\documentclass[prb,aps,amssymb,twocolumn,superscriptaddress,notitlepage]{revtex4-2}
\usepackage{amsmath}
\usepackage{amssymb}
\usepackage{amsthm}
\usepackage{amsfonts}
\usepackage{listings}
\usepackage{physics}
\usepackage{enumerate}
\usepackage{latexsym}
\usepackage{psfrag}
\usepackage{bm}
\usepackage{graphicx}
\usepackage[caption=false]{subfig}
\usepackage{blkarray}
\usepackage{array}
\usepackage{color}
\usepackage{dsfont}
\usepackage[com pat=1.1.0]{tikz-feynman}
\usepackage[normalem]{ulem}
\usepackage{hyperref}
\hypersetup{
     colorlinks = true,
     linkcolor = magenta,
     citecolor = magenta
}

\def\charge{\overline{e}}

\def\beq{\begin{equation}}
\def\eeq{\end{equation}}
\def\bal{\begin{aligned}}
\def\eal{\end{aligned}}

\begin{document}

\title{Charge Conservation Beyond Uniformity: Spatially Inhomogeneous Electromagnetic Response in Periodic Solids}

\author{Robert C.~McKay}
\affiliation{Department of Physics and Institute for Condensed Matter Theory, University of Illinois at Urbana-Champaign, Urbana IL, 61801-3080, USA}

\author{Fahad Mahmood}
\affiliation{Department of Physics, University of Illinois at Urbana-Champaign, Urbana, IL 61801, USA}
\affiliation{Materials Research Laboratory, University of Illinois at Urbana-Champaign, Urbana, IL 61801, USA}

\author{Barry Bradlyn}
\email{bbradlyn@illinois.edu}
\affiliation{Department of Physics and Institute for Condensed Matter Theory, University of Illinois at Urbana-Champaign, Urbana IL, 61801-3080, USA}

\begin{abstract}
Nonlinear electromagnetic response functions have reemerged as a crucial tool for studying quantum materials, due to recently appreciated connections between optical response functions, quantum geometry, and band topology. 
Most attention has been paid to responses to spatially uniform electric fields, relevant to low-energy optical experiments in conventional solid state materials.
However, magnetic and magnetoelectric phenomena are naturally connected by responses to spatially varying electric fields due to Maxwell's equations. 
Furthermore, in the emerging field of moir\'{e} materials, characteristic lattice scales are much longer, allowing spatial variation of optical electric fields to potentially have a measurable effect in experiments. 
In order to address these issues, we develop a formalism for computing linear and nonlinear responses to spatially inhomogeneous electromagnetic fields. 
Starting with the continuity equation, we derive an expression for the second-quantized current operator that is manifestly conserved and model independent. 
Crucially, our formalism makes no assumptions on the form of the microscopic Hamiltonian and so is applicable to model Hamiltonians derived from tight-binding or ab initio calculations. 
We then develop a diagrammatic Kubo formalism for computing the wavevector dependence of linear and nonlinear conductivities, using Ward identities to fix the value of the diamagnetic current order-by-order in the vector potential. 
We apply our formula to compute the magnitude of the Kerr effect at oblique incidence for a model of a moir\'{e} Chern insulator and demonstrate the experimental relevance of spatially inhomogeneous fields in these systems. 
We further show how our formalism allows us to compute the (orbital) magnetic multipole moments and magnetic susceptibilities in insulators. 
Turning to nonlinear response, we use our formalism to compute the second-order transverse response to spatially varying transverse electric fields in our moir\'{e} Chern insulator model, with an eye towards the next generation of experiments in these systems. 
\end{abstract}

\maketitle
\tableofcontents

\section{Introduction}

The study of optical properties of materials has long been one of the guiding themes in condensed matter physics. 
From the ability of X-ray crystallography to determine crystal structure to spectroscopic probes of band structure and collective dynamics, optical response experiments have served as a vital tool to learn about the structure of solid state systems. 
On the theoretical side, sum rules for optical response functions provide some of the few exact and experimentally relevant results in quantum many body physics~\cite{martin1988sum,nozieres1959electron,watanabe2020generalized,graf1995electromagnetic}.
Combined with first principles calculations, the Kubo formula for linear and nonlinear optical responses has allowed for an understanding of the electronic properties of materials~\cite{gross1985local,gross2005density,yao2004first,kuhne2020disordered,veithen2005nonlinear,baroni2001phonons}. 
More recently, it has come to be appreciated that optical response functions such as the linear and nonlinear Hall conductivity~\cite{thouless1982quantized,niu1985quantized,Xiao2019NonlinearFromClassical,Sodemann2015NonlinearHallEffect}, the photogalvanic effect~\cite{dejuan2017quantized,flicker2018chiral,alexandradinata2023anomalous}, the magnetoelectric polarizability~\cite{essin2009magnetoelectric,qi2008topological,wu2016quantized,dziom2017observation,ahn2022KerrEffectAxions}, and nonlinear magnetoresistance~\cite{nielsen1983adlerbelljackiw,huang2015observation,hirschberger2016chiral,zhang2016signatures,xiong2015evidence} all receive large topological contributions, and can be used to both diagnose and functionalize topological materials.

The fundamental objects of interest for optical responses are the (nonlinear) conductivity tensors of a material, which relate the measured current density $\langle \mathbf{j}(\mathbf{r},t)\rangle$ at position $\mathbf{r}$ and time $t$ that flow in a material in response to an electric field $\mathbf{E}(\mathbf{r},t)$. 
Typically, the electric field $\mathbf{E}(\mathbf{r},t)$ is introduced via a voltage source or an optical probe. 
The typical objects of interest are the experimentally-measurable (nonlinear) conductivities
\begin{equation}
\sigma_{(n)}^{\mu,\nu_1,\dots,\nu_n}(\mathbf{r},\mathbf{r}_1,\dots,\mathbf{r}_n,t,t_1,\dots, t_n),
\end{equation}
which determine the measured current via
\begin{widetext}
\begin{equation}
\label{eq:responsegeneral}
\langle j^\mu(\mathbf{r},t)\rangle = \sum_n \int \prod_i d\mathbf{r}_id t_i\sigma_{(n)}^{\mu,\nu_1,\dots,\nu_n}(\mathbf{r},\mathbf{r}_1,\dots,\mathbf{r}_n,t,t_1,\dots, t_n)E_{\nu_1}(\mathbf{r_1},t_1)\times\dots\times E_{\nu_n}(\mathbf{r_n},t_n).
\end{equation}
\end{widetext}
Here and throughout we use Greek indices $\mu,\nu=1,2,3$ to index spatial components of vectors, and the Einstein summation convention is used for repeated indices. 
Since we work entirely in flat space time, there is no functional distinction between upper and lower indices; we will try to choose the index arrangement that makes expressions easiest to parse.
Eq.~\eqref{eq:responsegeneral} parametrizes the current density order-by-order in powers of the external electric field. 
At order $n=1$ we recover the familiar generalized Ohm's law with $\sigma_1^{\mu\nu}(\mathbf{r},\mathbf{r}_1,t,t_1)=\sigma^{\mu\nu}(\mathbf{r},\mathbf{r}_1,t-t_1)$ being the linear conductivity tensor. 
Note that in writing Eq.~\eqref{eq:responsegeneral} we have not assumed that our system is translationally-invariant.

Because the speed of light is large compared with typical velocity scales in a solid state material, and because the typical energy scales of interest are in the microwave, terahertz, and optical regimes, the typical wavelength of the probe electric field (on the order of thousands of angstroms for visible light all the way up to the order of meters for AC fields) is significantly larger than the typical lattice spacing of crystals (on the order of nanometers). 
This means that for most experimentally-relevant situations, we can ignore the spatial variation of the electric field~\cite{basov2011electrodynamics}. 
In this approximation, we can introduce the uniform optical conductivities $\sigma_{n}^{\mu,\nu_1,\dots,\nu_n}(t,t_1,\dots,t_n)$ which depend on time and not space, and which determine the spatially averaged current density $\langle \mathbf{j}(t)\rangle$ via
\begin{equation}
\langle j^\mu(t)\rangle = \sum_n \int \prod_i dt_i \sigma^{\mu,\nu_1,\dots,\nu_n}_{n}(t,t_1,\dots,t_n)\prod_i E_{\nu_i}(t_i)\label{eq:spatuniform}.
\end{equation}
Eq.~\eqref{eq:spatuniform} follows from Eq.~\eqref{eq:responsegeneral} after assuming the electric field is spatially uniform. 

The uniform optical conductivities can be computed using generalizations of the Kubo formula, in terms of correlation functions of the electronic velocity operator $\mathbf{v}$. 
Recent theoretical work~\cite{parker2019diagrammatic,sipe2000secondorder,sipe2020ThirdOrderResponse,aversa1995nonlinear,Joao2020OpticalResponseTBWithDiagrams} has developed streamlined and efficient analytical machinery for computing $\sigma_{n}^{\mu,\nu_1,\dots,\nu_n}(t,t_1,\dots,t_n)$ for a variety of interesting materials, both for finite systems, continuum crystalline systems, and crystalline systems in the tight-binding approximation. 
For noninteracting electron systems, recent work has shown how the Kubo formula for the uniform optical conductivities can be interpreted in terms of the quantum geometry of electronic wavefunctions in the systems~\cite{ahn2022riemannian,alexandradinata2022topological,Yan2020TRBInCurvatureMetricAndMotion}. 
This theoretical work has been complemented by advances in nonlinear spectroscopy, which have seen signatures of intriguing topological and band-geometric effects in quantum materials~\cite{orenstein2021topology,wu2017giant}.

That said, there are limitations to focusing on the uniform optical conductivity. 
First, the uniform optical conductivity captures response to spatially uniform electric fields, but does not capture response to magnetic fields. 
To understand this, we can see from Faraday's law,
\begin{equation}\label{eq:faraday}
\curl\mathbf{E} = -\frac{\partial \mathbf{B}}{\partial t},
\end{equation}
that a spatially uniform and time-varying magnetic field generates a spatially inhomogeneous (transverse) electric field. 
Magnetic and magnetoelectric responses are thus encoded in the gradient expansion of $\sigma_{(n)}^{\mu,\nu_1,\dots,\nu_n}(\mathbf{r},\mathbf{r}_1,\dots,\mathbf{r}_n,t,t_1,\dots t_n)$. 
Such a gradient expansion has been carried out to low orders to obtain, for instance, the Streda formula relating the Hall conductivity to the magnetization density~\cite{streda1982theory,smrcka1977transport}, the Kubo formula for the magnetoelectric polarizability~\cite{malashevich2010theory}, and optical gyrotropy~\cite{zhong2016gyrotropic}. 
Similarly, a gradient expansion of the second order ($n=2$) longitudinal optical conductivity yielded diagrammatic Kubo formulas for electric quadrupole transitions in periodic crystals~\cite{Mele2023NonlinearOpticalRegLattice}. 
Lastly, the study of electron hydrodynamics has brought renewed attention to the connection between the viscosity tensor of an Galilean-invariant electron fluid and the (linear) optical conductivity expanded to second order in gradients~\cite{hoyos2012hall,bradlyn2012kubo,holder2019unified,lucas2018hydrodynamics,sulpizio2019visualizing,delacretaz2017transport,zhang2022GeometricSpinHallEffectFromInhomogeneity}. 
Attempts to generalize this relationship to periodic solids have yielded unintuitive results~\cite{kozii2021intrinsic}. 

Second, the explosion of interest in moir\'{e} materials~\cite{cao2018unconventional,bistritzer2011moire,ciarrocchi2022excitonic,andrei2021marvels,torma2022superconductivity,liu2020tunable,park2021tunable,zhang2021electronic,sung2020broken,wang2022one} gives cause to reevaluate the focus on the uniform optical conductivity. 
In systems such as twisted multilayer graphene, twisted transition metal dichalcogenides, and lattice-mismatched van der Waals interfaces, the effective moir\'{e} lattice constant can be several orders of magnitude larger than the atomic spacing in individual layers. 
In such a system, the wavelength of optical frequency light can be an appreciable fraction of the moir\'{e} lattice spacing. 
Thus, we can expect that optical properties such as Kerr and Faraday rotation, as well as nonlinear shift current and second harmonic generation may obtain measurable corrections from the spatial inhomogeneity of the applied optical electric field. 
Recent advances in ultrafast spectroscopy have put the study of nonlinear optical effects in moir\'{e} and van der Waals materials within reach of modern experiments~\cite{Liu2020AHEInBilayerGraphene,moon2013optical,zhang2022GeometricSpinHallEffectFromInhomogeneity}. 

Thus, in order to study both generalized magnetoelectric responses and optical properties of moir\'{e} materials, it would be desirable to have a complete theory of spatially inhomogeneous electromagnetic response applicable to periodic systems. 
From the Kubo formula, we know that $\sigma_{(n)}^{\mu,\nu_1,\dots,\nu_n}(\mathbf{r},\mathbf{r}_1,\dots,\mathbf{r}_n,t,t_1,\dots, t_n)$ can be computed from a retarded correlation function of $n$ current operators $\mathbf{j}(\mathbf{r})$. 
When the full microscopic Hamiltonian $H_0$ for a system of interest is known, the operator $\mathbf{j}(\mathbf{r})$ can be identified via minimal coupling, by replacing the momentum of each particle $\mathbf{p}_i$ by the covariant combination $\mathbf{p}_i - \charge\mathbf{A}(x_i)$, where $\mathbf{A}(\mathbf{x_i})$ is the electromagnetic vector potential as a function of the position operator $\mathbf{x}_i$ of particle $i$, and $\charge$ is the charge of particle $i$ \cite{Peres2018GuageCovarianceInNonlinearOpticalResponse, passos2018nonlinear}. 
Denoting by $H_A$ the Hamiltonian we obtain after minimal coupling, we can define the current operator as
\begin{equation}
\mathbf{j}_A(\mathbf{r}) = -\frac{\delta H_A}{\delta\mathbf{A}(\mathbf{r})},\label{eq:mincouplingcurrent}
\end{equation}
where $\delta/\delta\mathbf{A}(\mathbf{r})$ denotes a variational derivative with respect to the vector potential. 
This allows us to write
\begin{equation}
H_A = H_0 - \int d\mathbf{r} \mathbf{j}(\mathbf{r})\cdot\mathbf{A}(\mathbf{r}),
\end{equation}
from which the nonlinear conductivities can be extracted via standard response theory upon identifying (in this gauge \cite{Peres2018GuageCovarianceInNonlinearOpticalResponse,passos2018nonlinear}) the electric field $\mathbf{E} = -\partial \mathbf{A}/\partial t$. 
Crucially, the current defined via Eq.~\eqref{eq:mincouplingcurrent} manifestly obeys the continuity equation (charge conservation equation)
\begin{equation}
\partial_t\rho(\mathbf{r}) + \div \mathbf{j}_A(\mathbf{r}) = 0,\label{eq:continuityintro}
\end{equation}
where $\rho(\mathbf{r})$ is the charge density operator. 
Eq.~\eqref{eq:continuityintro} reflects the conservation of electric charge and is equivalent to gauge invariance.

While this gives an in-principle complete procedure for computing $\sigma_{(n)}^{\mu,\nu_1,\dots,\nu_n}(\mathbf{r},\mathbf{r}_1,\dots,\mathbf{r}_n,t,t_1,\dots ,t_n)$, it has one main drawback: in most situations in solid state physics we do not have access to the full microscopic Hamiltonian, but instead only an effective low-energy model for the degrees of freedom of interest. 
This can arise due to an approximate treatment of relativistic effects of core electrons in a solid (as in, e.g., a DFT calculation), or via the use of a tight-binding model that describes only the low-energy subset of states in the full Hilbert space of a material. 
When only an effective or approximate Hamiltonian is known, the minimal coupling substitution $H_0\rightarrow H_A$ cannot be carried out uniquely. 

One common approach~\cite{zhong2018linear,malashevich2010theory,zhong2016gyrotropic,Vignale1991OrbitalParamagnetism,Souza2023MidpointMethod} to circumvent this problem in nonrelativistic materials is to \emph{assume} that the full and unknown microscopic electronic Hamiltonian has the form
\begin{equation}
H_0\stackrel{!}{=} \sum_i \frac{|\mathbf{p}_i|^2}{2m} + V(\mathbf{x}_i) -\vec{\lambda}(\mathbf{x_i})\cdot (\mathbf{p}\times\vec{\sigma}) + \frac{1}{2} \sum_{i\neq j} U(\mathbf{x}_i-\mathbf{x}_j), \label{eq:assumedH}
\end{equation}
where $m$ is the electron mass, $V$ is the periodic potential of the ions in the system, $\vec{\lambda}$ is the approximate spin-orbit potential (usually expressed in terms of gradients of $V$), $\vec{\sigma}$ is a vector of Pauli matrices acting on electron spins, and $U$ is the Coulomb potential. 
With this assumption, the minimally coupled current operator can be evaluated exactly. 
In the absence of the external electromagnetic field, it is given by
\begin{align}
\mathbf{j}_\mathrm{min}(\mathbf{r}) &\equiv -\left.\frac{\delta H_A}{\delta\mathbf{A}(\mathbf{r})}\right|_{\mathbf{A}=0} = \sum_\mathbf{q} e^{i\mathbf{q}\cdot\mathbf{r}}\mathbf{j}_{\mathbf{q},\mathrm{min}}
\end{align}
where the Fourier components $\mathbf{j}_{\mathbf{q},\mathrm{min}}$ can be expressed in two equivalent and often used forms
\begin{align}
\mathbf{j}_{\mathbf{q},\mathrm{min}} &= \frac{\charge}{2}\sum_i\left\{e^{-i\mathbf{q}\cdot\mathbf{x}_i},\mathbf{v}_i\right\} =\mathbf{\tilde{j}}_\mathbf{q} \label{eq:trapfirsttime}\\
&= \charge\sum_i e^{-i\mathbf{q}\cdot\mathbf{x}_i/2}\mathbf{v_i}e^{-i\mathbf{q}\cdot\mathbf{x}_i/2} = \mathbf{j}_{\mathrm{mid},\mathbf{q}} \label{eq:midpointfirsttime},
\end{align}
where $\mathbf{v}_i = i\left[H,\mathbf{x}_i\right]$ is the velocity operator for particle $i$, and $\charge$ is the charge of the particles (which for convenience we assume to be the same for all particles throughout this work). 
The customary approach is to then take either Eq.~\eqref{eq:trapfirsttime} or Eq.~\eqref{eq:midpointfirsttime} as the definition of the current operator in the low-energy effective model, and use that to compute optical response functions. 

There are two problems with this approach. 
The first is that, although Eqs.~\eqref{eq:trapfirsttime} and \eqref{eq:midpointfirsttime} are equivalent for the microscopic Hamiltonian Eq.~\eqref{eq:assumedH}, they will in general give different results when applied to any effective model. 
This raises the question of which, if either, of them should be used to compute optical response functions. 
Additionally, neither Eq.~\eqref{eq:trapfirsttime} nor Eq.~\eqref{eq:midpointfirsttime} coincide with the minimally coupled current $\mathbf{j}_A(\mathbf{r})$ when relativistic corrections to the kinetic energy are incorporated, which may be important for core electrons in solid state systems. 
Even more severe, we will show below that generally neither Eq.~\eqref{eq:trapfirsttime} nor  Eq.~\eqref{eq:midpointfirsttime} satisfy the continuity equation Eq.~\eqref{eq:continuityintro} even when $\mathbf{A}=0$. 
When used in the context of effective low-energy models for even nonrelativistic systems, this means that Eqs.~\eqref{eq:trapfirsttime} and \eqref{eq:midpointfirsttime} represent non-number-conserving approximations to the minimally coupled current. 
Thus neither Eq.~\eqref{eq:trapfirsttime} nor  Eq.~\eqref{eq:midpointfirsttime} provide a satisfactory starting point for the computation of optical response functions in low-energy models of quantum materials. 

In what follows, we will develop a theory of spatially inhomogeneous linear and nonlinear optical response that manifestly obeys the continuity equation Eq.~\eqref{eq:continuityintro} when applied to arbitrary effective models of condensed matter systems.  
First in Section \ref{sec:jnew}, we show how to construct the position-dependent current operator for a generic Hamiltonian that manifestly respects the continuity equation.
Crucially, our construction makes no assumptions on the form of the
Hamiltonian, and is applicable to nonrelativistic, semi-relativistic, and approximately tight-binding Hamiltonians.
We show that our current agrees with Eqs.~\eqref{eq:trapfirsttime} and \eqref{eq:midpointfirsttime} to linear order in the wavevector $\mathbf{q}$, implying that our theory gives the same linear response to uniform electric and magnetic fields as in Refs.~\cite{zhong2018linear,zhong2016gyrotropic,malashevich2010theory,essin2009magnetoelectric}. 
We will also show how to define the diamagnetic current such that the continuity equation is satisfied order-by-order in the electromagnetic vector potential.  

Next in Section \ref{sec:DiagramsRuleAndSumRules}, we set out the Feynman diagrammatic rules for evaluating the spatially inhomogeneous conductivity in Fourier space as a function of frequency and wavevector (where the uniform optical conductivities correspond to the limit of zero wavevector). 
Focusing on the linear conductivity, we derive a modified f-sum rule relating the diamagnetic conductivity to the density-density response function.  
In Section \ref{sec:ApplicationOfTheLinearResponse}, we explore various observable phenomena that arises from our wavevector-dependent response theory at the linear response level, applied to non-interacting electron systems in a periodic potential. 
As a proof of concept, we examine the wavevector-dependence of the Hall conductivity in a time-reversal breaking Weyl semimetal model first.
We then show how our formulation of the wavevector dependent linear conductivity can be applied to compute the Kerr rotation and ellipticity in a model of a moir\'{e}-Chern insulator. 
We compare our predictions with an analogous calculation using the current operators in Eqs.~\eqref{eq:trapfirsttime} and \eqref{eq:midpointfirsttime}, showing that our gauge-invariant formulation predicts a measurably different Kerr response on the order of arcseconds to arcminutes for typical moir\'{e} length and energy scales. 
We round off this section be examining the magnetic properties of insulators. 
We analyze the relationship between magnetic susceptibility and transverse conductivity, present expressions for the magnetic quadrupole moment, and give a new derivation of the Streda formula for the Hall conductivity.

Then in Section \ref{sec:SecondOrderResponseInMoireMaterials}, we move on from phenomena derived from linear conductivity, and evaluate the diagrammatic Kubo formula for the second-order electromagnetic conductivity for noninteracting electrons in a periodic potential.
We then focus on the component of the second-order conductivity that is second harmonic generation in frequency, and self-focusing in wavevector, which is relevant to future ultrafast optical experiments. 
We show how to compute this component of the conductivity for our model of a moir\'{e}-Chern insulator. 
Finally, we conclude our paper by summarizing the key results and suggesting new directions in Section \ref{sec:Conclusion}.

\section{The Current Operator}\label{sec:jnew}
Motivated by recent efforts to obtain a complete spatially-inhomogeneous electromagnetic response theory \cite{kozii2021intrinsic,zhong2018linear, Mele2023NonlinearOpticalRegLattice, Souza2023MidpointMethod}, our goal in this section will be to derive the Fourier components of the conserved current operator in a model-independent fashion. 
Measurable phenomena such as conductivities, optical response, and magnetization can only be accurately modeled in terms of a charge-conserving current operator.
As we will show, the conventionally defined current densities Eq.~\eqref{eq:midpointfirsttime} and \eqref{eq:trapfirsttime} fail to satisfy the continuity equation~\eqref{eq:continuityintro} for generic Hamiltonians.

To find the conserved current density, let us consider a general Hamiltonian 
\begin{equation}
H_0=\sum_i T(\mathbf{p}_i) + V(\mathbf{x}_i,\mathbf{p}_i) + \frac{1}{2}\sum_{i\neq j} U(\mathbf{x}_i-\mathbf{x}_j) \label{eq:ham}
\end{equation}
for a (possibly interacting) system of electrons. 
Here, $i=1\dots N$ indexes the particles of the system, $\mathbf{p}_i$ is the momentum operator for particle $i$, and $\mathbf{x}_i$ is the position operator for particle $i$.
Also, $T(\mathbf{p}_i)$ is the kinetic energy operator for the $i$-th particle.
Although this kinetic operator is usually quadratic in momenta, it can extend to semi-relativistic systems.
For example, the kinetic energy could take the form
\begin{equation}
T(\mathbf{p}_i)\approx \frac{|\mathbf{p}_i|^2}{2m} + \frac{|\mathbf{p}_i|^4}{8m^3c^2} +\dots,
\end{equation}
which is relevant for all-electron ab initio modeling of spin-orbit coupled materials~\cite{blaha1990full}. 
Our results will hold for arbitrary $T(\mathbf{p}_i)$. 
The single particle potential  $V(\mathbf{x}_i,\mathbf{p}_i)$ includes both the momentum-independent external potential as well as momentum-dependent external potential terms such as the spin-orbit potential. 
Finally, $U(\mathbf{x}_i-\mathbf{x}_j)$ is the interaction potential, which we take to be momentum-independent for simplicity (we expect this is a good assumption for most models of interest; however, for long-range interacting models of Mott insulators that have recently attracted attention, this assumption must be checked~\cite{zhao2023failure}). 
We assume that the potential $V$ has discrete translation symmetry such that 
\begin{align}
T_\mathbf{R}^\dag H T_\mathbf{R} &= e^{i\mathbf{R}\cdot\sum_i\mathbf{p}_i} H e^{-i\mathbf{R}\cdot\sum_i\mathbf{p}_i} \nonumber \\
&= \sum_i T(\mathbf{p}_i) + V(\mathbf{x}_i+\mathbf{R},\mathbf{p}_i) + \frac{1}{2}\sum_{i\neq j} U(\mathbf{x}_i-\mathbf{x}_j) \nonumber \\
&= H\label{eq:dts}
\end{align}
where $\mathbf{R}$ is a Bravais lattice vector. 

The global $U(1)$ symmetry of $H_0$ implies that the density operator
\begin{equation}
\rho(\mathbf{r}) = \charge\sum_i \delta(\mathbf{r}-\mathbf{x}_i) \label{eq:density}
\end{equation}
satisfies the continuity equation~\eqref{eq:continuityintro}, where all particles have charge $\charge$. 
We can rewrite the continuity equation in the Heisenberg picture as 
\begin{equation}
\partial_t \rho(\mathbf{r}) = i\left[H_0,\rho(\mathbf{r})\right] = -\nabla\cdot\mathbf{j}_\mathrm{min}(\mathbf{r}),\label{eq:continuity}
\end{equation}
where we work in natural units where $\hbar=c=1$ unless stated otherwise. 
In the presence of an external electromagnetic field with scalar potential $A_0$ and vector potential $\mathbf{A}$, we have that the minimally coupled Hamiltonian becomes
\begin{align}
H_0\rightarrow H_A = \sum_i T(\mathbf{p}_i-\charge A(\mathbf{x}_i)) + V(\mathbf{x}_i,\mathbf{p}_i-\charge A(\mathbf{x}_i))& \nonumber
\\
  - \charge A_0(\mathbf{x}_i) + \frac{1}{2}\sum_{i\neq j} U(\mathbf{x}_i-\mathbf{x}_j)&\label{eq:mincouple} .
\end{align}
We can we take the functional derivative with respect to the scalar potential to arrive at
\begin{equation}
\rho_A(\mathbf{r}) = -\frac{\delta H}{\delta A_0(\mathbf{r})} = \rho(\mathbf{r}),
\end{equation}
where $\rho_A(\mathbf{r})$ indicates the density operator in the presence of the electromagnetic field $\rho(\mathbf{r})$ is defined in Eq.~\eqref{eq:density}, and we have made use of the identity
\begin{equation}
\frac{\delta A_0(\mathbf{x}_i)}{\delta A_0(\mathbf{r})} = \delta(\mathbf{r}-\mathbf{x}_i).
\end{equation}
Furthermore, we have that
\begin{equation}
\partial_t \rho_A(\mathbf{r}) = i\left[H_A,\rho_A(\mathbf{r})\right] = -\nabla\cdot\mathbf{j}_A(\mathbf{r}),
\end{equation}
where by evaluating the commutator we find 
\begin{equation}
j_A^\mu(\mathbf{r}) = -\frac{\delta H_A}{\delta A_\mu(\mathbf{r})}.\label{eq:jA}
\end{equation}

Eq.~\eqref{eq:jA} allows us to define the current in terms of the the Hamiltonian $H_A$ minimally coupled to the external electromagnetic field. 
In particular, we can find the unperturbed current operator $\mathbf{j}_{\mathrm{min}}(\mathbf{r})$ from Eq.~\eqref{eq:continuity} as
\begin{equation}
j_\mathrm{min}^\mu(\mathbf{r}) = -\left.\frac{\delta H_A}{\delta A_\mu(\mathbf{r})}\right|_{\mathbf{A}=0}.\label{eq:variational_current_def}
\end{equation}
However, in order to utilize Eq.~\eqref{eq:variational_current_def}, we need to know the explicit microscopic form of the Hamiltonian in Eq.~\eqref{eq:ham}. 
In many cases of interest, we do not have this information available. 
For instance, in modeling electronic systems we often only have access to Wannier-based tight-binding models, or other effective Hamiltonians, that are projected into a subset of bands of interest. 
For an effective Hamiltonian, it is not possible to implement the minimal coupling of Eq.~\eqref{eq:mincouple} in a model independent fashion. 
One possible way to circumvent this complication is to make use of a Peierls substitution to couple a Wannierized effective model to the vector potential, provided the Wannier orbitals are sufficiently well localized and the vector potential varies sufficiently slowly~\cite{graf1995electromagnetic}. 
However, defining the current via a Peierls substitution requires making choices for how to evaluate path integrals of the vector potential between Wannier centers that are not necessarily natural. 
It also requires assuming that the vector potential varies slowly on the length scale of the Wannier function localization length, which may not be appropriate for all models (especially for models of moir\'{e} or topologically nontrivial systems). 
Finally, the Peierls substitution cannot be used to define the current when a Wannierized tight-binding model has not been explicitly constructed. 

In order to circumvent these issues, we will derive an alternative expression for the current $\mathbf{j}(\mathbf{r})$ that is manifestly conserved and that does not require detailed knowledge of the microscopic Hamiltonian. 
To do so, let us first recast the continuity equation, Eq.~\eqref{eq:continuity}, in momentum space. 
We introduce the Fourier-transformed density
\begin{equation}
\rho_\mathbf{q} = \int d\mathbf{r}\; e^{-i\mathbf{q}\cdot\mathbf{r}} \rho(\mathbf{r}) = \charge \sum_i e^{-i\mathbf{q}\cdot\mathbf{x}_i}. \label{eq:rhoqdef}
\end{equation}
In Fourier space, the continuity equation reads
\begin{equation}
\partial_t \rho_\mathbf{q}= i\left[H_0,\rho_\mathbf{q}\right] =- i\mathbf{q}\cdot\mathbf{j}_\mathbf{q}, \label{eq:fouriercont}
\end{equation}
where
\begin{equation}
\mathbf{j}_\mathbf{q} = \int d\mathbf{r}\; e^{-i\mathbf{q}\cdot\mathbf{r}} \mathbf{j}(\mathbf{r}).
\end{equation}
Rather than attempt to evaluate the commutator in Eq.~\eqref{eq:fouriercont} directly, we will take a more general approach. 
In the Heisenberg picture, we see that $\mathbf{\rho}_\mathbf{q}$ in Eq.~\eqref{eq:rhoqdef} depends on time implicitly through the operators $\mathbf{x}_i$. 
Defining the single-particle velocity operators
\begin{equation}
\mathbf{v}_i \equiv \partial_t\mathbf{x}_i = i\left[H_0,\mathbf{x}_i\right],
\end{equation}
we can write
\begin{equation}
\rho_\mathbf{q}(t+\delta t) =\charge  e^{-i\mathbf{q}\cdot\mathbf{x}_i(t+\delta t)} \approx\charge  e^{-i\mathbf{q}\cdot\left(\mathbf{x}_i(t)+\delta t \mathbf{v}_i(t)\right)}.
\end{equation}
Thus, using the definition of the derivative, the rate of change of the density operator is
\begin{equation}
\partial_t\rho_\mathbf{q} =\charge  \lim_{\delta t\rightarrow 0}\frac{e^{-i\mathbf{q}\cdot\left(\mathbf{x}_i+\delta t \mathbf{v}_i\right)} - e^{-i\mathbf{q}\cdot\mathbf{x}_i}}{\delta t}. \label{eq:rhoderivexpand}
\end{equation}

To simplify this further, we make use of a general result of Karplus and Schwinger~\cite{karplus1948note}, who showed that for any operators $A$ and $B$,
\begin{equation}
e^{A+\delta t B} = e^A + \delta t\int_0^1 d\lambda e^{(1-\lambda)A}Be^{\lambda A} + \mathcal{O}[(\delta t)^2].\label{eq:kp}
\end{equation}
We prove this result in App.~\ref{app:karplus_proof}. 
Inserting Eq.~\eqref{eq:kp} with $A=-i\mathbf{q}\cdot\mathbf{x}_i$ and $B=-i\mathbf{q}\cdot\mathbf{v}_i$ into Eq.~\eqref{eq:rhoderivexpand}, we find that
\begin{equation}\label{eq:rhodot}
\partial_t\rho_\mathbf{q} = \charge \sum_i\int_0^1d\lambda e^{-i(1-\lambda)\mathbf{q}\cdot\mathbf{x}_i}(-i\mathbf{q}\cdot\mathbf{v}_i)e^{-i\lambda\mathbf{q}\cdot\mathbf{x}_i}.
\end{equation}
Comparing with the continuity equation in Eq.~\eqref{eq:fouriercont}, we can identify 
\begin{equation}
\mathbf{j}_\mathbf{q} = \charge \sum_i\int_0^1d\lambda e^{-i(1-\lambda)\mathbf{q}\cdot\mathbf{x}_i}\mathbf{v}_ie^{-i\lambda\mathbf{q}\cdot\mathbf{x}_i}. \label{eq:mainj}
\end{equation}

Eq.~\eqref{eq:mainj} has several desirable features. 
Most importantly, it is manifestly conserved. 
Second, for nonrelativistic Hamiltonians (where $\mathbf{v}(\mathbf{p},\mathbf{x})$ is a linear function of $\mathbf{p}$), a short calculation (see App.~\ref{app:jsr}) shows that $\mathbf{j}_\mathbf{q}$ coincides with the minimally coupled current.
Note that since the Hamiltonian $H_0$ commutes with the projection operator onto a set of low-energy bands, the projected low-energy density operator satisfies Eq.~\eqref{eq:continuity} with the conserved current given by the low-energy projection of Eq.~\eqref{eq:mainj}. 
Thus, for nonrelativistic systems with microscopic Hamiltonians of the form of Eq.~\eqref{eq:assumedH}, Eq.~\eqref{eq:mainj} is the approximation to the minimally coupled current in Eq.~\eqref{eq:mincouplingcurrent} that is manifestly conserved in any low-energy approximation to the Hamiltonian.
However, we emphasize that a choice was made in going from Eq.~\eqref{eq:rhodot} to Eq.~\eqref{eq:mainj}: we can add any operator $\delta\mathbf{j}_\mathbf{q}$ that is purely transverse ($\mathbf{q}\cdot\mathbf{\delta j}=0$)  to Eq.~\eqref{eq:mainj} without changing the continuity equation~\eqref{eq:continuity}. 
Different choices of $\delta\mathbf{j}$ correspond to different definitions for the (orbital) \emph{magnetization current}, which do not contribute to transport and are not constrained by gauge invariance. 
As such, it is in principle not possible to fix $\delta\mathbf{j}$ unambiguously from effective Hamiltonians alone. 
Even in the context of the microscopic Hamiltonian nonrelativistic approximations to the Dirac equation, nonminimal coupling to the electromagnetic field can modify the definition of the magnetization current. 
Faced with these challenges, we will take the natural choice $\delta\mathbf{j}=0$---corresponding to our definition Eq.~\eqref{eq:mainj} for the current operator---as our working, model-independent definition of the current. 

As we will now show, Eq.~\eqref{eq:mainj} resolves many of the problems of the non-conserved choices Eqs.~\eqref{eq:trapfirsttime} and \eqref{eq:midpointfirsttime} for the current. 
In Sec.~\ref{sec:jTildeCurrentOperator} we will show that both the longitudinal and transverse components of Eqs.~\eqref{eq:trapfirsttime} and \eqref{eq:midpointfirsttime} differ from those of Eq.~\eqref{eq:mainj} at second order in wavevector. 
Then, in Sec.~\ref{sec:SecondQuantizedOperator} we will compute the matrix elements of the conserved current Eq.~\eqref{eq:mainj} in the basis of Bloch eigenstates, allowing us to apply our formalism to compute electromagnetic responses. 
Finally, in Sec.~\ref{sec:HigherOrderCurrentOperators}, we will show that the formula of Karplus and Schwinger can be used to define the generalized wavevector-dependent diamagnetic contributions to the current operator in the presence of a nonzero electromagnetic field. 

\subsection{The Non-Conserved Current Operators}\label{sec:jTildeCurrentOperator}
Eq.~\eqref{eq:mainj} is our first main result. 
Let us compare it quantitatively with the (non-conserved) current operators Eqs.~\eqref{eq:trapfirsttime} and \eqref{eq:midpointfirsttime}, which have been used to compute spatially inhomogeneous electromagnetic response functions to low order in the wavevector~\cite{zhong2018linear,kozii2021intrinsic}.
Recall that Eq.~\eqref{eq:trapfirsttime} proposes the definition
\begin{equation}
\mathbf{\tilde{j}}_\mathbf{q} = \frac{\charge }{2}\sum_i\left\{e^{-i\mathbf{q}\cdot\mathbf{x}_i},\mathbf{v}_i\right\}\label{eq:jtildedef}
\end{equation}
for the current operator. 
We will refer to this as the ``trapezoidal'' current, since it is the trapezoidal approximation the integration over $\lambda$ in our conserved current Eq.~\eqref{eq:mainj}.

Similarly, Eq.~\eqref{eq:midpointfirsttime} proposes the commonly-used~\cite{kozii2021intrinsic, zhong2018linear, Souza2023MidpointMethod} definition
\begin{equation}
    \mathbf{j}_{\mathrm{mid},\mathbf{q}} = \charge \sum_i e^{-i\mathbf{q}\cdot\mathbf{x}_i/2}\mathbf{v_i}e^{-i\mathbf{q}\cdot\mathbf{x}_i/2} \label{eq:jmidpointdef}
\end{equation}
We will refer to this as the ``midpoint" current,  since it is the midpoint approximation the integration over $\lambda$ in our conserved current Eq.~\eqref{eq:mainj}.

To see how the definitions of current operator from Eqs.~\eqref{eq:jtildedef} and~\eqref{eq:jmidpointdef} compare with our Eq.~\eqref{eq:mainj}, it is helpful to write $\mathbf{v}_i=\mathbf{v}_i(\mathbf{p}_i,\mathbf{x}_i)$ to make explicit the $\mathbf{p}_i$- and $\mathbf{x}_i$-dependence of $\mathbf{v}_i$. 
Inserting this into Eq.~\eqref{eq:mainj} and using the fact that
\begin{equation}\label{eq:momtranslation}
e^{i\mathbf{\alpha}\cdot\mathbf{x}_i}f(\mathbf{p_i},\mathbf{x}_i) = f(\mathbf{p_i}-\mathbf{\alpha},\mathbf{x}_i)e^{i\mathbf{\alpha}\cdot\mathbf{x}_i},
\end{equation}
we find
\begin{align}
\mathbf{j}_\mathbf{q} &= \charge \sum_i\int_0^1d\lambda e^{-i(1-\lambda)\mathbf{q}\cdot\mathbf{x}_i}\mathbf{v}_i(\mathbf{p}_i,\mathbf{x}_i)e^{-i\lambda\mathbf{q}\cdot\mathbf{x}_i}\nonumber \\
&=\charge \sum_i \int_0^1 d\lambda e^{-i\mathbf{q}\cdot\mathbf{x}_i}\mathbf{v}_i(\mathbf{p}_i-\lambda \mathbf{q},\mathbf{x}_i). \label{eq:jinstandardform}
\end{align}
Similarly, using Eq.~\eqref{eq:momtranslation} to simplify Eq.~\eqref{eq:jtildedef} yields
\begin{equation}
\tilde{\mathbf{j}}_\mathbf{q} = \charge \frac{1}{2}\sum_ie^{-i\mathbf{q}\cdot\mathbf{x}_i}\left(\mathbf{v}_i(\mathbf{p}_i,\mathbf{x}_i) + \mathbf{v}_i(\mathbf{p}_i-\mathbf{q},\mathbf{x}_i)\right).\label{eq:jTildeInitial}
\end{equation}
Lastly, using Eq.~\eqref{eq:momtranslation} to simplify Eq.~\eqref{eq:jmidpointdef} yields
\begin{equation}
\mathbf{j}_{\text{mid}, \mathbf{q}} = \charge \sum_ie^{-i\mathbf{q}\cdot\mathbf{x}_i}\mathbf{v}_i(\mathbf{p}_i - \mathbf{q}/2, \mathbf{x}_i).\label{eq:jMidInitial}
\end{equation}

We thus generically have that $\mathbf{j}_\mathbf{q}\neq \mathbf{\tilde{j}}_\mathbf{q} \neq \mathbf{j}_{\text{mid}, \mathbf{q}}$. 
Note that if $\mathbf{v}_i$ is a linear function of momentum (which occurs when the kinetic energy includes no relativistic corrections and the spin-orbit potential is linear in momentum), then we can carry out the integral in Eq.~\eqref{eq:jinstandardform} to find that $\mathbf{j}_\mathbf{q} = \mathbf{\tilde{j}}_\mathbf{q} =  \mathbf{j}_{\text{mid}, \mathbf{q}}$. 
However, for Hamiltonians with relativistically-corrected kinetic energy and complicated spin-orbit potentials, this will not be the case. 
This is relevant not just for ab-initio calculations of charge transport in heavy elements, but also for effective models where integrating out high energy degrees of freedom can lead to a renormalization of the effective Hamiltonian away from the standard nonrelativistic form. 

We can compute the difference $\mathbf{j}_\mathbf{q}-\mathbf{\tilde{j}}_\mathbf{q}$ and $\mathbf{j}_\mathbf{q}-\mathbf{j}_{\text{mid}, \mathbf{q}}$ order-by-order in $\mathbf{q}$. 
We find that
\begin{align}
\mathbf{j}_\mathbf{q}- \mathbf{\tilde{j}}_\mathbf{q} =& -\frac{\charge }{12}\sum_iq_\mu q_\nu \frac{\partial^2 \mathbf{v}_i( \mathbf{p}_i,\mathbf{x}_i)}{\partial p_i^\mu \partial p_i^\nu} + \mathcal{O}(|\mathbf{q}|^3),\label{eq:discrepancy}
\\
\mathbf{j}_\mathbf{q}- \mathbf{j}_{\text{mid}, \mathbf{q}} =& \frac{\charge }{24}\sum_iq_\mu q_\nu \frac{\partial^2 \mathbf{v}_i( \mathbf{p}_i,\mathbf{x}_i)}{\partial p_i^\mu \partial p_i^\nu} + \mathcal{O}(|\mathbf{q}|^3).\label{eq:discrepancyMid}
\end{align}
We see from Eqs.~\eqref{eq:discrepancy} and~\eqref{eq:discrepancyMid} that $\mathbf{j}_\mathbf{q}$, $\mathbf{\tilde{j}}_\mathbf{q}$, and $\mathbf{j}_{\text{mid}, \mathbf{q}}$ all coincide to linear order in $\mathbf{q}$, with the first discrepancy at order $|\mathbf{q}|^2$. 
This suggests that while calculations of uniform magnetoelectric response functions (which requires knowing $\mathbf{j}_\mathbf{q}$ to linear order in $\mathbf{q}$, as in Refs.~\cite{zhong2018linear,malashevich2010theory,zhong2016gyrotropic}) can use $\mathbf{\tilde{j}}_\mathbf{q}$ or $\mathbf{j}_{\text{mid}, \mathbf{q}}$ in place of the conserved Eq.~\eqref{eq:mainj}, the calculations of the $\mathcal{O}(|\mathbf{q}|^2)$ corrections to the conductivity in Ref.~\cite{kozii2021intrinsic,Souza2023MidpointMethod} should be reexamined. 
We will show below that while $\mathbf{j}_\mathbf{q},\mathbf{\tilde{j}}_{\mathbf{q}},$ and $\mathbf{j}_{\mathrm{mid},\mathbf{q}}$ all yield the same Hall conductivity to order $\mathcal{O}(|\mathbf{q}|^2)$, they yield different predictions for the longitudinal conductivity at the same order.
Even further, we see from Eq.~\eqref{eq:discrepancy} that $\mathbf{q}\cdot\left(\mathbf{j}_\mathbf{q}- \mathbf{\tilde{j}}_\mathbf{q}\right)\neq 0$ and $\mathbf{q}\cdot\left(\mathbf{j}_\mathbf{q}- \mathbf{j}_{\text{mid}, \mathbf{q}}\right)\neq 0$, so that \emph{only} $\mathbf{j}_\mathbf{q}$ and \emph{not} $\mathbf{\tilde{j}}_\mathbf{q}$ \emph{nor} $\mathbf{j}_{\text{mid}, \mathbf{q}}$ satisfies the continuity equation Eq.~\eqref{eq:continuity}. 
Generically, we have
\begin{align}
\mathbf{q}\cdot\mathbf{j}_\mathbf{q} - \mathbf{q}\cdot\mathbf{j}_{\mathrm{min},\mathbf{q}} &= 0, \\
\mathbf{q}\cdot\mathbf{\tilde{j}}_\mathbf{q} - \mathbf{q}\cdot\mathbf{j}_{\mathrm{min},\mathbf{q}} &\neq 0,
\\
\mathbf{q}\cdot\mathbf{j}_{\text{mid}, \mathbf{q}} - \mathbf{q}\cdot\mathbf{j}_{\mathrm{min},\mathbf{q}} &\neq 0.
\end{align}

In App.~\ref{app:jsr} we give a direct analysis of how the conserved $\mathbf{j}_{\mathbf{q}}$ differs from non-conserved $\Tilde{\mathbf{j}}_{\mathbf{q}}$ and $\mathbf{j}_{\mathrm{mid},\mathbf{q}}$ for a concrete model of a semi-relativistic free-electron system.
We show explicitly that once the free electron is energetic enough for quartic or higher-order corrections to the kinetic energy to become appreciable, the conserved $\mathbf{j}_{\mathbf{q}}$ and non-conserved $\Tilde{\mathbf{j}}_{\mathbf{q}}$ and $\mathbf{j}_{\mathrm{mid},\mathbf{q}}$ will disagree at order $|\mathbf{q}|^2$.

Although we have shown how the conserved$\mathbf{j}_{\mathbf{q}}$ of Eq.~\eqref{eq:mainj} differs from the nonconserved $\Tilde{\mathbf{j}}_{\mathbf{q}}$ and $\mathbf{j}_{\mathrm{mid},\mathbf{q}}$, in the current form our results only apply to continuum first-quantized Hamiltonians of the form of Eq.~\eqref{eq:ham}.
To apply our same approach to formulate the current density operator for a wider variety of condensed matter problems, we will in Sec.~\ref{sec:SecondQuantizedOperator} calculate the second-quantized form of the current operator Eq.~\eqref{eq:mainj} by examining its matrix elements in a basis of single-particle Bloch eigenstates. 

\subsection{Matrix Elements of the Current Operator in the Bloch Basis with Second-Quantization}\label{sec:SecondQuantizedOperator}

In this section, we will evaluate the matrix elements of the conserved current operator Eq.~\eqref{eq:mainj}---as well as the non-conserved current operators Eqs.~\eqref{eq:jtildedef} and \eqref{eq:jmidpointdef}---in a basis of single-particle Bloch eigenstates. 
Consider a noninteracting Hamiltonian of the form of Eq.~\eqref{eq:ham} with $U=0$ with discrete translation symmetry as in Eq.~\eqref{eq:dts}. 
From Bloch's theorem we can introduce single-particle eigenstates
\begin{equation}\label{eq:eigenbasis}
\psi_{n\mathbf{k}}(r) = \frac{1}{\sqrt{N}}e^{i\mathbf{k}\cdot\mathbf{r}}u_{n\mathbf{k}}(r),
\end{equation}
where $N$ is the nominally infinite number of unit cells in the system. 
We denote the eigenstates with kets $\ket{\psi_{n\mathbf{k}}}$ with inner product
\begin{equation}
\bra{\psi_{n\mathbf{k}}}\ket{\psi_{m\mathbf{k'}}} = \int d\mathbf{r}\psi^*_{n\mathbf{k}}(\mathbf{r})\psi_{m\mathbf{k'}}(\mathbf{r}) = \delta_{\mathbf{k,k'}}\delta_{nm}.
\end{equation}
It will also be convenient to introduce kets $\ket{u_{n\mathbf{k}}}$ to represent the cell-periodic functions $u_{n\mathbf{k}}(\mathbf{r})$. 
The inner product of cell-periodic kets is given by
\begin{equation}
\bra{u_{n\mathbf{k}}}\ket{u_{m\mathbf{k'}}} = \int_{\mathrm{cell}} d\mathbf{r}u^*_{n\mathbf{k}}(\mathbf{r})u_{m\mathbf{k'}}(\mathbf{r}),
\end{equation}
where $\int_\mathrm{cell}$ denotes an integration over a single unit cell. 
We will work in the periodic gauge such that
\begin{equation}\label{eq:periodicgaugeconstraint}
\ket{\psi_{n\mathbf{k+G}}}=\ket{\psi_{n\mathbf{k}}}
\end{equation}
for any reciprocal lattice vector $\mathbf{G}$. 
Note that the periodic gauge always exists, even for topologically-nontrivial systems, as we discuss further at the end of this subsection.  
The periodic gauge constraint implies~\cite{vanderbilt2018berry}
\begin{equation}
\ket{u_{n\mathbf{k+G}}}=e^{-i\mathbf{G}\cdot\mathbf{x}}\ket{u_{n\mathbf{k}}}.
\end{equation}
Let us begin by computing the matrix elements of the density operator $\rho_\mathbf{q} = \charge \sum_i e^{-i\mathbf{q}\cdot\mathbf{x}_i}$ between two Bloch states. 
We have
\begin{align}
\bra{\psi_{n\mathbf{k}}}\rho_\mathbf{q}\ket{\psi_{m\mathbf{k'}}} &= \charge \int d\mathbf{r} \psi^*_{n\mathbf{k}}(\mathbf{r})e^{-i\mathbf{q}\cdot\mathbf{r}}\psi_{m\mathbf{k'}}(\mathbf{r}) \nonumber \\
&=\charge \frac{1}{N}\sum_\mathbf{R}e^{i\mathbf{R}\cdot(\mathbf{k'-k-q})}\nonumber
\\&
\qquad  \times \int_{\mathrm{cell}}d\mathbf{r}u_{n\mathbf{k}}^*(\mathbf{r})e^{i\mathbf{r}\cdot(k'-k-q)}u_{m\mathbf{k'}}(\mathbf{r}) \nonumber 
\\
&=\delta_{\mathbf{k',k+q}}\charge \bra{u_{n\mathbf{k}}}\ket{u_{m\mathbf{k+q}}}.\label{eq:densitymatrixelements}
\end{align}
Introducing creation and annihilation operators $c^\dag_{n\mathbf{k}}$ and $c_{n\mathbf{k}}$ that create or annihilate electrons in the state $\ket{\psi_{n\mathbf{k}}}$, Eq.~\eqref{eq:densitymatrixelements} implies that we can write the density operator in second-quantized notation as
\begin{align}
\rho_\mathbf{q} &= \sum_{\mathbf{k}\mathbf{k}^{\prime} n m }\bra{\psi_{n\mathbf{k}}}\rho_\mathbf{q}\ket{\psi_{m\mathbf{k'}}}c^\dag_{n\mathbf{k}}c_{m\mathbf{k'}} \nonumber \\
&= \charge \sum_{\mathbf{k}nm}\bra{u_{n\mathbf{k}}}\ket{u_{m\mathbf{k+q}}}c^\dag_{n\mathbf{k}}c_{m\mathbf{k+q}}.
\end{align}

Next, we can examine the second quantized form of our conserved current $\mathbf{j}_\mathbf{q}$ from Eq.~\eqref{eq:mainj}. 
Note first that
\begin{align}
T_\mathbf{R}\mathbf{j}_\mathbf{q}|\psi_{n\mathbf{k}} \rangle &=\charge T_\mathbf{R}\int_0^1 d\lambda e^{-i(1-\lambda)\mathbf{q}\cdot\mathbf{x}}\mathbf{v}e^{-i\lambda\mathbf{q}\cdot\mathbf{x}}\ket{\psi_{n\mathbf{k}}} \nonumber \\
&= \charge \int_0^1 d\lambda e^{-i(1-\lambda)\mathbf{q}\cdot(\mathbf{x-R})}\mathbf{v}e^{-i\lambda\mathbf{q}\cdot(\mathbf{x-R})}T_\mathbf{R}\ket{\psi_{n\mathbf{k}}} \nonumber \\
&=e^{-i(\mathbf{k-q})\cdot\mathbf{R}}\mathbf{j}_\mathbf{q}\ket{\psi_{n\mathbf{k}}}.\label{eq:jstatemomentum}
\end{align}
Eq.~\eqref{eq:jstatemomentum} indicates that $\mathbf{j}_\mathbf{q}\ket{\psi_{n\mathbf{k}}}$ has crystal momentum $\mathbf{k-q}$. 
Since Bloch states with different crystal momentum are orthogonal, we deduce that the only nonvanishing single-particle matrix elements of $\mathbf{j}_\mathbf{q}$ are
\begin{align}
\label{eq:mainjmatelem}
\bra{\psi_{n\mathbf{k}}}\mathbf{j}_\mathbf{q}&\ket{\psi_{m\mathbf{k+q}}} =\frac{\charge }{N}\int_0^1 d\lambda\int d\mathbf{r} u^*_{n\mathbf{k}}(\mathbf{r})e^{-i[(1-\lambda)\mathbf{q}+\mathbf{k}]\cdot\mathbf{r}}\nonumber
\\&
\qquad \qquad  \times \mathbf{v}e^{i[\mathbf{k+(1-\lambda\mathbf{q}})]\cdot\mathbf{r}} u_{m\mathbf{k+q}}(\mathbf{r}) \nonumber \\
&=\charge \int_0^1 d\lambda\int_{\mathrm{cell}}d\mathbf{r} u^*_{n\mathbf{k}}(\mathbf{r}) \mathbf{v}_{\mathbf{k}+(1-\lambda)\mathbf{q}} u_{m\mathbf{k+q}}(\mathbf{r}) \nonumber \\
&=\charge \int_0^1d\lambda \bra{u_{n\mathbf{k}}}\mathbf{v}_{\mathbf{k}+(1-\lambda)\mathbf{q}}\ket{u_{m\mathbf{k+q}}},
\end{align}
where we have defined
\begin{equation}
\mathbf{v}_\mathbf{k} = e^{-i\mathbf{k}\cdot\mathbf{x}}\mathbf{v}e^{i\mathbf{k}\cdot\mathbf{x}} = \partial_\mathbf{k} (e^{-i\mathbf{k}\cdot\mathbf{x}}He^{i\mathbf{k}\cdot\mathbf{x}}) \equiv \partial_\mathbf{k}H_\mathbf{k}.
\end{equation}
Using Eq.~\eqref{eq:mainjmatelem}, we conclude that the current $\mathbf{j}_\mathbf{q}$ has the second-quantized representation
\begin{equation}
\mathbf{j}_\mathbf{q} = \charge \sum_{\mathbf{k}nm}\int_0^1d\lambda \bra{u_{n\mathbf{k}}}\mathbf{v}_{\mathbf{k}+(1-\lambda)\mathbf{q}}\ket{u_{m\mathbf{k+q}}} c^\dag_{n\mathbf{k}}c_{m\mathbf{k+q}}.\label{eq:TBCurrentOperator}
\end{equation}
By the same logic, we find the second-quantized representation of the non-conserved ``trapezoid'' current $\tilde{\mathbf{j}}_\mathbf{q}$ from Eq.~\eqref{eq:jtildedef} is
\begin{equation}
\mathbf{\tilde{j}}_\mathbf{q} = \charge \sum_{\mathbf{k}nm}\frac{1}{2}\bra{u_{n\mathbf{k}}}\mathbf{v}_{\mathbf{k}+\mathbf{q}}+\mathbf{v}_{\mathbf{k}}\ket{u_{m\mathbf{k+q}}} c^\dag_{n\mathbf{k}}c_{m\mathbf{k+q}}.\label{eq:tildejmatrixelements}
\end{equation}
Finally, for the non-conserved ``midpoint'' current $\mathbf{j}_{\text{mid}, \mathbf{q}}$ in Eq.~\eqref{eq:jmidpointdef} we find the second-quantized representation
\begin{equation}
\mathbf{j}_{\text{mid}, \mathbf{q}} = \charge \sum_{\mathbf{k}nm}\bra{u_{n\mathbf{k}}}\mathbf{v}_{\mathbf{k}+\mathbf{q}/2}\ket{u_{m\mathbf{k+q}}} c^\dag_{n\mathbf{k}}c_{m\mathbf{k+q}}.\label{eq:midpointjmatrixelements}
\end{equation}

Eq.~\eqref{eq:TBCurrentOperator} is the main result of this section, and allows us to reformulate our conserved current operator Eq.~\eqref{eq:mainj} in terms of second-quantized Bloch orbitals. 
Eq.~\eqref{eq:TBCurrentOperator} can be used to compute the matrix elements of the conserved current for effective models that include only a subset of the energy bands of interest in a solid, in which the effective kinetic energy need not be a quadratic function of momentum. 
Additionally, Eq.~\eqref{eq:TBCurrentOperator} manifestly satisfies the continuity equation, and so gives a proper starting point for the evaluation of linear and nonlinear electromagnetic response coefficients. 

Although we supposed that the wavefunctions $\ket{u_{n\mathbf{k}}}$ were the (cell periodic parts of the) energy eigenstates for the non-interacting part of a Hamiltonian, we can follow the logic leading to Eq.~\eqref{eq:TBCurrentOperator} for any orthonormal Bloch-like basis set. 
In App.~\ref{sec:tbderivation} for example, we derive the second-quantized representation of the current Eq.~\eqref{eq:mainj} in a basis of ultra-localized tight-binding orbitals which will be useful for calculations in model systems. 
Additionally, provided that the interaction energy depends only on (unprojected) density operators, Eq.~\eqref{eq:TBCurrentOperator} will still give the second-quantized current operator.

Note that for systems with nonvanishing Chern numbers, the periodic gauge constraint for $\ket{u_{n\mathbf{k}}}$ in Eq.~\eqref{eq:periodicgaugeconstraint} requires that the phase of $\ket{u_{n\mathbf{k}}}$ be non-smooth in the Brillouin zone~\cite{thonhauser2006insulator,thouless1984wannier}. 
This presents no difficulty for our formalism, which will not involve derivatives or Taylor expansions of the wavefunctions themselves with respect to $\mathbf{k}$, but only derivatives of the manifestly-smooth operator $H_\mathbf{k}$. 
Additionally, when we work in a basis of tight-binding orbitals as in App.~\ref{sec:tbderivation}, we can include a sufficiently large number of bands to guarantee that the basis Wannier orbitals have smooth Fourier transforms. 
Finally, we note that in any numerical computation, the wave functions $\ket{u}_{n\mathbf{k}}$ are typically evaluated on a discrete grid of points in the Brillouin zone; we can then choose a periodic gauge such that any nonremovable singularities due to nonvanishing Chern number occur at points that are not included in the grid.

Having now expressed the current operator Eq.~\eqref{eq:mainj} in the second-quantized Bloch basis, we can now turn our attention to how the current density is modified in the presence of a background electromagnetic field. 
This will allow us to generalize the usual diamagnetic current to nonzero wavevector for systems with generic Hamiltonians. 
We will see that these generalized diamagnetic current operators---which are essential to maintain gauge-invariance of response functions---appear naturally in our formalism for $\mathbf{j}_\mathbf{q}$ defined in Eq.~\eqref{eq:mainj}.

\subsection{Diamagnetic Current Operators}\label{sec:HigherOrderCurrentOperators}

In order to compute electromagnetic response functions, we need to know the current $\mathbf{j}_A(\mathbf{r})$ order-by-order in the vector potential $\mathbf{A}$. 
Although we have primarily focused on $\mathbf{j}_{\mathbf{q},\mathrm{min}}$ in Eq.~\eqref{eq:variational_current_def}, for a general system, we can write
\begin{equation}
\label{eq:GeneraljADefinition}
j^\mu_A(\mathbf{r}) = j^\mu_{\mathrm{min}}(\mathbf{r}) + \int d\mathbf{r}' j^{\mu\nu}(\mathbf{r},\mathbf{r'})A_\nu(\mathbf{r}',t) + \dots,
\end{equation}
or in Fourier space
\begin{equation}
\label{eq:GeneraljADefinitionq}
j^\mu_{A,\mathbf{q}} = j^\mu_{\mathrm{min},\mathbf{q}} + \sum_{\mathbf{q'}} j^{\mu\nu}_{\mathbf{q},-\mathbf{q'}}A_{\nu,\mathbf{q}'}+\dots.
\end{equation}

While knowledge of $j^{\mu\nu}_{\mathbf{q},-\mathbf{q}'}$ requires knowledge of the minimally coupled Hamiltonian in Eq.~\eqref{eq:mincouple}, the continuity equation at Eq.~\eqref{eq:continuity} places constraints on the longitudinal components of $j^{\mu\nu}_{\mathbf{q},-\mathbf{q}'}$. 
In particular, note first that
\begin{align}
j^{\mu\nu}_{\mathbf{q},-\mathbf{q}'} &= -\left.\frac{\delta^2 H_A}{\delta A_{\mu,\mathbf{-q}} \delta A_{\nu,\mathbf{q'}}}\right|_{\mathbf{A}=0},\nonumber\\
&=\left.\frac{\delta j^\mu_{\mathbf{q},A}}{\delta A_{\nu,\mathbf{q'}}}\right|_{\mathbf{A}=0}.\label{eq:jmunuvardef}
\end{align}
Inserting Eq.~\eqref{eq:jmunuvardef} into the continuity equation~\eqref{eq:continuity}, we find that
\begin{align}
iq_\mu j^{\mu\nu}_{\mathbf{q},-\mathbf{q}'} &= \left.\frac{\delta}{\delta A_{\nu,\mathbf{q}'}}(iq_\mu j^\mu_{\mathbf{q},A})\right|_{\mathbf{A}=0}\nonumber \\
&=-i\left.\frac{\delta}{\delta A_{\nu,\mathbf{q}'}}\left[H_A,\rho_\mathbf{q}\right]\right|_{\mathbf{A}=0} \nonumber \\
&=i\left[j^\nu_{\mathbf{-q'},\mathrm{min}},\rho_\mathbf{q}\right],\label{eq:j2fromj1}
\end{align}
where we have used from Eq.~\eqref{eq:rhoqdef} that $\rho_\mathbf{q}$ is independent of the vector potential. 
Eq.~\eqref{eq:j2fromj1} shows that the longitudinal component of the diamagnetic current $j^{\mu\nu}_{\mathbf{q,-\mathbf{q}'}}$ can be expressed entirely in terms of the unperturbed current $j^\nu_{\mathbf{-q'},\mathrm{min}}$. 
Furthermore, since the longitudinal component of $j^\nu_{\mathbf{-q'},\mathrm{min}}$ is equal to the current in Eq.~\eqref{eq:mainj}, we find that the longitudinal component $q_\mu q'_\nu j^{\mu\nu}_{\mathbf{q,-q'}}$ of the diamagnetic current is entirely expressible in terms of the velocity operator.

Eq.~\eqref{eq:j2fromj1} is equivalent to the Ward identity from quantum field theory~\cite{peskin1995introduction}. 
We can iterate the logic leading to Eq.~\eqref{eq:j2fromj1} in order to obtain the longitudinal part of all higher-order current vertices $q_{\mu_1}q'_{\mu_2}q''_{\mu_3}\dots j^{\mu_1\mu_2\mu_3\dots}_{\mathbf{q},\mathbf{q'},\mathbf{q''}\dots}$. 
Since the operators $j^{\mu_1\mu_2\mu_3\dots}_{\mathbf{q},\mathbf{q'},\mathbf{q''}\dots}$ appear as vertex functions in calculations of higher-order electromagnetic responses~\cite{parker2019diagrammatic,Yan2023UnifyingSemiclassicsNonlinear}, we see that our expression for the current in Eq.~\eqref{eq:mainj} in terms of the velocity operator can be used to generate the longitudinal components of these higher-order vertices.

Using Eqs.~\eqref{eq:jmunuvardef} and \eqref{eq:j2fromj1}, we can recast this second order operator in terms of an integration over an auxiliary variable $\lambda$ via the Karplus-Schwinger relation. 
Namely, we show in App.~\ref{sec:GeneralizedKarplusSchwingerIntegrationDerivation}, that for any single-particle operator $O$,
\begin{align}
     i\left[ O, e^{-i \mathbf{q} \cdot \mathbf{x}} \right] = \int^{1}_{0} d\lambda e^{-i(1- \lambda)\mathbf{q}\cdot \mathbf{x}} (\mathbf{q}\cdot [O, \mathbf{x}]) e^{-i \lambda \mathbf{q}\cdot \mathbf{x}}.
     \label{eq:CommutatorKarplusSchwingerRelation}
\end{align}

Applying Eq.~\eqref{eq:CommutatorKarplusSchwingerRelation} to our Ward identity Eq.~\eqref{eq:j2fromj1} we find
\begin{align}
    iq_\mu q'_{\nu} j^{\mu\nu}_{\mathbf{q},-\mathbf{q}'} &= 
i\left[q'_{\nu}j^\nu_{\mathbf{-q'},\mathrm{min}}, \charge \sum_i e^{-i \mathbf{q}\cdot \mathbf{x}_i}\right] \nonumber
\\&= \charge  \int^{1}_{0} d\lambda e^{-i(1- \lambda)\mathbf{q}\cdot \mathbf{x}} (q_{\mu}q'_\nu [j^{\nu}_{-\mathbf{q}^{\prime}, \text{min}}, x^{\mu}]) e^{-i \lambda \mathbf{q}\cdot \mathbf{x}}. \nonumber \\
&=\charge  \int^{1}_{0} d\lambda e^{-i(1- \lambda)\mathbf{q}\cdot \mathbf{x}} (q_{\mu}q'_\nu [j^{\nu}_{-\mathbf{q}^{\prime}}, x^{\mu}]) e^{-i \lambda \mathbf{q}\cdot \mathbf{x}}\label{eq:j2LambdaIntegral},
\end{align}
where we have used the fact that both $\mathbf{j}_{\mathrm{min},\mathbf{q}}$ and $\mathbf{j}_{\mathbf{q}}$ of Eq.~\eqref{eq:mainj} are conserved. 
The advantage to using the Karplus-Schwinger relation is being able to strip off the $\mathbf{q}$ and $\mathbf{q'}$, thus allowing us to define $j^{\mu\nu}_{\mathbf{q},-\mathbf{q}'}$ that satisfies the Ward identity for generic systems.

We can rewrite Eq.~\eqref{eq:j2LambdaIntegral} in terms of the second-quantized creation and annihilation operators for Bloch eigenstates following the formalism of Sec.~\ref{sec:SecondQuantizedOperator} in order to define
\begin{widetext}
\begin{equation}
     j^{\mu\nu}_{\mathbf{q},-\mathbf{q}'} = -(\charge )^2 \int^{1}_{0} \int^{1}_{0} d\lambda d\lambda^{\prime} \left \langle u_{n \mathbf{k}} \right| \left[ \frac{\partial^2 H_{\mathbf{k}}}{\partial k^{\mu} \partial k^{\nu}} \right]_{\mathbf{k} \rightarrow \mathbf{k} - (1 - \lambda^{\prime}) \mathbf{q}^{\prime} + (1 - \lambda) \mathbf{q}}\left| u_{m \mathbf{k} - \mathbf{q}^{\prime} + \mathbf{q}} \right\rangle c^{\dagger}_{n \mathbf{k}} c_{m \mathbf{k} - \mathbf{q}^{\prime} + \mathbf{q}}.\label{eq:j2BlochLambdaIntegral}
\end{equation}

In a similar manner, we can iterate this procedure to derive expressions for the $N$-th order current operator
\begin{align}
   & j^{\mu \nu_1 \nu_2 \cdots \nu_N}_{\mathbf{q}, - \mathbf{q}_1, -\mathbf{q}_2, \ldots , - \mathbf{q}_N} = -\left.\frac{\delta^{N+1} H_A}{\delta A_{\mu,\mathbf{-q}} \delta A_{\nu_1,\mathbf{q}_1} \delta A_{\nu_2,\mathbf{q}_2} \cdots \delta A_{\nu_N,\mathbf{q}_N} }\right|_{\mathbf{A}=0}\label{eq:generalj},
\end{align}
which satisfies the generalized Ward identity
\begin{align}\label{eq:ward}
    &(iq_{\mu})(-iq_{1, \nu_1})(-i q_{2, \nu_2}) \cdots (-i q_{N-1, \nu_{N-1}})j^{\mu \nu_1 \nu_2 \cdots \nu_N}_{\mathbf{q}, - \mathbf{q}_1, -\mathbf{q}_2, \ldots, - \mathbf{q}_N}
    = i \left[ i \left[ i \left[j^{\nu_N}_{-\mathbf{q}_N, \text{min}}, \rho_{-\mathbf{q}_{N-1}} \right],\rho_{-\mathbf{q}_{N-2}}\right], \cdots , \rho_{\mathbf{q}} \right].
\end{align}
Applying the Karplus-Schwinger relation Eq.~\eqref{eq:CommutatorKarplusSchwingerRelation} iteratively, and using our expressions Eqs.~\eqref{eq:mainj} and \eqref{eq:TBCurrentOperator} we find that we can write the $n$-th order diamagnetic current operator in second-quantized notation as

\begin{align}
    j^{\mu \nu_1 \cdots \nu_N}_{\mathbf{q}, -\mathbf{q}_1, \cdots, -\mathbf{q}_N} =&-(-\charge )^{N+1} \int^1_0 d \lambda \int^1_0 d \lambda_1 \cdots \int^1_0 d \lambda_N\nonumber
    \\
    &\times
    \left \langle u_{n \mathbf{k}} \right| \left[ \partial_{\mathbf{k}^\mu} \partial_{\mathbf{k}^{\nu_1}} \cdots \partial_{\mathbf{k}^{\nu_N}} H_{\mathbf{k}} \right]_{\mathbf{k} \rightarrow \mathbf{k} + (1 - \lambda) \mathbf{q} - (1 - \lambda_1) \mathbf{q}_1 \cdots - (1 - \lambda_N )\mathbf{q}_N} \left|  u_{m \mathbf{k} + \mathbf{q} - \mathbf{q}_1  \cdots - \mathbf{q}_N}\right \rangle c^{\dagger}_{n \mathbf{k}} c_{m \mathbf{k} + \mathbf{q} - \mathbf{q}_1 \cdots - \mathbf{q}_N}.\label{eq:tight-bindingGeneralCurrentOperator}
\end{align}
Note, importantly, that Eq.~\eqref{eq:tight-bindingGeneralCurrentOperator} is explicitly symmetric under the exchange of any pair of indices, since partial derivatives commute.

For later convenience, we will define the $N$-th  order velocity vertex $v^{\nu_N}_{(N)}(\mathbf{k}, \mathbf{q}, -\mathbf{q}_1, \cdots ,-\mathbf{q}_N)$ as the operator appearing in the matrix element of Eq. \eqref{eq:tight-bindingGeneralCurrentOperator}, i.e.,
\begin{equation}\label{eq:vertexfuncgeneral}
    v^{\nu_N}_{(N)}(\mathbf{k},  \mathbf{q}, -\mathbf{q}_1, \cdots ,-\mathbf{q}_N) \equiv \int^1_0 d \lambda \int^1_0 d \lambda_1 \cdots \int^1_0 d \lambda_N \left[ \partial_{\mathbf{k}^\mu} \partial_{\mathbf{k}^{\nu_1}} \cdots \partial_{\mathbf{k}^{\nu_{N}}} H_{\mathbf{k}} \right]_{\mathbf{k} \rightarrow \mathbf{k}  + (1 - \lambda) \mathbf{q} - (1 - \lambda_1) \mathbf{q}_1 \cdots - (1 - \lambda_N )\mathbf{q}_N}.
\end{equation}
Therefore, applying Eq.~\eqref{eq:vertexfuncgeneral} to Eq.~\eqref{eq:tight-bindingGeneralCurrentOperator}, we can simply write
\begin{equation}
j^{\mu \nu_1 \cdots \nu_N}_{\mathbf{q}, -\mathbf{q}_1, \cdots, -\mathbf{q}_N} = -(-\charge )^{N+1} \sum_{\mathbf{k}}\left \langle u_{n \mathbf{k}} \right| v^{\nu_N}_{(N)}(\mathbf{k},  \mathbf{q}, -\mathbf{q}_1, \cdots ,-\mathbf{q}_N)  \left|  u_{m \mathbf{k} + \mathbf{q} - \mathbf{q}_1  \cdots - \mathbf{q}_N}\right \rangle c^{\dagger}_{n \mathbf{k}} c_{m \mathbf{k} + \mathbf{q} - \mathbf{q}_1 \cdots - \mathbf{q}_N}.
\end{equation} 
\end{widetext}

Note that since the ``trapezoid'' and ``midpoint'' currents [Eqs.~\eqref{eq:trapfirsttime} and \eqref{eq:midpointfirsttime}] do not satisfy the continuity equation, there is no general procedure for determining their corresponding diamagnetic currents. 
Nevertheless, for the sake of comparison, we will define ``trapezoid'' and ``midpoint'' diamagnetic current operators by approximating each auxiliary integral in Eq.~\eqref{eq:tight-bindingGeneralCurrentOperator} with the trapezoid or midpoint approximation, respectively.

Given Eq.~\eqref{eq:tight-bindingGeneralCurrentOperator} for the generalized diamagnetic current operators as a function of wavevector(s), we will now develop a diagrammatic formalism to compute linear and nonlinear conductivities as a function of frequency and wavevector. 
Our construction of Eq.~\eqref{eq:tight-bindingGeneralCurrentOperator} and Eq.~\eqref{eq:TBCurrentOperator} from generalized Ward identities ensures that these conductivities will give a current that respects the continuity equation.

\section{Feynman Diagrammatic Method and Sum Rules}\label{sec:DiagramsRuleAndSumRules}

Using the conserved current operator, Eq.~\eqref{eq:mainj} will now develop the formalism for computing the spatially-inhomogeneous linear and nonlinear conductivities defined in Eq.~\eqref{eq:responsegeneral}. 
We will build off the work of Ref.~\cite{parker2019diagrammatic}, which presented a simple yet powerful framework for calculating the spatially uniform ($\mathbf{q}=0$) nonlinear conductivities.
Continuing off of this framework, we look to expand this framework to accommodate spatially inhomogeneous fields and currents.
This framework utilizes diagrammatic perturbation theory in terms of non-interacting Matsubara Green's functions for electrons, with interactions with the external electromagnetic field given by the (noninteracting) vertex functions governed by Eq.~\eqref{eq:vertexfuncgeneral}. 
We will generalize this diagrammatic method of Ref.~\cite{parker2019diagrammatic, Joao2020OpticalResponseTBWithDiagrams} to allow for electromagnetic fields and currents with nonzero wavevectors.
As a consequence, we will be able to capture not just response to electric fields, but---via Faraday's law Eq.~\eqref{eq:faraday}---response to magnetic fields as well. 
We begin in Sec.~\ref{sec:FeynmanDiagramSetup} by establishing the foundation for diagrammatic evaluation of the Kubo formula. 
Then, in Sec.~\ref{sec:DiagramRules} we will give the rules for setting up and evaluating our Feynman diagrams. 
Finally, in Secs.~\ref{sec:linearResponse}--\ref{sec:SecondOrderResponseInMoireMaterials} we will apply our formalism to analyze the linear and second order response as a function of wavevector for 3D Weyl and 2D moire systems. 

\subsection{Feynman Diagram Setup}\label{sec:FeynmanDiagramSetup}
To write the Feynman diagram rules for evaluating the average current, we must first relate the conductivities to the generating function for correlations.
Since the Feynman diagrams are a direct interpretation of perturbative expansions of the generating function, this allows us to write the conductivities as a sum of diagrams, which may then be translated to a mathematical statement involving Green's functions and interaction vertices~\cite{coleman2015introduction, bruus2004manybody, Coleman2018QFTBook, parker2019diagrammatic, McKay2021CDWWeyl, altland2010condensed, fenton1983infrared, takano1982diagrammatical, Joao2020OpticalResponseTBWithDiagrams}.

First consider the generic Hamiltonian that has been coupled to a vector potential, as posed in Eq.~\eqref{eq:mincouple}.
Notice that the minimally coupled Hamiltonian can be expanded in powers of the vector potential as:
\begin{align}
\label{eq:expandedHamiltonian}
    H_A=& H_0 -\left( A_{\alpha_1, -\mathbf{q}_1} j^{\alpha_1}_{\mathbf{q}_1} + \frac{1}{2} A_{\alpha_1, -\mathbf{q}_1} A_{\alpha_2, \mathbf{q}_2} j^{\alpha_1 \alpha_2}_{\mathbf{q}_1, -\mathbf{q}_2}\right. \nonumber
    \\
    &+ \left.\frac{1}{6} A_{\alpha_1, -\mathbf{q}_1} A_{\alpha_2, \mathbf{q}_2} A_{\alpha_3, \mathbf{q}_3} j^{\alpha_1 \alpha_2 \alpha_3}_{\mathbf{q}_1, -\mathbf{q}_2, -\mathbf{q}_2}+ \cdots\right).
\end{align}
In the last term of Eq.~\eqref{eq:expandedHamiltonian}, we replaced the functional derivatives of the coupled Hamiltonian with the generalized current term described in Eq.~\eqref{eq:generalj}. 
We will now specify to studying systems of electrons, where we take $\charge = -e$, with $e$ being the (positive) elementary charge.
With this definition of the minimally coupled Hamiltonian, we can define the expectation values of the current operator to all orders in the vector potential \cite{parker2019diagrammatic}
\begin{equation}
\label{eq:masterEqnforJ}
    \langle j^{\mu} \rangle = \frac{1}{Z} \text{Tr} \left[ T_{t} e j^{\mu}_A(\mathbf{r}) e^{-i \int d t^{\prime}H_A(t^{\prime}, \mathbf{r})} \right],
\end{equation}
where $j^{\mu}_A(\mathbf{r})$ is defined via Eq.~\eqref{eq:GeneraljADefinition} in terms of our conserved current, Eq.~\eqref{eq:TBCurrentOperator}, and the diamagnetic currents, Eq.~\ref{eq:tight-bindingGeneralCurrentOperator}.
Also, $T_{t}$ is the time-ordering operator.

In Eq.~\eqref{eq:masterEqnforJ}, two levels of expansion may be implemented in orders of the vector potential: an expansion in $j^{\mu}_A(\mathbf{r})$ and an expansion in $H_A$.
We will assume the vector potential can be written as a superposition of plane waves, 
\begin{equation}
\mathbf{A}(\mathbf{r},t)=\sum_{\omega, \mathbf{q}}\mathbf{A}_{\mathbf{q}}(\omega) e^{i \mathbf{r} \cdot \mathbf{q} - i \omega t}.
\end{equation}
To obtain the average current, we can expand the left-hand-side of Eq.~\eqref{eq:masterEqnforJ} in terms of a Fourier-transformed response coefficient and the corresponding fields, as described in Eq.~\eqref{eq:responsegeneral}, taking the Fourier transform of every time and position argument.

Note that for systems of electrons in a periodic potential, Umklapp processes allow for the generation of currents that oscillate in space with wavevectors that differ from the wavevectors of the applied field by a reciprocal lattice vector. 
Concretely, expanding Eq.~\eqref{eq:masterEqnforJ} in Fourier space [or, equivalently, taking the Fourier transform of Eq.~\eqref{eq:responsegeneral}] yields
\begin{widetext}
\begin{equation}\label{eq:responsewithnonzeroG}
    \langle j^\mu_\mathbf{q}(t)\rangle = \sum_n\sum_\mathbf{G}\int\prod d\mathbf{q}_i dt_i \sigma_{(n)}^{\mu,\nu_1,\dots,\nu_n}(\mathbf{q},\mathbf{q}_1,\dots\mathbf{q}_n,t,t_1,\dots,t_n)\delta(\mathbf{q}-\sum_i\mathbf{q}_i-\mathbf{G})\prod_i E_{\nu_i,\mathbf{q}_i}(t_i),
\end{equation}
\end{widetext}
where $\mathbf{G}$ are the reciprocal lattice vectors. 
Since the spatial scale of variation of a current with wavevector $\mathbf{G}\neq 0$ is on the order of (fractions of) an angstrom, these Umklapp terms are typically not experimentally interesting. 
Therefore, we will focus the remaining work on computing the $\mathbf{G}=0$ component of the electromagnetic response. 
We emphasize, however, that our diagrammatic calculation can be easily generalized to compute the $\mathbf{G}\neq 0$ Umklapp components of the current as well.

By comparing Eq.~\eqref{eq:masterEqnforJ} with Eq.~\eqref{eq:responsewithnonzeroG}, we may extract the $n$-th order conductivity by using the following process: we first expand the average current to $n$-th order in the vector potential. 
Next, we Fourier transform with respect to time, and make use of the definition of the electric field
\begin{equation}
\mathbf{E}_{\mathbf{q}}(\omega) = i\omega \mathbf{A}_{\mathbf{q}}(\omega)
\end{equation}
in the $A_0=0$ gauge. 
Note that this produces a time-ordered, rather than a causal response function. 
To recover the usual causal response function that can be measured in experiment, we can analytically continue all frequencies into the complex plane, $\omega\rightarrow \omega+i\eta$ for a small infinitesimal $\eta$, following the procedure outlined in Refs.~\cite{parker2019diagrammatic,passos2018nonlinear,McKay2021CDWWeyl}.
In additional to preserving causality, a small positive finite $\eta$ can be used to approximate the electron self-energy due to impurity scattering~\cite{parker2019diagrammatic,passos2018nonlinear}.

Practically speaking, we will carry out the perturbative calculation using the imaginary-time Matsubara formalism, using Matsubara frequencies
$i\nu_n = \frac{(2 n + 1) i\pi}{\beta}$ for fermions and $i\omega_n =  \frac{2 n i\pi}{\beta}$ for bosons, where $\beta = (k_B T)^{-1}$ \cite{bruus2004manybody}. 
We will take the $T\rightarrow 0$ limit at the end of all computations, which has the effect of turning Fermi-Dirac distributions, $n_F$, into Heaviside step functions.
Moreover, the $T\rightarrow 0$ limit will still give a good approximation for low temperature transport coefficients in insulators.
Since the Matsubara frequencies become continuous in this limit, we will suppress the subscript $n$~\cite{parker2019diagrammatic, bruus2004manybody, altland2010condensed, fenton1983infrared}.

With this as our starting point, we can now outline the diagrammatic rules that allow us to express Eq.~\eqref{eq:responsewithnonzeroG} in terms of Matsubara Green's functions and the vertex functions of Eq.~\eqref{eq:tight-bindingGeneralCurrentOperator}.

\subsection{List of Feynman Diagram Rules}\label{sec:DiagramRules}
In this section, we will describe the rules for writing down and evaluating the Feynman diagrams for the wavevector- and frequency-dependent conductivity tensors.
These diagrams rely on the generalized current operator derived in Section \ref{sec:HigherOrderCurrentOperators}.
Our formalism will closely follow Ref.~\cite{parker2019diagrammatic}, contrasting only in the inclusion of the wavevector dependence of the external fields and the current operator.

First let us define the symbology that will compose our diagrams:
\begin{enumerate}
  \item The free fermion propagator (Matsubara Green's function), $G(k)$, is denoted by
  \begin{tikzpicture}[baseline={(current bounding box.center)}]
    \begin{feynman}
    \vertex (a) ;
    \vertex [       left=of a] (b) ;
    \diagram* {
        (b) -- [fermion]  (a),
    };
    \end{feynman}
    \end{tikzpicture}.
    \item The perturbing vector potential field is denoted by \begin{tikzpicture}[baseline={(current bounding box.center)}]
    \begin{feynman}
    \vertex (a) ;
    \vertex [       left=of a] (b) ;
    \diagram* {
        (b) -- [photon]  (a),
    };
    \end{feynman}
    \end{tikzpicture}, and carries with it a (Matsubara) frequency and a wavevector.
    \item Where the fermion and photon propagator meet, we place a vertex in the $\alpha$ direction. 
    When corresponding to the output current operator it is denoted by an outgoing photon vertex,
    \begin{tikzpicture}[baseline={(current bounding box.center)}]
    \begin{feynman}
    \vertex (a) ;
    \vertex [empty dot](b)[       left=1.5emof a]{}  ;
    \diagram* {
        (b) -- [photon]  (a),
    };
    \end{feynman}
    \end{tikzpicture}, and when corresponding to a perturbing field it is denoted by an incoming vertex, \begin{tikzpicture}[baseline={(current bounding box.center)}]
    \begin{feynman}
    \vertex (a) ;
    \vertex [dot](b)[       right=1.5emof a]{}  ;
    \diagram* {
        (b) -- [photon]  (a),
    };
    \end{feynman}
    \end{tikzpicture}.
\end{enumerate}
We have introduced a four vector notation for Matsubara frequencies, with $k = (i\nu, \mathbf{k})$ representing the fermionic Matsubara frequencies and momenta, and $q = (i\omega, \mathbf{q})$ representing the bosonic Matsubara frequencies and momenta. 
We will also use the shorthand $d k \equiv d \mathbf{k} d \nu_1$ to denote the integration measure, and $q_1 + q_2 \equiv q_{12}$ to denote the componentwise sum of two four vectors. 

With these components defined, we may write diagrams describing the contributions to $\langle j^\mu_\mathbf{q}\rangle$ order-by-order in perturbation theory. 
For concreteness, we will specialize at this point to a free-electron system. 
As such, the Feynman rules below require us to consider only diagrams with a single fermion loop. 
We note, however, that our diagrammatic rules can be extended in the presence of interactions, such as from phonons \cite{McKay2021CDWWeyl}, by including additional interaction vertices. To simplify the bookkeeping arising from the negative signs in Eq.~\eqref{eq:tight-bindingGeneralCurrentOperator} and from the negative sign of the electron charge, we will formulate our diagrammatic rules in terms of the velocity vertex functions in Eq.~\eqref{eq:vertexfuncgeneral}.

The diagrammatic rules are:
\begin{enumerate}
    \item Every loop has an output vertex, and most have input vertices. 
    Diagrams contributing the the $n$-th order response have $n+1$ photon lines. 
    All photon lines are incoming at input vertices. 
    At output vertices, exactly one photon line is outgoing.
    \item Every loop denotes an integral over both $\mathbf{k}$ space as well as a Matsubara sum (which can be made into an integral) in frequency $\nu$.
    \item Every closed loop conserves momentum and energy \footnote{Note that we can generalize this rule to allow for calculation of nonzero-$\mathbf{G}$ Umklapp response in Eq.~\eqref{eq:responsewithnonzeroG} by imposing momentum conservation only modulo a reciprocal lattice vector}. 
    By convention, incoming photon lines carry momentum out of the loop, and outgoing photon lines carry momentum into the loop.
    \item Each incoming field vector also contributes a factor of $\frac{e i}{\hbar i\omega_\alpha}$, where $\omega_\alpha$ is the Matsubara frequency for the $\alpha$-th photon line.
    \item To avoid double counting, only topologically unique diagrams should be considered. 
    In particular, if the exchange of two or more four-vector and index labels on photon lines (not including the photon corresponding to the output current) does not change a diagram, then the diagram should be divided by the appropriate multiplicity factor~\cite{parker2019diagrammatic}. 
    In practice, this means that $N$-th order input diamagnetic current vertices are accompanied by a factor of $1/N!$, and $N$-th order output diamagnetic vertices are accompanied by a factor of $1/(N-1)!$ We will see an application of this rule in Sec.~\ref{sec:SecondOrderResponseInMoireMaterials}.
    \item 
    A vertex with $N$ photon lines corresponds a factor of (a matrix elements of) $v^{\nu_N}_{(N)}(\mathbf{k},\mathbf{q}_1,\dots,\mathbf{q}_N)$ defined in Eq.~\eqref{eq:tight-bindingGeneralCurrentOperator},
    where $\mathbf{k}$ is the fermion momenta going into the vertex.
\end{enumerate}
After evaluating a particular set of diagrams, we can finally obtain a causal response function by analytically continuing each Matsubara frequency back to a real frequency, $i\omega\rightarrow \omega+i\eta$, for $\eta$ a positive infinitesimal.

\section{Linear Response}\label{sec:linearResponse}
In this section, we will apply the Feynman diagrammatic rules from Sec.~\ref{sec:DiagramRules} to calculate the linear conductivity as a function of frequency and wavevector for a general noninteracting periodic solid.
First, in Sec.~\ref{sec:FullLinearResponse} we will use our diagrammatic method to recover the Kubo formula for the conductivity, where we comment on the relationship between our approach and that of Refs.~\cite{kozii2021intrinsic,zhong2018linear,Souza2023MidpointMethod}. 
Then, in Sec.~\ref{sec:fsum} we will show how our formalism relates to the generalized f-sum rule~\cite{graf1995electromagnetic,husain2023pines}.

\subsection{Full Linear Response from Diagrams}\label{sec:FullLinearResponse}

\begin{figure}[h]
      \centering
\begin{minipage}{0.95\hsize}
\centering
\includegraphics[width=0.95\hsize]{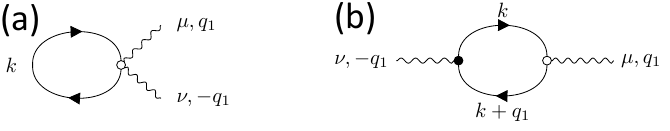}
\caption{Linear-order response diagrams}
\label{Fig:FeynmanDiagramsLinearOrder}
\end{minipage}
\end{figure}

The linear conductivity is given by the sum of the two diagrams in Fig.~\ref{Fig:FeynmanDiagramsLinearOrder}.
Using our Feynman rules, this becomes
\begin{align}
    &\sigma^{\mu \nu}(i\omega,\mathbf{q}) =  \frac{ie^2}{i\omega}\int dk \left[ G_{n_1}(k) v^{\mu \nu }_{(2), n_1 n_1}(\mathbf{k} , -\mathbf{q}, \mathbf{q}) \right. \nonumber
    \\&
    \left.+ G_{n_2 }(k + q) v^{\nu}_{(1) n_2 n_1}(\mathbf{k+q}, -\mathbf{q}) G_{n_1}(k )   v^{\mu}_{(1) n_1 n_2}(\mathbf{k}, \mathbf{q})\right],\label{eq:fullGeneralizedLinearConductivity}
\end{align}
where $G_n(k)=(\nu - \epsilon_{n \mathbf{k}})^{-1}$ is the Matsubara Green's function in the energy eigenbasis $\ket{\psi_{n\mathbf{k}}}$ of Eq.~\eqref{eq:eigenbasis}, $\epsilon_{n \mathbf{k}}$ is the corresponding energy, and we introduce the notation $v^{\mu\nu_1,\dots\nu_N}_{(N+1),n_1n_2}$ to denote the matrix elements of the velocity vertex Eq.~\eqref{eq:vertexfuncgeneral} in the basis of Bloch eigenstates $\ket{u_{n\mathbf{k}}}$.

The first term in Eq.~\eqref{eq:fullGeneralizedLinearConductivity}, corresponding to the diagram in Fig.~\ref{Fig:FeynmanDiagramsLinearOrder}(a), is the diamagnetic conductivity.
We can evaluate Eq.~\eqref{eq:fullGeneralizedLinearConductivity} and analytically continue to real frequency to find
\begin{widetext}
\begin{align}
\label{eq:diamagnetic}
    \sigma^{\mu \nu}_{\text{dia}}(\omega)=&  \frac{ie^2}{\omega^+} \int dk \sum_{n_1}G_{n_1}(k) v^{\nu}_{(2),n_1 n_1}(\mathbf{k}, -\mathbf{q}, \mathbf{q})\nonumber
    \\
    =&  \frac{ie^2}{\omega^+} \int d\nu d \mathbf{k} d \lambda d \lambda_1 \sum_{n_1}\frac{
    \left\langle u_{n_1 \mathbf{k}} \right|\left( \partial_{\mathbf{k}^\mu} \partial_{\mathbf{k}^{\nu}} H_{\mathbf{k}}\right)|_{\mathbf{k} \rightarrow \mathbf{k} - (1 - \lambda_1) \mathbf{q} +  (1 - \lambda) \mathbf{q}}\left| u_{n_1 \mathbf{k}} \right\rangle
    }{i\nu- \epsilon_{n_1 \mathbf{k}}}  \nonumber
    \\
    =&  \frac{ie^2}{\omega^+} \int d \mathbf{k} d\lambda d\lambda_1 \sum_{n_1} n_F\left(\epsilon_{n_1}(\mathbf{k})\right) \left\langle u_{n_1 \mathbf{k}} \right|\left( \partial_{\mathbf{k}^\mu} \partial_{\mathbf{k}^{\nu}} H_{\mathbf{k}}\right)|_{\mathbf{k} \rightarrow \mathbf{k} + (\lambda_1 - \lambda) \mathbf{q}}\left| u_{n_1 \mathbf{k}} \right\rangle,
\end{align}
\end{widetext}
where $n_F$ is the Fermi-Dirac distribution and $\omega^+=\omega+i\eta$. 
In the limit $\mathbf{q}\rightarrow 0$ this reproduces the generalized diamagnetic conductivity of Ref.~\cite{parker2019diagrammatic}. 
Here, however, we see that when $H_\mathbf{k}$ contains non-quadratic momentum dependence---as it will for any semirelativistic system or for any system modeled by a low-energy effective Hamiltonian---the diamagnetic conductivity acquires a nontrivial $\mathbf{q}$ dependence. 
The diamagnetic conductivity serves to regularize the conductivity in the limit $\omega\rightarrow 0$. 
We can gain some intuition for this by noting that Fig.~\ref{Fig:FeynmanDiagramsLinearOrder}(a) gives the diamagnetic conductivity in terms of the average
\begin{equation}\label{eq:diamagneticintermsofj2}
\sigma^{\mu \nu}_{\text{dia}}(\omega) = \frac{i}{\omega^+}\langle j^{\mu\nu}_{\mathbf{-q,q}}\rangle
\end{equation}
of the diamagnetic current Eq.~\eqref{eq:j2BlochLambdaIntegral}. 
The generalized Ward identity from Eq.~\eqref{eq:j2fromj1} constrains the diamagnetic conductivity.

The bubble diagram in Fig.~\ref{Fig:FeynmanDiagramsLinearOrder}(b) gives the paramagnetic conductivity, corresponding to the second term in Eq.~\eqref{eq:fullGeneralizedLinearConductivity}. 
By evaluating the Matsubara integrals in Eq.~\eqref{eq:fullGeneralizedLinearConductivity} and analytically continuing back to real frequencies, we find that the paramagnetic conductivity is given by
\begin{equation}
\label{eq:kubo}
    \sigma^{\alpha \beta}_{\text{para}}(\omega, \mathbf{q})=\frac{-ie^2}{ \omega^+}Q^{\alpha \beta}( \omega, \mathbf{q}) ,
\end{equation}
where $Q^{\alpha \beta}( t - t^{\prime}, \mathbf{q})\equiv i\left\langle\left[j^{\alpha}_\mathbf{q}(t), j^{\beta}_{-\mathbf{q}}(t^{\prime})\right]\right\rangle \theta(t - t^\prime)$ is the retarded current-current correlation function in the time domain. 
In the frequency domain, we can use our expression Eq.~\eqref{eq:TBCurrentOperator} for the current operator to find
\begin{align}
     &Q^{\alpha \beta}( \omega, \mathbf{q} )  \nonumber
     \\
     =&
    \sum_{\mathbf{k}} v^{\alpha}_{a  b}(\mathbf{k}, \mathbf{q})   v^{\beta}_{b a}(\mathbf{k}+\mathbf{q}, -\mathbf{q}) \frac{n_{F}(\epsilon_{a \mathbf{k}}) - n_{F}(\epsilon_{b \mathbf{k} + \mathbf{q} })}{ \epsilon_{a \mathbf{k} } -\epsilon_{b \mathbf{k} + \mathbf{q} }  + \omega^+ }\nonumber
    \\ \equiv &
    \sum_{\mathbf{k}}  F_{\alpha \beta}^{a b}(\mathbf{k}, \mathbf{q}) \frac{n_{F}(\epsilon_{a \mathbf{k}}) - n_{F}(\epsilon_{b \mathbf{k} + \mathbf{q} })}{ \epsilon_{a \mathbf{k} } -\epsilon_{b \mathbf{k} + \mathbf{q} }  + \omega^+ },\label{eq:currentcurrent}
\end{align}
where $v^{\alpha}_{a  b}(\mathbf{k}, \mathbf{q})$ is the charge-conserving velocity, given by
\begin{equation}
v^{\alpha}_{a  b}(\mathbf{k}, \mathbf{q}) = \int^{1}_{0} d\lambda \langle u_{a \mathbf{k}} | (\partial_{\mathbf{k}^{\alpha}} H_{\mathbf{k}})|_{\mathbf{k} \rightarrow \mathbf{k} + (1 + \lambda)\mathbf{q}} | u_{b \mathbf{k} + \mathbf{q}} \rangle 
\end{equation}
 as indicated in Eq.~\eqref{eq:tight-bindingGeneralCurrentOperator}.

Our improved definition for the conserved current operator enters into the matrix elements $F^{ab}_{\alpha\beta}(\mathbf{k},\mathbf{q})$, while the ratio of Fermi functions to energy differences is universal and arises from the evaluation of the Green's function integral in Eq.~\eqref{eq:fullGeneralizedLinearConductivity}. 
We can compare our result for the paramagnetic conductivity to that of Refs.~\cite{kozii2021intrinsic, Souza2023MidpointMethod} which uses the midpoint current from Eq.~\eqref{eq:jmidpointdef} instead of the conserved current. 
This leads to the appearance of a modified matrix element
\begin{align}
F_{\text{mid}; \alpha \beta}^{a b}(\mathbf{k}, \mathbf{q}) = &\left\langle u_{a \mathbf{k}}\left|\hat{v}_{\mathbf{k}+\frac{\mathbf{q}}{2}}^{\alpha}\right| u_{b \mathbf{k} + \mathbf{q}}\right\rangle \nonumber
\\
\times & \left\langle u_{b \mathbf{k}+\mathbf{q}}\left|\hat{v}_{\mathbf{k}+\frac{\mathbf{q}}{2}}^{\beta}\right| u_{a \mathbf{k}}\right\rangle,
\label{eq:midpointHallFormula}
\end{align}
in place of $F_{\alpha \beta}^{a b}(\mathbf{k}, \mathbf{q})$, with $\hat{v}_{\mathbf{k}}^{\alpha} \equiv\frac{\partial \hat{H}_\mathbf{k}}{\partial k_{\alpha}}$.
The subscript ``mid" is assigned to this quantity to remind the reader that it was derived using the midpoint definition of the current operator.

Similarly, if the (non-conserved) trapezoid current $\Tilde{\mathbf{j}}$ from Eq.~\eqref{eq:jtildedef} were used in place of $\mathbf{j}$ to derive the response, then we would find for the matrix element
\begin{align}
    \Tilde{F}_{\alpha \beta}^{a b}(\mathbf{k}, \mathbf{q}) = &\left\langle u_{a \mathbf{k}}\left|\frac{\hat{v}_{\mathbf{k}+\mathbf{q}}^{\alpha} + \hat{v}_{\mathbf{k}}^{\alpha}}{2}\right| u_{b \mathbf{k}+\mathbf{q}}\right\rangle \nonumber
    \\
    \times & \left\langle u_{b \mathbf{k}+\mathbf{q}}\left|\frac{\hat{v}_{\mathbf{k}+\mathbf{q}}^{\beta} + \hat{v}_{\mathbf{k}}^{\beta}}{2} \right| u_{a \mathbf{k}}\right\rangle.
    \label{eq:tildeHallFormula}
\end{align}
We emphasize that since Eq.~\eqref{eq:currentcurrent} is computed using the conserved current in Eq.~\eqref{eq:TBCurrentOperator}, it gives the physically-meaningful conductivity. 
We will use Eqs.~\eqref{eq:midpointHallFormula} and \eqref{eq:tildeHallFormula} in Sec.~\ref{sec:ApplicationOfTheLinearResponse} to show how the trapezoid and midpoint currents give quantitatively different predictions for the conductivity as compared to Eq.~\eqref{eq:currentcurrent}.

In particular, Ref.~\cite{kozii2021intrinsic} used the (non-conserved) Eq.~\eqref{eq:midpointHallFormula} to evaluate the $\mathbf{q}$-dependent Hall response for a two-band tight-binding model, where we expect the use of the conserved current Eq.~\eqref{eq:TBCurrentOperator} to be important to obtain reliable results. 
We now argue that, since the current operators Eqs.~\eqref{eq:mainj}, \eqref{eq:jtildedef}, and \eqref{eq:jmidpointdef} agree to linear order in $\mathbf{q}$, they yield identical calculations of the paramagnetic conductivity Eq.~\eqref{eq:kubo} to order $|\mathbf{q}|^2$. 
To see this, let us rewrite the current-current correlator in Eq.~\eqref{eq:kubo} as
\begin{align}\label{eq:correlator-rewrite}
\left\langle\left[j^\alpha_\mathbf{q}(t),j^\beta_\mathbf{-q}(t')\right]\right\rangle &= \left\langle\left[j^\alpha_\mathbf{0}(t),j^\beta_\mathbf{-q}(t')\right]\right\rangle\nonumber\\
&+\left\langle\left[j^\alpha_\mathbf{q}(t),j^\beta_\mathbf{0}(t')\right]\right\rangle\nonumber\\
&+\left\langle\left[\delta j^\alpha_\mathbf{q}(t),\delta j^\beta_\mathbf{-q}(t')\right]\right\rangle\nonumber\\
&-\left\langle\left[j^\alpha_\mathbf{0}(t),j^\beta_\mathbf{0}(t')\right]\right\rangle,
\end{align}
where we have introduced 
\begin{equation}\label{eq:deltaj}
\delta j^\alpha_\mathbf{q} = j^\alpha_\mathbf{q}-j^\alpha_\mathbf{0}.
\end{equation}
To proceed, note that for crystalline systems, we know from conservation of (crystal) momentum that the ground state operator of any operator with nonzero momentum (modulo reciprocal lattice vectors) must vanish. 
We can apply this observation to conclude that the first two terms on the right hand side of Eq.~\eqref{eq:correlator-rewrite} vanish unless $\mathbf{q}=0$. 
Thus, we have
\begin{align}\label{eq:simplified-para}
\left\langle\left[j^\alpha_\mathbf{q}(t),j^\beta_\mathbf{-q}(t')\right]\right\rangle &= \left\langle\left[\delta j^\alpha_\mathbf{q}(t),\delta j^\beta_\mathbf{-q}(t')\right]\right\rangle \nonumber \\
&-\left\langle\left[j^\alpha_\mathbf{0}(t),j^\beta_\mathbf{0}(t')\right]\right\rangle.
\end{align}

Finally, Taylor expanding Eq.~\eqref{eq:simplified-para} for small $\mathbf{q}$ shows that the $\mathcal{O}(\mathbf|q|^2)$ contribution to the paramagnetic conductivity is determined by the $\mathcal{O}(|\mathbf{q}|)$ term in $\delta j^\alpha_\mathbf{q}$. 
Since Eqs.~\eqref{eq:mainj}, \eqref{eq:jtildedef}, and \eqref{eq:jmidpointdef} agree to linear order in $\mathbf{q}$, they yield the same prediction for the paramagnetic (and hence the Hall) conductivity.

Note, however, that the diamagnetic vertex Eq.~\eqref{eq:diamagneticintermsofj2} derived from the conserved current in Eq.~\eqref{eq:mainj} differs from the naive $\mathbf{q}$-independent diamagnetic current derived from the nonrelativistic Hamiltonian in Eq.~\eqref{eq:assumedH} and used in Refs.~\cite{zhong2018linear,Souza2023MidpointMethod,Vignale1991OrbitalParamagnetism} to calculate the symmetric part of the conductivity from tight-binding models. 
In particular, since the conserved two-photon vertex $j^{\mu\nu}_{-\mathbf{q},\mathbf{q}}$, defined in Eq.~\eqref{eq:j2BlochLambdaIntegral}, is an even and symmetric function of $\mathbf{q}$, there will be $\mathcal{O}(|\mathbf{q}|^2)$ contributions to the (symmetric) diamagnetic conductivity for tight binding models that are necessary to ensure charge conservation.

\subsection{The f-Sum Rule and the Diamagnetic Conductivity}\label{sec:fsum}
In this section, we will show how our definition for the conserved current $\mathbf{j}_\mathbf{q}$ allows us to derive a generalized f-sum rule valid for any effective model. 
While the derivations in this section do not make use of the explicit form of the current operator, they highlight constraints on the conductivity imposed by charge conservation that are satisfied only when the conserved current operator is used. 
We begin with a short review of density-density response and the derivation of the f-sum rule~\cite{kadanoff1963hydrodynamic}. 
Recall that the f-sum rule constrains the spectral weight of the density-density response function
\begin{equation}
\chi(\omega, \mathbf{q}) = -i\int_0^\infty dt e^{i\omega^+t}\langle\left[\rho_\mathbf{q}(t),\rho_{-\mathbf{q}}(0)\right]\rangle.
\end{equation}
We start by defining the spectral density
\begin{equation}\label{eq:spectraldensitydef}
\chi{''}(\omega, \mathbf{q}) = \frac{1}{2}\int_{-\infty}^\infty dt e^{i\omega t}\langle\left[\rho_\mathbf{q}(t),\rho_{-\mathbf{q}}(0)\right]\rangle
\end{equation}
which satisfies the Kramers-Kronig relation
\begin{equation}\label{eq:kkreln}
\chi(\omega, \mathbf{q})=\frac{1}{\pi} \int_{-\infty}^\infty d\omega' \frac{\chi{''}(\omega', \mathbf{q})}{\omega^+-\omega'}.
\end{equation}
The spectral density $\chi{''}(\omega, \mathbf{q})$ can also be expressed as the imaginary part of $\chi(\omega, \mathbf{q})$, and can be directly measured through absorption spectroscopy. 

From Eq.~\eqref{eq:kkreln} we can deduce an expansion for the large-$\omega$ asymptotics of $\chi(\omega, \mathbf{q})$ in terms of moments of $\chi''(\omega, \mathbf{q})$. 
By Taylor expanding the denominator in the integral in Eq.~\eqref{eq:kkreln} we obtain the asymptotic expansion
\begin{equation}
\chi(\omega\rightarrow\infty, \mathbf{q})\sim\sum_{n=0}^{\infty} \frac{\chi_n''(\mathbf{q})}{(\omega)^{n+1}}\label{eq:asymptotic}
\end{equation}
where
\begin{equation}\label{eq:momentdef}
\chi_n''(\mathbf{q})=\frac{1}{\pi}\int_{-\infty}^{\infty} d\omega (\omega)^n\chi{''}(\omega, \mathbf{q})
\end{equation}
is ($1/\pi$ times) the $n$-th frequency moment of $\chi{''}(\omega, \mathbf{q})$. 
Note that since the (unprojected) density operators commute, i.e. $\left[\rho_{\mathbf{q}},\rho_{\mathbf{q}'}\right]=0$, we have
\begin{align}
\chi''_0(\mathbf{q}) &= \frac{1}{\pi}\int_{-\infty}^\infty d\omega \chi''(\omega, \mathbf{q}) \\
&=\frac{1}{2\pi}\int_{-\infty}^\infty d\omega\int_{-\infty}^\infty dt e^{i\omega t}\langle\left[\rho_\mathbf{q}(t),\rho_{-\mathbf{q}}(0)\right]\rangle\label{eq:0moment1} \\
&=\int_{-\infty}^\infty dt \delta(t) \langle\left[\rho_\mathbf{q}(t),\rho_{-\mathbf{q}}(0)\right]\rangle \label{eq:0moment2}\\
&=\langle\left[\rho_\mathbf{q}(0),\rho_{-\mathbf{q}}(0)\right]\rangle \\
&=0,\label{eq:0momentfinal}
\end{align}
where in going from Eq.~\eqref{eq:0moment1} to Eq.~\eqref{eq:0moment2} we exchanged the order of the $\omega$ and $t$ integrals and used
\begin{equation}
\frac{1}{2\pi}\int d\omega e^{i\omega t} = \delta(t).
\end{equation}
This means that the $n=0$ term in the asymptotic expansion Eq.~\eqref{eq:asymptotic} vanishes. 
Using Eqs.~\eqref{eq:asymptotic} and \eqref{eq:0momentfinal} we thus have asymptotically to leading order as $\omega\rightarrow \infty$,
\begin{equation}
\lim_{\omega\rightarrow \infty}\omega^2\chi(\omega, \mathbf{q}) = \chi''_1(\mathbf{q}) = \frac{1}{\pi}\int_{-\infty}^\infty d\omega (\omega)\chi''(\omega, \mathbf{q}).
\end{equation}
We can go further and evaluate $\chi''_1(\mathbf{q})$ using the continuity equation to arrive at a general form of the f-sum rule. 
In particular, inserting the definition Eq.~\eqref{eq:spectraldensitydef} into the definition Eq.~\eqref{eq:momentdef} of $\chi''_1(\mathbf{q})$ and integrating by parts we find
\begin{align}\label{eq:sumruleintermediate}
\chi''_1(\mathbf{q}) &= \frac{1}{\pi}\int_{-\infty}^\infty d\omega (\omega)\chi''(\omega, \mathbf{q}) \nonumber
\\
&=\frac{1}{2\pi}\int_{-\infty}^\infty d\omega \int_{-\infty}^{\infty}dt e^{i\omega t}\omega\langle\left[\rho_\mathbf{q}(t),\rho_{-\mathbf{q}}(0)\right]\rangle \nonumber
\\
&=\frac{-i}{2\pi}\int_{-\infty}^\infty d\omega \int_{-\infty}^{\infty}dt \frac{d}{dt}(e^{i\omega t})\langle\left[\rho_\mathbf{q}(t),\rho_{-\mathbf{q}}(0)\right]\rangle \nonumber
\\
&=\frac{1}{2\pi}\int_{-\infty}^\infty d\omega \int_{-\infty}^{\infty}dt e^{i\omega t}\langle\left[i\partial_t\rho_\mathbf{q}(t),\rho_{-\mathbf{q}}(0)\right]\rangle \nonumber
\\
&=\int_{\infty}^{\infty} dt \delta(t) \langle\left[i\partial_t\rho_\mathbf{q}(t),\rho_{-\mathbf{q}}(0)\right]\rangle \nonumber
\\
&=\langle\left[i\partial_t\rho_\mathbf{q},\rho_{-\mathbf{q}}\right]\rangle,
\end{align}
where we have suppressed the time arguments in the equal-time commutator in the last line. 
Using the continuity equation, Eq.~\eqref{eq:fouriercont}, along with the Karplus-Schwinger relationship to simplify commutators of the density operator, we find
\begin{align}
\chi''_1(\mathbf{q}) &=\langle\left[i\partial_t\rho_\mathbf{q},\rho_{-\mathbf{q}}\right]\rangle \nonumber
\\ 
&=q_\mu\langle\left[j^\mu_\mathbf{q},\rho_{-\mathbf{q}}\right]\rangle \nonumber
\\
&=q_\mu q_\nu \langle j^{\mu\nu}_{\mathbf{-q,q}}\rangle, \nonumber
\\
&=-i\lim_{\omega\rightarrow 0}\omega q_\mu q_\nu \sigma_{\mathrm{dia}}^{\mu\nu}(\mathbf{q,\omega}),\label{eq:fsum}
\end{align}
where we have used Eq.~\eqref{eq:diamagneticintermsofj2} to express the sum rule in terms of the average of the two-photon velocity vertex, which is the coefficient of the diamagnetic conductivity $\sigma_{\mathrm{dia}}^{\mu\nu}(\mathbf{q,\omega})$.
For Hamiltonians with quadratic momentum dependence, the diamagnetic conductivity is independent of $\mathbf{q}$ and this reproduces the usual f-sum rule.
Note that since Eq.~\eqref{eq:sumruleintermediate} involves an integral over all frequencies, the sum rule of Eq.~\eqref{eq:fsum} in the form we have written it relates to the full (unprojected) density operator and diamagnetic conductivity. 
We defer the exploration of restricted sum rules over a limited frequency range to future work (though see the recent Ref.~\cite{mendez2023theory} for progress along these lines).

Our derivation shows how the f-sum rule generalizes to nonzero wavevector for general systems.
In particular, we have shown in Sec.~\ref{sec:HigherOrderCurrentOperators} that the two-photon velocity vertex, $j^{\mu\nu}_{\mathbf{q',-q}}$, which uses the conserved current $\mathbf{j}_\mathbf{q}$ is generally given by Eq.~\eqref{eq:j2LambdaIntegral}.
We have also shown how the correlation-correlation operator through the spectral density function relates back to conductivity in Eq.~\eqref{eq:fsum}, and, in App.~\ref{app:ConsistencyWithPlasmonDispersion}, we explore this connection even further through the plasmon dispersion.

For general Hamiltonians, such as those arising in tight-binding approximations to solid state systems, the $\mathbf{q}$ dependence of $j^{\mu\nu}_{\mathbf{q',-q}}$ will lead to a nontrivial $\mathbf{q}$-dependent diamagnetic conductivity, and hence a modification to the f-sum rule according to Eq.~\eqref{eq:fsum}. 
In Ref.~\cite{graf1995electromagnetic} the deviation between the generalized f-sum rule [in the form of Eq.~\eqref{eq:sumruleintermediate}] and the electron filling was taken as a measure of goodness of fit for tight-binding parameters. 
Here, we take a different point of view: the deviation between the generalized f-sum rule and the electron filling is a consequence of gauge invariance and quantifies a combination of semirelativistic effects as well as information about the truncation of the Hilbert space in an effective low-energy model. 
Since the spectral density at nonzero $\mathbf{q}$ is measurable via absorption spectroscopy, Eq.~\eqref{eq:fsum} gives an experimental probe of the two-photon velocity vertex, and hence of the conserved current $\mathbf{j}_\mathbf{q}$~\cite{husain2023pines}.

\section{Applications of the Linear Response}\label{sec:ApplicationOfTheLinearResponse}
Having now derived the complete expression for the $\mathbf{q}$-dependent linear conductivity in Eq.~\eqref{eq:fullGeneralizedLinearConductivity}, we will now use it to analyze electromagnetic response in insulators and semimetals. 
We will begin in Sec.~\ref{sec:weylhall} by computing the frequency- and wavevector-dependent Hall conductivity in a model of a Weyl semimetal. 
We will see that for large wavevectors the use of the conserved current Eq.~\eqref{eq:TBCurrentOperator} in the Kubo formula yields quantitatively different predictions for the Hall response as compared to the trapezoid or midpoint approximations prevalent in the literature. 
This analysis is applicable to studying the response of the system to spatially-modulated AC electromagnetic fields, such as can be applied using standing waves or gate potentials.
Next, in Sec.~\ref{sec:KerrRotation} we turn our attention to moir\'{e} materials, where the wavevector dependence of $\sigma^{\mu\nu}(\omega,\mathbf{q})$ has implications even for optical response due to the large effective lattice constant. 
We will compute the Kerr angle and ellipticity at oblique incidence as a function of frequency for a model of a moir\'{e} Chern insulator in two dimensions, showing that spatial inhomogeneous electric fields can lead to experimentally relevant modifications to the Kerr effect. 
Finally, in Sec.~\ref{sec:MagneticPropertiesOfInsulators} we use our formalism for the conserved current to analyze the magnetic moment and magnetic susceptibility of insulators.

\subsection{Linear Hall Effect in a Weyl Semimetal}\label{sec:weylhall}
As a proof of principle, we will apply our definition for the conserved current Eq.~\eqref{eq:TBCurrentOperator} to compute the wavevector-dependent Hall conductivity
\begin{equation}
    \sigma_H^{xy}(\omega,\mathbf{q}) = \frac{1}{2}\left(\sigma^{xy}(\omega,\mathbf{q}) - \sigma^{yx}(\omega,\mathbf{q})\right)
\end{equation}
for a toy model of a time-reversal symmetry breaking, inversion symmetric Weyl semimetal with two Weyl points first presented in Ref.~\cite{Weyl1929OriginalWeylFermionPaper}.
We will also compare our predictions with analogous calculations of the Hall conductivity using the (non-conserved) midpoint and trapezoid currents via Eqs.~\eqref{eq:midpointHallFormula}. and \eqref{eq:tildeHallFormula}.
In the Weyl semimetal, Hall conductivity arises due to the Berry curvature texture of the occupied bands; each Weyl point is a source of Berry curvature of charge $|C|=1$. 
For two Weyl points of opposite charge separated by momentum $2k_0\hat{z}$, the Berry curvature, when the chemical potential is exactly at the two Weyl nodes, leads to a spatially uniform anomalous Hall conductivity~\cite{Steiner2017AnomalousHallWSM, burkov2014anomalous, Huang2017MultifoldWSMAnomHall,Yan2020TRBInCurvatureMetricAndMotion}
\begin{equation}
\sigma_H^{xy}(\omega\rightarrow 0, \mathbf{q}\rightarrow 0) = 2e^2 k_0.
\end{equation}
For future convenience, we define the wavevector-dependent anomalous Hall conductivity $\sigma^{xy}_{\text{anom}}(\mathbf{q})$ as
\begin{equation}
\label{eq:anomalousDefinition}
    \sigma_H^{xy}(\omega\rightarrow 0, \mathbf{q}) \equiv \sigma^{xy}_{\text{anom}}(\mathbf{q}).
\end{equation}

We will use our formalism to compute how this topological Hall response is modified in the presence of inhomogeneous electric fields.
At this stage, we do not assume that our fields are optical---that is, we do not require that $\omega = c|\mathbf{q}|$.
As such, our analysis is primarily applicable to the response of the Weyl semimetal to spatially dispersive, slowly varying AC electromagnetic fields. 
Note that the diamagnetic conductivity Eq.~\eqref{eq:diamagnetic} is explicitly symmetric, owing to the symmetry of the diamagnetic current vertex in Eq.~\eqref{eq:j2fromj1}. 
Therefore, when we go to calculate the Hall conductivity (which is antisymmetric), it will not contribute. 
We can thus focus on the paramagnetic conductivity Eqs.~\eqref{eq:kubo} and \eqref{eq:currentcurrent}. 

\subsubsection{The Weyl Semimetal Model}

\begin{figure}[h]
      \centering
\begin{minipage}{0.95\hsize}
\centering
\includegraphics[width=0.95\hsize]{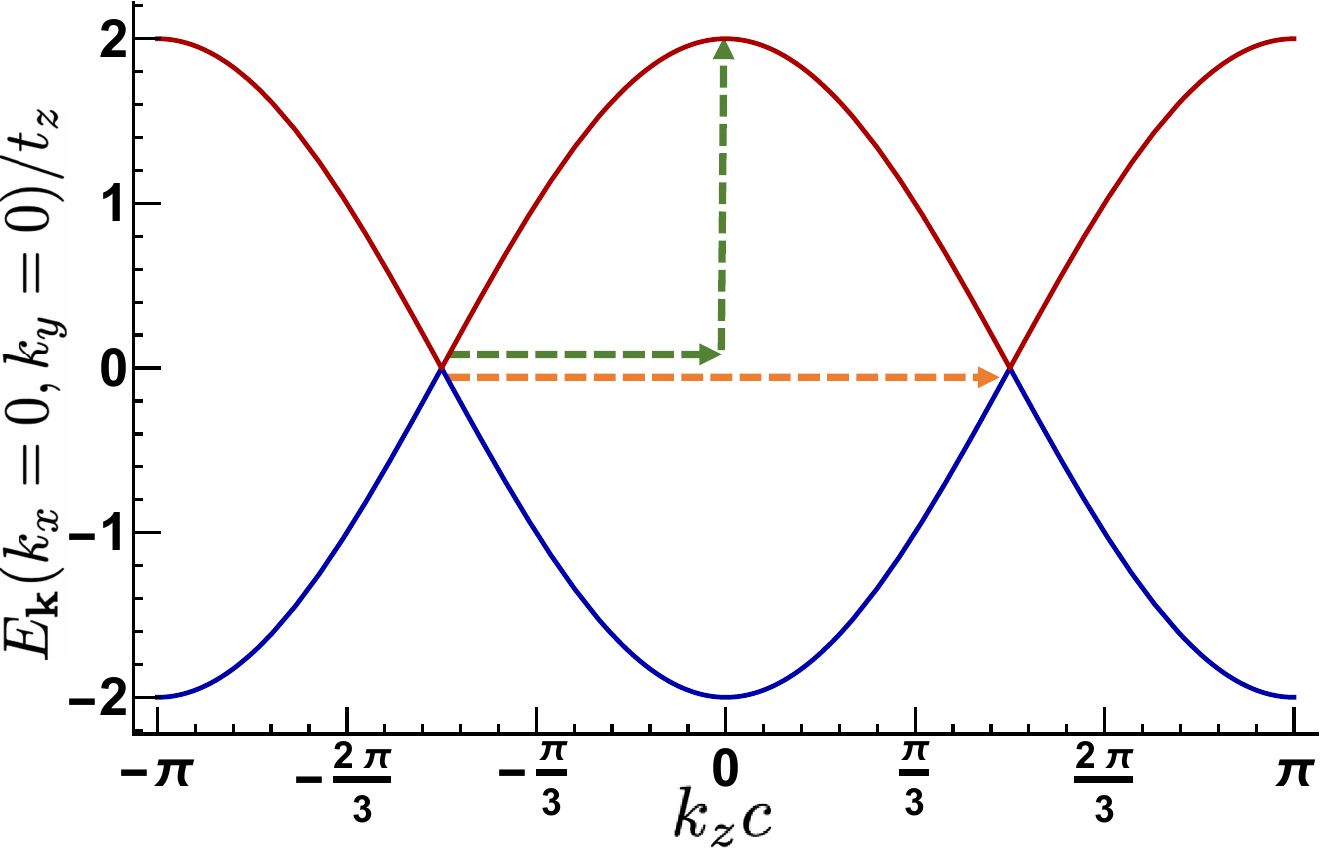}
\end{minipage}
\caption{Energy spectrum of the Weyl semimetal Hamiltonian Eq.~\eqref{eq:weylHam} along the $k_x = k_y = 0$ line in the Brillouin zone, 
with parameter values $m=4 t_z$, $t_x = t_y = t_z $, $\gamma=0$, $\gamma_z=0$, and $\mu = 0$.
The dotted lines indicate two types of indirect transitions that can be driven by a time- and space-dependent electric field: the green dotted line shows a jump of of $|\mathbf{q}|=\pi/2$ in momentum and $\omega=2 t_z$ in energy, and the dotted yellow line show an jump of $|\mathbf{q}|=\pi$ in momentum space with zero energy transfer $\omega=0$.}
\label{Fig:weylSpectrumPlot}
\end{figure}

We will consider a model for a time-reversal symmetry breaking, inversion symmetric Weyl semimetal with tight-binding Hamiltonian given by~\cite{mccormick2017minimal, mccormick2017tiltedweyl, Ran2011WeylModel, Tewari2016WeylMagnetothermalConductivity, Bradlyn2020AxionicWSMCDW, Cano2019MultifoldWeyl}:
\begin{align}
    H =& \sum^4_{i=0} d_i(\mathbf{k}) \sigma^i,\label{eq:weylHam}
    \\
    d_0(\mathbf{k}) =& \gamma (\cos(k_z) - \cos(k_0)) - \mu,\nonumber
    \\
    d_1(\mathbf{k}) =& -2 t_x \sin(k_x),\nonumber
    \\
    d_2(\mathbf{k}) =& -2 t_y \sin(k_y),\nonumber
    \\
    d_3(\mathbf{k}) =& -\left(2 t_z \left( \cos(k_z) - \cos(k_0) \right) \right.\nonumber
    \\&
    \left. + m \left( 2 - \cos(k_x) - \cos(k_y) \right)\right.\nonumber
    \\&
    \left.+ \gamma_z \left( \cos(3 k_z) - \cos(3 k_0) \right) \right),\nonumber    
\end{align}
where $\sigma^i$ are the $2\times 2$ Pauli matrices with $\sigma^0$ defined to be the identity matrix. 
All values of $\mathbf{k}$ are measured in reduced coordinates, i.e. $\mathbf{k}_i \in [-\pi, \pi]$ for $i=x, y, z$.
Here $k_0$ quantifies the separation of the Weyl nodes along the $\hat{z}$ direction. 
When $|m|>2t_z$, this model has a gapped spectrum everywhere in the Brillouin zone except at two Weyl points with (reduced) coordinates $(0,0,\pm k_0)$; we will focus on this regime for our analysis.
The energy parameters $t_x$, $t_y$, and $t_z$ determine the velocities close to the Weyl node in each principal direction.
The quantities $\gamma$ and $\gamma_z$ are tilting parameters that allow for type-II Weyl semimetals to occur.
\cite{mccormick2017minimal}.
A plot of the spectrum of Eq. \eqref{eq:weylHam} with
$k_x = k_y = 0$ is shown in Fig.~\ref{Fig:weylSpectrumPlot}.

Using this model along with our Kubo formula Eq.~\eqref{eq:currentcurrent}, we will examine how the Hall conductivity $\sigma^{xy}_H(\omega,\mathbf{q})$ deviates from its topological value as a function of both $\omega$ and $\mathbf{q}$. 

\subsubsection{Hall Response when $\mathbf{q} = (q_0, 0,  0)$}\label{sec:HallResponseAt000}

\begin{figure*}[t]
      \centering
\begin{minipage}{0.98\hsize}
\centering
\includegraphics[width=0.98\hsize]{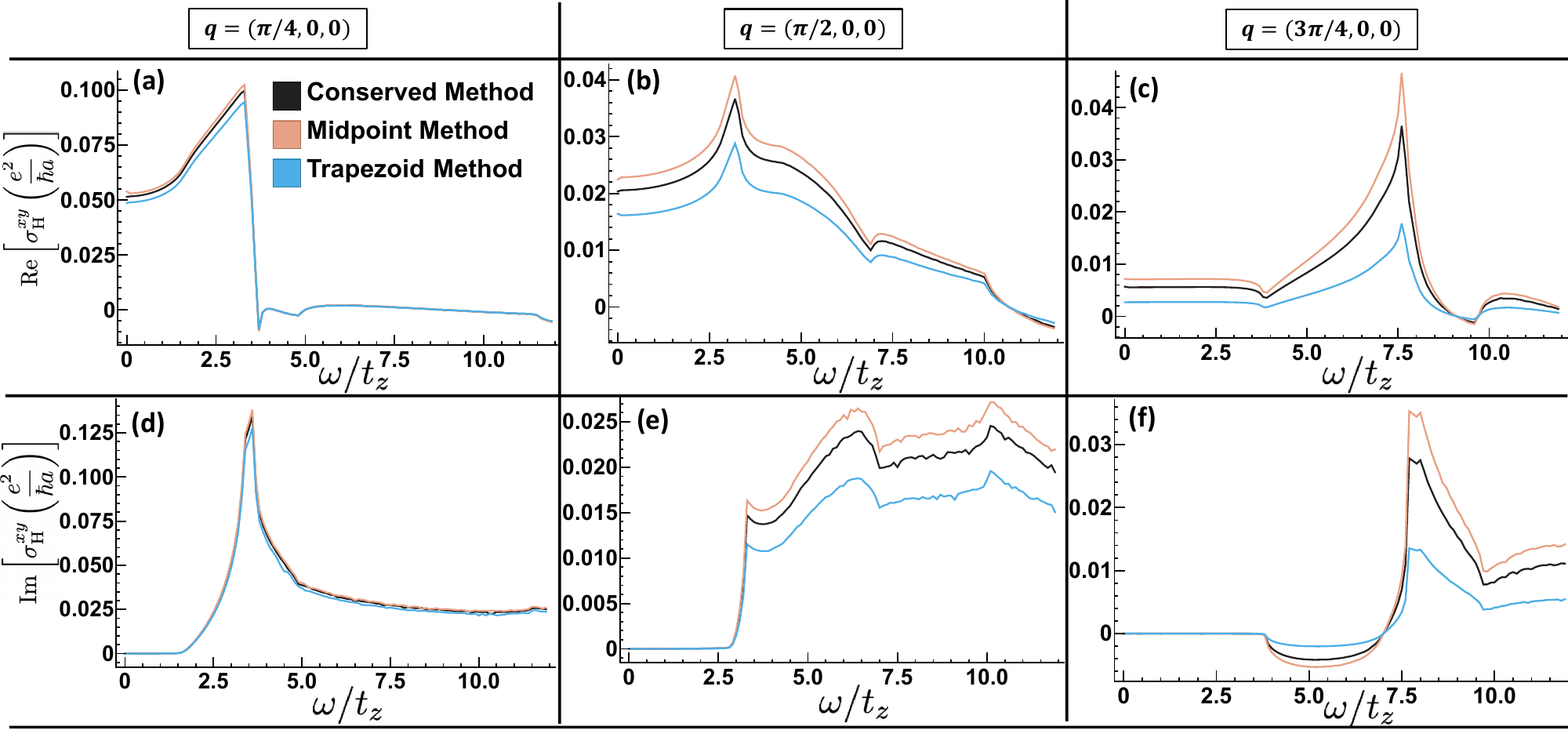}
\end{minipage}
\caption{ The Hall conductivity versus the frequency of the perturbing electric field at different values of the wavevector ($|q| = \pi/4, \pi/2, 3 \pi/4$) in the $\hat{x}$ direction. 
(a-c) show the real parts of the Hall conductivity for each wavevector while plots (d-f) show the imaginary part.
The parameters $m=4 t_z$, $t_x = t_y = t_z$, $\gamma=0$, $\gamma_z=0$, and $\mu = 0$ were used for these calculations. 
Black curves correspond to the conductivity computed with the conserved current $\mathbf{j}_\mathbf{q}$ using Eq.~\eqref{eq:currentcurrent}. 
For comparison, we also show the Hall conductivity computed with the non-conserved trapezoid current $\tilde{j}_{\mathbf{q}}$ (blue) and the non-conserved midpoint current (pink). 
We see that at large wavevectors, the non-conserved currents give quantitatively different predictions for the Hall conductivity as compared to the conserved current. 
All plots are generated with natural units (i.e. all energy parameters in terms of $t_z$ and $e=1$).}
\label{Fig:differenceInMethodsAlongQxWeyl}
\end{figure*}

Using Eq.~\eqref{eq:currentcurrent}, we first compute the Hall response in this Weyl semimetal model when the electric field is parallel to the $\hat{y}$ direction and the wavevector is oriented in the $\hat{x}$ direction with a magnitude $q_0$. 
The Hall current then flows along the $\hat{x}$ direction, parallel to the wavevector. 
Note that in this case, the electric field is transverse, while the measured current is longitudinal.
We examine this response because it illustrates the differences between the Hall conductivity calculated with the conserved current and the Hall conductivity calculated with the non-conserved currents commonly used in the literature.  

We first illustrate the difference in the three definitions as outlined in Equations \eqref{eq:currentcurrent}, \eqref{eq:midpointHallFormula}, and \eqref{eq:tildeHallFormula}: the conserved current, the (non-conserved) midpoint definition, and the (non-conserved) trapezoidal definition respectively.

\begin{figure*}[t]
      \centering
\begin{minipage}{0.98\hsize}
\centering
\includegraphics[width=0.98\hsize]{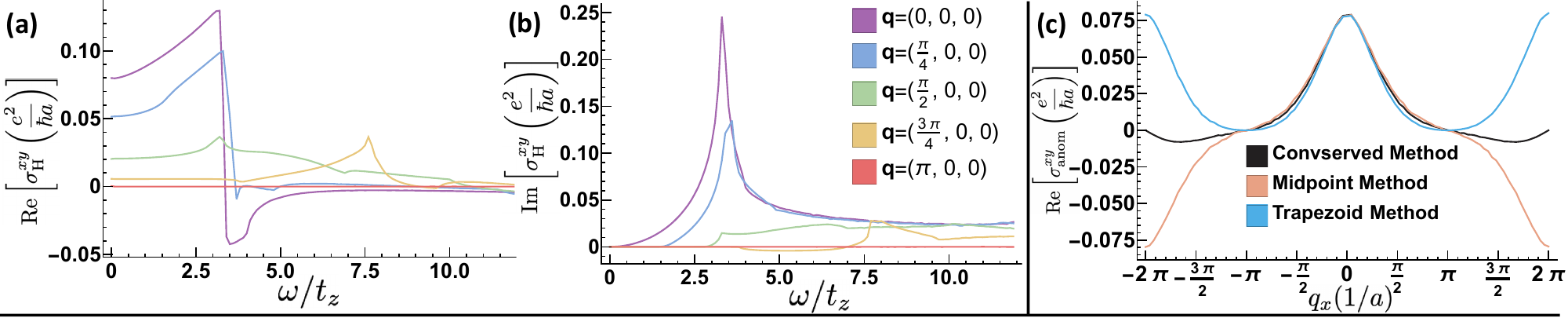}
\end{minipage}
\caption{The real [(a)] and imaginary [(b)] parts of the Hall conductivity as a function of frequency for several values of the wavevector in the $\hat{x}$ direction.
The conductivities in (a) and (b) were computed using the conserved current Eq.~\eqref{eq:TBCurrentOperator} and the Kubo formula Eq.~\eqref{eq:currentcurrent}.
(c) shows the real part of the anomalous Hall conductivity [as defined in Eq.~\eqref{eq:anomalousDefinition}] as a function of the wavevector changing in the $\hat{x}$ direction.
The parameters $m=4 t_z$, $t_x = t_y = t_z$, $\gamma=0$, $\gamma_z=0$, and $\mu = 0$ were used for these calculations.}
\label{Fig:sigmaxyVsOmegaAndQx}
\end{figure*}

Each of these three different definitions, yielding different predictions for the Hall conductivity, is shown in Fig.~\ref{Fig:differenceInMethodsAlongQxWeyl}.
Increasing the wavevector generally leads to a more pronounced difference among the three predictions, which makes sense since the midpoint and trapezoid currents are approximations to the $\lambda$ integration in the definition Eq.~\eqref{eq:TBCurrentOperator} of the conserved current; these integral approximations become less exact as the wavevector increases. 
This is further demonstrated in Fig.~\ref{Fig:sigmaxyVsOmegaAndQx}(c) where we can see that the discrepancies between the anomalous Hall conductivity $\sigma_\mathrm{anom}^{xy}(\mathbf{q})$ [defined in Eq.~\eqref{eq:anomalousDefinition}], computed using each of the three methods increase as $\mathbf{q}$ increases. 
We note also that due to the integration over $\lambda$ in the definition of the conserved current Eq.~\eqref{eq:TBCurrentOperator}, the anomalous Hall conductivity computed using the conserved current decays as $1/|\mathbf{q}|^2$ for asymptotically large $\mathbf{q}$.
Hence, the continuity equation has an influence on the Hall response that becomes more apparent with increasing wavevector. 
On the contrary, the non-conserved trapezoidal and midpoint definitions of the current fail to capture this feature, and instead predict an anomalous Hall conductivity that is periodic in $q_x$ with period $2\pi$ and $4\pi$, respectively.
This point is illustrated in Fig. \ref{Fig:sigmaxyVsOmegaAndQx}(c).

We also find, as expected, that the three current operators predict the same Hall conductivity in the limit $\mathbf{q}\rightarrow 0$ for all frequencies. 
Furthermore, from Fig.~\ref{Fig:sigmaxyVsOmegaAndQx}(c) we see that for small $|\mathbf{q}|$, deviations between the conserved and nonconserved predictions for the anomalous Hall conductivity vanish faster than $|\mathbf{q}|^2$. 
We see that for small $\mathbf{q}$, the anomalous Hall conductivity behaves as
\begin{equation}
    \sigma^{xy}_{\text{anom}}(q_x\hat{x}) \sim \sigma^{xy}_{\text{anom}}(0) - |\alpha| q_x^2.
\end{equation}
This is consistent with the results of Ref.~\cite{kozii2021intrinsic} who found that to quadratic order, the non-conserved midpoint approximation Eq.~\eqref{eq:midpointHallFormula} predicts a negative $\mathcal{O}(q^2)$ correction to the anomalous Hall conductivity arising from band-geometric effects (recall that to quadratic order the midpoint current and the conserved current yield the same Hall conductivity).

Focusing now on the physically-meaningful conserved current, we show in Figs.~\ref{Fig:sigmaxyVsOmegaAndQx}(a) and (b) the real and imaginary part of the Hall conductivity as a function of frequency for five different wavevectors in the $\hat{x}$ direction. 
Strikingly, we see that the Hall conductivity at $\mathbf{q}=(\pi,0,0)$ is identically zero, i.e. $\sigma^{xy}_H(\omega, \pi\hat{x})=0$.
This is true for computations based on both the conserved and non-conserved currents, and arises due to the simplicity of the model, and can be most clearly understood in terms of the trapezoid method: since $v^x_\mathbf{k}$ depends only on $k_x$ and includes only nearest-neighbor hopping, we have $\tilde{F}_{xy}(\mathbf{k},\pi\hat{x})=0$ identically. 
In a more complicated model with longer-range hopping, we would expect $\sigma^{xy}_H(\omega, \pi\hat{x})\neq 0$ generically.
Examining Figs.~\ref{Fig:sigmaxyVsOmegaAndQx}(a-b), we see that as the magnitude of $\mathbf{q}=q\hat{x}$ is increased, the magnitude of the Hall conductivity decreases. 
We also see peaks in the Hall response at frequencies corresponding to indirect (nonzero momentum transfer) transitions between occupied and unoccupied states induced by the external electromagnetic field.

\subsubsection{Hall Response when $\mathbf{q} = (0, 0, q_0)$}
Now we take the orientation of the incoming field to be in the $\hat{z}$ direction, which is parallel to the Weyl node separation.
Unlike when $\mathbf{q}$ was in the $\hat{x}$ orientation in the Sec. \ref{sec:HallResponseAt000}, the three different definitions of the current operators yield identical predictions for the Hall conductivity in this case.
This is because for our toy model Eq.~\eqref{eq:weylHam}, only the matrix elements Eqs. \eqref{eq:currentcurrent}, \eqref{eq:midpointHallFormula}, and \eqref{eq:tildeHallFormula} change in the Kubo formula for the conductivity.
Since our model has no diagonal hopping, and since the Hall conductivity depends only on the $x$ and $y$ components of the current, we have that $F^{ab}_{xy}(\mathbf{k}, \mathbf{q}) = F^{ab}_{\text{mid}; xy}(\mathbf{k}, \mathbf{q}) = \Tilde{F}^{ab}_{xy}(\mathbf{k}, \mathbf{q}) = F^{ab}_{xy}(\mathbf{k},0) $.
To change this, we could include diagonal hoppings of the form $\sin(k_i) \rightarrow \sin(k_i) f(k_z)$ or $\cos(k_i) \rightarrow \cos(k_i) f(k_z)$ for $i \in \{ x, y \}$ and $f(k_z)$ being a general periodic function.
Mixing the $k_x$ or $k_y$ dependence with $k_z$ gives rise to the possibility of a Hall conductance where each of the three mentioned methods would then yield different results.

\begin{figure*}[t]
      \centering
\begin{minipage}{0.98\hsize}
\centering
\includegraphics[width=0.98\hsize]{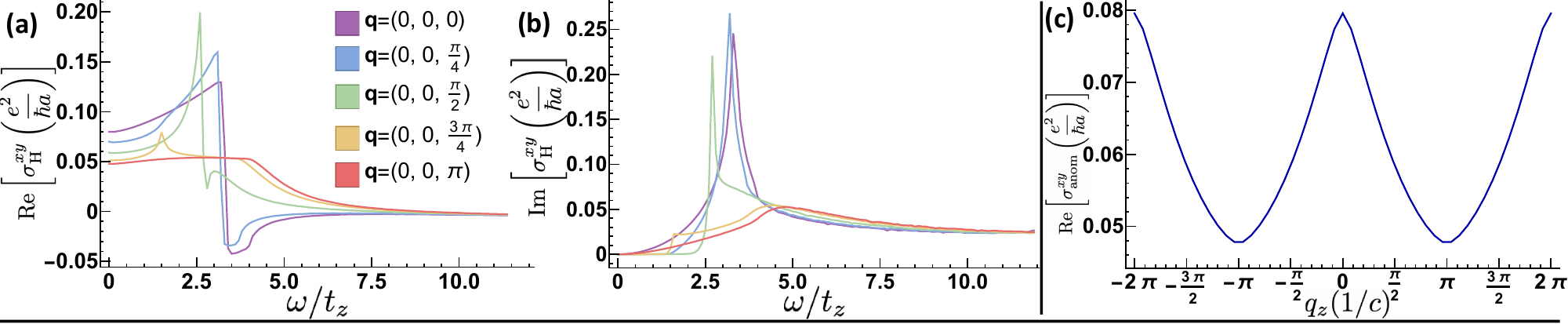}
\end{minipage}
\caption{The real [(a)] and imaginary [(b)] parts of the Hall conductivity as a function of frequency for several values of the wavevector in the $\hat{z}$ direction.
(c) shows the real part of the anomalous Hall conductivity [as defined in Eq.~\eqref{eq:anomalousDefinition}] as a function of the wavevector changing in the $\hat{z}$ direction.
The parameters $m=4 t_z$, $t_x = t_y = t_z $, $\gamma=0$, $\gamma_z=0$, and $\mu = 0$ were used for these calculations.}
\label{Fig:sigmaxyVsOmegaAndQz}
\end{figure*}

In Figs.~~\ref{Fig:sigmaxyVsOmegaAndQz}(a-b) we show the real and imaginary part of the Hall conductivity $\sigma^{xy}_H(\mathbf{q}= q_z \hat{z}, \omega)$ for five values of $q_z$. 
We point out two relevant wavevector values: $q_z  = \pi/2$ (the momentum transfer of the transition from one Weyl node to the highest energy eigenstate) and $q_z = \pi$ (momentum transfer of the transition transition between Weyl nodes at zero frequency).
The $q_z  = \pi/2$ transition is shown in Fig.~\ref{Fig:weylSpectrumPlot} with the green dotted arrow and the $q_z = \pi$ is illustrated in that same figure with a yellow dotted line.
At $q_z  = \pi/2$, we would expect a large Hall response, since the density of states at the band maximum is large.
To examine the Hall effect in this orientation, we inspect Fig.~\ref{Fig:sigmaxyVsOmegaAndQz}(a-b) which plots the Hall coefficient as a function of frequency for several wavevectors.
At the value $q_z  = \pi/2$, we notice the real part of the Hall effect achieves its maximal value at $\omega\sim 2.63 t_z$.

The other interesting value of $q_z$ is at $\pi$.
At $q_z=\pi$ a zero-frequency electric field can excite transitions between the two Weyl nodes.
This is, an electron can directly hop from one Weyl node to the other without having to absorb energy.
This type of excitation at $\mathbf{q} = \pi \hat{z}$ puts the electron in strongest differential of Berry curvature possible, hopping from one topologically charged Fermi point to the Fermi point with opposite charge.
We find that at low frequencies, this leads to a Hall conductivity $\sigma^{xy}_H(\mathbf{q}= \pi \hat{z}, \omega)$ with real part that is nearly constant over a wide frequency range, before decaying to zero at high frequencies. 
We see this in Fig.~\ref{Fig:sigmaxyVsOmegaAndQz}(a).
In Fig.~\ref{Fig:sigmaxyVsOmegaAndQz}(b), see that the imaginary part of  $\sigma^{xy}_H(\mathbf{q}= \pi \hat{z}, \omega)$ the conductivity that is nearly linear at lower frequencies and then decays to zero at high frequencies.

We next consider the anomalous Hall conductivity $\sigma^{xy}_{\text{anom}}(\mathbf{q} = q_z \hat{z})$ as a function of the wavevector.
We plot the anomalous Hall conductivity as a function of the $\mathbf{q} = q_z \hat{z}$, in Fig.~\ref{Fig:sigmaxyVsOmegaAndQz}(c). 
We see that the anomalous Hall conductivity for this model is periodic as a function of $q_z$ due to the absence of diagonal hopping. 
We also see that the Hall conductivity decreases quadratically at small $q_z$, consistent with previous investigations into $\mathcal{O}(\mathbf{q}^2)$ corrections to the Hall conductivity~\cite{hoyos2012hall,bradlyn2012kubo,hoyos2014hall}.

Lastly, to study the effect of a finite Fermi surface on the Hall conductivity, we study $\sigma^{xy}_H(\omega,\mathbf{q}=q_z\hat{z})$ for our Weyl semimetal as a function of chemical potential. 
When the Fermi surface volume is finite, we expect a diverse range of intra- and inter-Fermi surface indirect transitions to contribute to the $\mathbf{q}$-dependence of the Hall conductivity.

First, we focus on the anomalous Hall conductivity as a function of wavevector for several different values of the chemical potential.
For $|\mu|<|2t_z|$, we see from Fig~\ref{Fig:weylSpectrumPlot} that the Fermi surface will consist of two disjoint pockets centered on each Weyl node. 
At $|\mu|=|2t_z|$ there is a Lifshitz transition, where the two Fermi pockets meet; for $|\mu|>|2t_z|$ there is a single Fermi pocket. 
Since the density of states and topological charge of the Fermi surface changes drastically at the Lifshitz transition, we expect to see a drastic change in the Hall conductivity near $\mu=2t_z$ for all $\mathbf{q}$. 

\begin{figure*}[t]
      \centering
\begin{minipage}{0.98\hsize}
\centering
\includegraphics[width=0.98\hsize]{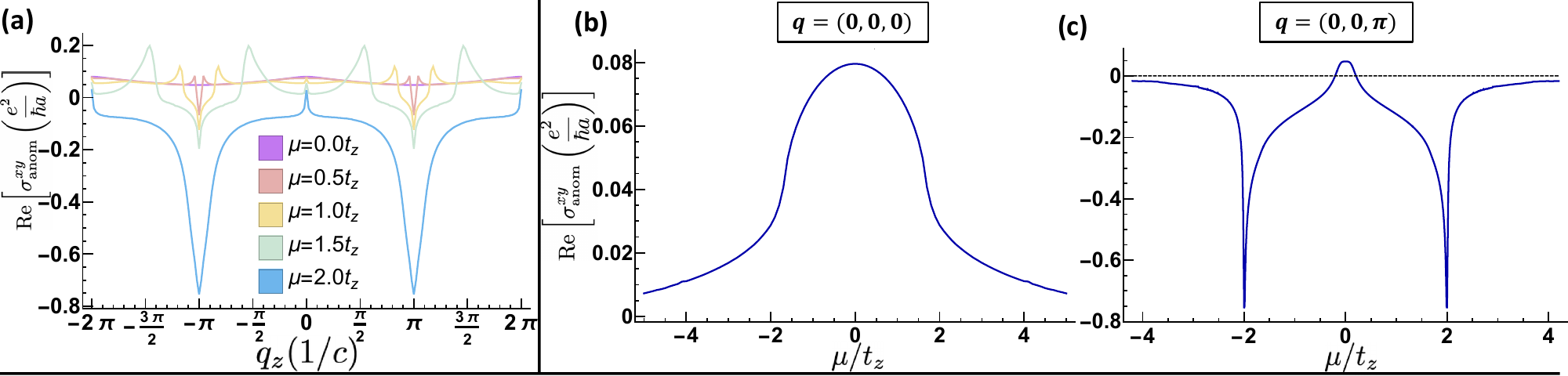}
\end{minipage}
\caption{Chemical potential dependence of the anomalous Hall conductivity [as defined in Eq.~\eqref{eq:anomalousDefinition}] for the Weyl semimetal model. (a) shows the real part of the anomalous Hall response as a function of the wavevector in the $\hat{z}$ direction, for several different values of $\mu$ up to the Lifshitz transition at $\mu = 2 t_z$.
(b) shows the real part of the anomalous Hall response as a function of chemical potential at $q_z = 0$, and (c) shows the real part of the anomalous Hall response as a function of chemical potential at $q_z = \pi$. 
The parameters $m=4 t_z$, $t_x = t_y = t_z $, $\gamma=0$, and $\gamma_z=0$ were used in these calculations.}
\label{Fig:sigmaxyForDifferentMu}
\end{figure*}

We show this in Fig.~\ref{Fig:sigmaxyForDifferentMu}(a), where we plot the anomalous Hall conductivity $\sigma^{xy}_{\text{anom}}(\mathbf{q}=q_z\hat{z})$ as a function of $q_z$ for five different values of the chemical potential. 
We see that when $\mu=2t_z$,
the anomalous Hall conductivity is positive at $q_z=0$ but decreases dramatically as a function of $q_z$. 
For larger $q_z$ the anomalous Hall conductivity reverses sign, and achieves a large negative value at $q_z=\pi$; the Hall response at $q_z=\pi$ and $\mu=2t_z$ is almost an order of magnitude larger than the topological anomalous Hall response $\sigma^{xy}_{\text{anom}}(\mathbf{q}=0)$.

To investigate this further, we compute the anomalous Hall conductivity $\sigma^{xy}_{\text{anom}}(\mathbf{q}=q_z\hat{z})$ at $\omega=0$ and at fixed $q_z$ as a function of $\mu$, shown in Fig.~\ref{Fig:sigmaxyForDifferentMu}(b) for $q_z=0$, and we see that $\sigma^{xy}_{\text{anom}}(q_z = 0)$ decreases quadratically as $\mu \rightarrow 0$, consistent with previous results~\cite{sharma2016topological, messica2022anomalousWSM, kanagaraj2022topologicalWSM, Francesco2022QuadradicConductanceVsChemPot}.
Importantly, the anomalous Hall response at $q_z=0$ is proportional to the Berry curvature at the Fermi surface leading to this quadratic dependence on the chemical potential~\cite{sharma2016topological, messica2022anomalousWSM, kanagaraj2022topologicalWSM, Francesco2022QuadradicConductanceVsChemPot}.
Furthermore, we see in Fig.~\ref{Fig:sigmaxyForDifferentMu}(b) that $\sigma^{xy}_{\text{anom}}(\mathbf{q}=0)$ is always positive for any value of $\mu$.
Therefore, since we know from previous literature~\cite{mccormick2017tiltedweyl,Yan2020TRBInCurvatureMetricAndMotion} that this contribution is proportional to the Berry curvature, we can establish both the positivity and quadratic behavior as indicators of a Berry curvature-dominated influence at the Fermi surface.

However, as noted previously, $\sigma^{xy}_{\text{anom}}(\mathbf{q}=\pi\hat{z})$ demonstrates very different behavior as a function of $\mu$, as shown in Fig.~\ref{Fig:sigmaxyForDifferentMu}(c).
We see that the anomalous response is positive for $\mu = 0$, and then, as $|\mu|$ grows, eventually switches sign and achieves a large negative value at the Lifshitz transition. 
The behavior of $\sigma^{xy}_{\text{anom}}(\mathbf{q}=\pi\hat{z})$ near the $\mu=2t_z$ arises due to the divergence of the joint density of states (essentially the velocity-independent universal factor in our Kubo formula Eq.~\eqref{eq:currentcurrent}) at the Lifshitz transition.

\subsection{Kerr Rotation in Moir\'e Materials}\label{sec:KerrRotation}

While the wavevector dependence of the optical conductivity can usually be ignored in most solids, the large length scales present in moir\'{e} lattice systems mean that spatial inhomogeneities in optical electromagnetic fields may have an appreciable effect on transport. 
In this section, we will focus on applying our formalism to compute the wavevector dependence of the Kerr effect in 2D systems. 
Recall that the Kerr effect describes the change in relative angle and ellipticity from the light reflected from a material~\cite{Osterloh1994MOKE, ahn2022KerrEffectAxions}. and is a direct probe of time-reversal symmetry breaking~\cite{armitage2014constraints}. 
As such, the polar Kerr effect has been used as a direct probe of time-reversal symmetry breaking in unconventional superconductors and charge-ordered systems~\cite{xia2006high,xia2007polar,schemm2014observation,kapitulnik2009polar,saykin2023high,gong2017time}. 
Additionally, reflectivity studies on 2D materials allow us to probe the frequency- and wavevector- dependence of response functions off the light cone, since for a fixed frequency, the magnitude of the in-plane wavevector can be varied by changing the incidence angle.  
Due to the intrinsic time-reversal symmetry breaking of states in twisted graphene and TMD materials, a study of the Kerr effect in these systems is a natural avenue for future experiments~\cite{Liu2020AHEInBilayerGraphene}. 

To this end, we will compute the magnitude of the Kerr effect in a model of a moir\'{e}-Chern insulator. 
First, in Sec.~\ref{sec:haldanemodel} we introduce a toy model for a Chern insulator in a moir\'{e} system. 
Then in Sec.~\ref{sec:kerrresults} we compute the Kerr angle and ellipticity for the model, using Eq.~\eqref{eq:fullGeneralizedLinearConductivity}. 
Unlike in our analysis of the Hall effect in Sec.~\ref{sec:weylhall}, here both the paramagnetic and diamagnetic conductivities will play a role. 
We will compare our results with approximate calculations using the non-conserved trapezoid [Eq.~\eqref{eq:jtildedef}] and midpoint [Eq.~\eqref{eq:jmidpointdef}] definitions of the current prevalent in the literature, showing that they yield quantitatively distinct predictions for the Kerr angle and ellipticity that could be distinguished in experiment.

\subsubsection{Moir\'{e} Haldane Chern Insulator Model}\label{sec:haldanemodel}

For our toy model of a moir\'e Chern insulator, we start with the spinless Haldane model on a honeycomb lattice~\cite{haldane1988HaldaneModel, Manna2020StrainHaldaneModel, Filusch2023FlatHaldaneModel, wright2013BuckledHaldaneModel}.
The Bravais lattice vectors connecting next-nearest-neighbor honeycomb lattice sites (in the same sublattice) are, in Cartesian coordinates,
\begin{align}
    \mathbf{b}_1 =& \frac{a}{2}\begin{bmatrix}
        -\sqrt{3} \\
        3
    \end{bmatrix},
    \\
     \mathbf{b}_2 =& \frac{a}{2}\begin{bmatrix}
        -\sqrt{3} \\
        -3
    \end{bmatrix},
    \\
     \mathbf{b}_3 =& a\begin{bmatrix}
        \sqrt{3} \\
        0
    \end{bmatrix},
\end{align}
where $a$ is the moir\'{e} lattice constant.
The nearest neighbor vectors connecting honeycomb lattice sites in opposite sublattices can similarly be written as
\begin{align}
    \mathbf{a}_1 =& \frac{a}{2}\begin{bmatrix}
        \sqrt{3} \\
        1
    \end{bmatrix},
    \\
     \mathbf{a}_2 =& \frac{a}{2}\begin{bmatrix}
        -\sqrt{3} \\
        1
    \end{bmatrix},
    \\
     \mathbf{a}_3 =& a\begin{bmatrix}
        0 \\
        -1
    \end{bmatrix}.
\end{align}
We construct a tight-binding model with nearest and next-nearest-neighbor  hoppings, as well as a staggered on-site potential and an orbital magnetic flux. 
We can write the Hamiltonian as $H = \sum^3_{i = 0} h_i(\mathbf{k}) \sigma_i$, with
\begin{align}
    h_0(\mathbf{k}) =& 2 t_2 \sum^3_{j=1} \cos(\mathbf{k} \cdot  \mathbf{b}_j) - \mu + 3 t_2 \cos(\phi)\label{eq:modifiedHaldaneTerm},
    \\
    h_1(\mathbf{k}) = & t \sum^3_{j=1}\cos(\mathbf{k} \cdot \mathbf{a}_j),
    \\
    h_2(\mathbf{k}) = & t \sum^3_{j=1}\sin(\mathbf{k} \cdot \mathbf{a}_j),
    \\
    h_3(\mathbf{k}) = & M - 2 t_2 \sin(\phi) \sum^3_{j=1}\sin(\mathbf{k} \cdot \mathbf{b}_j)\label{eq:lastHaldaneTerm}.
\end{align}
Here $M$ is the staggered on-site potential, $t$ is the nearest-neighbor hopping amplitude, and $t_2$ is the next-nearest neighbor hopping amplitude.
The parameter $\mu$ is the chemical potential.
Also, the last term in Eq. \eqref{eq:modifiedHaldaneTerm} is modified from Haldane's original derivation by the extra term, $3 t_2 \cos(\phi)$.
We include this extra term for the convenience of putting the zero of energy in the band gap.
Finally, $\phi$ is the time-reversal symmetry breaking magnetic flux per plaquette. 

\begin{figure}[t]
      \centering
\begin{minipage}{0.95\hsize}
\centering
\includegraphics[width=0.95\hsize]{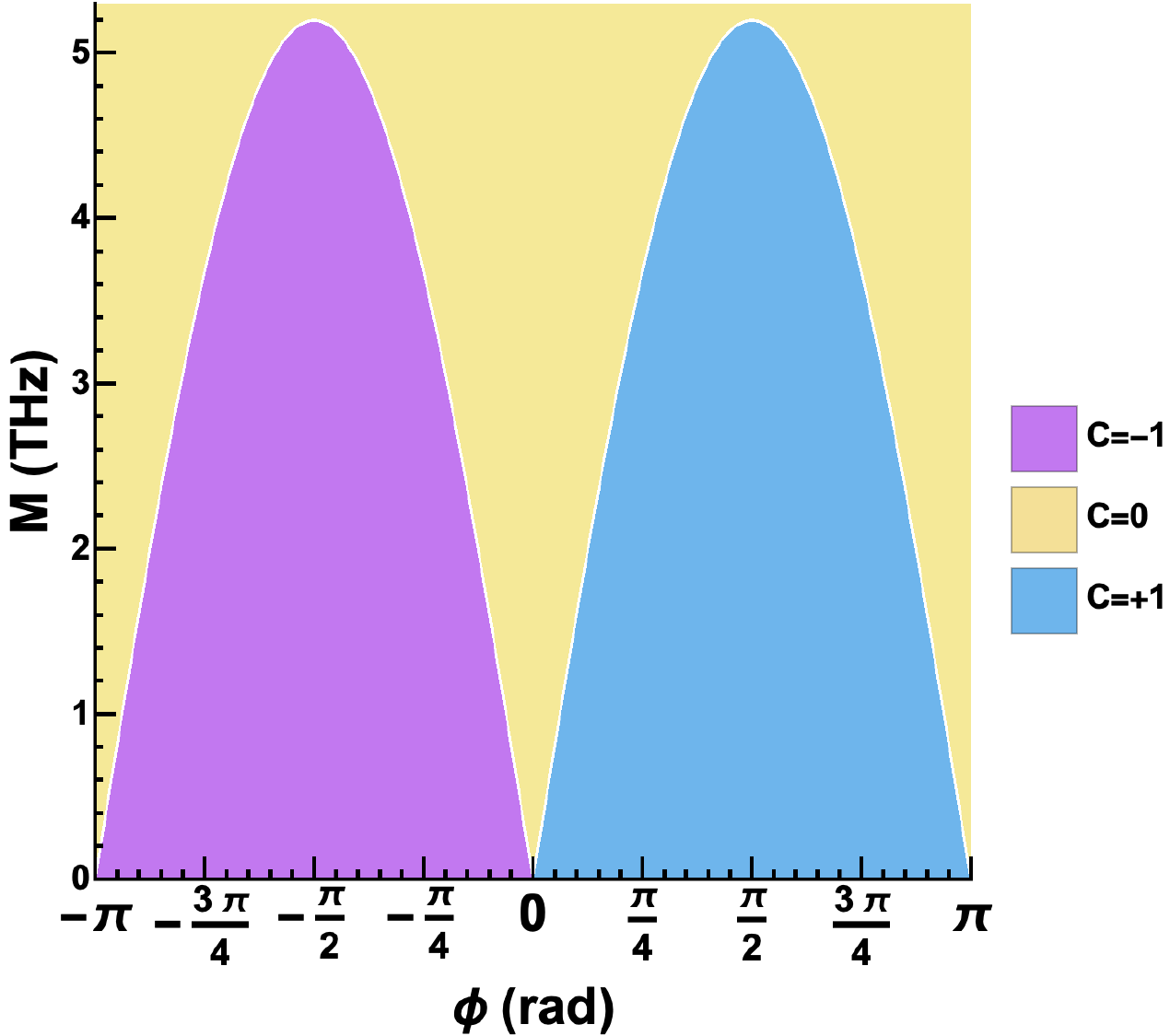}
\includegraphics[width=0.95\hsize]{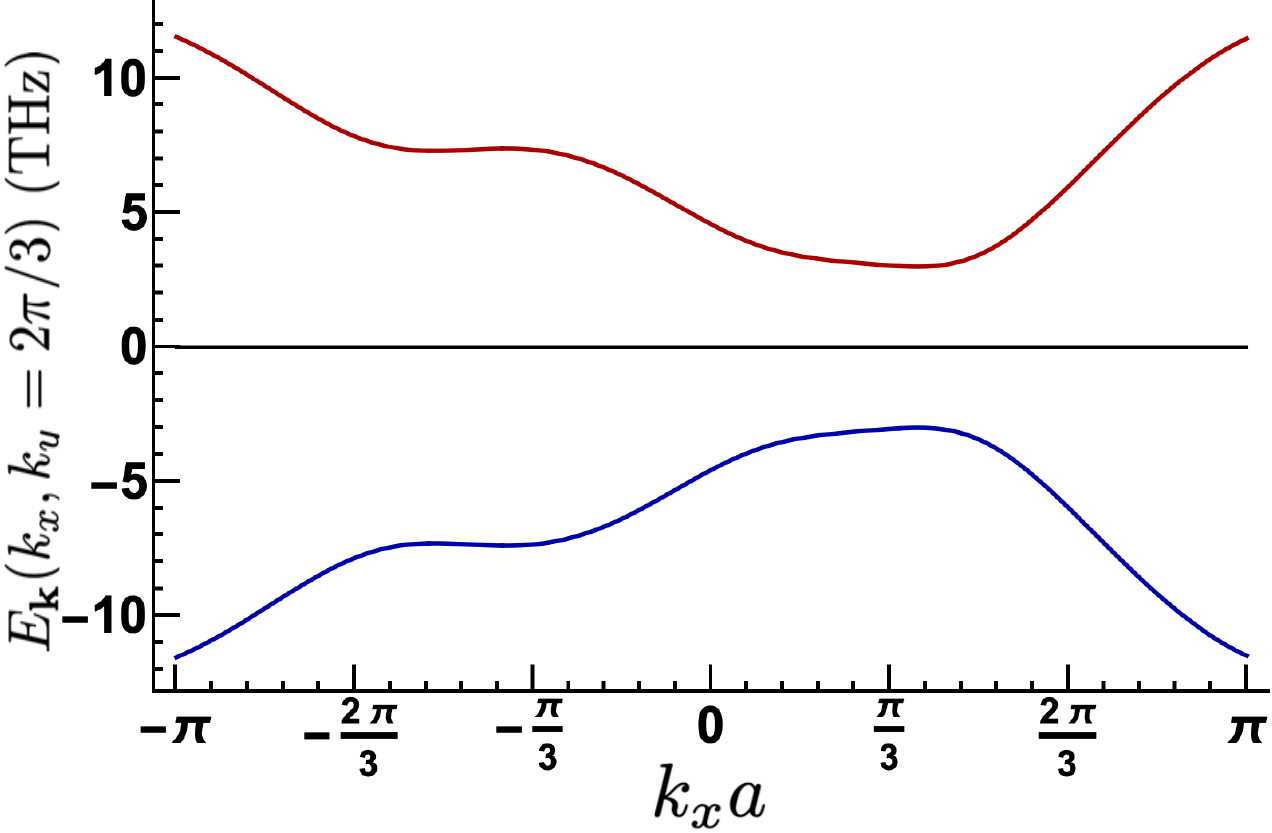}
\end{minipage}
\caption{
The top is the topological phase diagram of the Haldane Chern insulator as described by Eqs.~\eqref{eq:modifiedHaldaneTerm}--\eqref{eq:lastHaldaneTerm}.
We set $t=4 t_2$ and $t_2 = 1 \text{ THz}$ to adhere to the non-overlap condition of $|t_2/t| < 1/3$.
The Chern number, as illustrated in this figure, is calculated by integrating the Berry curvature across the Brillouin zone.
The bottom is the spectrum for the Haldane Chern insulator, capturing both $K$ and $K^{\prime}$ points when $k_y = 2 \pi/3$, which is at $k_x = \pm 2 \pi / 3 \sqrt{3}$.
At these points, the separation in energies follows as $2 | M \pm 3 \sqrt{3} t_2 \sin(\phi)|$, which take on separation values of $6 \text{ THz}$ and $14.8 \text{ THz}$.
}
\label{Fig:HaldaneChernInsulatorPhaseDiagram}
\end{figure}

In this model, we will take $|t_2 / t| < 1/3$, which forces the bands to not overlap.
Additionally, to be in a topologically nontrivial state with Chern number $C = \pm 1$, our choice of parameters must satisfy $|M/ t_2| < 3 \sqrt{3} | \sin(\phi) |$.
The phase diagram for this model is shown in Fig.~\ref{Fig:HaldaneChernInsulatorPhaseDiagram}.

\begin{figure}[b]
      \centering
\begin{minipage}{0.95\hsize}
\centering
\includegraphics[width=0.95\hsize]{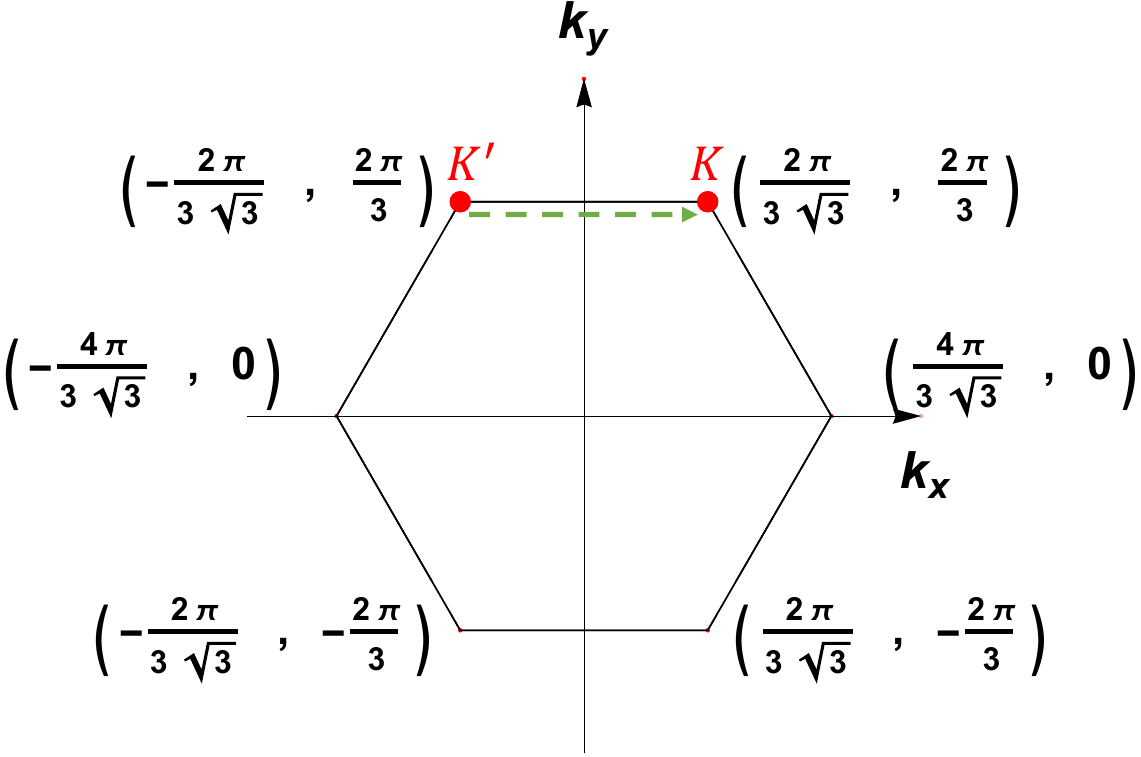}
\end{minipage}
\caption{Brillouin zone for the Chern insulator model in Eqs.~\eqref{eq:modifiedHaldaneTerm}--\eqref{eq:lastHaldaneTerm} measured in natural units of inverse lattice constants.
The dotted green line denotes the momentum transfer between the $K^{\prime}$ and $K$ points.
An excitation between these two points can be excited by adjusting the wavevector of the incoming electromagnetic field.}
\label{Fig:HaldaneLatticeDiagram}
\end{figure}

Furthermore, we note this model contains two gapped Dirac points at points $K$ and $K^{\prime}$, with coordinates $\frac{2 \pi}{3} (\frac{1}{\sqrt{3}}, 1)$ and $\frac{2 \pi}{3} (-\frac{1}{\sqrt{3}}, 1)$ in units of the inverse moire lattice constant, respectively.
The $K$ and $K^{\prime}$ points are shown in the Brillouin zone in Fig.~\ref{Fig:HaldaneLatticeDiagram}.
By adjusting $M$, $t_2$, and $\phi$, the Dirac nodes can be gapped out.
A gap closes at a single Dirac node at each phase boundary as illustrated in Fig.~\ref{Fig:HaldaneChernInsulatorPhaseDiagram}.
At the $K$ and $K^{\prime}$ points, the band gaps are $2 |M -3 \sqrt{3} t_2 \sin(\phi) |$ and $2 |M + 3 \sqrt{3} t_2 \sin(\phi) |$ respectively; we refer to these as the topological gaps. 
We show the band structure of the model for a representative set of parameters $t_2=1$ THz, $t=4$ THz, $M=2.2$ THz, and $\phi = \pi/2$.

\subsubsection{The Momentum-Dependent  Kerr Effect of the Haldane Chern Insulator}\label{sec:kerrresults}

We will now consider the influence of an optical electric field on the moir\'{e} Chern insulator, focusing on the Kerr effect. 
When the moir\'{e} lattice constant is large, the in-plane wavevector $|\mathbf{q}| \sim \omega / c$ may be an appreciable fraction of the inverse lattice spacing $1/a$, at frequencies near the topological gaps. 
To model this, we will choose the moir\'{e} length and energy scales of our model to be comparable to those of twisted bilayer graphene reported in the literature~\cite{Po2019tightbindingModelsForBilayerGraphene, Liu2020AHEInBilayerGraphene, brey2020PlasmonicsInBilayerGraphene, Kang2018WannierStatesInBilayerGraphene, Faulstich2022QuantumChemistryBilayerGraphene}. 
In particular, we take the lattice constant $a=200$\AA, which is achievable at small twist angles in graphene. 
Additionally, we take $t_2=1$THz, $t=4$THz, and $M\approx 4.45$THz, such that the relevant energy scale for optical excitations is on the order of $10$THz. 
Additionally, by differentiating the Hamiltonian $H$ defined in Eqs.~\eqref{eq:modifiedHaldaneTerm}--\eqref{eq:lastHaldaneTerm} with respect to $\mathbf{k}$, we find that the matrix elements of the velocity operator are all on the order of $t a \sim 2.6\times 10^{-4} c$ or smaller, where $c$ is the speed of light; this is comparable to the Fermi velocity in a typical metal.

\begin{figure}[t]
      \centering
\begin{minipage}{0.95\hsize}
\centering
\includegraphics[width=0.95\hsize]{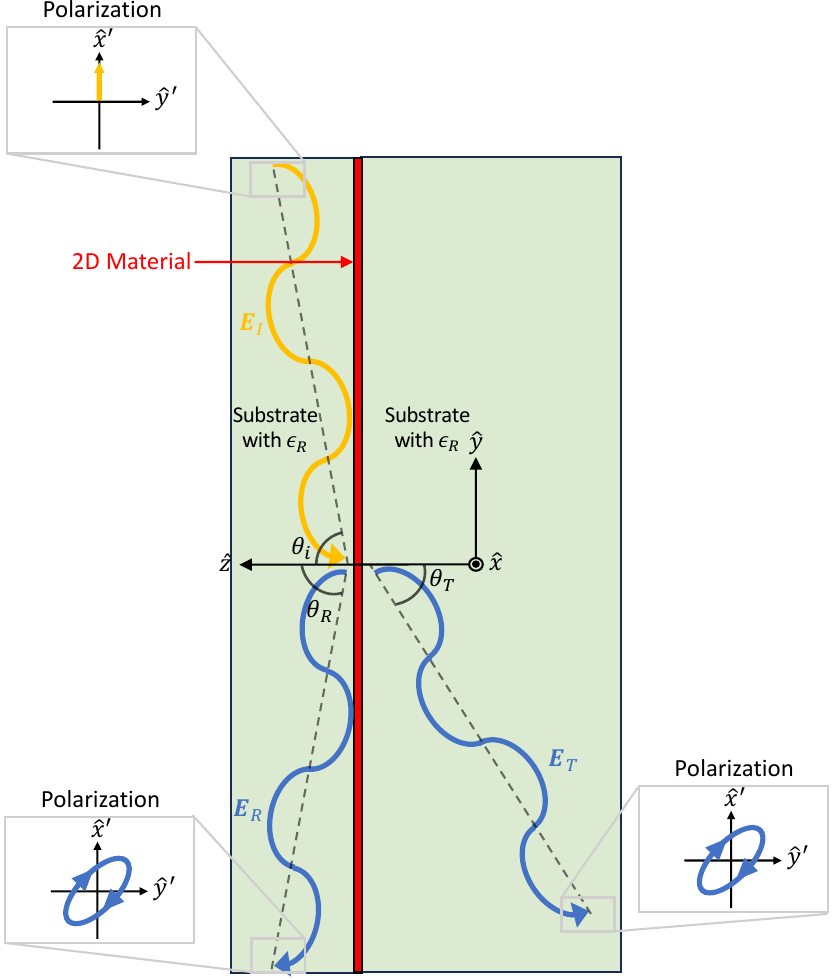}
\end{minipage}
\caption{Diagram of the polarization and configuration of the Kerr effect setup. 
The 2D sample (red) is encapsulated in a high-$\epsilon_R$ dielectric (green). 
The yellow curve is the incident light with incidence angle $\theta_i$ and incident field $\mathbf{E}_I$.
The left blue curve is the reflected wave with reflected angle $\theta_R$ and reflected electric field $\mathbf{E}_R$.
The right blue curve is the transmitted light with transmitted angle $\theta_T$ and transmitted electric field $\mathbf{E}_T$.
Each of the reflected and transmitted polarizations can be described in terms of an ellipticity and angle of rotation that describe the Kerr (reflected) and Faraday (transmitted) effects.
}
\label{Fig:KerrEffectDiagram}
\end{figure}

To compute the Kerr response, we consider the experimental geometry shown in Fig.~\ref{Fig:KerrEffectDiagram}. 
We consider an infinite sample oriented in the $xy$ plane, encapsulated within a dielectric substrate with large index of refraction $n_R=\sqrt{\epsilon_R}=38$~\cite{Choi2011HighRefractiveIndex}. 
We illuminate the system with linearly polarized light at oblique incidence $\theta_i = 7 \pi/16$, $\mathbf{E}_{I} = E_0 \hat{x} e^{-i \omega t + i \mathbf{q} \cdot \mathbf{r}}$. 
Since the in-plane component of the electric field is continuous across the interface, this means that the in-plane component $q_y$ of the wavevector relevant for scattering is, in dimensionless units
\begin{equation}\label{eq:inplaneq}
q_y a = a n_R\sin(\theta_I) \omega/(c) \sim (0.0025\mathrm{THz}^{-1}) \omega
\end{equation}
for our choice of moir\'{e} lattice constant.
This is several orders of magnitude larger than the typical scale $q_y a_0 \sim (10^{-6} \mathrm{THz}^{-1}) \omega$ for crystalline systems with typical lattice constants $a_0\sim2$\AA. 
Taking the incident photon frequency to be on the order of $\sim10 \text{ THz}$, which should be reasonable with recent breakthroughs in terahertz spectroscopy experiments, we see $\mathbf{q}$ in the moir\'{e} system is comparable enough to the lattice spacing to make finite wavevector corrections to optical response large enough to be experimentally measurable; although the effects of nonzero $\mathbf{q}$ are still small, they are experimentally non-negligible in moir\'{e} systems.
Although we use a very large $n_R$ substrate appropriate to metamaterials, we see from Eq.~\eqref{eq:inplaneq} that a lower dielectric constant can be used provided the moir\'{e} lattice constant is similarly increased (e.g., by decreasing the twist angle).
Additionally, the effect of the wavevector correction will become more noticeable as the angle of incidence becomes closer to being oblique.

We can now compute the Kerr angle (rotation of the plane of polarization of the reflected wave) and ellipticity (imaginary part of the Kerr angle) for light reflected off our model of a moir\'{e} Chern insulator. 
Since our sample is two-dimensional, its primary influence on the propagation of light is through the boundary conditions that enter Maxwell's equations. 
In particular, the conductivity tensor $\sigma^{\mu\nu}(\omega,\mathbf{q}=q_y)$ determines the surface current at the sample, which determines the discontinuity in the magnetic field across the sample and thus the reflection coefficient. 
We give a full derivation of the Kerr angle and ellipticity in terms of the conductivity tensor in App.~\ref{app:KerrDerivations}.
We can thus use our Kubo formula Eq.~\eqref{eq:fullGeneralizedLinearConductivity} in terms of the conserved current operator in Eq.~\eqref{eq:TBCurrentOperator} to compute the Kerr angle and ellipticity.
Since the derivation for the Kerr angle and ellipticity involves both the longitudinal and Hall components of the conductivity tensor, we need to include both the diamagnetic and paramagnetic parts of the conductivity, as presented in Section \ref{sec:FullLinearResponse}.

\begin{figure*}[t]
      \centering
\begin{minipage}{0.98\hsize}
\centering
\includegraphics[width=0.98\hsize]{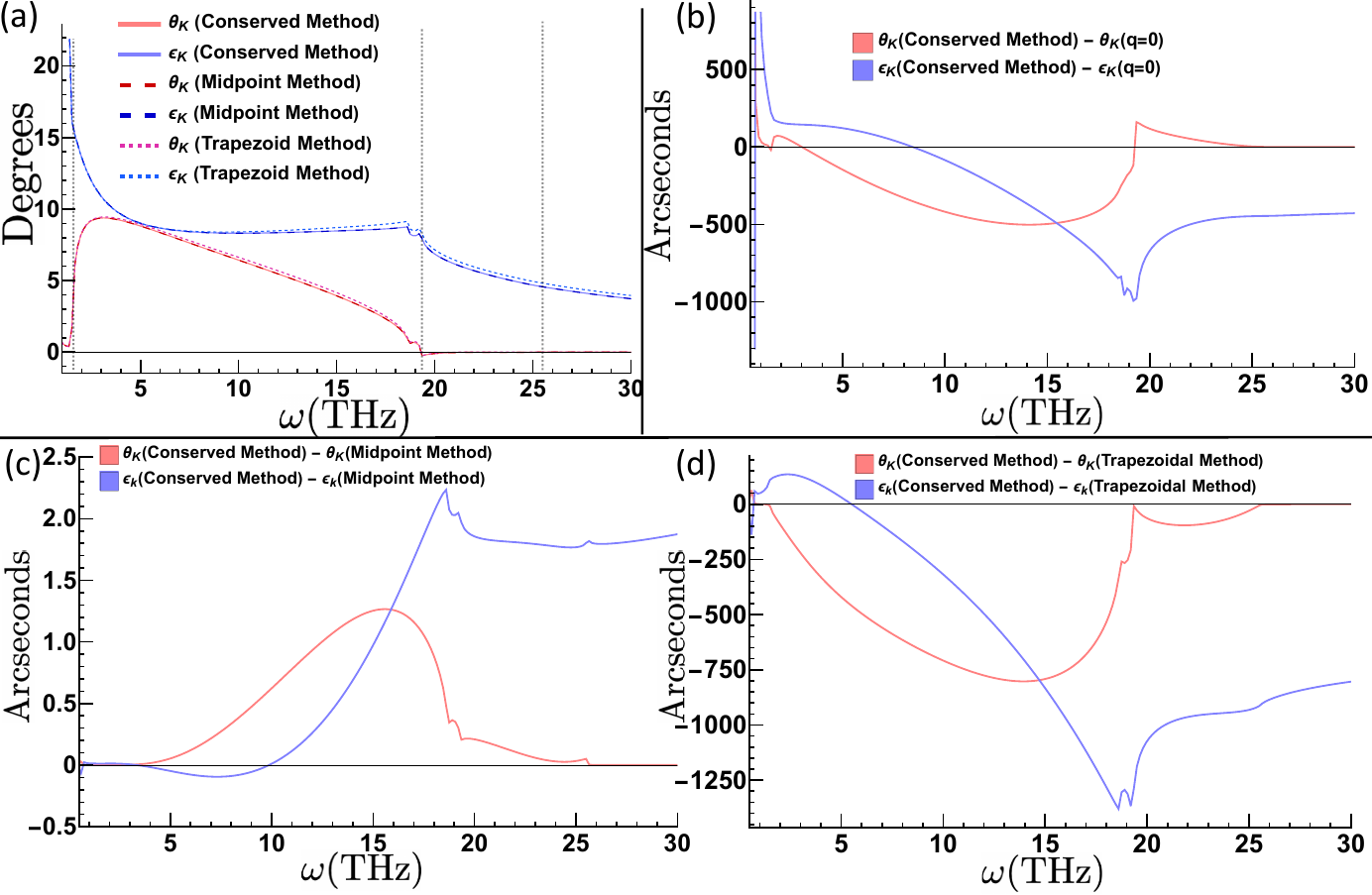}
\end{minipage}
\caption{Kerr effect in the moir\'{e} Chern insulator model. (a) shows the Kerr angle $\theta_K$ and ellipticity $\epsilon_K$ as a function of frequency computed using the conserved current, the non-conserved midpoint current, and the nonconserved trapezoid current.
(b) shows the difference between the angle and ellipticity computed using conserved current compared to the uniform approximation $\sigma_{\mu\nu}(\mathbf{q},\omega) \approx \sigma_{\mu\nu}(\mathbf{0},\omega)$, effectively showing the importance of $\mathbf{q}$-dependent modifications to the optical response.
In (c) we show the differences between the conserved and midpoint current predictions for the Kerr angle and ellipticity. 
Similarly, in (d) we show the differences between the conserved and trapezoid current predictions for the Kerr angle and ellipticity. 
These figures used model parameters of $M = (3 \sqrt{3} -3/4)t _2$, $t = 4 t_2$, $t_2=1$ THz and $\phi = \pi/2$.
Vertical dashed lines denote the two topological gaps and the highest allowed transition frequency.
}
\label{Fig:HaldaneModelKerrEffect}
\end{figure*}

In Fig.~\ref{Fig:HaldaneModelKerrEffect}(a) we show the Kerr angle, $\theta_K$, and Kerr ellipticity, $\epsilon_K$, as functions of frequency, using the Kubo formula in Eq.~\eqref{eq:fullGeneralizedLinearConductivity}. 
For comparison, we also include the analogous calculation using the non-conserved trapezoid and midpoint definitions of the current operator. 
As noted at the end of Sec.~\ref{sec:jnew}, to use the non-conserved currents we introduced an approximate diamagnetic current in order to compute the corresponding $\mathbf{q}$-dependent diamagnetic conductivity.
For the parameter values considered here, the topological gaps in the band structure occur at $\omega = 1.5 \text{ THz}$ and another at $\omega \approx 19.28 \text{ THz}$, which are indicated by vertical dotted lines in Fig.~\ref{Fig:HaldaneModelKerrEffect}(a). 
As expected, we see that the Kerr angle rises rapidly (and the ellipticity falls rapidly) near the smaller topological gap, and the Kerr angle nearly vanishes above the larger topological gap. 
Note also that while the ellipticity and Kerr angles also grow at very low frequencies ($\omega < 0.5$THz), the intensity of the reflected light goes rapidly to zero at low frequencies. 
We also indicate with a vertical dotted line the frequency $\omega=25.60 \text{ THz}$, which corresponds to the highest possible excitation energy in this system.

To highlight the non-negligiblity of including wavevector-dependent corrections to the optical response in moir\'{e} systems, in Fig.~\ref{Fig:HaldaneModelKerrEffect}(b) we show the difference between the $\mathbf{q}$-dependent conserved current calculation of the Kerr angle and ellipticity and the Kerr angle and ellipticity computed using the common $\mathbf{q}\rightarrow 0$ uniform approximation to the conductivity tensor. 
This shows the calculational error arising from neglecting the wavevector dependence in the Kerr effect.
We see that the difference varies on the order of $10^3$ arcseconds, which shows that including spatial inhomogeneity when modeling light scattering off the material surface has quantitative effects that cannot be ignored in order to accurately describe the Kerr effect in moir\'{e} systems.
Moreover, this magnitude of difference is detectable using existing experimental techniques.
Thus, this further demonstrates that optical response in moir\'{e} materials is sensitive to spatial inhomogeneities in electromagnetic fields.

At the scale of Fig.~\ref{Fig:HaldaneModelKerrEffect}(a), it is difficult to distinguish between the predicted value of $\theta_K$ and $\epsilon_K$ from the conserved and non-conserved currents.
According to our analysis of the dimensionless in-plane wavevector in Eq.~\eqref{eq:inplaneq}, we expect the difference in prediction between the three definitions of the current operator to be small but measurable.
To demonstrate this, in Fig.~\ref{Fig:HaldaneModelKerrEffect}(c) and (d) we show the difference between $\theta_K$ and $\epsilon_K$ calculated with the conserved current and the midpoint [(c)] and trapezoid [(d)] approximations to the current. 
We see that the differences are on the order of $1$ arcsecond for the midpoint current, to as large as almost $1250$ arcseconds (or about $20$ arcminutes) for the trapezoid current, which are both within reach of experimental detection.
Thus, we expect that the conserved current in Eq.~\eqref{eq:TBCurrentOperator} will provide a better fit to Kerr effect experiments, especially for terahertz frequencies in moir\'{e} lattices.

Next, we turn our attention to the dependence of the Kerr angle on the magnetic flux at fixed frequency.
In our tight-binding model, the parameter $\phi$ controls the magnetic flux through each plaquette; varying $\phi$ at fixed $M$ allows us to tune between the topological phases with Chern numbers $-1$, $0$, and $+1$ as indicated in Fig.~\ref{Fig:HaldaneChernInsulatorPhaseDiagram}
We will examine signatures of the topological phase transition on the Kerr angle and ellipticity, restricting our attention only to predictions using the conserved current of Eq.~\eqref{eq:TBCurrentOperator}.

\begin{figure*}[t]
      \centering
\begin{minipage}{0.98\hsize}
\centering
\includegraphics[width=0.98\hsize]{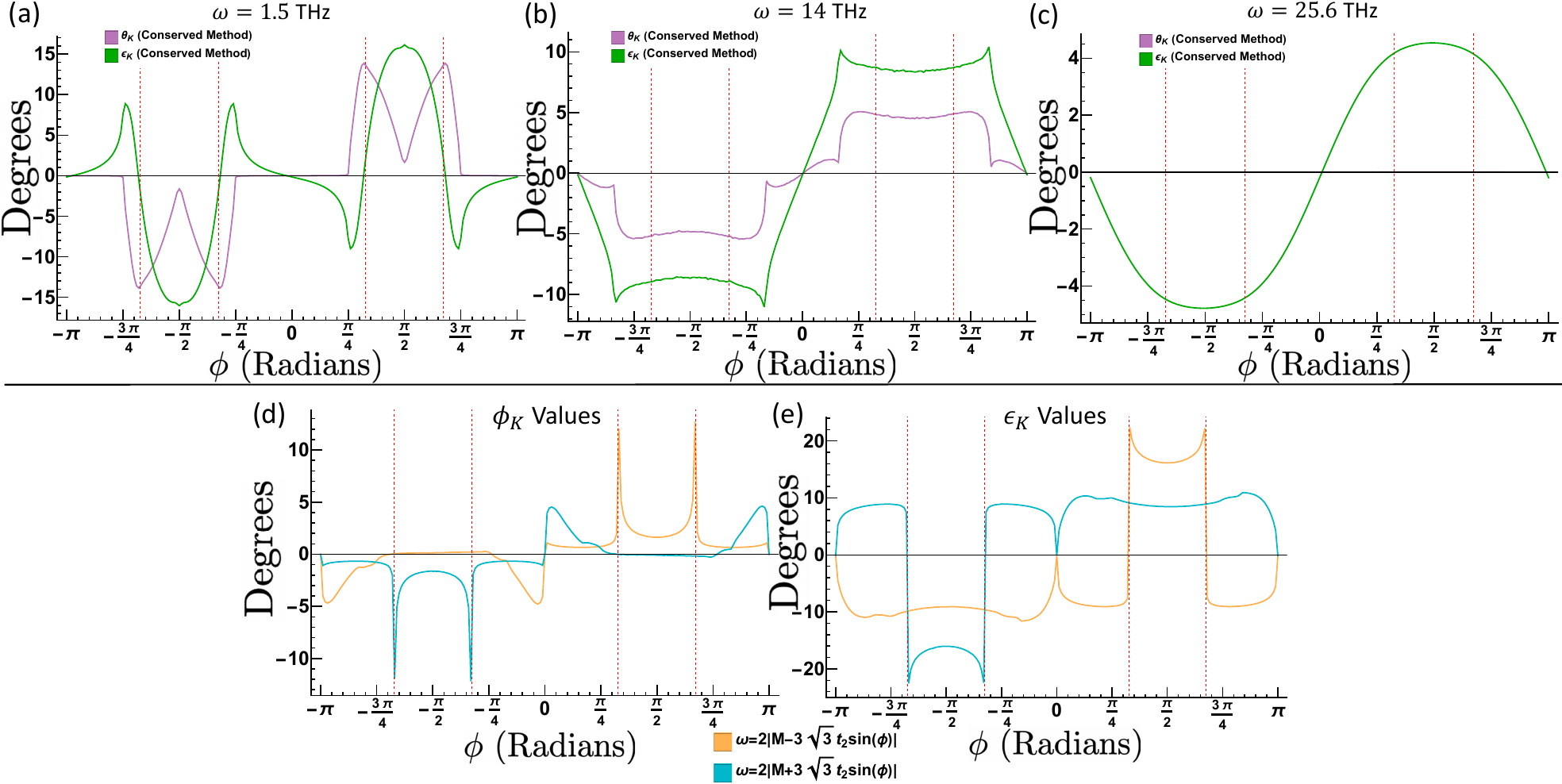}
\end{minipage}
\caption{
Kerr angle $\theta_K$ and ellipticity $\epsilon_K$ as a function of magnetic flux $\phi$ per plaquette in the moir\'{e} Chern insulator model, computed using our conserved current in Eq.~\eqref{eq:TBCurrentOperator}. $\theta_K$ and ellipticity $\epsilon_K$ are shown at fixed frequency $\omega=1.5$\text{ THz} in (a), $14\text{ THz}$ in (b), and $25.6\text{ THz}$ in (c).
(d) and (e) show the Kerr angles [(d)] and Kerr ellipticities [(e)] as a function of flux when $\omega$ is pinned to the $\phi$-dependent topological band gap $\omega(\phi) = 2|M \pm 3 \sqrt{3} t_2 \sin(\phi)|$ at the $K$ (for positive $\phi$) or $K^\prime$ (for negative $\phi$) point in the Brillouin zone.
The dotted red vertical lines indicate values of $\phi$ corresponding to the topological phase boundaries shown in Fig.~\ref{Fig:HaldaneChernInsulatorPhaseDiagram}.
All quantities are computed using $M = (3 \sqrt{3} -3/4) t_2$, $t = 4 t_2$, and $t_2=1$THz.
}
\label{Fig:HaldaneModelKerrEffectVsPhi}
\end{figure*}

In general, we expect the $\theta_K$ and $\epsilon_K$ to both be odd functions of $\phi$, since all of $\theta_K,\epsilon_K$, and $\phi$ are odd under time-reversal.
Additionally, since $\phi=0$ and $\phi=\pi$ are time-reversal invariant values of the magnetic flux per plaquette, we expect the Kerr angle and ellipticity to vanish for these value of $\phi$.
We see this reflected in Fig.~\ref{Fig:HaldaneModelKerrEffectVsPhi} where we show the Kerr angles and Kerr ellipticities as functions of $\phi$ for various values of $\omega$. 
Fig.~ (a-c) show $\theta_K$ and $\epsilon_K$ for $\omega=1.5$THz (the frequency corresponding to the topological gap), $14$THz (an intermediate frequency scale in the model), and $25.6$THz (the highest possible excitation energy in the model), respectively. 
We show with vertical dashed lines the values of $\phi$ corresponding to the topological phase boundaries in Fig.~\ref{Fig:HaldaneChernInsulatorPhaseDiagram}. 
We see that when the incident photon frequency is on resonance with the topological gap [Fig.~\ref{Fig:HaldaneModelKerrEffectVsPhi}(a)], the Kerr angle and ellipticity peak near the value of $\phi$ corresponding to the topological phase transition. 
On the other hand, when the photon frequency is off-resonance [Fig.~\ref{Fig:HaldaneModelKerrEffectVsPhi}(b,c)]  the evolution of the Kerr angle and ellipticity as a function of $\phi$ becomes more smooth; for very large frequencies [Fig.~\ref{Fig:HaldaneModelKerrEffectVsPhi}(c)], we see that the Kerr angle is approximately zero for all $\phi$, and the dependence of the ellipticity angle on $\phi$ resembles a steepened sinusoidal function.

To further explore the sensitivity of the Kerr angle and ellipticity to the topological phase transition in our model, we compute $\theta_K$ and $\epsilon_K$ as functions of $\phi$ at frequency $\omega(\phi) = 2|M \pm 3 \sqrt{3} t_2 \sin(\phi)|$, which is on resonance with the smaller of the two topological band gaps for every $\phi$ (shown in Fig.~\ref{Fig:HaldaneChernInsulatorPhaseDiagram}); we show these results in Figs.~\ref{Fig:HaldaneModelKerrEffectVsPhi}(d) and (e).
In these figures, we examine the case when the frequency is varied in step with the size of the Dirac points' topological gaps at $\omega = 2|M \pm 3 \sqrt{3} t_2 \sin(\phi)|$. 
We see that both the Kerr angle and ellipticity on-resonance is large and sharply peaked at the phase boundary, with maximum values $|\theta_K|\gtrsim 40^\circ$ and $|\epsilon_K|\gtrsim 40^\circ$.

Thus, in order to fully understand and quantitatively model optical experiments being performed on moir\'{e} materials, it is necessary to account for wavevector dependence of the optical conductivity.
In particular, as  moir\'{e} systems become larger and as fabrication techniques allow for smaller twist angles and longer-wavelength superlattice potentials, the need to understand how $\mathbf{q}$ affects the conductivity will become more imperative for explaining the physics.

\subsection{Magnetic Properties of Insulators}\label{sec:MagneticPropertiesOfInsulators}

While we have derived our expression for the conductivity tensor in terms of response to an electric field, gauge invariance and Maxwell's equations imply that response to electric and magnetic fields are inextricably linked. 
In particular, Faraday's law Eq.~\eqref{eq:faraday} implies that a transverse electric field is always accompanied by a magnetic field. 
Rewriting Eq.~\eqref{eq:faraday} in Fourier space, we have that
\begin{equation}\label{eq:faradayq}
B_{\mu,\mathbf{q}}(\omega) = \frac{1}{\omega}\epsilon_{\mu\nu\lambda} q_\nu E_{\lambda,\mathbf{q}}(\omega),
\end{equation}
where $\epsilon_{\mu\nu\lambda}$ is the totally antisymmetric Levi-Civita symbol.
Additionally, the momentum-dependent conductivity tensor $\sigma^{\mu\nu}(\omega, \mathbf{q})$ can be used to derive equilibrium susceptibilities. 
Recall that in the limit $\omega\rightarrow 0$ for generic, nonzero $\mathbf{q}$, our perturbing fields $\mathbf{A}_{\mathbf{q}}(\omega)$ and $A_{0\mathbf{q}}(\omega)$ become time-independent.
The electric and magnetic fields corresponding to these potentials are time-independent, bounded, and oscillate in space with wavevector $\mathbf{q}$.
Note that in this limit, the static electric and magnetic fields interacting with the material are purely dependent on the wavevector, $\mathbf{q}$, and therefore offer a promising experimental setup to probe the physics discussed in this manuscript.
Thus, the Hamiltonian in the presence of a static $\omega\rightarrow 0$ electromagnetic perturbation at nonzero $\mathbf{q}$ has a static ground state. 
It follows that taking $\omega\rightarrow 0$ at fixed nonzero $\mathbf{q}$ leaves the system in an equilibrium state~\cite{luttinger1964theory}.
This means that there can be no charge transport, and so the ground state current is purely a \emph{magnetization current}~\cite{cooper1997thermoelectric}
\begin{equation}\label{eq:magcurrent}
\langle \mathbf{j}_\mathbf{q}\rangle = i\mathbf{q}\times \mathbf{M}_\mathbf{q},
\end{equation}
where $\mathbf{M}_\mathbf{q}$ is the magnetization density (i.e. magnetic dipole moment per unit volume).

Combining Eqs.~\eqref{eq:faradayq} and \eqref{eq:magcurrent}, we will be able to use our formalism for the nonuniform conductivity $\sigma^{\mu\nu}(\omega, \mathbf{q})$ to calculate magnetic properties of insulators. 
First, in Sec.~\ref{sec:magneticSusceptabilityInInsulators}, we will derive a formula for the magnetic susceptibility tensor in insulators. 
Next, in Sec.~\ref{sec:QuadrupoleAndOctopoleMagneticMoments}, we will derive expressions for the magnetic quadrupole moment in finite systems. 
Finally, in Sec.~\ref{sec:streda} we will show that our formalism is consistent with the Streda formula, for which we will provide a new derivation. 
We note that the results of this section are completely general, relying only on gauge invariance and the assumption that any two-particle interactions are independent of momentum; they apply equally well to interacting and noninteracting systems.

\subsubsection{Magnetic Susceptibility in Insulators}\label{sec:magneticSusceptabilityInInsulators}

We will derive an expression for the magnetic susceptibility starting from the defining relation for the linear conductivity,
\begin{equation}\label{eq:conductivitydef}
\langle j^\mu_\mathbf{q}(\omega)\rangle = \sigma^{\mu\nu}(\omega, \mathbf{q})E_{\nu,\mathbf{q}}(\omega).
\end{equation}
While the derivations in this section do not make use of the explicit form of the current operator, they highlight constraints on the conductivity imposed by charge conservation that are satisfied only when the conserved current operator is used. 
We can rewrite Eq.~\eqref{eq:conductivitydef} in the $A_0=0$ gauge (which we have used for our derivation of the diagrammatic response in Sec.~\ref{sec:DiagramsRuleAndSumRules}) as
\begin{equation}
\langle j^\mu_\mathbf{q}(\omega)\rangle =i\omega\sigma^{\mu\nu}(\omega, \mathbf{q})A_{\nu,\mathbf{q}}(\omega).
\end{equation}
Now let us suppose that our perturbing field $A_{\nu,\mathbf{q}}$ is time-independent, which in Fourier space implies that it consists of only an $\omega=0$ component. 
Defining
\begin{equation}
R^{\mu\nu}(\mathbf{q}) = \lim_{\omega\rightarrow 0}i\omega\sigma^{\mu\nu}(\omega, \mathbf{q}),
\end{equation}
we have that the time-independent current response is given by
\begin{equation}\label{eq:rresp}
\langle j^\mu_\mathbf{q}\rangle = R^{\mu\nu}(\mathbf{q})A_{\nu,\mathbf{q}},
\end{equation}
where the current and vector potential are taken at $\omega=0$. 
From Eq.~\eqref{eq:magcurrent}, we know that $R^{\mu\nu}(\mathbf{q})A_{\nu,\mathbf{q}}$ must be expressible as the cross product of $\mathbf{q}$ with a vector. 
This allows us to write
\begin{equation}\label{eq:rdef}
R^{\mu\nu}(\mathbf{q})A_{\nu,\mathbf{q}} = i\epsilon_{\mu\lambda\rho} q_\lambda \alpha^{\rho\nu}(\mathbf{q})A_{\nu,\mathbf{q}} 
\end{equation}
for some tensor $\alpha^{\rho\nu}(\mathbf{q})$. 
Furthermore, gauge invariance restricts the form of $\alpha^{\rho\nu}(\mathbf{q})$; the average current can only depend on $A_{\nu,\mathbf{q}}$ through the magnetic field $B_{\nu,\mathbf{q}} = i\epsilon_{\nu\lambda\rho}q_\lambda A_{\nu,\mathbf{q}}$, and so
\begin{equation}\label{eq:chidef}
\alpha^{\rho\nu}(\mathbf{q}) = i\chi^{\rho\lambda}(\mathbf{q})\epsilon_{\lambda\mu\nu}q_\mu.
\end{equation}
\begin{figure*}[t!]
      \centering
\begin{minipage}{0.55\hsize}
\centering
\includegraphics[width=0.98\hsize]{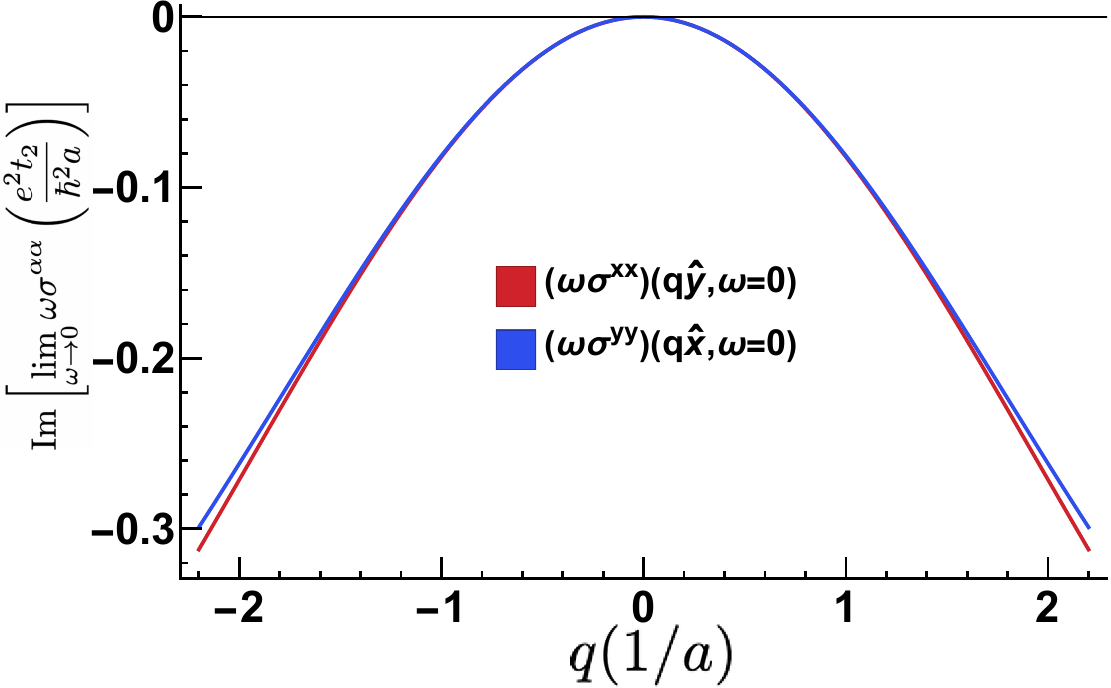}
\end{minipage}
\caption{Plot showing the imaginary part of the diagonal components of the transverse conductivity scaled by the frequency, in the limit of zero frequency for the moir\'{e} Chern insulator model introduced in Eqs.~\eqref{eq:modifiedHaldaneTerm}--\eqref{eq:lastHaldaneTerm}.
The real part vanishes.
Notice the two curves are similarly quadratic at leading order, and higher-order curvatures separate the curves with increasing $\mathbf{q}$.
We used model parameters of $\phi = \pi/2$, $M = (3 \sqrt{3} -3/4) t_2$, and $t = 4 t_2$, which puts the ground state in the topological phase at $C=1$ as shown in Fig.~\ref{Fig:HaldaneChernInsulatorPhaseDiagram}.
}
\label{Fig:LinearConductivityVsq}
\end{figure*}

Combining Eqs.~\eqref{eq:magcurrent}, \eqref{eq:rresp} \eqref{eq:rdef}, and \eqref{eq:chidef}, we have
\begin{align}
i\epsilon_{\mu\nu\lambda} q_\nu M_{\lambda,\mathbf{q}} &= -\epsilon_{\mu\lambda\rho} q_\lambda \chi^{\rho\sigma}(\mathbf{q})\epsilon_{\sigma\delta\nu}q_\delta A_{\mathbf{q},\nu} \nonumber \\ 
&=i\epsilon_{\mu\lambda\rho} q_\lambda \chi^{\rho\sigma}(\mathbf{q})B_{\mathbf{q},\sigma}.
\end{align}
and hence,
\begin{equation}
M_{\lambda,\mathbf{q}} = \chi^{\lambda\sigma}(\mathbf{q})B_{\mathbf{q},\sigma}.
\end{equation}
We thus see that $\chi^{\lambda\sigma}(\mathbf{q})$ is the wavevector dependent magnetic susceptibility. 
Furthermore, we see from Eqs.~\eqref{eq:rresp}, \eqref{eq:rdef} and \eqref{eq:chidef} that the magnetic susceptibility is related to the conductivity tensor via
\begin{equation}\label{eq:conductivity-susceptibility}
\lim_{\omega\rightarrow 0} i\omega \sigma^{\mu\nu}(\omega, \mathbf{q}) = -\epsilon_{\mu\lambda\rho} q_\lambda \chi^{\rho\sigma}(\mathbf{q})\epsilon_{\sigma\delta\nu}q_\delta.
\end{equation}

From Eq.~\eqref{eq:conductivity-susceptibility} we can deduce several general features of the conductivity tensor in the $\omega\rightarrow 0$ limit. 
First, we see that if the magnetic susceptibility $\chi^{\mu\nu}$ is finite, then the conductivity tensor is singular in the $\omega\rightarrow 0$ limit. 
In particular, Eq.~\eqref{eq:conductivity-susceptibility} shows that the magnetic susceptibility determines the weight of the $1/\omega$ pole in the conductivity tensor as $\omega\rightarrow 0$ at nonzero $\mathbf{q}$. 
This singularity appears only in the transverse component of the conductivity (i.e, the transverse current response to a transverse field). 
Counterintuitively, this shows that the transverse conductivity can be divergent as $\omega\rightarrow 0$ even for an insulator. 
We also deduce for any system with a finite uniform magnetic susceptibility
\begin{equation}
\chi^{\mu\nu} = \lim_{\mathbf{q}\rightarrow 0} \chi^{\mu\nu}(\mathbf{q}),
\end{equation}
the low-frequency conductivity satisfies
\begin{equation}
\lim_{\omega\rightarrow 0} \sigma^{\mu\nu}(\omega, \mathbf{q}) \sim \frac{i}{\omega}\epsilon_{\mu\lambda\rho} q_\lambda \chi^{\rho\sigma}\epsilon_{\sigma\delta\nu}q_\delta + \mathcal{O}(q^3),
\end{equation}
and so the singular part of the low-frequency transverse conductivity depends quadratically on $\mathbf{q}$ and vanishes as $|\mathbf{q}|\rightarrow 0$. 

To support these conclusions, we use our formalism to compute $\lim_{\omega\rightarrow 0}\sigma^{\mu\nu}(\omega, \mathbf{q})$ for our model of a moir\'{e} Chern insulator described in Eqs.~\eqref{eq:modifiedHaldaneTerm}-~\eqref{eq:lastHaldaneTerm}. 
We show the results in Fig.~\ref{Fig:LinearConductivityVsq}.
We see that, in accordance with Eq.~\eqref{eq:conductivity-susceptibility}, we have that
\begin{equation}
\lim_{\omega\rightarrow 0}\mathrm{Im} \left[ \omega\sigma^{xx}(\omega, q\hat{y}) \right] = \lim_{\omega\rightarrow 0}\mathrm{Im} \left[ \omega\sigma^{yy}(\omega, q\hat{x}) \right] \sim -q^2\chi^{zz}
\end{equation}
for small $q$.
This figure not only supplicates the $q$ dependence but also the singularity that is linear in $\omega$.
Notably, this numerical computation does \textit{not} assume a form of the magnetic susceptibility, since it just faithfully carries out the conductivity calculation; equality of $\lim_{\omega\rightarrow 0}\omega\sigma^{xx}(q\hat{y},\omega)$ and $\lim_{\omega\rightarrow 0}\omega\sigma^{yy}(q\hat{x},\omega)$ in the limit of small $q$ is a reflection of gauge invariance alone. 

We can also invert Eq.~\eqref{eq:conductivity-susceptibility} to derive an expression for the uniform magnetic susceptibility $\chi^{\mu\nu}$. 
First, we note that for a system that conserves energy, the uniform magnetic susceptibility must be a symmetric tensor. 
To derive this, we follow the logic of Ref.~\cite{scheibner2019odd}, which proved an analogous result for the tensor of elastic moduli. 
We can consider the change in (free) energy for a system as we slowly move the magnetic field through a closed cycle,
\begin{align}
\Delta \langle H\rangle &= -\oint \mathbf{B}\cdot d\mathbf{M} \nonumber \\
&=-\oint B_\mu \chi^{\mu\nu} dB_\nu \nonumber \\
&=-\frac{1}{2}\oint B_\mu \left(\chi^{\mu\nu}-\chi^{\nu\mu}\right) dB_\nu.
\end{align}
Since the total energy is a state function, its change over every closed cycle must be zero. 
Thus, we deduce that $\chi^{\mu\nu}=\chi^{\nu\mu}$. 

Taking two derivatives of Eq.~\eqref{eq:conductivity-susceptibility} with respect to $\mathbf{q}$ and making use of the antisymmetry of the Levi-Civita symbol, we find after making use of the symmetry of $\chi^{\mu\nu}$ that
\begin{equation}\label{eq:chifinal}
\chi^{\mu\nu} = \lim_{\omega\rightarrow 0}\left.\frac{-i\omega}{3}\epsilon_{\mu\lambda\rho}\epsilon_{\nu\sigma\delta}\frac{\partial^2 \sigma^{\lambda\sigma}}{\partial q_\rho\partial q_\delta}\right|_{\mathbf{q}= 0}.
\end{equation}
Eq.~\eqref{eq:chifinal} is consistent with and generalizes expressions for the orbital susceptibility that appear in the literature~\cite{Louie1996MagneticSusceptibilityOfInsulators, Vignale1991OrbitalParamagnetism, Bouchiat2021DiamagnetismInDirac, Loktev2021OrbitalSuscpInGraphene}. 
It allows us to compute the magnetic susceptibility of insulators using our Kubo formula Eq.~\eqref{eq:fullGeneralizedLinearConductivity} in terms of the conserved current operator.

\subsubsection{Magnetic Quadrupole Moments}\label{sec:QuadrupoleAndOctopoleMagneticMoments}
In this section, we will derive an expression for the magnetic quadrupole moment in the ground state of a finite system with zero external field using our
manifestly-conserved operator $\mathbf{j}_\mathbf{q}$ defined in Eq.~\eqref{eq:mainj}.
Using Eq.~\eqref{eq:magcurrent}, we see that evaluating the average magnetic moment will involve expanding our expression for the current operator in powers of $\mathbf{q}$; we thus expect our conserved current operator will give different predictions for magnetic multipole moments as compared to the non-conserved currents in Eqs.~\eqref{eq:trapfirsttime} or \eqref{eq:midpointfirsttime}. 
Since we showed in Sec.~\ref{sec:jnew} that all definitions of the current operator agree to linear order in $\mathbf{q}$, we begin our analysis with the magnetic quadrupole moment. 
In the interest of generality, we will here return to consider particles of charge $\charge$

We start by following Refs.~\cite{hughes2022quadrupolemagneticmoment,ceresoli2006orbital,vanderbilt2022surfaceorbitalmagnetization}, and take the average of Eq.~\eqref{eq:magcurrent} to find in components, 
\begin{equation}
    \left \langle j^{\mu}_{\mathbf{q}} \right\rangle = i \epsilon^{\mu \nu \gamma} q_{\nu} M_{ \gamma \mathbf{q}}.
\end{equation}
To find the higher-order moments (quadrupole, octupole, etc.), we expand both sides of the equation in orders of the wavevector,
\begin{align}
\label{eq:expandedMagneticMoment}
    &\left \langle j^{\mu}_{0}\right\rangle + \left \langle j^{\mu \nu}_{1}\right\rangle q_{\nu} + \left \langle j_{2}^{\mu \nu \gamma}\right\rangle q_{\nu} q_{\gamma} + \cdots \nonumber
    \\&
    \qquad = i \epsilon^{\mu \nu \gamma} q_{\nu} \left( M_{\gamma 0} + M^{\alpha}_{\gamma  1} q_\alpha + M^{\alpha  \beta}_{\gamma  2} q_\alpha q_{\beta} + \cdots \right).
\end{align}
We point out that in the thermodynamic limit, $\left \langle j^{\mu}_{0}\right\rangle = 0$ by Bloch's theorem~\cite{bohm1949BlochsTheorem,watanabe2022bloch}.
We now match up the left and right hand sides by orders of $\mathbf{q}$. 
The magnetic dipole moment $M_{\gamma 0}$ has been extensively considered in the literature~\cite{thonhauser2006insulator,ceresoli2006orbital}. 
We focus instead first on the quadrupolar term, $\left. \langle j^{\mu \nu \alpha}_2 \right\rangle= i \epsilon^{\mu \nu \gamma} M^{\alpha}_{\gamma 1}$.
Using the antisymmetry of the Levi-Civita symbol, we can rewrite this as 
\begin{equation}
-\frac{i}{2} \epsilon_{\mu \nu \Delta} \left \langle j_{2}^{\mu \nu \alpha} \right\rangle = M^{\alpha}_{\Delta 1}.
\end{equation}
Since 
\begin{equation}
\left \langle j^{\mu \nu \alpha}_2 \right\rangle = \left. \frac{1}{2} \partial_{\mathbf{q}^\nu} \partial_{\mathbf{q}^{\alpha}} \left\langle j^{\mu}_{\mathbf{q}} \right\rangle \right|_{\mathbf{q}=0}\label{eq:quadrupolemomentdef}
\end{equation}
we can use our definition of the conserved current in Eq.~\eqref{eq:mainj} to find
\begin{widetext}
\begin{align}
\label{eq:quadrupolejOperator}
    \left \langle j^{\mu \nu \alpha}_2 \right\rangle =& -\frac{\charge}{2} \sum_i \int^1_0 d\lambda \left\langle (1- \lambda)^2 x^\nu_i x^{\alpha}_i v^{\mu}_i(\mathbf{p}_i,\mathbf{x}_i) + \lambda^2  v^{\mu}_i(\mathbf{p}_i,\mathbf{x}_i) x^\nu_i x^{\alpha}_i  +  \lambda (1- \lambda) x^\nu_i v^{\mu}_i(\mathbf{p}_i,\mathbf{x}_i) x^{\alpha}_i +  \lambda (1- \lambda) x^{\alpha}_i v^{\mu}_i(\mathbf{p}_i,\mathbf{x}_i) x^\nu_i \right\rangle\nonumber
    \\
    =& -\frac{\charge}{6} \sum_i  \left\langle \{ x^\nu_i x^{\alpha}_i, v^{\mu}_i(\mathbf{p}_i,\mathbf{x}_i) \} +  \frac{1}{2} x_i^{\nu} v^{\mu}_i(\mathbf{p}_i, \mathbf{x}_i) x_i^\alpha +  \frac{1}{2}x_i^{\alpha} v^{\mu}_i(\mathbf{p}_i, \mathbf{x}_i) x_i^\nu \right\rangle.
\end{align}
\end{widetext}
This implies from Eq.~\eqref{eq:quadrupolejOperator} that the magnetic quadrupole moment takes the form
\begin{align}
\label{eq:quadrupoleMomentj}
    M^{\alpha}_{\Delta 1} =& -\frac{i\charge}{12} \epsilon_{\mu \nu \Delta} \sum_i  \Big\langle \{ x^\nu_i x^{\alpha}_i, v^{\mu}_i(\mathbf{p}_i,\mathbf{x}_i) \} \nonumber
    \\& + \frac{1}{2} x_i^{\nu} v^{\mu}_i(\mathbf{p}_i, \mathbf{x}_i) x_i^{\alpha} +  \frac{1}{2}x_i^{\alpha} v^{\mu}_i(\mathbf{p}_i, \mathbf{x}_i) x_i^{\nu }\Big\rangle.
\end{align}

Given that we know our conserved current in Eq.~\eqref{eq:mainj} differs from the non-conserved trapezoid [\eqref{eq:jtildedef}] and midpoint [\eqref{eq:jmidpointdef}] at order $\mathbf{q}^2$, we expect Eq.~\eqref{eq:quadrupoleMomentj} to differ from conventional expressions for the magnetic quadrupole moment that have appeared in the literature. 
Indeed, if we insert Eq.~\eqref{eq:jtildedef} for the non-conserved midpoint current $\tilde{j}$ operator into Eq.~\eqref{eq:quadrupolemomentdef}, we find the alternative expression
\begin{equation}
    \left \langle \Tilde{j}^{\mu \nu \alpha}_2 \right\rangle = -\frac{\charge}{4} \sum_i  \left\langle \{ x^\nu_i x^{\alpha}_i, v^{\mu}_i(\mathbf{p}_i,\mathbf{x}_i) \}  \right\rangle,
\end{equation}
which also gives a quadrupole moment of
\begin{align}
\label{eq:quadrupoleMomentjTilde}
    \Tilde{M}^{\alpha}_{\Delta 1} =& -\frac{i\charge}{8} \epsilon_{\mu \nu \Delta} \sum_i  \Big\langle \{ x^\nu_i x^{\alpha}_i, v^{\mu}_i(\mathbf{p}_i,\mathbf{x}_i) \} \Big\rangle.
\end{align}
Finally, if we use instead the midpoint current $\mathbf{j}_{\mathrm{mid},\mathbf{q}}$ of Eq.~\eqref{eq:jmidpointdef}, we find the alternative expression
\begin{align}
      \left \langle j^{\mu \nu \alpha}_{\mathrm{mid},2} \right\rangle&= -\frac{\charge}{8} \sum_i  \left\langle \{ x^\nu_i x^{\alpha}_i, v^{\mu}_i(\mathbf{p}_i,\mathbf{x}_i) \}\right\rangle \nonumber \\
     &+ \left\langle x_i^{\nu} v^{\mu}_i(\mathbf{p}_i, \mathbf{x}_i) x_i^\alpha +  x_i^{\alpha} v^{\mu}_i(\mathbf{p}_i, \mathbf{x}_i) x_i^\nu \right\rangle,
\end{align}
yielding a quadrupole moment of 
\begin{align}
\label{eq:quadrupoleMomentmid}
    M^{\alpha}_{\mathrm{mid},\Delta 1} =& -\frac{i\charge}{16} \epsilon_{\mu \nu \Delta} \sum_i  \Big\langle \{ x^\nu_i x^{\alpha}_i, v^{\mu}_i(\mathbf{p}_i,\mathbf{x}_i) \} \nonumber
    \\& +  x_i^{\nu} v^{\mu}_i(\mathbf{p}_i, \mathbf{x}_i) x_i^{\alpha} + x_i^{\alpha} v^{\mu}_i(\mathbf{p}_i, \mathbf{x}_i) x_i^{\nu }\Big\rangle.
\end{align}
By comparing Eqs.~\eqref{eq:quadrupoleMomentj}, \eqref{eq:quadrupoleMomentjTilde}, and \eqref{eq:quadrupoleMomentmid}, we see that the conserved current predicts a different value for the ground state magnetic quadrupole moment compared to the nonconserved currents conventionally used in the literature. 
This suggests that care must be exercised when computing the magnetic quadrupole moment in effective low-energy models, for which the non-conserved currents may not correspond to physically relevant observables. 
Furthermore, similar discrepancies will appear in the octupole magnetic moment, as well as all higher moments.

\subsubsection{Gauge invariance and the Streda formula}\label{sec:streda}
Continuing with our study of response to time-independent fields, we will now specialize to two dimensions, and show how the Streda formula~\cite{streda1982theory,smrcka1977transport} arises as a consequence of gauge invariance and Eq.~\eqref{eq:magcurrent}. 
While this result is somewhat tangential to our main argument, it highlights the importance of defining the $\mathbf{q}$-dependent conductivity using the conserved current operator. 
To begin, recall that although we derived the Kubo formula for the conductivity in the $A_0=0$ gauge, gauge invariance requires that the conductivity tensor $\emph{also}$ governs the response to gradients of $A_0$. 
In particular, let us consider response to a longitudinal electric field $\mathbf{E}_{\mathbf{q}}(\omega) = -i\mathbf{q}A_{0,\mathbf{q}}(\omega)$. 
We have from Eq.~\eqref{eq:conductivitydef} that
\begin{equation}
\langle j^\mu_\mathbf{q}(\omega)\rangle = -i\sigma^{\mu\nu}(\omega, \mathbf{q})q_\nu A_{0,\mathbf{q}}(\omega) 
\end{equation}
We now take the $\omega\rightarrow 0$ limit at fixed nonzero $\mathbf{q}$. 
Recall from our discussion preceding Eq.~\eqref{eq:magcurrent} that in this limit, the perturbing electric field is static, bounded, and spatially periodic. 
Thus, the system remains in a perturbed equilibrium state and the only currents that flow are magnetization currents. 
Hence, we have
\begin{equation}\label{eq:maglong}
i\epsilon^{\mu\nu\lambda}q_\nu M_{\lambda,\mathbf{q}} =  -i\lim_{\omega\rightarrow 0}\sigma^{\mu\nu}(\mathbf{q},\omega)q_\nu A_{0,\mathbf{q}}.
\end{equation}
We can expand Eq.~\eqref{eq:maglong} to lowest order in $\mathbf{q}$ to obtain
\begin{equation}
\epsilon^{\mu\nu\lambda} M_{\lambda,0} = -\sigma^{\mu\nu}(0,0) A_0,
\end{equation}
where we have introduced
\begin{equation}\label{eq:sigma00def}
\sigma^{\mu\nu}(0,0) = \lim_{\mathbf{q}\rightarrow 0}\lim_{\omega\rightarrow 0}  \sigma^{\mu\lambda}(\omega, \mathbf{q})\frac{q_\lambda q_\nu}{|\mathbf{q}|^2}
\end{equation}
Finally, we note that a time and space independent scalar potential $A_0$ is indistinguishable from (the negative) of a uniform variation of the chemical potential $-\delta\mu$. 
Taking derivatives on both sides then yields the generalized Streda relation
\begin{equation}
\sigma^{\mu\nu}(0,0) = \epsilon^{\mu\nu\lambda} \frac{\partial M_{\lambda}}{\partial \mu}.
\end{equation}
Finally, for an insulator, we know that the longitudinal component $\sigma^{\mu\nu}(\omega, \mathbf{q})q_\nu$ must be analytic and vanishing as $\mathbf{q}$ and $\omega$ both go to zero~\cite{bradlyn2012kubo}. 
This allows us to interchange the order of limits in Eq.~\eqref{eq:sigma00def}, and identify the susceptibility $\sigma^{\mu\nu}(0,0)$ with the DC Hall conductivity.  
We see then that for an insulator, the only nonvanishing component of the response to a longitudinal electric field at zero frequency is the Hall conductivity, and the components of the Hall conductivity are equal to the derivatives of the magnetization with respect to chemical potential. 
We note that this derivation makes clear the importance of considering an insulating system: it is only for an insulator that we can guarantee the regularity of $\sigma^{\mu\nu}(\omega, \mathbf{q})q_\nu$.

\section{Second-Order Response in Moir\'{e} Materials}\label{sec:SecondOrderResponseInMoireMaterials}
Impressive experimental advances in recent years have spurred a renewed interest in nonlinear electromagnetic responses in (topological) materials. 
Theoretical and experimental investigations into the nonlinear optical response of electrons in crystals has yielded new insights into band topology and geometry that can be probed using spatially uniform ($\mathbf{q}\rightarrow 0$) optical fields~\cite{sipe2000secondorder, parker2019diagrammatic, sipe2020ThirdOrderResponse, Sodemann2015NonlinearHallEffect, Tewari2021NonlinearBerryCurvature, Xiao2019NonlinearFromClassical,dejuan2017quantized,flicker2018chiral,rees2020helicitydependent,wu2017giant,kim2023observation,ahn2022riemannian,Peres2018GuageCovarianceInNonlinearOpticalResponse,passos2018nonlinear,Yan2020TRBInCurvatureMetricAndMotion,Yan2023UnifyingSemiclassicsNonlinear}. 
So far, relatively little attention has been paid to the nonlinear response to finite $\mathbf{q}$ electromagnetic fields.   
As we have shown in Sec.~\ref{sec:KerrRotation}, in moir\'{e} materials even the wavevector dependence of optical fields may play an important part in determining the electromagnetic response. 

Recently, Ref.~\cite{Mele2023NonlinearOpticalRegLattice} examined the wavevector-dependent longitudinal nonlinear conductivity at low order in wavevector.   Here, however, we will show how our diagrammatic perturbation theory of Sec.~\ref{sec:DiagramsRuleAndSumRules} can be used in conjunction with our definition of the conserved current in Eq.~\eqref{eq:TBCurrentOperator} to compute longitudinal and transverse nonlinear conductivities for arbitrary wavevector. 
We will focus primarily on the second-order response. 
In Sec.~\ref{sec:secondorderformulas} we will write down and evaluate the Feynman diagrams for the second-order nonlinear conductivity $\sigma^{\mu\gamma\nu}(\omega_1,\omega_2,\mathbf{q}_1,\mathbf{q}_2)$ that determines the second-order response via
\begin{widetext}
\begin{equation}\label{eq:secondorderdef}
\langle\mathbf{j}^\mu_\mathbf{q}(\omega)\rangle = \int d\omega_1 d\omega_2d\mathbf{q}_1 d\mathbf{q}_2\delta(\omega-\omega_1-\omega_2)\delta(\mathbf{q}-\mathbf{q}_1-\mathbf{q}_2)\sigma^{\mu\gamma\nu}(\omega_1,\omega_2,\mathbf{q}_1,\mathbf{q}_2) E_{\gamma,\mathbf{q}_1}(\omega_1)E_{\nu,\mathbf{q}_2}(\omega_2).
\end{equation}
\end{widetext}
This will require using all of the diagrammatic rules of Sec.~\ref{sec:DiagramRules}, and will involve a careful treatment of the $\mathbf{q}$-dependent diamagnetic current vertices from Sec.~\ref{sec:HigherOrderCurrentOperators}. 
Next, in Sec.~\ref{sec:HarmonicGenAndSelfFocusing} we will apply our results to compute the second order response of our toy model of a moir\'{e} Chern insulator from Eqs.~\eqref{eq:modifiedHaldaneTerm}--\eqref{eq:lastHaldaneTerm}. 
Motivated by experimental considerations, we will focus on spatially rectified ($\mathbf{q}_1=-\mathbf{q}_2$) second harmonic generation ($\omega_1=\omega_2$). 
This component of the nonlinear conductivity can be probed using transient grating spectroscopy techniques~\cite{weber2007nondiffusive,koralek2009emergence,torchinsky2013fluctuating,rouxel2021hard,bencivenga2019nanoscale}. 
In such measurements, a pair of coherent lasers are made to interfere at the sample to generate a sinusoidal electric field, whose wavevector can be tuned by changing the wavelength of the beams and the angle between the beams. 
The nonlinear response of the system to such a finite wavevector perturbation is then determined using a probe pulse. 
By using light in the visible to extreme ultraviolet range, grating wavevectors on the order $0.001$--$0.1$\AA$^{-1}$ can be achieved. 
For moir\'{e} systems with lattice constants $a\sim 100$\AA, this means that the nonlinear response can be measured over a range of wavevectors spanning multiple moir\'{e} Brillouin zones; even for conventional materials with lattice constants on the order of $1$\AA, the transient grating wavevector can span a large portion of the first Brillouin zone. 
In both cases, a formalism such as ours is necessary to compute the nonlinear response of the manifestly conserved current at large $\mathbf{q}$.

\subsection{Second-Order Response}\label{sec:secondorderformulas}
\begin{figure}[h]
      \centering
\begin{minipage}{0.95\hsize}
\centering
\includegraphics[width=0.95\hsize]{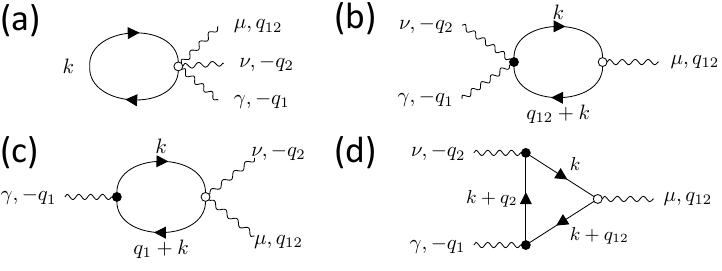}
\caption{Second-order response diagrams}
\label{Fig:FeynmanDiagramsSecondOrder}
\end{minipage}
\end{figure}
We may now turn to higher-order conductivities like the second-order response.
We will proceed in a similar fashion as in the linear response case, by using the rules and conventions set out in Secs. \ref{sec:HigherOrderCurrentOperators} and \ref{sec:DiagramRules}.

To compute the second-order conductivity, we begin by using the diagrammatic rules in Sec.~\ref{sec:DiagramRules} to write down the four relevant Feynman diagrams, shown in Fig.~\ref{Fig:FeynmanDiagramsSecondOrder}. 
Note that in all cases, $q_1$ and $q_2$ flow ``into'' the fermion loop, while $q_{12}$ flows ``out'' of the fermion loop.
We see that diagrams Figs.~\ref{Fig:FeynmanDiagramsSecondOrder}(b) and (c) involve the two-photon diamagnetic current vertex, while diagram Fig.~\ref{Fig:FeynmanDiagramsSecondOrder}(a) involves the three photon vertex. 
Using rule \#5 from the Feynman diagram rules in Sec.~\ref{sec:DiagramRules}, we see that since Figs.~\ref{Fig:FeynmanDiagramsSecondOrder}(b) and (c) are invariant under the exchange of $(\gamma,q_1)$ with $(\nu, q_2)$, these diagrams enter into the expression for the conductivity with a multiplicity factor of $1/2$.

To this end, we will examine a representative example of the input and output vertices in Fig.~\ref{Fig:FeynmanDiagramsSecondOrder}(c) in applying rules \#5 and \#6 before writing the full mathematical expression from the diagrams.
First, the input vertex has a fermion line going into the solid dot vertex, so that $\mathbf{k}^{\prime} = \mathbf{k} + \mathbf{q}_1$.
The photon line associated with this line carries a value of $-\mathbf{q}_1$ as well, so this input vertex goes as
\begin{align}
    \int & d\lambda_1  \langle u_{n_1 \mathbf{k} + \mathbf{q}_1}| \left[ \partial_{\mathbf{k}^{\gamma}}  H_{\mathbf{k}} \right]_{\mathbf{k} \rightarrow \mathbf{k} + \mathbf{q}_1 - (1 - \lambda_1) \mathbf{q}_1 } | u_{n_2 \mathbf{k}} \rangle.
\end{align}

Similarly, we can apply Feynman diagram rules \#5 and \#6 to the output vertex of the same diagram.
Since the fermion line coming into the vertex has momentum $\mathbf{k}$, then $\mathbf{k}^{\prime} = \mathbf{k}$.
We also see there are two vector potential lines emanating from the open dot vertex: one is going ``in" with momentum $-\mathbf{q}_2$, and the other is going ``out" with $\mathbf{q}_{12}$.
So the output velocity vertex goes as
\begin{align}
    \int & d\lambda^{\prime}_2 d\lambda_2 \langle u_{n_2 \mathbf{k}} |\nonumber
    \\
    & \times \left[ \partial_{\mathbf{k}^{\nu}} \partial_{\mathbf{k}^{\mu}} H_{\mathbf{k}} \right]_{\mathbf{k} \rightarrow \mathbf{k} - (1 - \lambda^{\prime}_2) \mathbf{q}_2 + (1 - \lambda_2) \mathbf{q}_{12}} | u_{n_1 \mathbf{k} + \mathbf{q}_1} \rangle.
\end{align}

Similarly applying the Feynman rules to each diagram in Fig.~\ref{Fig:FeynmanDiagramsSecondOrder} and summing the results, we can write the second-order conductivity in terms of Matsubara frequencies as
\begin{widetext}
\begin{align}\label{eq:translatingsecondorderdiags}
    \sigma^{\mu \gamma\nu}(i\omega_1,i\omega_2,\mathbf{q}_1,\mathbf{q}_2) &= \frac{-e^3}{(i\omega_1) (i\omega_2)}\int d k \Bigg[ \frac{1}{2}  G_{n_1}(k) v_{(3), n_1  n_1}^{\mu \nu \gamma}(\mathbf{k}, -\mathbf{q}_1, -\mathbf{q}_2, \mathbf{q}_{12})\nonumber \\
    &+\frac{1}{2} G_{n_1}(k + q_{12}) v^{\gamma \nu}_{(2), n_1  n_2}(\mathbf{k} + \mathbf{q}_{12}, -\mathbf{q}_1 , -\mathbf{q}_2) G_{n_2}(k)  v^{\mu}_{(1), n_2  n_1}(\mathbf{k}, \mathbf{q}_{12})\nonumber
    \\
    &+G_{n_1}(k + q_1) v^{\gamma}_{(1), n_1  n_2}(\mathbf{k} + \mathbf{q}_1, -\mathbf{q}_1) G_{n_2 }(k)v^{\nu \mu}_{(2), n_2  n_1}(\mathbf{k}, -\mathbf{q}_2, \mathbf{q}_{12})\nonumber
    \\&
    + G_{n_1}(k + q_{12}) v^{\gamma}_{(1), n_1  n_2}(\mathbf{k} + \mathbf{q}_{12}, -\mathbf{q}_1 ) G_{n_2}(k + q_2 ) v^{\nu}_{(1), n_2  n_3}(\mathbf{k} + \mathbf{q}_2, -\mathbf{q}_2) G_{n_3}(k  ) v^{\mu}_{(1), n_3  n_1}(\mathbf{k}, \mathbf{q}_{12})\nonumber \\
    &+ (\gamma, q_1) \longleftrightarrow (\nu, q_2) \Bigg].
\end{align}
Each of these diagrams must be symmetrized under the exchange of $(\gamma,\mathbf{q}_1)$ with $(\nu,\mathbf{q}_2)$, since as we see from Eq.~\eqref{eq:secondorderdef}, these are merely labels for components of the incident electromagnetic field. 
We indicate this explicit symmetrization in the last line of Eq.~\eqref{eq:translatingsecondorderdiags}.  
Feynman rule \#5 in Sec.~\ref{sec:DiagramRules} ensures that this explicit symmetrization does not overcount diagrams, by including multiplicity factors like the $\frac{1}{2}$ in the second line of Eq.~\eqref{eq:translatingsecondorderdiags} from Fig.~\ref{Fig:FeynmanDiagramsSecondOrder}(b).

In this format, we can carry out the Matsubara frequency integrals and analytically continue back to real frequencies (using the prescription of Refs.~\cite{passos2018nonlinear,parker2019diagrammatic,McKay2021CDWWeyl} that $i\omega_1\rightarrow\omega_1^+$, $i\omega_2\rightarrow \omega_2^+$, and $\omega_{12}^+\equiv \omega_1+\omega_2+2i\eta$) in order to find
\begin{align}
\label{eq:GeneralSecondOrderResponse}
    &\sigma^{\mu  \gamma\nu}(\omega_1,\omega_2,\mathbf{q}_1,\mathbf{q}_2)= \frac{-e^3}{\omega^+_1 \omega^+_2}\int d \mathbf{k} d \lambda_1 d\lambda_2 d\lambda_3\Bigg[\frac{1}{2} n_F(\epsilon_{n_1 \mathbf{k}}) \langle u_{n_1 \mathbf{k}}|  \left[ \partial_{\mathbf{k}^\mu} \partial_{\mathbf{k}^\nu} \partial_{\mathbf{k}^\gamma} H_{\mathbf{k}} \right]_{\mathbf{k} \rightarrow \mathbf{k} - (1 - \lambda_1) \mathbf{q}_1  - (1 - \lambda_2) \mathbf{q}_2 +  (1 - \lambda_3) \mathbf{q}_{12} } | u_{n_1 \mathbf{k}} \rangle\nonumber
    \\&
    +\frac{1}{2} \frac{n_F(\epsilon_{ n_2 \mathbf{k}  }) - n_F(\epsilon_{n_1 \mathbf{k} + \mathbf{q}_{12}} ) }{\omega^+_{12} + \epsilon_{n_2 \mathbf{k}} - \epsilon_{n_1 \mathbf{k} + \mathbf{q}_{12}}} \langle u_{n_1 \mathbf{k}  + \mathbf{q}_{12}} |\left[ \partial_{\mathbf{k}^{\gamma}} \partial_{\mathbf{k}^{\nu}}  H_{\mathbf{k} } \right]_{\mathbf{k} \rightarrow \mathbf{k} + \mathbf{q}_{12} - (1 - \lambda_1) \mathbf{q}_1 - (1 - \lambda_2) \mathbf{q}_2} | u_{n_2 \mathbf{k}} \rangle \langle u_{n_2 \mathbf{k} } | \left[ \partial_{\mathbf{k}^{\mu}} H_{\mathbf{k}} \right]_{\mathbf{k} \rightarrow \mathbf{k} +(1 - \lambda_3) \mathbf{q}_{12}} | u_{n_1 \mathbf{k} + \mathbf{q}_{12} } \rangle \nonumber
    \\&
    + \frac{n_F(\epsilon_{n_2 \mathbf{k}}) - n_F(\epsilon_{n_1 \mathbf{k} + \mathbf{q}_1}) }{\omega^+_1 + \epsilon_{n_2 \mathbf{k}} - \epsilon_{n_1 \mathbf{k} + \mathbf{q}_1}}  \langle u_{n_1 \mathbf{k}  + \mathbf{q}_1}|\left[ \partial_{\mathbf{k}^{\gamma}} H_{\mathbf{k}} \right]_{\mathbf{k} \rightarrow \mathbf{k}  + \mathbf{q}_1 - (1 - \lambda_1) \mathbf{q}_1} | u_{n_2 \mathbf{k} } \rangle \langle u_{n_2 \mathbf{k}} |\left[ \partial_{\mathbf{k}^{\nu}} \partial_{\mathbf{k}^{\mu}} H_{\mathbf{k} } \right]_{\mathbf{k} \rightarrow \mathbf{k}  - (1 - \lambda_2) \mathbf{q}_2 + (1 - \lambda_3) \mathbf{q}_{12}} | u_{n_1 \mathbf{k}  + \mathbf{q}_1} \rangle \nonumber
    \\&
    +\left( \frac{n_F(\epsilon_{n_1 \mathbf{k} + \mathbf{q}_{12}} ) }{\left( -\omega^+_1 + \epsilon_{n_1 \mathbf{k} + \mathbf{q}_{12}} - \epsilon_{n_2 \mathbf{k} + \mathbf{q}_2} \right) \left( -\omega^+_{12} + \epsilon_{n_1 \mathbf{k} + \mathbf{q}_{12}} - \epsilon_{n_3 \mathbf{k}} \right)} \right. 
    \left.  -\frac{n_F(\epsilon_{n_2 \mathbf{k} + \mathbf{q}_2 } ) }{\left( -\omega^+_1 + \epsilon_{n_1 \mathbf{k} + \mathbf{q}_{12}} - \epsilon_{n_2 \mathbf{k} + \mathbf{q}_2} \right) \left( -\omega^+_{2} + \epsilon_{n_2 \mathbf{k} + \mathbf{q}_2} - \epsilon_{n_3 \mathbf{k}} \right)} \right. \nonumber
    \\&
    \left. + \frac{n_F(\epsilon_{n_3 \mathbf{k}} ) }{\left( -\omega^+_2 + \epsilon_{n_2 \mathbf{k} + \mathbf{q}_2 } - \epsilon_{n_3 \mathbf{k}} \right) \left( -\omega^+_{12} + \epsilon_{n_1 \mathbf{k} + \mathbf{q}_{12}} - \epsilon_{n_3 \mathbf{k}} \right)} \right) \nonumber
    \\&
     \times \langle u_{n_1 \mathbf{k} + \mathbf{q}_{12}} | \left[ \partial_{\mathbf{k}^{\gamma}} H_{\mathbf{k} }\right]_{\mathbf{k} \rightarrow \mathbf{k} +\mathbf{q}_{12} - (1 - \lambda_1) \mathbf{q}_1} | u_{n_2 \mathbf{k} + \mathbf{q}_2} \rangle \langle u_{n_2 \mathbf{k} + \mathbf{q}_2} | \left[ \partial_{\mathbf{k}^{\nu}} H_{\mathbf{k}}\right]_{\mathbf{k} \rightarrow \mathbf{k} + \mathbf{q}_2 - (1 - \lambda_2) \mathbf{q}_2} | u_{n_3 \mathbf{k} } \rangle \nonumber
    \\&
     \times \langle u_{n_3 \mathbf{k} } | \left[ \partial_{\mathbf{k}^{\mu}} H_{\mathbf{k}}\right]_{\mathbf{k} \rightarrow \mathbf{k} + (1 - \lambda_3)  \mathbf{q}_{12}} | u_{n_1 \mathbf{k} + \mathbf{q}_{12} } \rangle \nonumber
    \\& + (\gamma, q_1) \longleftrightarrow (\nu, q_2) \Bigg] .
\end{align}
\end{widetext}

This analytic expression in Eq.~\eqref{eq:GeneralSecondOrderResponse} also allows us to comment on the regularity of the conductivity at low frequencies. 
In the $\mathbf{q}_1,\mathbf{q}_2\rightarrow 0$ limit, it was shown in Refs.~\cite{parker2019diagrammatic, McKay2021CDWWeyl,Mele2023NonlinearOpticalRegLattice} that the diamagnetic vertices in Figs.~\ref{Fig:FeynmanDiagramsSecondOrder} (a--c) serve a crucial role in regulating the low frequency conductivity; in particular, proper definitions of the diamagnetic current is necessary to ensure that the second-order conductivity for a band insulator satisfies
\begin{equation}
\lim_{\omega_1\rightarrow 0}\lim_{\omega_2\rightarrow 0}\sigma^{\mu\gamma\nu}(\omega_1,\omega_2,0,0) = 0.
\end{equation}
Using our formalism, we can now examine the behavior of the static ($\omega_1,\omega_2\rightarrow 0$) second-order conductivity as a function of $\mathbf{q}_1$ and $\mathbf{q}_2$. 
Since we defined our diamagnetic current vertices to satisfy the generalized Ward identity in Eq.~\eqref{eq:ward}, we are guaranteed that for a band insulator the longitudinal components of the static second order conductivity go to zero as the wavevectors tend to zero. 
On the other hand, as we showed for the linear conductivity in Sec.~\ref{sec:magneticSusceptabilityInInsulators}, the transverse components of the second order conductivity can have singularities as we take $\omega\rightarrow 0$ at fixed $\mathbf{q}_1$ and $\mathbf{q}_2$. 
These singularities correspond to second order magnetic and magnetoelectric responses, and are not unphysical.

\subsection{Second-Order Response in a Moir\'{e} Chern Insulator: Harmonic Generation in Frequency and Self-Focusing in Wavevector}\label{sec:HarmonicGenAndSelfFocusing}
\begin{figure*}[t]
      \centering
\begin{minipage}{0.98\hsize}
\centering
\includegraphics[width=0.98\hsize]{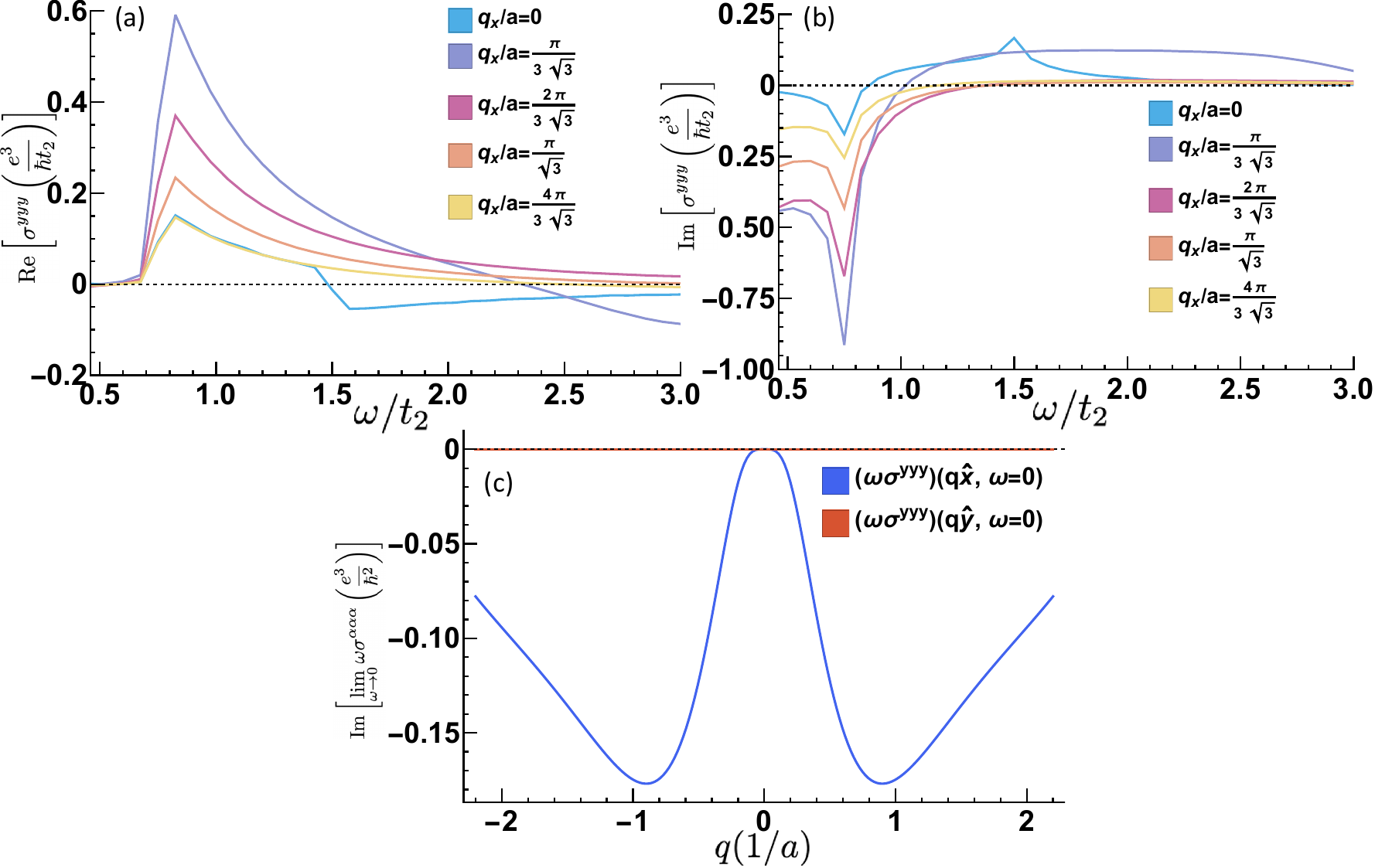}
\end{minipage}
\caption{
Plots showing the real (a) and imaginary (b) parts of the second-order transverse conductivities as functions of the frequency, $\sigma^{yyy}(\omega, q_x, \omega, -q_x; 2 \omega, 0)$.
The chosen direction for the conductivities' indices are $yyy$ and the wavevector is varied in $q_x \hat{x}$.
Plot (c) shows the imaginary parts of the transverse and longitudinal wavevectors of $\lim_{\omega \rightarrow 0}\omega \sigma^{yyy}(\omega, \mathbf{q}, \omega, -\mathbf{q}; 2 \omega, 0)$.
These figures used model parameters of $\phi = \pi/2$, $M = (3 \sqrt{3} -3/4) t_2$, and $t = 4 t_2$, which puts the topological phase at $C=1$ as shown in Fig.~\ref{Fig:HaldaneChernInsulatorPhaseDiagram}.
}
\label{Fig:SecondOrderConductivityVsOmega}
\end{figure*}

Using Eq.~\eqref{eq:GeneralSecondOrderResponse}, we can numerically compute the second-order conductivity as a function of frequency and wavevector for any noninteracting electron system.
Typically, calculations of the second-order conductivity have focused on sum frequency generation (where the measured current oscillates as the sum of applied electric field frequencies) or difference frequency generation responses (where the measured current oscillates as the difference of applied electric field frequencies) \cite{parker2019diagrammatic, Moore2021SecondHarmonicGenerationinWeyl, Gong2020HigherHarmonicGeneration, Benfatto2016ThirdHarmonicsInSuperconductors, 
Joao2020OpticalResponseTBWithDiagrams,McKay2021CDWWeyl}.
When we allow for spatially inhomogeneous electric fields, however, the response at second order becomes more complicated. 
In particular, when the applied electric field varies sinusoidally in space with wavevector $\mathbf{q}$ and in time with frequency $\omega$, we can
consider sum frequency generation (measured current oscillating at frequency $2\omega$) and ``difference wavevector generation''  in wavevector (measured current is spatially uniform with wavevector $\mathbf{q}-\mathbf{q}=0$). 
Experimentally, this could be measured in a transient grating experiment by looking at the second harmonic generation signal.

To see an example of this in practice, we return to our model of a moir\'{e} Chern insulator from Eqs.~\eqref{eq:modifiedHaldaneTerm}--\eqref{eq:lastHaldaneTerm}. 
We can consider the ``sum-in-frequency, difference-in-wavevector'' response. 
Since previous works~\cite{Mele2023NonlinearOpticalRegLattice} have presented a formalism for computing the longitudinal components of the response, we primarily focus on the transverse response $\sigma^{yyy}(\omega,\omega,q\hat{x},-q\hat{x})$ that quantifies the spatially uniform current $j^y(2\omega)$ that flows in the $y$ direction in response to a transverse $y$-polarized electric field. 
Such a response could be probed using in-plane polarized THz radiation at oblique incidence. 
Referring to Fig.~\ref{Fig:HaldaneLatticeDiagram}, we see that the wavevector of incident light is parallel to the vector $\Delta K=\frac{4\pi}{3\sqrt{3}a}$ separating the $K$ and $K'$ points in the moir\'{e} Brillouin zone. 
In Fig.~\ref{Fig:SecondOrderConductivityVsOmega} (a) and (b) we plot the real and imaginary parts of  $\sigma^{yyy}(\omega,\omega,q\hat{x},-q\hat{x})$ respectively, as a function of frequency for five different values of $q$ between $q=0$ and $q=|\Delta K|$. 
We see that for all wavevectors the real part of the response initially grows as a function of frequency, peaks at $\omega\sim 0.8t_2=0.8$THz---such that $2\omega$ is approximately equal to the topological gap---and then decreases at larger frequency. 
Intriguingly, as a function of wavevector we see that for small values of $q$ the magnitude of the current at $\omega\sim 0.8t_2$ initially grows as $q$ increases, before again decreasing: the peak value of  $\sigma^{yyy}(\omega,\omega,q\hat{x},-q\hat{x})$ is similar for $q=0$ and $q=|\Delta K|$.

To conclude, we can also study magnetoelectric response by examining the low frequency behavior of $\sigma^{yyy}(\omega,\omega,q\hat{x},-q\hat{x})$. 
In line with our discussion in Sec.~\ref{sec:magneticSusceptabilityInInsulators}, we expect that as $\omega\rightarrow 0$ for fixed $\mathbf{q}$, the there will be a $1/\omega$ singularity in $\sigma^{yyy}(\omega,\omega,q\hat{x},-q\hat{x})$ whose weight disperses at least quadratically at small $\mathbf{q}$; the weight of this pole quantifies the magnetization current that flows in response to the external magnetic and electric fields (with the magnetic field determined by Eq.~\eqref{eq:faraday}). 
At the same time, we expect the longitudinal response $\omega \sigma^{yyy}(\omega,\omega,q\hat{y},-q\hat{y})$ to be regular as $\omega\rightarrow 0$ for any fixed $\mathbf{q}$. 
We can see both properties in Fig.~\ref{Fig:SecondOrderConductivityVsOmega}(c).
We show the quantities $\lim_{\omega\rightarrow 0}\omega\sigma^{yyy}(\omega,\omega,q\hat{x},-q\hat{x})$ (in blue) and $\lim_{\omega\rightarrow 0}\omega\sigma^{yyy}(\omega,\omega,q\hat{y},-q\hat{y})$ (in red) computed for our moir\'{e} Chern insulator model. 
Thus, our formalism for computing transverse nonlinear electromagnetic responses correctly captures the low-energy behavior dictated by Maxwell's equations and gauge invariance discussed in Sec.~\ref{sec:magneticSusceptabilityInInsulators} for the linear response.

\section{Conclusion}\label{sec:Conclusion}
In this work, we have developed a formalism for computing spatially nonuniform (wavevector dependent) linear and nonlinear electromagnetic response functions for condensed matter systems. 
In Sec.~\ref{sec:jnew} we introduced a definition of the current operator that can be defined using only the velocity and position operators, independent of detailed knowledge of the microscopic form of the Hamiltonian. 
Furthermore, unlike the approximations Eqs.~\eqref{eq:midpointfirsttime} and \eqref{eq:trapfirsttime}, our current in Eq.~\eqref{eq:mainj} is manifestly conserved independent of the microscopic details of the Hamiltonian. 
It is important to emphasize that while Eqs.~\eqref{eq:midpointfirsttime}, \eqref{eq:trapfirsttime} and \eqref{eq:mainj} coincide for nonrelativistic Hamiltonians of the form of Eq.~\eqref{eq:assumedH}, only our conserved current in Eq.~\eqref{eq:mainj} remains conserved when Eq.~\eqref{eq:assumedH} is approximated by truncating the Hilbert space. 
For nonrelativistic systems, we can thus view Eq.~\eqref{eq:mainj} as a conserving approximation to the total current applicable to effective models such as Wannier-based tight-binding models. 
This is crucial for calculations involving approximate models of non-superconducting systems with truncated Hilbert spaces, since physically relevant approximations must conserve charge if the total Hamiltonian also conserves charge. 
Additionally, unlike Eqs.~\eqref{eq:midpointfirsttime} and \eqref{eq:trapfirsttime}, our conserved current \eqref{eq:mainj} also remains conserved when relativistic corrections to the kinetic energy are taken into account.

Next, we showed how Ward identities could be iteratively applied to determine the diamagnetic current operator order-by-order in the electromagnetic vector potential.
Using our conserved current as a starting point, we continued in Sec.~\ref{sec:DiagramsRuleAndSumRules} to develop a diagrammatic perturbation theory for computing spatially nonuniform linear and nonlinear conductivities, focusing on the case of noninteracting electrons for simplicity. 
Focusing first on the linear conductivity, we showed how our fully charge-conserving approach implies a generalized f-sum rule relating the density-density response function to the diamagnetic conductivity as a function of wavevector. 

We also applied our formalism to compute the wavevector dependent Hall conductivity in a toy model of Weyl semimetal.
To connect our formalism to experimentally relevant systems, In Sec.~\ref{sec:KerrRotation} we introduced a model for a 2D Chern insulator in a moir\'{e} superlattice, such that the wavelength of THz radiation can be a non-negligible fraction of the moir\'{e} lattice scale in certain geometries. 
We used this model to compute the Kerr angle and ellipticity as a function of frequency for oblique incidence, showing that the effects of spatial nonuniformity are potentially measurable in the next generation of experiments. 
We also showed how the low frequency transverse conductivity yields insights into the magnetic susceptibility, magnetic quadrupole moment, and Streda formula for insulating systems.

Finally, we applied our formalism to study second-order response to spatially nonuniform, time varying electric fields in two-dimensional systems. 
We calculated the experimentally relevant spatially uniform second-harmonic generation current that flows in response to a transverse, spatially varying AC electric field for a model of a moir\'{e} Chern insulator, which points towards future experimental work on transient grating nonlinear spectroscopy in moir\'{e} materials. 

Our work opens up several avenues for future theoretical and experimental studies. 
First, while our explicit calculations in Secs.~\ref{sec:ApplicationOfTheLinearResponse} and \ref{sec:SecondOrderResponseInMoireMaterials} were carried out for noninteracting systems, our Feynman diagram formalism in Sec.~\ref{sec:DiagramsRuleAndSumRules} can naturally accommodate a treatment of interacting systems by including additional interaction vertices. 
Our manifestly charge conserving approach to transverse linear and nonlinear conductivity would thus allow a consistent quantitative treatment of magnetoelectric response in models for candidate axionic charge density wave materials such as (TaSe$_4$)$_2$I~\cite{gooth2019axionic,shi2021chargedensitywave}. 
Along the same lines, our formalism can be systematically applied to systems with disorder. 
In the lowest order approximation, the effect of disorder on the nonlinear conductivities may be phenomenologically accounted for by replacing $\omega^+\rightarrow \omega+i/\tau$, where $\tau$ is the quasiparticle lifetime~\cite{parker2019diagrammatic}. 
Our formalism, however, can treat disorder scattering more formally by including vertices for scattering of electrons by the disorder potential, and standard diagrammatic techniques~\cite{bruus2004manybody} for averaging over disorder may be applied. 
In particular, we expect the (nonlinear) conductivities to depend on both the current vertices defined in Sec.~\ref{sec:jnew} as well as (nonlinear generalizations of) the diffuson propagator obtained by averaging over disorder. 
Additionally, our formalism implies additional generalized sum rule relations between nonlinear density response functions and diamagnetic current vertices, generalizing our results of Sec.~\ref{sec:fsum} and making contact with Ref.~\cite{watanabe2020generalized}. 

As we showed that commonly used approximations for the current operator do not conserve charge within the context of effective models, our work also prompts a reexamination of wavevector-dependent quantities computed using those approximations, such as the magnetic quadrupole moment in models of higher-order topological insulators~\cite{hughes2022quadrupolemagneticmoment,vanderbilt2022surfaceorbitalmagnetization}. 
Furthermore, in calculations of (nonlinear) X-ray scattering processes where both core and valence electronic states must be treated on equal footing, our response formalism based on the conserved current of Eq.~\eqref{eq:mainj} can be used to ensure Ward identities are obeyed even when relativistic corrections to the kinetic energy cannot be ignored~\cite{ament2011resonant}. 
Our work also motivates experimental studies of nonlinear optics in moir\'{e} systems, where the effect of spatial inhomogeneity of optical fields may be nonnegligible provided samples are large enough.

Finally, recent advances in superlattice and gate engineering open the door to experimentally studying (nonlinear) response to spatially inhomogeneous electromagnetic field outside of optics. 
In particular, gate tunable electronic superlattice potentials~\cite{forsythe2018band,li2021anisotropic,barcons2022engineering} could be modulated in time to create time-dependent electromagnetic fields. 
While recent theoretical~\cite{zeng2024gate} and experimental~\cite{sun2023signature,ghorashi2023topological} progress along these lines has focused on tunable superlattices in untwisted bilayer graphene systems, applications of these techniques to twisted systems would allow the study of response to time-dependent electromagnetic fields with wavevectors comparable to the moir\'{e} lattice spacing. 
We thus emphasize that our results will be directly applicable to the next generation of transport experiments in gate-tunable superlattice devices.

Let us conclude by reemphasizing the importance of the conserved current from Eq.~\eqref{eq:mainj}. 
Unlike the approximations of Eqs.~\eqref{eq:trapfirsttime} and \eqref{eq:midpointfirsttime} to the minimally coupled current in nonrelativistic systems commonly used in the literature, Eq.~\eqref{eq:mainj} is manifestly conserved for any model Hamiltonian. 
Since most Hamiltonians of interest in condensed matter systems arise as low energy approximations to (i.e. truncations of the Hilbert space of) complicated many-body semi-relativistic systems, electromagnetic response functions computed within these models will only be faithful approximations to what is experimentally measured if the current used in the calculation is conserved \emph{within the Hilbert space of the model}. 
All of Eqs.~\eqref{eq:trapfirsttime}, \eqref{eq:midpointfirsttime}, and \eqref{eq:mainj} accomplish this task to linear order in the wavevector, ensuring that they yield the same approximations in (paramagnetic) conductivities to quadratic order in wavevector. 
However, only Eq.~\eqref{eq:mainj} is generally conserved at quadratic order in wavevector and beyond, and only Eq.~\eqref{eq:mainj} allows for the determination of diamagnetic current vertices via Ward identities, which are essential for properly regularizing the low-frequency conductivity. 
Thus, we expect that Eq.~\eqref{eq:mainj} is a necessary starting point for obtaining consistent approximations to response functions beyond quadratic order in wavevector.
\begin{acknowledgments}
The authors thank G.~Monteiro, P.~Rao and M.~Trigo for helpful discussions. 
The theoretical work of R.C.M. and B.B. in the development of the response theory formalism was supported by the Alfred P.~Sloan Foundation and the National Science Foundation under Grant No.~DMR-1945058. 
The numerical computation of response functions and the development of experimental proposals was supported by the U.S.~DOE, Office of Basic Energy Sciences, Energy Frontier Research Center for Quantum Sensing and Quantum Materials through Grant No.~DE-SC0021238. 
This work made use of the Illinois Campus Cluster, a computing resource that is operated by the Illinois Campus Cluster Program (ICCP) in conjunction with the National Center for Supercomputing Applications (NCSA) and which is supported by funds from the University of Illinois at Urbana-Champaign.

\end{acknowledgments}

\appendix

\section{Proof of the Karplus-Schwinger Relation}\label{app:karplus_proof}

In this Appendix section, we will review the proof of the Karplus-Schwinger relation, Eq.~\eqref{eq:kp}, which was first presented in Ref.~\cite{karplus1948note}. 
We consider two operators $A$ and $B$. 
Let us define a function
\begin{equation}
F(\tau) = e^{\tau ( A+\delta t B)}\label{eq:fdef}.
\end{equation}
We will develop an expansion for $F(1)=\exp(A+\delta t B)$ as a power series in $\delta t$. 
To do so, first note that
\begin{equation}
F'(\tau) = (A+\delta tB)F(\tau).
\end{equation}
Defining
\begin{equation}
G(\tau) \equiv e^{-\tau A}F(\tau),\label{eq:gdef}
\end{equation}
we find that
\begin{align}
G'(\tau) &= -AG(\tau) + e^{-\tau A}F'(\tau) \nonumber \\ 
&=-AG(\tau) + e^{-\tau A}(A+\delta t B)F(\tau) \nonumber \\
&= \delta t e^{-\tau A}Be^{\tau A}G(\tau). \label{eq:gdiffeq}
\end{align}
Note that from Eqs.~\eqref{eq:fdef} and \eqref{eq:gdef}, we also have that $G(0)=1$. 
Introducing 
\begin{equation}
C(\tau) =e^{-\tau A}Be^{\tau A},\label{eq:cdef}
\end{equation}
and recasting Eq.~\eqref{eq:gdiffeq} as an integral equation, we have 
\begin{equation}
G(\tau) = 1 + \int_0^\tau d\lambda \delta t C(\lambda) G(\lambda).
\end{equation}
This integral equation can be solved iteratively by means of a Dyson series. 
We find that
\begin{align}
G(\tau) =& \mathcal{P} e^{\int_0^\tau\delta t C(\lambda)d\lambda} \nonumber
\\
=& 1 + \delta t\int_0^\tau C(\lambda) d\lambda \nonumber
\\& + (\delta t)^2\int_0^\tau d\lambda \int_0^\lambda d\lambda' C(\lambda) C(\lambda') + \mathcal{O}[(\delta t)^3], \label{eq:gdyson}
\end{align}
where $\mathcal{P}$ represents the path ordering (in $\lambda$) of the exponential.

Setting $\tau=1$, multiplying by $e^A$, and using the definition of Eq.~\eqref{eq:cdef} for $C(\tau)$ we have
\begin{align}
F(1)=e^AG(1) =& e^{A+\delta t B} \nonumber \\
 =&e^A + \delta t \int_0^1e^AC(\lambda) d\lambda \nonumber
\\& + (\delta t)^2\int_0^1 d\lambda \int_0^\lambda d\lambda' e^AC(\lambda) C(\lambda') + \mathcal{O}[(\delta t)^3] \nonumber \\
=&e^A + \delta t \int_0^1 e^{(1-\lambda)A}Be^{\lambda A} d\lambda +\mathcal{O}[(\delta t)^2],
\end{align}
which is equal to Eq.~\eqref{eq:kp}.

\section{Example: Longitudinal and Transverse Current Operator for Semi-Relativistic Free Electrons}\label{app:jsr}
As an example and verification of our expression for the current operator derived from Eq.~\eqref{eq:mainj}, we consider the semi-relativistic free electron Hamiltonian
\begin{equation}
H_{SR} = \sum_i\frac{|\mathbf{p_i}|^2}{2m} + \frac{|\mathbf{p_i}|^4}{8m^3c^2}. \label{eq:hsr}
\end{equation}
By minimally coupling $H_{SR}$ to a background vector potential, we can derive from Eq.~\eqref{eq:variational_current_def} that the current operator is given by
\begin{widetext}
\begin{align}
\mathbf{j}_{\mathrm{min},\mathbf{q}} &= -\charge\int d\mathbf{r} e^{-i\mathbf{q}\cdot\mathbf{r}}\frac{\delta H_{SR}(\mathbf{A})}{\delta \mathbf{A}(\mathbf{r})} \nonumber \\
&=\charge \sum_i \frac{1}{2m}\left(\mathbf{p}_i e^{-i\mathbf{q}\cdot\mathbf{x_i}}+ e^{-i\mathbf{q}\cdot\mathbf{x_i}}\mathbf{p}_i\right) + \frac{1}{8m^3c^2}\left(\mathbf{p}_i|\mathbf{p}_i|^2 e^{-i\mathbf{q}\cdot\mathbf{x_i}} + \mathbf{p}_i e^{-i\mathbf{q}\cdot\mathbf{x_i}}|\mathbf{p}_i|^2 +|\mathbf{p}_i|^2 e^{-i\mathbf{q}\cdot\mathbf{x_i}}\mathbf{p}_i +  e^{-i\mathbf{q}\cdot\mathbf{x_i}} \mathbf{p}_i|\mathbf{p}_i|^2\right)\nonumber
\\
&=\charge\sum_i e^{-i\mathbf{q}\cdot\mathbf{x}_i}\left[ \frac{1}{2m}\left(2\mathbf{p}_i-\mathbf{q}\right) + \frac{1}{8m^3c^2}\left(|\mathbf{p}_i|^2\mathbf{p_i} +|\mathbf{p_i}|^2(\mathbf{p}_i-\mathbf{q}) + |\mathbf{p}_i-\mathbf{q}|^2\mathbf{p}_i +|\mathbf{p}_i-\mathbf{q}|^2(\mathbf{p}_i-\mathbf{q})\right)   \right].\label{eq:jsrmin}
\end{align}
\end{widetext}

We can alternatively use our Eq.~\eqref{eq:mainj} to obtain the current $\mathbf{j}_\mathbf{q}$. 
The single-particle velocity operator for $H_{SR}$ is
\begin{equation}
\mathbf{v}_i = i\left[H_{SR},\mathbf{x}_i\right] = \frac{1}{m}\mathbf{p}_i + \frac{1}{2m^3c^2}|\mathbf{p}_i|^2\mathbf{p}_i. \label{eq:srvel}
\end{equation}
Inserting Eq.~\eqref{eq:srvel} into Eq.~\eqref{eq:mainj}, we find that our formalism  for the current operator yields
\begin{alignat}{2}
\mathbf{j}_\mathbf{q} &= \charge\sum_i\int_0^1 d\lambda e^{-i(1-\lambda)\mathbf{q}\cdot\mathbf{x}_i}\left(\frac{1}{m}\mathbf{p}_i \right. \nonumber
\\&
\left. \qquad \qquad \qquad  \qquad  \qquad \quad + \frac{1}{2m^3c^2}|\mathbf{p}_i|^2\mathbf{p}_i\right)e^{-i\lambda\mathbf{q}\cdot\mathbf{x}_i} \nonumber \\
&=\charge\sum_i\int_0^1 d\lambda e^{-i\mathbf{q}\cdot\mathbf{x}_i}\left[\frac{1}{m}(\mathbf{p}_i-\lambda\mathbf{q}) \right.\nonumber
\\&
\quad\quad\qquad\qquad\qquad\quad\left.+ \frac{1}{2m^3c^2}|\mathbf{p}_i-\lambda\mathbf{q}|^2(\mathbf{p}_i-\lambda\mathbf{q}) \right]. \label{eq:jsrnew}
\end{alignat}
Finally, using Eqs.~\eqref{eq:jtildedef} and \eqref{eq:jmidpointdef}, we have that
\begin{align}
\mathbf{\tilde{j}}_{\mathbf{q}} &= \frac{\charge}{2}\sum_i e^{-i\mathbf{q}\cdot\mathbf{x_i}}\left(\mathbf{v}_i(\mathbf{p_i},\mathbf{x}_i) + \mathbf{v}_i(\mathbf{p_i}-\mathbf{q},\mathbf{x}_i)\right) \nonumber
\\
&=\sum_{i}e^{-i\mathbf{q}\cdot\mathbf{x_i}} \left[
\frac{1}{2m}(2\mathbf{p}_i-\mathbf{q}) \right.\nonumber
\\&
\qquad \qquad \qquad \left. + \frac{1}{4m^3c^2}(\mathbf{p}_i|\mathbf{p}_i|^2 + (\mathbf{p_i}-\mathbf{q})|\mathbf{p}_i-\mathbf{q}|^2)
\right],\label{eq:jsrold}
\end{align}
and
\begin{align}
\mathbf{j}_{\text{mid}, \mathbf{q}} &=\charge \sum_i e^{-i\mathbf{q}\cdot\mathbf{x_i}}\mathbf{v}_i( \mathbf{v}_i(\mathbf{p_i}-\mathbf{q}/2,\mathbf{x}_i) \nonumber
\\
&=\sum_{i}e^{-i\mathbf{q}\cdot\mathbf{x_i}} \left[
\frac{1}{2m}(2\mathbf{p}_i-\mathbf{q}) \right.\nonumber
\\&
\qquad \qquad \qquad \left. + \frac{1}{16 m^3c^2} (2\mathbf{p_i}-\mathbf{q})|2\mathbf{p}_i-\mathbf{q}|^2
\right]\label{eq:jsrmidold}
\end{align}

Let us examine the longitudinal and transverse components of the current using the three possible definitions: Eqs.~\eqref{eq:jsrmin}, \eqref{eq:jsrnew} and \eqref{eq:jsrold}. 
For the longitudinal current, we find
\begin{align}
\mathbf{q}\cdot&\mathbf{j}_{\mathbf{q},\mathrm{min}} =\mathbf{q}\cdot\mathbf{j}_{\mathbf{q}}  = \charge\sum_i e^{-i\mathbf{q}\cdot\mathbf{x}_i}\left[\frac{1}{2m}(2\mathbf{p}_i\cdot\mathbf{q}-|\mathbf{q}|^2) \right.\nonumber
\\&
\left.+\frac{1}{8m^3c^2}(2\mathbf{p}_i\cdot\mathbf{q}-|\mathbf{q}|^2)(2|\mathbf{p}_i|^2-2\mathbf{p}_i\cdot\mathbf{q}+|\mathbf{q}|^2)
\right], \label{eq:correctlong}
\end{align}
\begin{align}
 \mathbf{q}\cdot&\mathbf{\tilde{j}}_\mathbf{q}  = \charge\sum_i e^{-i\mathbf{q}\cdot\mathbf{x}_i}\left[\frac{1}{2m}(2\mathbf{p}_i\cdot\mathbf{q}-|\mathbf{q}|^2)\right.\nonumber
 \\&
 \left. +\frac{1}{4m^3c^2}(2\mathbf{p}_i\cdot\mathbf{q}-|\mathbf{q}|^2)(|\mathbf{p}_i|^2-\mathbf{p}_i\cdot\mathbf{q}+|\mathbf{q}|^2)
\right],\label{eq:tildelong}
\end{align}
\begin{align}
 \mathbf{q}\cdot&\mathbf{j}_{\text{mid}, \mathbf{q}}  = \charge\sum_i e^{-i\mathbf{q}\cdot\mathbf{x}_i}\left[\frac{1}{2m}(2\mathbf{p}_i\cdot\mathbf{q}-|\mathbf{q}|^2)\right.\nonumber
 \\&
 \left. +\frac{1}{16 m^3c^2}(2\mathbf{p}_i\cdot\mathbf{q}-|\mathbf{q}|^2)(4 |\mathbf{p}_i|^2-4\mathbf{p}_i\cdot\mathbf{q}+|\mathbf{q}|^2)
\right],\label{eq:midlong}
\end{align}
Eq.~\eqref{eq:correctlong} shows that the longitudinal component of the current $\mathbf{j}_\mathbf{q}$, as defined in Eq.~\eqref{eq:mainj}, agrees with the longitudinal component of the current $\mathbf{j}_{\mathbf{q},\mathrm{min}}$ defined through minimal coupling via Eq.~\eqref{eq:jA}; both $\mathbf{j}_\mathbf{q}$ and $\mathbf{j}_{\mathbf{q},\mathrm{min}}$ satisfy the continuity equation, Eq.~\eqref{eq:continuity}, as mentioned in the main text. 
The longitudinal components of $\tilde{\mathbf{j}}_\mathrm{q}$ and $\mathbf{j}_{\text{mid}, \mathrm{q}}$ from Eqs.~\eqref{eq:tildelong} and~\eqref{eq:midlong} are distinct, implying that the (non-conserved) $\tilde{\mathbf{j}}_\mathrm{q}$ and $\mathbf{j}_{\text{mid}, \mathrm{q}}$ as defined in Eqs.~\eqref{eq:jtildedef} and~\eqref{eq:jmidpointdef} \emph{do not} satisfy the continuity equation. 
Subtracting Eq.~\eqref{eq:correctlong} from Eqs.~\eqref{eq:tildelong} and~\eqref{eq:midlong} we find
\begin{align}
\mathbf{q}\cdot\left(\tilde{\mathbf{j}}_\mathrm{q}-{\mathbf{j}}_{\mathrm{q},\mathrm{min}}\right) &= \frac{\charge}{8m^3c^4}\sum_ie^{-i\mathbf{q}\cdot\mathbf{x}_i}|\mathbf{q}|^2\left(2\mathbf{p}_i\cdot\mathbf{q}-|\mathbf{q}|^2\right) \nonumber
\\&
\neq 0,
\end{align}
and
\begin{align}
&\mathbf{q}\cdot\left(\mathbf{j}_{\text{mid}, \mathrm{q}}-{\mathbf{j}}_{\mathrm{q},\mathrm{min}}\right) = \nonumber
\\&-\frac{\charge}{16 m^3c^2}\sum_ie^{-i\mathbf{q}\cdot\mathbf{x}_i}|\mathbf{q}|^2\left(2\mathbf{p}_i\cdot\mathbf{q}-|\mathbf{q}|^2\right)
\neq 0.
\end{align}
This explicitly shows that $\tilde{\mathbf{j}}_\mathbf{q}$ and $\mathbf{j}_{\text{mid}, \mathbf{q}}$, as defined in Eqs.~\eqref{eq:jtildedef} and~\eqref{eq:jmidpointdef}, are not conserved for the semi-relativistic Hamiltonian from Eq.~\eqref{eq:hsr}.

Let us now examine the transverse components of the currents Eqs.~\eqref{eq:jsrmin}, \eqref{eq:jsrnew}, and $\eqref{eq:jsrold}$. 
Taking the cross product with $\mathbf{q}$, we find that the minimally coupled current satisfies
\begin{align}
&\mathbf{q}\times\mathbf{j}_{\mathbf{q},\mathrm{min}} =\nonumber
\\ &
\charge\sum_i e^{-i\mathbf{q}\cdot\mathbf{x}_i}\mathbf{q}\times\mathbf{p}_i\left[\frac{1}{m}\right. \left.+\frac{1}{4m^3c^2}(|\mathbf{p}_i|^2+|\mathbf{p}_i-\mathbf{q}|^2)
\right].
\end{align}
On the other hand, for our $\lambda$ integral definition of the current from Eqs.~\eqref{eq:mainj} and \eqref{eq:jsrnew} we find
\begin{align}
\mathbf{q}\times\mathbf{j}_{\mathbf{q}} &=\charge \sum_i e^{-i\mathbf{q}\cdot\mathbf{x}_i}\mathbf{q}\times\mathbf{p}_i\left[\frac{1}{m} + \frac{1}{2m^3c^2}\int_0^1 d\lambda |\mathbf{p}_i-\lambda\mathbf{q}|^2\right] \nonumber \\
&=\mathbf{q}\times\mathbf{j}_{\mathbf{q},\mathrm{min}} +\frac{1}{3m^3c^2}\sum_i e^{-i\mathbf{q}\cdot\mathbf{x}_i}|\mathbf{q}|^2(\mathbf{p}_i\times\mathbf{q}).
\end{align}
This means that although the current $\mathbf{j}_\mathbf{q}$ is conserved, its transverse components differ from that of the minimally coupled current for the semi-relativistic Hamiltonian in Eq.~\eqref{eq:hsr}, implying a different definition for the magnetization current.

On the other hand, the transverse current for the midpoint definition of the current Eq.~\eqref{eq:jmidpointdef} is
\begin{align}
\mathbf{q}\times\mathbf{j}_{\text{mid}, \mathbf{q}} &= \charge\sum_i e^{-i\mathbf{q}\cdot\mathbf{x}_i}\mathbf{q}\times\bigg[
\frac{1}{2m}(2\mathbf{p}_i-\mathbf{q}) \nonumber
\\&
\qquad\qquad\qquad+ \frac{1}{16m^3c^2} (2 \mathbf{p_i}-\mathbf{q})| 2\mathbf{p}_i-\mathbf{q}|^2
\bigg] \nonumber \\
 &=\sum_i e^{-i\mathbf{q}\cdot\mathbf{x}_i}\mathbf{q}\times\mathbf{p}_i\left[\frac{1}{m}+\frac{1}{4m^3c^2}(|\mathbf{p}_i|^2\right.\nonumber
 \\
 & \left.\qquad\qquad\qquad +|\mathbf{p}_i-\mathbf{q}|^2 - \frac{1}{2} |\mathbf{q}|^2)
 \right] \nonumber \\
&=\mathbf{q}\times \mathbf{j}_{\mathbf{q},\mathrm{min}} -\frac{1}{8 m^3 c^2} \sum_i e^{-i\mathbf{q}\cdot\mathbf{x}_i}|\mathbf{q}|^2 \mathbf{q}\times\mathbf{p}_i  .\label{eq:srtransversemid}
\end{align}

Curiously, if we examine the transverse component of the non-conserved trapezoid current, $\tilde{\mathbf{j}}_\mathbf{q}$, from Eq.~\eqref{eq:jtildedef}, we find that for the semi-relativistic system
\begin{align}
\mathbf{q}\times\tilde{\mathbf{j}}_{\mathbf{q}} &= \charge\sum_i e^{-i\mathbf{q}\cdot\mathbf{x}_i}\mathbf{q}\times\bigg[
\frac{1}{2m}(2\mathbf{p}_i-\mathbf{q}) + \frac{1}{4m^3c^2}(\mathbf{p}_i|\mathbf{p}_i|^2 \nonumber
\\&
\qquad\qquad\qquad\qquad+ (\mathbf{p_i}-\mathbf{q})|\mathbf{p}_i-\mathbf{q}|^2)
\bigg] \nonumber \\
 &=\sum_i e^{-i\mathbf{q}\cdot\mathbf{x}_i}\mathbf{q}\times\mathbf{p}_i\left[\frac{1}{m}+\frac{1}{4m^3c^2}(|\mathbf{p}_i|^2+|\mathbf{p}_i-\mathbf{q}|^2)
 \right] \nonumber \\
&=\mathbf{q}\times \mathbf{j}_{\mathbf{q},\mathrm{min}}.\label{eq:srtransverse}
\end{align}
Thus, for the semi-relativistic Hamiltonian, the transverse component of the non-conserved current $\tilde{\mathbf{j}}_\mathbf{q}$ is equal to the transverse component of the minimally coupled current. 
This means that for the semi-relativistic Hamiltonian, Eq.~\eqref{eq:hsr}, we have
\begin{equation}
\mathbf{j}^\mu_{\mathbf{q},\mathrm{min}}= \frac{q^\mu q^\nu}{|\mathbf{q}|^2}j^\nu_\mathbf{q} + \left(\delta_{\mu\nu}-\frac{q^\mu q^\nu}{|\mathbf{q}|^2}\right)\tilde{j}^\nu_\mathbf{q}.\label{eq:srfullcurrent}
\end{equation}
In other words, for the semi-relativistic Hamiltonian in Eq.~\eqref{eq:hsr}, the minimally coupled current in Eq.~\eqref{eq:variational_current_def} can be written entirely in terms of the velocity operator via Eqs.~\eqref{eq:mainj} and \eqref{eq:jtildedef}. 
We could, if we desired a model-dependent formulation of the current, apply Eq.~\eqref{eq:srfullcurrent} to the semirelativistic Hamiltonian which is applicable to heavy elements,
\begin{align}
H_{SOC} =& \sum_i\frac{|\mathbf{p}_i|^2}{2m} + \frac{|\mathbf{p}_i|^4}{8m^3c^2} + V(\mathbf{x}_i) + \frac{1}{8m^2c^2} \nonumber
\\&
+ \frac{1}{4m^2c^2} \vec{\sigma}\cdot(\mathbf{p}_i \times\nabla V(\mathbf{x}_i)) +\mathcal{O}(1/c^3),\label{eq:socham}
\end{align}
where $\vec{\sigma}$ is a vector of Pauli matrices acting on electron spin; $\vec{\sigma}\cdot(\mathbf{p}_i \times\nabla V(\mathbf{x}_i))$ represents the spin-orbit coupling energy. 

If we can assume that our Hamiltonians under study have the form of $H_{SOC}$ in Eq.~\eqref{eq:socham}, then the current defined in Eq.~\eqref{eq:srfullcurrent} entirely in terms of the velocity operator can be used to compute both longitudinal and transverse responses.
If we cannot assume that Eq.~\eqref{eq:socham} holds for our system of interest, then we cannot determine the transverse component of the minimally coupled current entirely in terms of the velocity operator without detailed knowledge of the full microscopic Hamiltonian. 
In such a situation, the only guiding principle is that we define a conserved current, such as Eq.~\eqref{eq:mainj}. 
In addition, we note that the nonrelativistic approximation to the Dirac equation that yields Eq.~\eqref{eq:socham} in the absence of an external electromagnetic field does \emph{not} give a minimally coupled Hamiltonian in the presence of a nonzero vector potential~\cite{messiah2014quantum}. 
Furthermore, Eq.~\eqref{eq:srfullcurrent} fails when higher-order relativistic corrections are taken into account. 
To see this, we can consider a toy model with the next highest power in momentum according to the semi-relativistic kinetic term:
\begin{equation}
H'=\frac{1}{6}\sum_i|\mathbf{p}_i|^6.\label{eq:h6}
\end{equation}
Minimally coupling $H'$ to a vector potential via Eq.~\eqref{eq:variational_current_def} gives the minimally coupled current
\begin{widetext}
\begin{align}
\mathbf{j}_{\mathbf{q},\mathrm{min}} &= \frac{\charge}{6}\sum_i \left[ e^{-i\mathbf{q}\cdot\mathbf{x}_i}\mathbf{p}_i|\mathbf{p}_i|^4 + \mathbf{p}_i e^{-i\mathbf{q}\cdot\mathbf{x}_i} |\mathbf{p}_i|^4 + |\mathbf{p}_i|^2e^{-i\mathbf{q}\cdot\mathbf{x}_i}\mathbf{p}_i|\mathbf{p}_i|^2 + |\mathbf{p}_i|^2\mathbf{p}_ie^{-i\mathbf{q}\cdot\mathbf{x}_i}|\mathbf{p}_i|^2 + |\mathbf{p}_i|^4e^{-i\mathbf{q}\cdot\mathbf{x}_i}\mathbf{p}_i + e^{-i\mathbf{q}\cdot\mathbf{x}_i}\mathbf{p}_i|\mathbf{p}_i|^4\right] \nonumber \\
&=\frac{\charge}{6}\sum_i e^{-i\mathbf{q}\cdot\mathbf{x}_i}\left[\mathbf{p}_i|\mathbf{p}_i|^4 + (\mathbf{p}_i-\mathbf{q})|\mathbf{p}_i|^4 + |\mathbf{p}_i-\mathbf{q}|^2\mathbf{p}_i|\mathbf{p}_i|^2 +|\mathbf{p}_i-\mathbf{q}|^2(\mathbf{p}_i-\mathbf{q})|\mathbf{p}_i|^2+|\mathbf{p}_i-\mathbf{q}|^4\mathbf{p}_i + |\mathbf{p}_i-\mathbf{q}|^4(\mathbf{p}_i-\mathbf{q})
\right].\label{eq:jmin6}
\end{align}
\end{widetext}
Similarly, applying Eqs.~\eqref{eq:jtildedef} to the Hamiltonian $H'$ in Eq.~\eqref{eq:h6} results in the non-conserved current operator
\begin{align}
\tilde{\mathbf{j}}_\mathbf{q} =& \frac{\charge}{2}\sum_ie^{-i\mathbf{q}\cdot\mathbf{x}_i}\left[|\mathbf{p}_i|^4\mathbf{p}_i + |\mathbf{p}_i-\mathbf{q}|^4(\mathbf{p}_i-\mathbf{q})\right],\label{eq:jold6}
\end{align}

Decomposing Eqs.~\eqref{eq:jmin6} and \eqref{eq:jold6} into longitudinal and transverse components, we find
\begin{align}
\mathbf{q}\cdot\mathbf{\tilde{j}}_\mathbf{q} \neq \mathbf{q}\cdot\mathbf{j}_{\mathbf{q},\mathrm{min}},
\end{align}
such that Eq.~\eqref{eq:jold6} is not conserved. 
Importantly, using Eq.~\eqref{eq:jold6} we also find that
\begin{align}
\mathbf{q}\times\mathbf{\tilde{j}}_\mathbf{q} = &\mathbf{q}\times\mathbf{j}_{\mathbf{q},\mathrm{min}} + \frac{1}{6}\sum_ie^{-i\mathbf{q}\cdot\mathbf{x}_i}(\mathbf{q}\times\mathbf{p}_i)(|\mathbf{q}|^2-2\mathbf{q}\cdot\mathbf{p}_i)^2\nonumber
\\ \neq&  0.
\end{align}
Thus, for a general Hamiltonian, neither the longitudinal nor the transverse components of the conventional current $\tilde{\mathbf{j}}_\mathbf{q}$ will correctly reproduce the minimally coupled current $\mathbf{j}_{\mathbf{q},\mathrm{min}}$.

\section{Current Operator in the Tight-Binding Basis}\label{sec:tbderivation}

In many cases, we will be interested in operators and dynamics projected into a (tight-binding) basis of Loewdin orbitals $\phi_\alpha(\mathbf{r}-\mathbf{R}-\mathbf{r}_\alpha)$. 
Let us introduce the Bloch-Loewdin basis functions
\begin{equation}
\chi_{\alpha\mathbf{k}}(\mathbf{r})=\frac{1}{\sqrt{N}}\sum_\mathbf{R} e^{i\mathbf{k}\cdot(\mathbf{R})}\phi_\alpha(\mathbf{r}-\mathbf{R}-\mathbf{r}_\alpha).\label{eq:blochlowdin}
\end{equation}
We can equally well use $\ket{\chi_{\alpha\mathbf{k}}}$ as a basis for expanding the current operators. 
Taking matrix elements of the conserved current from Eq.~\eqref{eq:mainj} in the basis of Eq.~\eqref{eq:blochlowdin}, we find
\begin{align}
\mathbf{j}_\mathbf{q} &= \charge\sum_{\mathbf{k}\alpha\beta}\int_0^1d\lambda \bra{\chi_{\alpha\mathbf{k}}}e^{-i(1-\lambda)\mathbf{q}\cdot\mathbf{x}}\mathbf{v}e^{-i\lambda\mathbf{q}\cdot\mathbf{x}}\ket{\chi_{\beta\mathbf{k+q}}}c^\dag_{\alpha\mathbf{k}}c_{\beta\mathbf{k+q}} \nonumber
\\
&=\sum_{\mathbf{k}\mathbf{k'}\mathbf{k''}\alpha\beta\gamma\eta}\int_0^1d\lambda \bra{\chi_{\alpha\mathbf{k}}}e^{-i(1-\lambda)\mathbf{q}\cdot\mathbf{x}}\ket{\chi_{\gamma\mathbf{k'}}} \nonumber
\\ &\quad \times\bra{\chi_{\gamma\mathbf{k'}}}\mathbf{v}\ket{\chi_{\eta\mathbf{k''}}}\bra{\chi_{\eta\mathbf{k''}}}e^{-i\lambda\mathbf{q}\cdot\mathbf{x}}\ket{\chi_{\beta\mathbf{k+q}}}c^\dag_{\alpha\mathbf{k}}c_{\beta\mathbf{k+q}}.\label{eq:jbasisint}
\end{align}
Introducing the cell-periodic basis functions
\begin{equation}
\tilde{\chi}_{\alpha\mathbf{k}}(\mathbf{r}) = \sqrt{N}e^{-i\mathbf{k}\cdot\mathbf{r}}\chi_{\alpha\mathbf{k}}(\mathbf{r}) 
\end{equation}
and using Eq.~\eqref{eq:densitymatrixelements}, we observe that Eq.~\eqref{eq:jbasisint} reduces to
\begin{align}
\mathbf{j}_\mathbf{q}=\charge\sum_{\mathbf{k}\alpha\beta\gamma\eta} \int_0^1d\lambda & \bra{\tilde{\chi}_{\alpha\mathbf{k}}}\ket{\tilde{\chi}_{\gamma\mathbf{k+(1-\lambda)q}}} \nonumber
\\ 
\times
&\bra{{\chi}_{\gamma\mathbf{k+(1-\lambda)q}}}\mathbf{v}\ket{\chi_{\eta\mathbf{k+(1-\lambda)q}}} \nonumber
\\ 
\times&
\bra{\tilde{\chi}_{\eta\mathbf{k+(1-\lambda)q}}}\ket{\tilde{\chi}_{\beta\mathbf{k+q}}}c^\dag_{\alpha\mathbf{k}}c_{\beta\mathbf{k+q}}. \label{eq:tbj1} 
\end{align}
We can simplify Eq.~\eqref{eq:tbj1} by rewriting $\bra{{\chi}_{\gamma\mathbf{k+(1-\lambda)q}}}\mathbf{v}\ket{\chi_{\eta\mathbf{k+(1-\lambda)q}}}$ in terms of the Bloch Hamiltonian. 
First, note that
\begin{align}
\bra{\chi_{\alpha\mathbf{k}}}\mathbf{v}\ket{\chi_{\beta\mathbf{k}}} =& \bra{\tilde{\chi}_{\alpha\mathbf{k}}}\partial_\mathbf{k}H_\mathbf{k}\ket{\tilde{\chi}_{\beta\mathbf{k}}} \nonumber \\
=&\partial_\mathbf{k}\bra{\tilde{\chi}_{\alpha\mathbf{k}}}H_\mathbf{k}\ket{\tilde{\chi}_{\beta\mathbf{k}}} \nonumber
\\& 
-\bra{\partial_\mathbf{k}\tilde{\chi}_{\alpha\mathbf{k}}}H_\mathbf{k}\ket{\tilde{\chi}_{\beta\mathbf{k}}} -\bra{\tilde{\chi}_{\alpha\mathbf{k}}}H_\mathbf{k}\ket{\partial_\mathbf{k}\tilde{\chi}_{\beta\mathbf{k}}} \nonumber \\
=&\partial_\mathbf{k}H^{\alpha\beta}_\mathbf{k} - i\left[\mathcal{A}_\mathbf{k},H_\mathbf{k}\right]_{\alpha\beta},\label{eq:velcovderiv}
\end{align}
where we have introduced
\begin{align}
H^{\alpha\beta}_{\mathbf{k}} &= \bra{\tilde{\chi}_{\alpha\mathbf{k}}}H_\mathbf{k}\ket{\tilde{\chi}_{\beta\mathbf{k}}}, \\
\mathcal{A}^{\alpha\beta}_\mathbf{k} & = i\bra{\tilde{\chi}_{\alpha\mathbf{k}}}\ket{\partial_\mathbf{k}\tilde{\chi}_{\beta\mathbf{k}}}, \\
\left[\mathcal{A}_\mathbf{k},H_\mathbf{k}\right] & = \sum_\gamma \mathcal{A}_{\mathbf{k}}^{\alpha\gamma}H_{\mathbf{k}}^{\gamma\beta} - \mathcal{H}_{\mathbf{k}}^{\alpha\gamma}A_{\mathbf{k}}^{\gamma\beta}.
\end{align}
In the tight-binding limit where 
\begin{equation}
\int d\mathbf{r} \phi_\alpha(\mathbf{r-R-r_\alpha})\mathbf{r}\phi_\beta(\mathbf{r-R'-r_\beta}) = \mathbf{r}_\alpha\delta_{\alpha\beta}\delta_{\mathbf{RR'}},\label{eq:tblimit}
\end{equation}
we can simplify Eq.~\eqref{eq:velcovderiv} further as
\begin{align}
\partial_\mathbf{k}H^{\alpha\beta}_\mathbf{k} - i\left[\mathcal{A}_\mathbf{k},H_\mathbf{k}\right]_{\alpha\beta}&\rightarrow \partial_\mathbf{k}H^{\alpha\beta}-i\mathbf{r}_\alpha H_\mathbf{k}^{\alpha\beta} + iH_\mathbf{k}^{\alpha\beta}\mathbf{r}_\beta \nonumber \\
&\mkern-36mu =e^{i\mathbf{k}\cdot\mathbf{r}_\alpha}\partial_\mathbf{k}\left(e^{-i\mathbf{k}\cdot\mathbf{r}_\alpha}H_\mathbf{k}^{\alpha\beta}e^{i\mathbf{k}\cdot\mathbf{r}_\beta}\right)e^{-i\mathbf{k}\cdot\mathbf{r}_\beta} \nonumber \\
&\mkern-36mu = \left[V(\mathbf{k})(\partial_\mathbf{k}h_\mathbf{k})V^\dag(\mathbf{k})\right]_{\alpha\beta}\label{eq:veltb}
\end{align}
where 
\begin{equation}
V^{\alpha\beta}(\mathbf{k}) = e^{i\mathbf{k}\cdot\mathbf{r}_\alpha}\delta_{\alpha\beta}
\end{equation}
is the tight-binding embedding matrix, and 
\begin{equation}
h^{\alpha\beta}_\mathbf{k} =\sum_{\gamma\eta} V_{\alpha\gamma}^\dag(\mathbf{k})H^{\gamma\eta}_\mathbf{k}V_{\eta\beta}(\mathbf{k})
\end{equation}
is the matrix Bloch Hamiltonian in the ``non-periodic'' gauge with $h_{\mathbf{k+G}}=V^\dag(\mathbf{G})h_\mathbf{k}V(\mathbf{G})$ for any reciprocal lattice vector $\mathbf{G}$.  

Furthermore, in the tight-binding limit we also have
\begin{align}
\bra{\tilde{\chi}_{\alpha\mathbf{k}}}\ket{\tilde{\chi}_{\beta\mathbf{k'}}} &= \sum_\mathbf{RR'}\int_{\mathrm{cell}}d\mathbf{r} e^{i\mathbf{k}\cdot(\mathbf{r-R})}e^{-i\mathbf{k'}\cdot(\mathbf{r-R'})} \nonumber
\\&
\qquad \qquad \times\phi_\alpha^*(\mathbf{r-R-r_\alpha})\phi_\beta(\mathbf{r-R'-r_\beta}) \nonumber
\\
&\rightarrow \delta_{\alpha\beta}\int d\mathbf{r} e^{i(\mathbf{k-k'})\cdot\mathbf{r}}\phi_\alpha^*(\mathbf{r-r_\alpha})\phi_\beta(\mathbf{r-r_\beta}) \nonumber
\\
&\rightarrow V^{\alpha\beta}(\mathbf{k-k'}).\label{eq:basisblochoverlap}
\end{align}
Notice in going from the second to third line in Eq.~\eqref{eq:basisblochoverlap}, we used an even stricter form of the tight-binding limit than in Eq.~\eqref{eq:tblimit}: we require not just the first moment but \emph{all} moments of the position operator to be diagonal in the tight-binding basis. 
Nevertheless, in this strict tight-binding limit we can combine Eqs.~\eqref{eq:tbj1}, \eqref{eq:veltb} and \eqref{eq:basisblochoverlap} to find
\begin{widetext}
\begin{align}
\mathbf{j}_\mathbf{q}&\rightarrow\charge\sum_{\mathbf{k}\alpha\beta}\int_0^1 d\lambda\left[V((\lambda-1)\mathbf{q})V(\mathbf{k}+(1-\lambda)\mathbf{q})\partial_\mathbf{k}h_{\mathbf{k+(1-\lambda)q}}V^\dag(\mathbf{k}+(1-\lambda)\mathbf{q})V(-\lambda\mathbf{q})\right]_{\alpha\beta}c^\dag_{\alpha\mathbf{k}}c_{\beta\mathbf{k+q}} \nonumber \\
&=\charge\sum_{\mathbf{k}\alpha\beta}\int_0^1 d\lambda \left[V(\mathbf{k})\partial_\mathbf{k}h_{\mathbf{k+(1-\lambda)q}} V^\dag(\mathbf{k+q})\right]c^\dag_{\alpha\mathbf{k}}c_{\beta\mathbf{k+q}} \nonumber \\
&=\charge\sum_{\mathbf{k}\alpha\beta}\int_0^1 d\lambda (\partial_\mathbf{k}h_{\mathbf{k+(1-\lambda)q}})^{\alpha\beta}\bar{c}^\dag_{\alpha\mathbf{k}}\bar{c}_{\beta\mathbf{k+q}},\label{eq:jtbfinal}
\end{align}
\end{widetext}
where we have introduced 
\begin{equation}
\bar{c}_{\alpha\mathbf{k}} = V^\dag_{\alpha\beta}(\mathbf{k})c_{\beta\mathbf{k}}
\end{equation}
as the annihilation operator for the non-periodic basis states,
\begin{align}\label{eq:nonperbasisstates}
\bar{\chi}_{\alpha\mathbf{k}}(\mathbf{r}) =& \frac{1}{\sqrt{N}}\sum_\mathbf{R} e^{i\mathbf{k}\cdot(\mathbf{R}+\mathbf{r}_\alpha)}\phi_\alpha(\mathbf{r}-\mathbf{R}-\mathbf{r}_\alpha) \nonumber
\\
=& V_{\alpha\beta}(\mathbf{k})\chi_{\beta\mathbf{k}}(\mathbf{r}).
\end{align}

Eq.~\eqref{eq:jtbfinal} allows us to compute matrix elements of the conserved current $\mathbf{j}_\mathbf{q}$ in the tight-binding basis entirely in terms of the tight-binding Bloch Hamiltonian $h^{\alpha\beta}(\mathbf{k})$.
Note that Eq.~\eqref{eq:jtbfinal} could have been anticipated from Eqs.~\eqref{eq:tbj1} and \eqref{eq:velcovderiv} by noting that the nonperiodic basis states in Eq.~\eqref{eq:nonperbasisstates} have vanishing Berry connection in the strict tight binding limit. 

Following a completely analogous set of steps, we find for the nonconserved trapezoid current in Eq.~\eqref{eq:jtildedef} that 
\begin{equation}
\mathbf{\tilde{j}}_\mathbf{q} = \frac{\charge}{2}\sum_{\mathbf{k}\alpha\beta}\left(\partial_\mathbf{k}h^{\alpha\beta}_{\mathbf{k+q}}+\partial_\mathbf{k}h^{\alpha\beta}_{\mathbf{k}}\right)\bar{c}^\dag_{\alpha\mathbf{k}}\bar{c}_{\beta\mathbf{k+q}},\label{eq:jtildetb}
\end{equation}
and similarly for the non-conserved midpoint current in Eq.~\eqref{eq:jmidpointdef},
\begin{equation}
\mathbf{{j}}_{\mathrm{mid},\mathbf{q}} = \charge\sum_{\mathbf{k}\alpha\beta}\partial_\mathbf{k}h^{\alpha\beta}_{\mathbf{k+q/2}}\bar{c}^\dag_{\alpha\mathbf{k}}\bar{c}_{\beta\mathbf{k+q}}.\label{eq:jmidtb}
\end{equation}
Eqs.~\eqref{eq:jtbfinal}, \eqref{eq:jtildetb}, and \eqref{eq:jmidtb} serve as our starting point for computing linear and nonlinear response coefficients from tight-binding models in the main text.

\section{Generalized Integration Formulation of an Operator Using the Karplus-Schwinger Relation}\label{sec:GeneralizedKarplusSchwingerIntegrationDerivation}
In this Appendix, we apply the Karplus-Schwinger relation from Eq.~\eqref{eq:CommutatorKarplusSchwingerRelation} to the specific case of the density operator $\rho_\mathbf{q}$ in order to derive an iterative formula for the diamagnetic current vertices.

First, consider the commutator between a general operator $A$ and the exponential $e^{-i \mathbf{q} \cdot \mathbf{x}}$ that appears in the Fourier component $\rho_{q}$. 
By introducing an auxiliary parameter $\tau$, we can write
\begin{align}
    i\left[ A, e^{-i \mathbf{q} \cdot \mathbf{x}} \right] =&\left. \frac{d}{d \tau}\left( e^{i A \tau} e^{-i \mathbf{q}\cdot \mathbf{x} } e^{-i A \tau} \right) \right|_{\tau \rightarrow 0} \nonumber
    \\
    =& \left. \frac{d}{d \tau} e^{ e^{i A \tau} (-i \mathbf{q} \cdot \mathbf{x}) e^{-i A \tau} } \right|_{\tau \rightarrow 0}\nonumber
    \\
    =& \left. \frac{d}{d \tau} e^{-i \mathbf{q} \cdot \mathbf{x}(\tau) } \right|_{\tau \rightarrow 0},
\end{align}
where we have defined
\begin{equation}
\mathbf{x}(\tau)\equiv e^{iA\tau}\mathbf{x}e^{-iA\tau}.
\end{equation}
From here we may follow our derivation in App.~\ref{app:karplus_proof} of the Karplus-Schwinger relation to find
\begin{align}
     i\left[ A, e^{-i \mathbf{q} \cdot \mathbf{x}} \right] =& \left.\lim_{\delta \tau \rightarrow 0} \frac{e^{-i \mathbf{q}\cdot \mathbf{x}(\tau + \delta \tau)} - e^{-i \mathbf{q}\cdot \mathbf{x}(\tau )} }{\delta \tau}\right|_{\tau \rightarrow 0} \nonumber
     \\
     \approx&\left. \lim_{\delta \tau \rightarrow 0} \frac{e^{-i \mathbf{q}\cdot (\mathbf{x}(\tau) + \delta \tau i[A, \mathbf{x}])} - e^{-i \mathbf{q}\cdot \mathbf{x}(\tau )} }{\delta \tau}\right|_{\tau \rightarrow 0}\nonumber
     \\
     =& \int^{1}_{0} d\lambda e^{-i(1- \lambda)\mathbf{q}\cdot \mathbf{x}} (\mathbf{q}\cdot [A, \mathbf{x}]) e^{-i \lambda \mathbf{q}\cdot \mathbf{x}}.
     \label{eq:CommutatorGeneralKarplusSchwingerRelation}
\end{align}
Eq.~\eqref{eq:CommutatorGeneralKarplusSchwingerRelation} allows us to rewrite commutators with the density operator that appear in the generalized Ward Identity of Eq.~\eqref{eq:ward} in terms of iterated integrals over auxiliary variables $\lambda$.

We can also use Eq.~\eqref{eq:CommutatorGeneralKarplusSchwingerRelation} in conjunction with the tight-binding expression in App.~\ref{sec:tbderivation}  to express the generalized diamagnetic current operators in  Sec.~\ref{sec:HigherOrderCurrentOperators} in the tight-binding basis. 
In keeping with notation from Sec.~\ref{sec:tbderivation}, we will let an overline indicate the operator in the orbital basis. 
We have
\begin{widetext}
\begin{align}
    \overline{j}^{\mu \nu_1 \cdots \nu_N}_{\mathbf{q}, -\mathbf{q}_1, \cdots, -\mathbf{q}_N} =& \int^1_0 d \lambda \int^1_0 d \lambda_1 \cdots \int^1_0 d \lambda_N\nonumber
    \\
    &\times
    \left[ \partial_{\mathbf{k}^\mu} \partial_{\mathbf{k}^{\nu_1}} \cdots \partial_{\mathbf{k}^{\nu_N}} h_{\mathbf{k}} \right]_{\mathbf{k} \rightarrow \mathbf{k} + (1 - \lambda) \mathbf{q} - (1 - \lambda_1) \mathbf{q}_1 \cdots - (1 - \lambda_N )\mathbf{q}_N} \overline{c}^{\dagger}_{n \mathbf{k}} \overline{c}_{m \mathbf{k} + \mathbf{q} - \mathbf{q}_1 \cdots - \mathbf{q}_N}\label{eq:orbitalGeneralCurrentOperator}.
\end{align}
Furthermore, in analogy with Eq.~\eqref{eq:vertexfuncgeneral} we can define the tight-binding velocity vertex
\begin{equation}
    \overline{v}^{\nu_N}_{(N)}(\mathbf{k}, \mathbf{q}_1, \cdots ,\mathbf{q}_N) \equiv \left[ \partial_{\mathbf{k}^\mu} \partial_{\mathbf{k}^{\nu_1}} \cdots \partial_{\mathbf{k}^{\nu_{N-1}}} h_{\mathbf{k}} \right]_{\mathbf{k} \rightarrow \mathbf{k}  + (1 - \lambda_1) \mathbf{q}_1 \cdots + (1 - \lambda_N )\mathbf{q}_N}.
\end{equation}
\end{widetext}
Note that we can express any of the diagrams and subsequent conductivities in Sec.~\ref{sec:DiagramRules} in either the orbital or tight-binding bases.

\section{The Kerr Effect Derivation}\label{app:KerrDerivations}
In this Appendix, we present a derivation of the Kerr effect for systems with an anisotropic dielectric tensor. 
In App.~\ref{app:KerrDerivationIn3D} we start by analyzing electromagnetic scattering from a 3D interface. 
Then, in App.~\ref{app:KerrDerivationIn2D} we apply these results to the experimentally-relevant situation of scattering from an encapsulated 2D sample considered in Sec.~\ref{sec:KerrRotation}.

\subsection{Kerr Effect Derivation in 3D Materials}\label{app:KerrDerivationIn3D}

Consider an interface between vacuum and a 3D semi-infinite slab of material, with the interface normal to the $z$-axis.
We take the material dielectric tensor $\epsilon$ to have the form~\cite{Catarina2020KerrIn2DDirac}
\begin{equation}
    \epsilon = \begin{bmatrix}
    \epsilon_{xx} & \epsilon_{xy} & 0
    \\
    -\epsilon_{xy} & \epsilon_{yy} & 0
    \\
    0 & 0 & \epsilon_{zz}
    \end{bmatrix}\label{eq:epsilonTensor}.
\end{equation}
We will also make the approximation  $\mu \sim 1$, which is tantamount to including all magnetic response in the transverse dielectric tensor~\cite{Kargarian2015KerrAndFaradayInWeyl, Oppeneer1999MOKEinFerromagnets, Ghosh2023KerrInWeyl}.

We now consider an incident light beam with electric field of the form
\begin{equation}\label{eq:Ein}
    \mathbf{E}_{\pm I} = {E_0} R(\theta_i) \cdot \begin{bmatrix}
        1
        \\
        0
        \\ 0
    \end{bmatrix} e^{-i \omega t + i k_i (y \sin(\theta_i) + z \cos(\theta_i))}.
\end{equation}
Here, $R(\theta_i)$ denotes a rotation matrix about the $x$-axis by angle $\theta_i$ (while this has no effect on $x$-polarized light such as in Eq.~\eqref{eq:Ein}, we retain it for ease of comparison with later steps in the derivation).
Notice that this electric field is linearly polarized and propagating in the direction of $\theta_i$, measured from the normal of the plane of the material.
The wavevector $k_i$ and frequency $\omega$ will be related later and $E_0$ is the real and positive amplitude.

We next consider the reflected field \cite{Creed2020MethodsforKerrDetection, Ghosh2023KerrInWeyl}:
\begin{equation}
    \mathbf{E}_{R} = E_0\begin{bmatrix}
        r_{x}
        \\
        r_{y}
        \\ r_{z}
    \end{bmatrix} e^{-i \omega t + i k_R (y \sin(\theta_R) + z \cos(\theta_R))}.
\end{equation}
Before considering the form of the transmitted field, we consider the differential equations it must satisfy.
Consider Maxwell's equations in a material~\cite{Jackson1999ClassicalEMBook, Oppeneer1999MOKEinFerromagnets, Wolski2014theoryOfEMFields}:
\begin{align}
    \nabla \cdot \mathbf{D}_T =& \rho_f\label{eq:Maxwell1},
    \\
    \nabla \times \mathbf{H}_T =& \mathbf{J}_f + \frac{\partial \mathbf{D}_T}{\partial t}\label{eq:equationForWave1},
    \\
    \nabla \cdot \mathbf{B}_T =& 0\label{eq:Maxwell2},
    \\
    \nabla \times \mathbf{E}_T =& -\frac{\partial \mathbf{B}_T}{\partial t}\label{eq:equationForWave2}.
\end{align}
Here, with the assumption $\mu \sim 1$, then $\mathbf{H}_T = \mathbf{B}_T$.
Also, $\mathbf{D}_T = \epsilon \mathbf{E}_T $, and we take $\rho_f=\mathbf{J}_f=0$.
We can now combine Eqs.~\eqref{eq:equationForWave1} and \eqref{eq:equationForWave2} to obtain a wave equation,
\begin{equation}
    \nabla\left( \nabla \cdot \mathbf{E}_T \right)-\nabla^2 \mathbf{E}_T = -\frac{\partial \mathbf{J}_f}{\partial t} - \frac{\partial^2 \epsilon \mathbf{E}_T}{\partial t^2}.\label{eq:waveEquationInMaterial}
\end{equation}

The fields that satisfy Eq.~\eqref{eq:waveEquationInMaterial} are of the form $t_{\ell} \mathbf{E}_{\ell} e^{-i \omega + i \mathbf{k}_{\ell} \cdot \mathbf{r}}$ where $\mathbf{E}_{\ell}$ is the $\ell$-th eigenvector of $\epsilon$ and $|\mathbf{k}_{\ell}|$ is proportional to the corresponding eigenvalue.
Given the form of the dielectric tensor in equation \eqref{eq:epsilonTensor}, we can solve the eigenvalue equation $\epsilon \mathbf{E} = n^2 \mathbf{E}$ \cite{Ebert1996MOKEInTransitionMetals}, to obtain the eigenvalues

\begin{align}
n^2_{p} &= \frac{1}{2}\left( \epsilon_{xx} + \epsilon_{yy} + \sqrt{\left( \epsilon_{xx} - \epsilon_{yy} \right)^2 - 4 \epsilon^2_{xy}} \right), \nonumber \\
n^2_{m} &= \frac{1}{2}\left( \epsilon_{xx} + \epsilon_{yy} - \sqrt{\left( \epsilon_{xx} - \epsilon_{yy} \right)^2 - 4 \epsilon^2_{xy}} \right),\nonumber \\
n^2_3 &= \epsilon_{zz}
\end{align}
 with normalized (in units of $E_0$) eigenvectors, $\mathbf{E}_{p}$, $\mathbf{E}_{m}$, and $\mathbf{E}_{3}$ respectively~\cite{Higo2018LargeMOKE, McGee1991MOKETheory}.
The meaning of $n_{\ell}^2$ is the refractive index (squared) in the material.

Next, let $\theta_{T \ell}$ be the angle of propagation of each eigenmode of the transmitted wave, measured with respect to the normal.
Satisfying the wave equation implies $|\mathbf{k}_{\ell}|^2 = n^2_{\ell} \omega^2 / c^2$.
Note that when we assign an angle, this also changes our eigenvalue equation to $\left[ R(\theta_{T\ell}) \epsilon R^{-1}(\theta_{T\ell}) \right] R(\theta_{T\ell}) \mathbf{E}_{\ell} = n_{\ell}^2 R(\theta_{T\ell}) \mathbf{E}_{\ell}$ but the eigenvalues remain the same even if the eigenvectors are now angle-dependent \cite{Ghosh2023KerrInWeyl}.
Given the required form of $|\mathbf{k}_{\ell}|^2$ to be a solution to Eq. \eqref{eq:waveEquationInMaterial}, we can now write down the generalized form of the solution of the transmitted electric field \cite{Ghosh2023KerrInWeyl, Higo2018LargeMOKE, Wolski2014theoryOfEMFields}:
\begin{align}
    \mathbf{E}_{T} = &t_p \mathbf{E}_p  e^{-i \omega t + i k_{T1} (y \sin(\theta_{T1}) + z \cos(\theta_{T1}))}\nonumber
    \\
    +& t_m \mathbf{E}_m  e^{-i \omega t + i k_{T2} (y \sin(\theta_{T2}) + z \cos(\theta_{T2}))} \nonumber
    \\
    +& t_3 \mathbf{E}_3  e^{-i \omega t + i k_{T3} (y \sin(\theta_{T3}) + z \cos(\theta_{T3}))},
\end{align}
where $t_{\ell}$ and $\mathbf{E}_{\ell}$ are angle dependent.

To now solve this electromagnetic scattering problem, we can first derive Snell's law for each of the transmitted modes.
Snell's law states that the argument in the exponential of all the electric fields must be equal at the boundary.
Therefore, we will now have a set of four equations that results in:
\begin{align}
    k_i \sin(\theta_i) &= k_R \sin(\theta_R),
    \\
    k_i \sin(\theta_i) &= k_{T1} \sin(\theta_{T1}),
    \\
    k_i \sin(\theta_i) &= k_{T2} \sin(\theta_{T2}),
    \\
    k_i \sin(\theta_i) &= k_{T3} \sin(\theta_{T3}).
\end{align}
It should be noted by the nature of reflection that $k _R= - k_i$ and $\theta_R= - \theta_i$.

We may now consider the remaining boundary conditions derived by requiring that Eqs.~\eqref{eq:Maxwell1}, \eqref{eq:equationForWave1}, \eqref{eq:Maxwell2}, and \eqref{eq:equationForWave2} are satisfied by components of the incident, reflected, and transmitted waves at the boundary.
The Snell's law equations ensure the position-dependent arguments in the exponential are the same, guaranteeing the reflection and transmission coefficients are position (and time) independent.
Following Ref.~\cite{Jackson1999ClassicalEMBook}, the boundary equations are~\cite{Creed2020MethodsforKerrDetection}
\begin{align}
    \left [ \epsilon_{\text{vacuum}} \cdot \left( \mathbf{E}_{\pm i} + \mathbf{E}_R \right) - \mathbf{D}_{T} \right] \cdot \mathbf{n} =& 0,
    \\
    \left [ \nabla \times \left( \mathbf{E}_{\pm i} + \mathbf{E}_R \right) - \nabla \times \mathbf{E}_T \right] \cdot \mathbf{n} =& 0,
    \\
    \left [  \mathbf{E}_{\pm i} + \mathbf{E}_R  - \mathbf{E}_{T} \right] \times \mathbf{n} =& 0,
    \\
    \left [ \nabla \times \left( \mathbf{E}_{\pm i} + \mathbf{E}_R \right) - \nabla \times \mathbf{E}_T \right] \times \mathbf{n} =& 0,
\end{align}
where $\mathbf{n} = [0, 0, 1]$, the normal to the 2D material plane.
We also observe, as good check, that $\nabla \cdot \mathbf{E}_R = 0$ and $\nabla \cdot \mathbf{E}_T = 0$ after resolving the coefficients through the boundary conditions.
Solving the transmitted coefficients in this basis results in $t_3 = 0$.

To solve for the Kerr angles, we first rotate our local coordinate system to the frame of the reflected wave, $R(\theta_R) \mathbf{E}_R$. 
Then we redefine the coefficients as $r_p = (r_x + i r_y)/2  $ and $r_m = (r_y - i r_y) / 2$, which describes the reflection coefficients for the right-hand and left-hand circular polarizations.
Each of these coefficients is, in general, complex so we may break up each according to complex polar coordinates, that is $r_p = |r_p| e^{i \alpha_p}$ and $r_m = |r_m| e^{i \alpha_m}$.
Kerr rotation results in a phase shift between right- and left-hand circularly polarized components of the reflected wave.
Furthermore the reflected wave can have an amplitude difference between right- and left-hand circularly polarized components.
This allows us to define\cite{Catarina2020KerrIn2DDirac, McGee1991MOKETheory, Oppeneer1999MOKEinFerromagnets, ahn2022KerrEffectAxions}
\begin{equation}
     \frac{r_p}{r_m} =\frac{|r_p|}{|r_m|} e^{i (\alpha_p - \alpha_m)} \equiv \frac{|r_p|}{|r_m|} e^{i (2 (\theta_{K} + i \epsilon_K))},\label{eq:kerrRotationDefinition}
\end{equation}
where $\theta_K$ is the Kerr rotation angle and $\epsilon_K$ is the ellipticity,
\begin{equation}
    \tan(\epsilon_K) = \frac{|r_p| - |r_m|}{|r_p| + |r_m|},\label{eq:kerrEllipticityDefinition}
\end{equation}
Our derivation expresses the Kerr angle and ellipticity in terms of components of the dielectric tensor. 
This can be related to the frequency and wavevector-dependent conductivity using~\cite{Catarina2020KerrIn2DDirac, Oppeneer1999MOKEinFerromagnets} 
\begin{equation}
\epsilon= \mathds{I} + \frac{4 \pi i}{\omega} \sigma,
\end{equation}
where $\mathds{I}$ is the $3\times 3$ identity matrix.

\subsection{Kerr Effect Derivation in 2D Materials with a Substrate}\label{app:KerrDerivationIn2D}
In this section we will extend our analysis of Kerr effect to 2D materials.
Most of the derivation will remain the same, but there are a few subtle changes.
In this setup there will be a vacuum, followed by a 2D thin film, and followed by a substrate which will have the permittivity $\epsilon_R$.
We will take the $\hat{z}$ axis to be normal to our 2D film, in the direction of the vacuum.

In this setup, the conductivity of the 2D materials now enters into Maxwell's equations as a boundary condition at the interface.
Therefore, the propagation speed of the transmitted wave will not be influenced by the film, only by the substrate in which it propagates.
That is, the the eigenvalue equation for the transmitted wave must satisfy $n^2_R \mathds{I}_{3} \mathbf{E}_{T,\text{2D}} = \epsilon_R \mathds{I}_{3} \mathbf{E}_{T,\text{2D}}$ \cite{Ebert1996MOKEInTransitionMetals, Catarina2020KerrIn2DDirac}. 
We can thus write the transmitted wave as
\begin{align}
    \mathbf{E}_{T,\text{2D}} = &R(\theta_T)\cdot \left( t_1 \mathbf{e}_1 + t_2 \mathbf{e}_2 + t_3 \mathbf{e}_3 \right) \nonumber
    \\
    &\times e^{ - i \omega t + i (R(\theta_T) \cdot (0, 0, \sqrt{\epsilon_R} k_{i})) \cdot \mathbf{r}},
\end{align}
where $\mathbf{e}_1 = (1, 0, 0)$, $\mathbf{e}_2 = (0, 1, 0)$, and $\mathbf{e}_3 = (0, 0, 1)$ are the eigenmodes of $\epsilon_R \mathds{I}_{3}$.
The form of the incident and reflected waves remain the same respectively
\begin{align}
    \mathbf{E}_{\pm I, \text{2D}} =& \frac{E_0}{\sqrt{2}} R(\theta_i) \cdot \begin{bmatrix}
        1
        \\
        0
        \\ 0
    \end{bmatrix} e^{-i \omega t + i (R(\theta_i) \cdot (0, 0, k_i)) \cdot \mathbf{r}},
    \\
    \mathbf{E}_{R, \text{2D}} =& R(-\theta_i) \cdot \begin{bmatrix}
        r_{x}
        \\
        r_{y}
        \\ r_{z}
    \end{bmatrix} e^{-i \omega t + i (R(-\theta_i) \cdot (0, 0, -k_i)) \cdot \mathbf{r}}.
\end{align}
The Snell's law found by comparing the exponentials is also similar to before
\begin{align}
    k_i \sin(\theta_i) &= -k_i \sin(\theta_R),
    \\
    \sin(\theta_i) &= \sqrt{\epsilon_R} \sin(\theta_{T}).
\end{align}

Since we have a 2D material with nonzero conductivity tensor, a surface current is generated at the interface. 
As such, the boundary conditions get modified to include this surface current by way of the surface conductivity, $\mathbf{j}_s = \sigma^{\text{2D}} \cdot \mathbf{E}_{T, \text{2D}}$, which enters into Maxwell's equations as a free current on the surface.
Since $\nabla \cdot \mathbf{E}_{R, \text{2D}} = 0$ and $\nabla \cdot \mathbf{E}_{T, \text{2D}} = 0$ force $t_3 = r_z = 0$, we arrive at four equations arising from two boundary conditions~\cite{Kargarian2015KerrAndFaradayInWeyl, Shah2021FaradayAndKerrInThinFilms},
\begin{align}
    \left [  \mathbf{E}_{\pm i, \text{2D}} + \mathbf{E}_{R, \text{2D}}  - \mathbf{E}_{T} \right] \times \mathbf{n} = 0&,
    \\
    \frac{c}{i \omega}\left [ \nabla \times \left( \mathbf{E}_{\pm i, \text{2D}} + \mathbf{E}_{R, \text{2D}} \right) - \nabla \times \mathbf{E}_{T, \text{2D}} \right] \times \mathbf{n} &\nonumber
    \\
    = \sigma^{\text{2D}} \cdot \mathbf{E}_{T, \text{2D}}&.
\end{align}
Since $\sigma^{\text{2D}}$ is a property of the 2D sample, it takes the form
\begin{equation}
    \sigma^{\text{2D}} = \begin{bmatrix}
        \sigma^{\text{2D}}_{xx} & \sigma^{\text{2D}}_{xy} & 0
        \\
        \sigma^{\text{2D}}_{yx} & \sigma^{\text{2D}}_{yy} & 0
        \\ 0 & 0 & 0
    \end{bmatrix},
\end{equation}
where the bulk conductivity is related to the surface conductivity by $\sigma^{\text{2D}} = d \sigma^{\text{3D}}$ \cite{Catarina2020KerrIn2DDirac}. 
The Kerr rotation and ellipticity can be extracted using Eqs. \eqref{eq:kerrRotationDefinition} and \eqref{eq:kerrEllipticityDefinition} respectively.

\section{Consistency with Plasmon Dispersion}\label{app:ConsistencyWithPlasmonDispersion}
In this section, we examine the implications of Eqs.~\eqref{eq:mainj} and~\eqref{eq:fullGeneralizedLinearConductivity} for the dispersion of collective plasmon modes.
Since the plasmon dispersion depends on the $\mathbf{q}$ dependence of the density-density calculation, it serves as a consistency check on Eqs.~\eqref{eq:mainj} and the sum rule from Eq.~\eqref{eq:fsum}.

To begin, we will treat the Coulomb interaction between electrons in the random phase approximation (RPA).
That is, in writing the dielectric function, only the first bubble diagram is considered, such that 
\begin{equation}
    \epsilon(\omega, \mathbf{q}) = 1 - v_c(q) \Pi(\omega, \mathbf{q}).\label{eq:plasmonDielectricStart}
\end{equation}
Here, $v_c(\mathbf{q})$ is the Fourier transformed Coulomb interaction (in 2D, $v_c(q) = 2 \pi e^2 / \kappa q$ and  $v_c(q) = 4 \pi e^2 / \kappa q^2$ in 3D) \cite{Hwang2007PlasmonDispersionDirac, Mishra2012PlasmonDispersionDirac}.
The function $\Pi(\omega, \mathbf{q})$ is the polarizability or the electric susceptibility for noninteracting electrons.
The electric susceptibility is related to the conductivity through the continuity equation.
First consider a longitudinal electric field, $\mathbf{E}_\mathbf{q}=-i\mathbf{q}A_0(\omega,\mathbf{q})$, defined in terms of a scalar potential $A_0$.
The current that flows in response to this field is $\mathbf{j}(\omega, \mathbf{q}) = -i \sigma(\omega, \mathbf{q}) \cdot \mathbf{q} A_0(\omega, \mathbf{q})$ 
Additionally, from the continuity equation, $ -e\partial_t n(t, \mathbf{x}) + \nabla \cdot \mathbf{j}(t, \mathbf{r}) = 0$. 
Expressing the density in terms of the response function $\Pi$, we find \cite{Jani2003DensityDensityPlasmon}:
\begin{equation}
    i \omega e^2 \Pi(\omega, \mathbf{q})  + \mathbf{q} \cdot \sigma(\omega, \mathbf{q}) \cdot \mathbf{q}= 0.\label{eq:continuityForDensityDensity}
\end{equation}

Now Eq. \ref{eq:continuityForDensityDensity} may be substituted into Eq. \ref{eq:plasmonDielectricStart} to obtain the relation
\begin{equation}
    \epsilon(\omega, \mathbf{q}) = 1 + \frac{i}{\omega e^2}v_c(\mathbf{q}) \left( \mathbf{q} \cdot \sigma(\omega, \mathbf{q}) \cdot \mathbf{q} \right).\label{eq:plasmonDielectric}
\end{equation}
Note that since our conserved current in Eq.~\eqref{eq:mainj} satisfies the continuity equation, our formalism ensures that Eq.~\eqref{eq:plasmonDielectric} is obeyed.
To solve for the plasmon dispersion, we set Eq. \eqref{eq:plasmonDielectric} equal to zero and solve for $\omega(\mathbf{q})$, the plasma frequency as a function of $\mathbf{q}$. 

Since our conserved current Eq.~\eqref{eq:mainj} obeys the continuity equation, we can use the conductivity Eq.~\eqref{eq:fullGeneralizedLinearConductivity} to determine the RPA plasmon dispersion in generic models. 
By contrast, since the midpoint and trapezoid definitions of the current, Eqs.~\eqref{eq:trapfirsttime} and \eqref{eq:midpointfirsttime}, do not obey the continuity equation at large $\mathbf{q}$, they will not give reliable estimates of the plasmon dispersion.

\bibliography{refs}

\begin{thebibliography}{165}%
\makeatletter
\providecommand \@ifxundefined [1]{%
 \@ifx{#1\undefined}
}%
\providecommand \@ifnum [1]{%
 \ifnum #1\expandafter \@firstoftwo
 \else \expandafter \@secondoftwo
 \fi
}%
\providecommand \@ifx [1]{%
 \ifx #1\expandafter \@firstoftwo
 \else \expandafter \@secondoftwo
 \fi
}%
\providecommand \natexlab [1]{#1}%
\providecommand \enquote  [1]{``#1''}%
\providecommand \bibnamefont  [1]{#1}%
\providecommand \bibfnamefont [1]{#1}%
\providecommand \citenamefont [1]{#1}%
\providecommand \href@noop [0]{\@secondoftwo}%
\providecommand \href [0]{\begingroup \@sanitize@url \@href}%
\providecommand \@href[1]{\@@startlink{#1}\@@href}%
\providecommand \@@href[1]{\endgroup#1\@@endlink}%
\providecommand \@sanitize@url [0]{\catcode `\\12\catcode `\$12\catcode
  `\&12\catcode `\#12\catcode `\^12\catcode `\_12\catcode `\%12\relax}%
\providecommand \@@startlink[1]{}%
\providecommand \@@endlink[0]{}%
\providecommand \url  [0]{\begingroup\@sanitize@url \@url }%
\providecommand \@url [1]{\endgroup\@href {#1}{\urlprefix }}%
\providecommand \urlprefix  [0]{URL }%
\providecommand \Eprint [0]{\href }%
\providecommand \doibase [0]{https://doi.org/}%
\providecommand \selectlanguage [0]{\@gobble}%
\providecommand \bibinfo  [0]{\@secondoftwo}%
\providecommand \bibfield  [0]{\@secondoftwo}%
\providecommand \translation [1]{[#1]}%
\providecommand \BibitemOpen [0]{}%
\providecommand \bibitemStop [0]{}%
\providecommand \bibitemNoStop [0]{.\EOS\space}%
\providecommand \EOS [0]{\spacefactor3000\relax}%
\providecommand \BibitemShut  [1]{\csname bibitem#1\endcsname}%
\let\auto@bib@innerbib\@empty
\bibitem [{\citenamefont {Martin}(1988)}]{martin1988sum}%
  \BibitemOpen
  \bibfield  {author} {\bibinfo {author} {\bibfnamefont {P.~A.}\ \bibnamefont
  {Martin}},\ }\bibfield  {title} {\bibinfo {title} {Sum rules in charged
  fluids},\ }\href@noop {} {\bibfield  {journal} {\bibinfo  {journal} {Reviews
  of Modern Physics}\ }\textbf {\bibinfo {volume} {60}},\ \bibinfo {pages}
  {1075} (\bibinfo {year} {1988})}\BibitemShut {NoStop}%
\bibitem [{\citenamefont {Nozieres}\ and\ \citenamefont
  {Pines}(1959)}]{nozieres1959electron}%
  \BibitemOpen
  \bibfield  {author} {\bibinfo {author} {\bibfnamefont {P.}~\bibnamefont
  {Nozieres}}\ and\ \bibinfo {author} {\bibfnamefont {D.}~\bibnamefont
  {Pines}},\ }\bibfield  {title} {\bibinfo {title} {Electron interaction in
  solids. characteristic energy loss spectrum},\ }\href@noop {} {\bibfield
  {journal} {\bibinfo  {journal} {Physical Review}\ }\textbf {\bibinfo {volume}
  {113}},\ \bibinfo {pages} {1254} (\bibinfo {year} {1959})}\BibitemShut
  {NoStop}%
\bibitem [{\citenamefont {Watanabe}\ and\ \citenamefont
  {Oshikawa}(2020)}]{watanabe2020generalized}%
  \BibitemOpen
  \bibfield  {author} {\bibinfo {author} {\bibfnamefont {H.}~\bibnamefont
  {Watanabe}}\ and\ \bibinfo {author} {\bibfnamefont {M.}~\bibnamefont
  {Oshikawa}},\ }\bibfield  {title} {\bibinfo {title} {Generalized f-sum rules
  and kohn formulas on nonlinear conductivities},\ }\href@noop {} {\bibfield
  {journal} {\bibinfo  {journal} {Physical Review B}\ }\textbf {\bibinfo
  {volume} {102}},\ \bibinfo {pages} {165137} (\bibinfo {year}
  {2020})}\BibitemShut {NoStop}%
\bibitem [{\citenamefont {Graf}\ and\ \citenamefont
  {Vogl}(1995)}]{graf1995electromagnetic}%
  \BibitemOpen
  \bibfield  {author} {\bibinfo {author} {\bibfnamefont {M.}~\bibnamefont
  {Graf}}\ and\ \bibinfo {author} {\bibfnamefont {P.}~\bibnamefont {Vogl}},\
  }\bibfield  {title} {\bibinfo {title} {Electromagnetic fields and dielectric
  response in empirical tight-binding theory},\ }\href@noop {} {\bibfield
  {journal} {\bibinfo  {journal} {Physical Review B}\ }\textbf {\bibinfo
  {volume} {51}},\ \bibinfo {pages} {4940} (\bibinfo {year}
  {1995})}\BibitemShut {NoStop}%
\bibitem [{\citenamefont {Gross}\ and\ \citenamefont
  {Kohn}(1985)}]{gross1985local}%
  \BibitemOpen
  \bibfield  {author} {\bibinfo {author} {\bibfnamefont {E.}~\bibnamefont
  {Gross}}\ and\ \bibinfo {author} {\bibfnamefont {W.}~\bibnamefont {Kohn}},\
  }\bibfield  {title} {\bibinfo {title} {Local density-functional theory of
  frequency-dependent linear response},\ }\href@noop {} {\bibfield  {journal}
  {\bibinfo  {journal} {Physical review letters}\ }\textbf {\bibinfo {volume}
  {55}},\ \bibinfo {pages} {2850} (\bibinfo {year} {1985})}\BibitemShut
  {NoStop}%
\bibitem [{\citenamefont {Gross}\ \emph {et~al.}(2005)\citenamefont {Gross},
  \citenamefont {Dobson},\ and\ \citenamefont {Petersilka}}]{gross2005density}%
  \BibitemOpen
  \bibfield  {author} {\bibinfo {author} {\bibfnamefont {a.~E.}\ \bibnamefont
  {Gross}}, \bibinfo {author} {\bibfnamefont {J.}~\bibnamefont {Dobson}},\ and\
  \bibinfo {author} {\bibfnamefont {M.}~\bibnamefont {Petersilka}},\ }\bibfield
   {title} {\bibinfo {title} {Density functional theory of time-dependent
  phenomena},\ }\href@noop {} {\bibfield  {journal} {\bibinfo  {journal}
  {Density Functional Theory II: Relativistic and Time Dependent Extensions}\
  ,\ \bibinfo {pages} {81}} (\bibinfo {year} {2005})}\BibitemShut {NoStop}%
\bibitem [{\citenamefont {Yao}\ \emph {et~al.}(2004)\citenamefont {Yao},
  \citenamefont {Kleinman}, \citenamefont {MacDonald}, \citenamefont {Sinova},
  \citenamefont {Jungwirth}, \citenamefont {Wang}, \citenamefont {Wang},\ and\
  \citenamefont {Niu}}]{yao2004first}%
  \BibitemOpen
  \bibfield  {author} {\bibinfo {author} {\bibfnamefont {Y.}~\bibnamefont
  {Yao}}, \bibinfo {author} {\bibfnamefont {L.}~\bibnamefont {Kleinman}},
  \bibinfo {author} {\bibfnamefont {A.}~\bibnamefont {MacDonald}}, \bibinfo
  {author} {\bibfnamefont {J.}~\bibnamefont {Sinova}}, \bibinfo {author}
  {\bibfnamefont {T.}~\bibnamefont {Jungwirth}}, \bibinfo {author}
  {\bibfnamefont {D.-s.}\ \bibnamefont {Wang}}, \bibinfo {author}
  {\bibfnamefont {E.}~\bibnamefont {Wang}},\ and\ \bibinfo {author}
  {\bibfnamefont {Q.}~\bibnamefont {Niu}},\ }\bibfield  {title} {\bibinfo
  {title} {First principles calculation of anomalous hall conductivity in
  ferromagnetic bcc fe},\ }\href@noop {} {\bibfield  {journal} {\bibinfo
  {journal} {Physical review letters}\ }\textbf {\bibinfo {volume} {92}},\
  \bibinfo {pages} {037204} (\bibinfo {year} {2004})}\BibitemShut {NoStop}%
\bibitem [{\citenamefont {K{\"u}hne}\ \emph {et~al.}(2020)\citenamefont
  {K{\"u}hne}, \citenamefont {Heske},\ and\ \citenamefont
  {Prodan}}]{kuhne2020disordered}%
  \BibitemOpen
  \bibfield  {author} {\bibinfo {author} {\bibfnamefont {T.~D.}\ \bibnamefont
  {K{\"u}hne}}, \bibinfo {author} {\bibfnamefont {J.}~\bibnamefont {Heske}},\
  and\ \bibinfo {author} {\bibfnamefont {E.}~\bibnamefont {Prodan}},\
  }\bibfield  {title} {\bibinfo {title} {Disordered crystals from first
  principles ii: Transport coefficients},\ }\href@noop {} {\bibfield  {journal}
  {\bibinfo  {journal} {Annals of Physics}\ }\textbf {\bibinfo {volume}
  {421}},\ \bibinfo {pages} {168290} (\bibinfo {year} {2020})}\BibitemShut
  {NoStop}%
\bibitem [{\citenamefont {Veithen}\ \emph {et~al.}(2005)\citenamefont
  {Veithen}, \citenamefont {Gonze},\ and\ \citenamefont
  {Ghosez}}]{veithen2005nonlinear}%
  \BibitemOpen
  \bibfield  {author} {\bibinfo {author} {\bibfnamefont {M.}~\bibnamefont
  {Veithen}}, \bibinfo {author} {\bibfnamefont {X.}~\bibnamefont {Gonze}},\
  and\ \bibinfo {author} {\bibfnamefont {P.}~\bibnamefont {Ghosez}},\
  }\bibfield  {title} {\bibinfo {title} {Nonlinear optical susceptibilities,
  raman efficiencies, and electro-optic tensors from first-principles density
  functional perturbation theory},\ }\href@noop {} {\bibfield  {journal}
  {\bibinfo  {journal} {Physical Review B}\ }\textbf {\bibinfo {volume} {71}},\
  \bibinfo {pages} {125107} (\bibinfo {year} {2005})}\BibitemShut {NoStop}%
\bibitem [{\citenamefont {Baroni}\ \emph {et~al.}(2001)\citenamefont {Baroni},
  \citenamefont {De~Gironcoli}, \citenamefont {Dal~Corso},\ and\ \citenamefont
  {Giannozzi}}]{baroni2001phonons}%
  \BibitemOpen
  \bibfield  {author} {\bibinfo {author} {\bibfnamefont {S.}~\bibnamefont
  {Baroni}}, \bibinfo {author} {\bibfnamefont {S.}~\bibnamefont
  {De~Gironcoli}}, \bibinfo {author} {\bibfnamefont {A.}~\bibnamefont
  {Dal~Corso}},\ and\ \bibinfo {author} {\bibfnamefont {P.}~\bibnamefont
  {Giannozzi}},\ }\bibfield  {title} {\bibinfo {title} {Phonons and related
  crystal properties from density-functional perturbation theory},\ }\href@noop
  {} {\bibfield  {journal} {\bibinfo  {journal} {Reviews of modern Physics}\
  }\textbf {\bibinfo {volume} {73}},\ \bibinfo {pages} {515} (\bibinfo {year}
  {2001})}\BibitemShut {NoStop}%
\bibitem [{\citenamefont {Thouless}\ \emph {et~al.}(1982)\citenamefont
  {Thouless}, \citenamefont {Kohmoto}, \citenamefont {Nightingale},\ and\
  \citenamefont {{den Nijs}}}]{thouless1982quantized}%
  \BibitemOpen
  \bibfield  {author} {\bibinfo {author} {\bibfnamefont {D.~J.}\ \bibnamefont
  {Thouless}}, \bibinfo {author} {\bibfnamefont {M.}~\bibnamefont {Kohmoto}},
  \bibinfo {author} {\bibfnamefont {M.~P.}\ \bibnamefont {Nightingale}},\ and\
  \bibinfo {author} {\bibfnamefont {M.}~\bibnamefont {{den Nijs}}},\ }\bibfield
   {title} {\bibinfo {title} {Quantized {{Hall}} conductance in a
  two-dimensional periodic potential},\ }\href@noop {} {\bibfield  {journal}
  {\bibinfo  {journal} {Physical review letters}\ }\textbf {\bibinfo {volume}
  {49}},\ \bibinfo {pages} {405} (\bibinfo {year} {1982})}\BibitemShut
  {NoStop}%
\bibitem [{\citenamefont {Niu}\ \emph {et~al.}(1985)\citenamefont {Niu},
  \citenamefont {Thouless},\ and\ \citenamefont {Wu}}]{niu1985quantized}%
  \BibitemOpen
  \bibfield  {author} {\bibinfo {author} {\bibfnamefont {Q.}~\bibnamefont
  {Niu}}, \bibinfo {author} {\bibfnamefont {D.~J.}\ \bibnamefont {Thouless}},\
  and\ \bibinfo {author} {\bibfnamefont {Y.-S.}\ \bibnamefont {Wu}},\
  }\bibfield  {title} {\bibinfo {title} {Quantized {{Hall}} conductance as a
  topological invariant},\ }\href@noop {} {\bibfield  {journal} {\bibinfo
  {journal} {Physical Review B}\ }\textbf {\bibinfo {volume} {31}},\ \bibinfo
  {pages} {3372} (\bibinfo {year} {1985})}\BibitemShut {NoStop}%
\bibitem [{\citenamefont {Xiao}\ \emph {et~al.}(2019)\citenamefont {Xiao},
  \citenamefont {Du},\ and\ \citenamefont
  {Niu}}]{Xiao2019NonlinearFromClassical}%
  \BibitemOpen
  \bibfield  {author} {\bibinfo {author} {\bibfnamefont {C.}~\bibnamefont
  {Xiao}}, \bibinfo {author} {\bibfnamefont {Z.~Z.}\ \bibnamefont {Du}},\ and\
  \bibinfo {author} {\bibfnamefont {Q.}~\bibnamefont {Niu}},\ }\bibfield
  {title} {\bibinfo {title} {Theory of nonlinear hall effects: Modified
  semiclassics from quantum kinetics},\ }\bibfield  {journal} {\bibinfo
  {journal} {Physical Review B}\ }\textbf {\bibinfo {volume} {100}},\ \href
  {https://doi.org/10.1103/physrevb.100.165422} {10.1103/physrevb.100.165422}
  (\bibinfo {year} {2019})\BibitemShut {NoStop}%
\bibitem [{\citenamefont {Sodemann}\ and\ \citenamefont
  {Fu}(2015)}]{Sodemann2015NonlinearHallEffect}%
  \BibitemOpen
  \bibfield  {author} {\bibinfo {author} {\bibfnamefont {I.}~\bibnamefont
  {Sodemann}}\ and\ \bibinfo {author} {\bibfnamefont {L.}~\bibnamefont {Fu}},\
  }\bibfield  {title} {\bibinfo {title} {Quantum nonlinear {Hall} effect
  induced by {Berry} curvature dipole in time-reversal invariant materials},\
  }\href {https://doi.org/10.1103/PhysRevLett.115.216806} {\bibfield  {journal}
  {\bibinfo  {journal} {Physical Review Letters}\ }\textbf {\bibinfo {volume}
  {115}},\ \bibinfo {pages} {216806} (\bibinfo {year} {2015})},\ \bibinfo
  {note} {arXiv: 1508.00571}\BibitemShut {NoStop}%
\bibitem [{\citenamefont {De~Juan}\ \emph {et~al.}(2017)\citenamefont
  {De~Juan}, \citenamefont {Grushin}, \citenamefont {Morimoto},\ and\
  \citenamefont {Moore}}]{dejuan2017quantized}%
  \BibitemOpen
  \bibfield  {author} {\bibinfo {author} {\bibfnamefont {F.}~\bibnamefont
  {De~Juan}}, \bibinfo {author} {\bibfnamefont {A.~G.}\ \bibnamefont
  {Grushin}}, \bibinfo {author} {\bibfnamefont {T.}~\bibnamefont {Morimoto}},\
  and\ \bibinfo {author} {\bibfnamefont {J.~E.}\ \bibnamefont {Moore}},\
  }\bibfield  {title} {\bibinfo {title} {Quantized circular photogalvanic
  effect in {{Weyl}} semimetals},\ }\href@noop {} {\bibfield  {journal}
  {\bibinfo  {journal} {Nature communications}\ }\textbf {\bibinfo {volume}
  {8}},\ \bibinfo {pages} {15995} (\bibinfo {year} {2017})}\BibitemShut
  {NoStop}%
\bibitem [{\citenamefont {Flicker}\ \emph {et~al.}(2018)\citenamefont
  {Flicker}, \citenamefont {De~Juan}, \citenamefont {Bradlyn}, \citenamefont
  {Morimoto}, \citenamefont {Vergniory},\ and\ \citenamefont
  {Grushin}}]{flicker2018chiral}%
  \BibitemOpen
  \bibfield  {author} {\bibinfo {author} {\bibfnamefont {F.}~\bibnamefont
  {Flicker}}, \bibinfo {author} {\bibfnamefont {F.}~\bibnamefont {De~Juan}},
  \bibinfo {author} {\bibfnamefont {B.}~\bibnamefont {Bradlyn}}, \bibinfo
  {author} {\bibfnamefont {T.}~\bibnamefont {Morimoto}}, \bibinfo {author}
  {\bibfnamefont {M.~G.}\ \bibnamefont {Vergniory}},\ and\ \bibinfo {author}
  {\bibfnamefont {A.~G.}\ \bibnamefont {Grushin}},\ }\bibfield  {title}
  {\bibinfo {title} {Chiral optical response of multifold fermions},\
  }\href@noop {} {\bibfield  {journal} {\bibinfo  {journal} {Physical Review
  B}\ }\textbf {\bibinfo {volume} {98}},\ \bibinfo {pages} {155145} (\bibinfo
  {year} {2018})}\BibitemShut {NoStop}%
\bibitem [{\citenamefont {Alexandradinata}\ and\ \citenamefont
  {Zhu}(2023)}]{alexandradinata2023anomalous}%
  \BibitemOpen
  \bibfield  {author} {\bibinfo {author} {\bibfnamefont {A.}~\bibnamefont
  {Alexandradinata}}\ and\ \bibinfo {author} {\bibfnamefont {P.}~\bibnamefont
  {Zhu}},\ }\bibfield  {title} {\bibinfo {title} {Anomalous shift and optical
  vorticity in the steady photovoltaic current},\ }\href@noop {} {\bibfield
  {journal} {\bibinfo  {journal} {arXiv preprint arXiv:2308.08596}\ } (\bibinfo
  {year} {2023})}\BibitemShut {NoStop}%
\bibitem [{\citenamefont {Essin}\ \emph {et~al.}(2009)\citenamefont {Essin},
  \citenamefont {Moore},\ and\ \citenamefont
  {Vanderbilt}}]{essin2009magnetoelectric}%
  \BibitemOpen
  \bibfield  {author} {\bibinfo {author} {\bibfnamefont {A.~M.}\ \bibnamefont
  {Essin}}, \bibinfo {author} {\bibfnamefont {J.~E.}\ \bibnamefont {Moore}},\
  and\ \bibinfo {author} {\bibfnamefont {D.}~\bibnamefont {Vanderbilt}},\
  }\bibfield  {title} {\bibinfo {title} {Magnetoelectric polarizability and
  axion electrodynamics in crystalline insulators},\ }\href
  {https://doi.org/10.1103/PhysRevLett.102.146805} {\bibfield  {journal}
  {\bibinfo  {journal} {Phys. Rev. Lett.}\ }\textbf {\bibinfo {volume} {102}},\
  \bibinfo {pages} {146805} (\bibinfo {year} {2009})}\BibitemShut {NoStop}%
\bibitem [{\citenamefont {Qi}\ \emph {et~al.}(2008)\citenamefont {Qi},
  \citenamefont {Hughes},\ and\ \citenamefont {Zhang}}]{qi2008topological}%
  \BibitemOpen
  \bibfield  {author} {\bibinfo {author} {\bibfnamefont {X.-L.}\ \bibnamefont
  {Qi}}, \bibinfo {author} {\bibfnamefont {T.~L.}\ \bibnamefont {Hughes}},\
  and\ \bibinfo {author} {\bibfnamefont {S.-C.}\ \bibnamefont {Zhang}},\
  }\bibfield  {title} {\bibinfo {title} {Topological field theory of
  time-reversal invariant insulators},\ }\href
  {https://doi.org/10.1103/PhysRevB.78.195424} {\bibfield  {journal} {\bibinfo
  {journal} {Phys. Rev. B}\ }\textbf {\bibinfo {volume} {78}},\ \bibinfo
  {pages} {195424} (\bibinfo {year} {2008})}\BibitemShut {NoStop}%
\bibitem [{\citenamefont {Wu}\ \emph {et~al.}(2016)\citenamefont {Wu},
  \citenamefont {Salehi}, \citenamefont {Koirala}, \citenamefont {Moon},
  \citenamefont {Oh},\ and\ \citenamefont {Armitage}}]{wu2016quantized}%
  \BibitemOpen
  \bibfield  {author} {\bibinfo {author} {\bibfnamefont {L.}~\bibnamefont
  {Wu}}, \bibinfo {author} {\bibfnamefont {M.}~\bibnamefont {Salehi}}, \bibinfo
  {author} {\bibfnamefont {N.}~\bibnamefont {Koirala}}, \bibinfo {author}
  {\bibfnamefont {J.}~\bibnamefont {Moon}}, \bibinfo {author} {\bibfnamefont
  {S.}~\bibnamefont {Oh}},\ and\ \bibinfo {author} {\bibfnamefont
  {N.}~\bibnamefont {Armitage}},\ }\bibfield  {title} {\bibinfo {title}
  {Quantized faraday and kerr rotation and axion electrodynamics of a 3d
  topological insulator},\ }\href@noop {} {\bibfield  {journal} {\bibinfo
  {journal} {Science}\ }\textbf {\bibinfo {volume} {354}},\ \bibinfo {pages}
  {1124} (\bibinfo {year} {2016})}\BibitemShut {NoStop}%
\bibitem [{\citenamefont {Dziom}\ \emph {et~al.}(2017)\citenamefont {Dziom},
  \citenamefont {Shuvaev}, \citenamefont {Pimenov}, \citenamefont {Astakhov},
  \citenamefont {Ames}, \citenamefont {Bendias}, \citenamefont {B{\"o}ttcher},
  \citenamefont {Tkachov}, \citenamefont {Hankiewicz}, \citenamefont
  {Br{\"u}ne} \emph {et~al.}}]{dziom2017observation}%
  \BibitemOpen
  \bibfield  {author} {\bibinfo {author} {\bibfnamefont {V.}~\bibnamefont
  {Dziom}}, \bibinfo {author} {\bibfnamefont {A.}~\bibnamefont {Shuvaev}},
  \bibinfo {author} {\bibfnamefont {A.}~\bibnamefont {Pimenov}}, \bibinfo
  {author} {\bibfnamefont {G.}~\bibnamefont {Astakhov}}, \bibinfo {author}
  {\bibfnamefont {C.}~\bibnamefont {Ames}}, \bibinfo {author} {\bibfnamefont
  {K.}~\bibnamefont {Bendias}}, \bibinfo {author} {\bibfnamefont
  {J.}~\bibnamefont {B{\"o}ttcher}}, \bibinfo {author} {\bibfnamefont
  {G.}~\bibnamefont {Tkachov}}, \bibinfo {author} {\bibfnamefont
  {E.}~\bibnamefont {Hankiewicz}}, \bibinfo {author} {\bibfnamefont
  {C.}~\bibnamefont {Br{\"u}ne}}, \emph {et~al.},\ }\bibfield  {title}
  {\bibinfo {title} {Observation of the universal magnetoelectric effect in a
  3d topological insulator},\ }\href@noop {} {\bibfield  {journal} {\bibinfo
  {journal} {Nature communications}\ }\textbf {\bibinfo {volume} {8}},\
  \bibinfo {pages} {15197} (\bibinfo {year} {2017})}\BibitemShut {NoStop}%
\bibitem [{\citenamefont {Ahn}\ \emph {et~al.}(2022{\natexlab{a}})\citenamefont
  {Ahn}, \citenamefont {Xu},\ and\ \citenamefont
  {Vishwanath}}]{ahn2022KerrEffectAxions}%
  \BibitemOpen
  \bibfield  {author} {\bibinfo {author} {\bibfnamefont {J.}~\bibnamefont
  {Ahn}}, \bibinfo {author} {\bibfnamefont {S.-Y.}\ \bibnamefont {Xu}},\ and\
  \bibinfo {author} {\bibfnamefont {A.}~\bibnamefont {Vishwanath}},\ }\bibfield
   {title} {\bibinfo {title} {Theory of optical axion electrodynamics and
  application to the {Kerr} effect in topological antiferromagnets},\ }\href
  {https://doi.org/10.1038/s41467-022-35248-8} {\bibfield  {journal} {\bibinfo
  {journal} {Nature Communications}\ }\textbf {\bibinfo {volume} {13}},\
  \bibinfo {pages} {7615} (\bibinfo {year} {2022}{\natexlab{a}})}\BibitemShut
  {NoStop}%
\bibitem [{\citenamefont {Nielsen}\ and\ \citenamefont
  {Ninomiya}(1983)}]{nielsen1983adlerbelljackiw}%
  \BibitemOpen
  \bibfield  {author} {\bibinfo {author} {\bibfnamefont {H.~B.}\ \bibnamefont
  {Nielsen}}\ and\ \bibinfo {author} {\bibfnamefont {M.}~\bibnamefont
  {Ninomiya}},\ }\bibfield  {title} {\bibinfo {title} {The
  {{Adler-Bell-Jackiw}} anomaly and {{Weyl}} fermions in a crystal},\
  }\href@noop {} {\bibfield  {journal} {\bibinfo  {journal} {Physics Letters
  B}\ }\textbf {\bibinfo {volume} {130}},\ \bibinfo {pages} {389} (\bibinfo
  {year} {1983})}\BibitemShut {NoStop}%
\bibitem [{\citenamefont {Huang}\ \emph {et~al.}(2015)\citenamefont {Huang},
  \citenamefont {Zhao}, \citenamefont {Long}, \citenamefont {Wang},
  \citenamefont {Chen}, \citenamefont {Yang}, \citenamefont {Liang},
  \citenamefont {Xue}, \citenamefont {Weng}, \citenamefont {Fang},
  \citenamefont {Dai},\ and\ \citenamefont {Chen}}]{huang2015observation}%
  \BibitemOpen
  \bibfield  {author} {\bibinfo {author} {\bibfnamefont {X.}~\bibnamefont
  {Huang}}, \bibinfo {author} {\bibfnamefont {L.}~\bibnamefont {Zhao}},
  \bibinfo {author} {\bibfnamefont {Y.}~\bibnamefont {Long}}, \bibinfo {author}
  {\bibfnamefont {P.}~\bibnamefont {Wang}}, \bibinfo {author} {\bibfnamefont
  {D.}~\bibnamefont {Chen}}, \bibinfo {author} {\bibfnamefont {Z.}~\bibnamefont
  {Yang}}, \bibinfo {author} {\bibfnamefont {H.}~\bibnamefont {Liang}},
  \bibinfo {author} {\bibfnamefont {M.}~\bibnamefont {Xue}}, \bibinfo {author}
  {\bibfnamefont {H.}~\bibnamefont {Weng}}, \bibinfo {author} {\bibfnamefont
  {Z.}~\bibnamefont {Fang}}, \bibinfo {author} {\bibfnamefont {X.}~\bibnamefont
  {Dai}},\ and\ \bibinfo {author} {\bibfnamefont {G.}~\bibnamefont {Chen}},\
  }\bibfield  {title} {\bibinfo {title} {Observation of the
  {{Chiral-Anomaly-Induced Negative Magnetoresistance}} in {{3D Weyl Semimetal
  TaAs}}},\ }\href {https://doi.org/10.1103/PhysRevX.5.031023} {\bibfield
  {journal} {\bibinfo  {journal} {Phys. Rev. X}\ }\textbf {\bibinfo {volume}
  {5}},\ \bibinfo {pages} {031023} (\bibinfo {year} {2015})}\BibitemShut
  {NoStop}%
\bibitem [{\citenamefont {Hirschberger}\ \emph {et~al.}(2016)\citenamefont
  {Hirschberger}, \citenamefont {Kushwaha}, \citenamefont {Wang}, \citenamefont
  {Gibson}, \citenamefont {Liang}, \citenamefont {Belvin}, \citenamefont
  {Bernevig}, \citenamefont {Cava},\ and\ \citenamefont
  {Ong}}]{hirschberger2016chiral}%
  \BibitemOpen
  \bibfield  {author} {\bibinfo {author} {\bibfnamefont {M.}~\bibnamefont
  {Hirschberger}}, \bibinfo {author} {\bibfnamefont {S.}~\bibnamefont
  {Kushwaha}}, \bibinfo {author} {\bibfnamefont {Z.}~\bibnamefont {Wang}},
  \bibinfo {author} {\bibfnamefont {Q.}~\bibnamefont {Gibson}}, \bibinfo
  {author} {\bibfnamefont {S.}~\bibnamefont {Liang}}, \bibinfo {author}
  {\bibfnamefont {C.~A.}\ \bibnamefont {Belvin}}, \bibinfo {author}
  {\bibfnamefont {B.~A.}\ \bibnamefont {Bernevig}}, \bibinfo {author}
  {\bibfnamefont {R.~J.}\ \bibnamefont {Cava}},\ and\ \bibinfo {author}
  {\bibfnamefont {N.~P.}\ \bibnamefont {Ong}},\ }\bibfield  {title} {\bibinfo
  {title} {The chiral anomaly and thermopower of {{Weyl}} fermions in the
  half-{{Heusler GdPtBi}}},\ }\href@noop {} {\bibfield  {journal} {\bibinfo
  {journal} {Nature Materials}\ }\textbf {\bibinfo {volume} {15}},\ \bibinfo
  {pages} {1161} (\bibinfo {year} {2016})}\BibitemShut {NoStop}%
\bibitem [{\citenamefont {Zhang}\ \emph {et~al.}(2016)\citenamefont {Zhang},
  \citenamefont {Xu}, \citenamefont {Belopolski}, \citenamefont {Yuan},
  \citenamefont {Lin}, \citenamefont {Tong}, \citenamefont {Bian},
  \citenamefont {Alidoust}, \citenamefont {Lee}, \citenamefont {Huang} \emph
  {et~al.}}]{zhang2016signatures}%
  \BibitemOpen
  \bibfield  {author} {\bibinfo {author} {\bibfnamefont {C.-L.}\ \bibnamefont
  {Zhang}}, \bibinfo {author} {\bibfnamefont {S.-Y.}\ \bibnamefont {Xu}},
  \bibinfo {author} {\bibfnamefont {I.}~\bibnamefont {Belopolski}}, \bibinfo
  {author} {\bibfnamefont {Z.}~\bibnamefont {Yuan}}, \bibinfo {author}
  {\bibfnamefont {Z.}~\bibnamefont {Lin}}, \bibinfo {author} {\bibfnamefont
  {B.}~\bibnamefont {Tong}}, \bibinfo {author} {\bibfnamefont {G.}~\bibnamefont
  {Bian}}, \bibinfo {author} {\bibfnamefont {N.}~\bibnamefont {Alidoust}},
  \bibinfo {author} {\bibfnamefont {C.-C.}\ \bibnamefont {Lee}}, \bibinfo
  {author} {\bibfnamefont {S.-M.}\ \bibnamefont {Huang}}, \emph {et~al.},\
  }\bibfield  {title} {\bibinfo {title} {Signatures of the adler--bell--jackiw
  chiral anomaly in a weyl fermion semimetal},\ }\href@noop {} {\bibfield
  {journal} {\bibinfo  {journal} {Nature communications}\ }\textbf {\bibinfo
  {volume} {7}},\ \bibinfo {pages} {1} (\bibinfo {year} {2016})}\BibitemShut
  {NoStop}%
\bibitem [{\citenamefont {Xiong}\ \emph {et~al.}(2015)\citenamefont {Xiong},
  \citenamefont {Kushwaha}, \citenamefont {Liang}, \citenamefont {Krizan},
  \citenamefont {Hirschberger}, \citenamefont {Wang}, \citenamefont {Cava},\
  and\ \citenamefont {Ong}}]{xiong2015evidence}%
  \BibitemOpen
  \bibfield  {author} {\bibinfo {author} {\bibfnamefont {J.}~\bibnamefont
  {Xiong}}, \bibinfo {author} {\bibfnamefont {S.~K.}\ \bibnamefont {Kushwaha}},
  \bibinfo {author} {\bibfnamefont {T.}~\bibnamefont {Liang}}, \bibinfo
  {author} {\bibfnamefont {J.~W.}\ \bibnamefont {Krizan}}, \bibinfo {author}
  {\bibfnamefont {M.}~\bibnamefont {Hirschberger}}, \bibinfo {author}
  {\bibfnamefont {W.}~\bibnamefont {Wang}}, \bibinfo {author} {\bibfnamefont
  {R.~J.}\ \bibnamefont {Cava}},\ and\ \bibinfo {author} {\bibfnamefont
  {N.~P.}\ \bibnamefont {Ong}},\ }\bibfield  {title} {\bibinfo {title}
  {Evidence for the chiral anomaly in the dirac semimetal na3bi},\ }\href@noop
  {} {\bibfield  {journal} {\bibinfo  {journal} {Science}\ }\textbf {\bibinfo
  {volume} {350}},\ \bibinfo {pages} {413} (\bibinfo {year}
  {2015})}\BibitemShut {NoStop}%
\bibitem [{\citenamefont {Basov}\ \emph {et~al.}(2011)\citenamefont {Basov},
  \citenamefont {Averitt}, \citenamefont {Van Der~Marel}, \citenamefont
  {Dressel},\ and\ \citenamefont {Haule}}]{basov2011electrodynamics}%
  \BibitemOpen
  \bibfield  {author} {\bibinfo {author} {\bibfnamefont {D.~N.}\ \bibnamefont
  {Basov}}, \bibinfo {author} {\bibfnamefont {R.~D.}\ \bibnamefont {Averitt}},
  \bibinfo {author} {\bibfnamefont {D.}~\bibnamefont {Van Der~Marel}}, \bibinfo
  {author} {\bibfnamefont {M.}~\bibnamefont {Dressel}},\ and\ \bibinfo {author}
  {\bibfnamefont {K.}~\bibnamefont {Haule}},\ }\bibfield  {title} {\bibinfo
  {title} {Electrodynamics of correlated electron materials},\ }\href@noop {}
  {\bibfield  {journal} {\bibinfo  {journal} {Reviews of Modern Physics}\
  }\textbf {\bibinfo {volume} {83}},\ \bibinfo {pages} {471} (\bibinfo {year}
  {2011})}\BibitemShut {NoStop}%
\bibitem [{\citenamefont {Parker}\ \emph {et~al.}(2019)\citenamefont {Parker},
  \citenamefont {Morimoto}, \citenamefont {Orenstein},\ and\ \citenamefont
  {Moore}}]{parker2019diagrammatic}%
  \BibitemOpen
  \bibfield  {author} {\bibinfo {author} {\bibfnamefont {D.~E.}\ \bibnamefont
  {Parker}}, \bibinfo {author} {\bibfnamefont {T.}~\bibnamefont {Morimoto}},
  \bibinfo {author} {\bibfnamefont {J.}~\bibnamefont {Orenstein}},\ and\
  \bibinfo {author} {\bibfnamefont {J.~E.}\ \bibnamefont {Moore}},\ }\bibfield
  {title} {\bibinfo {title} {Diagrammatic approach to nonlinear optical
  response with application to {{Weyl}} semimetals},\ }\href
  {https://doi.org/10.1103/PhysRevB.99.045121} {\bibfield  {journal} {\bibinfo
  {journal} {Phys. Rev. B}\ }\textbf {\bibinfo {volume} {99}},\ \bibinfo
  {pages} {045121} (\bibinfo {year} {2019})}\BibitemShut {NoStop}%
\bibitem [{\citenamefont {Sipe}\ and\ \citenamefont
  {Shkrebtii}(2000)}]{sipe2000secondorder}%
  \BibitemOpen
  \bibfield  {author} {\bibinfo {author} {\bibfnamefont {J.~E.}\ \bibnamefont
  {Sipe}}\ and\ \bibinfo {author} {\bibfnamefont {A.~I.}\ \bibnamefont
  {Shkrebtii}},\ }\bibfield  {title} {\bibinfo {title} {Second-order optical
  response in semiconductors},\ }\href
  {https://doi.org/10.1103/PhysRevB.61.5337} {\bibfield  {journal} {\bibinfo
  {journal} {Physical Review B}\ }\textbf {\bibinfo {volume} {61}},\ \bibinfo
  {pages} {5337} (\bibinfo {year} {2000})}\BibitemShut {NoStop}%
\bibitem [{\citenamefont {Cheng}\ \emph {et~al.}(2020)\citenamefont {Cheng},
  \citenamefont {Sipe},\ and\ \citenamefont {Wu}}]{sipe2020ThirdOrderResponse}%
  \BibitemOpen
  \bibfield  {author} {\bibinfo {author} {\bibfnamefont {J.~L.}\ \bibnamefont
  {Cheng}}, \bibinfo {author} {\bibfnamefont {J.~E.}\ \bibnamefont {Sipe}},\
  and\ \bibinfo {author} {\bibfnamefont {S.~W.}\ \bibnamefont {Wu}},\
  }\bibfield  {title} {\bibinfo {title} {Third order optical nonlinearity of
  three dimensional massless {Dirac} fermions},\ }\bibfield  {journal}
  {\bibinfo  {journal} {arXiv:2005.13693 [cond-mat, physics:physics]}\ }\href
  {https://doi.org/10.1021/acsphotonics.0c00836} {10.1021/acsphotonics.0c00836}
  (\bibinfo {year} {2020}),\ \bibinfo {note} {arXiv: 2005.13693}\BibitemShut
  {NoStop}%
\bibitem [{\citenamefont {Aversa}\ and\ \citenamefont
  {Sipe}(1995)}]{aversa1995nonlinear}%
  \BibitemOpen
  \bibfield  {author} {\bibinfo {author} {\bibfnamefont {C.}~\bibnamefont
  {Aversa}}\ and\ \bibinfo {author} {\bibfnamefont {J.~E.}\ \bibnamefont
  {Sipe}},\ }\bibfield  {title} {\bibinfo {title} {Nonlinear optical
  susceptibilities of semiconductors: {{Results}} with a length-gauge
  analysis},\ }\href {https://doi.org/10.1103/PhysRevB.52.14636} {\bibfield
  {journal} {\bibinfo  {journal} {Physical Review B}\ }\textbf {\bibinfo
  {volume} {52}},\ \bibinfo {pages} {14636} (\bibinfo {year}
  {1995})}\BibitemShut {NoStop}%
\bibitem [{\citenamefont {João}\ and\ \citenamefont
  {Lopes}(2019)}]{Joao2020OpticalResponseTBWithDiagrams}%
  \BibitemOpen
  \bibfield  {author} {\bibinfo {author} {\bibfnamefont {S.~M.}\ \bibnamefont
  {João}}\ and\ \bibinfo {author} {\bibfnamefont {J.~M. V.~P.}\ \bibnamefont
  {Lopes}},\ }\bibfield  {title} {\bibinfo {title} {Basis-independent spectral
  methods for non-linear optical response in arbitrary tight-binding models},\
  }\href {https://doi.org/10.1088/1361-648X/ab59ec} {\bibfield  {journal}
  {\bibinfo  {journal} {Journal of Physics: Condensed Matter}\ }\textbf
  {\bibinfo {volume} {32}},\ \bibinfo {pages} {125901} (\bibinfo {year}
  {2019})}\BibitemShut {NoStop}%
\bibitem [{\citenamefont {Ahn}\ \emph {et~al.}(2022{\natexlab{b}})\citenamefont
  {Ahn}, \citenamefont {Guo}, \citenamefont {Nagaosa},\ and\ \citenamefont
  {Vishwanath}}]{ahn2022riemannian}%
  \BibitemOpen
  \bibfield  {author} {\bibinfo {author} {\bibfnamefont {J.}~\bibnamefont
  {Ahn}}, \bibinfo {author} {\bibfnamefont {G.-Y.}\ \bibnamefont {Guo}},
  \bibinfo {author} {\bibfnamefont {N.}~\bibnamefont {Nagaosa}},\ and\ \bibinfo
  {author} {\bibfnamefont {A.}~\bibnamefont {Vishwanath}},\ }\bibfield  {title}
  {\bibinfo {title} {Riemannian geometry of resonant optical responses},\
  }\href@noop {} {\bibfield  {journal} {\bibinfo  {journal} {Nature Physics}\
  }\textbf {\bibinfo {volume} {18}},\ \bibinfo {pages} {290} (\bibinfo {year}
  {2022}{\natexlab{b}})}\BibitemShut {NoStop}%
\bibitem [{\citenamefont
  {Alexandradinata}(2022)}]{alexandradinata2022topological}%
  \BibitemOpen
  \bibfield  {author} {\bibinfo {author} {\bibfnamefont {A.}~\bibnamefont
  {Alexandradinata}},\ }\bibfield  {title} {\bibinfo {title} {A topological
  principle for photovoltaics: Shift current in intrinsically polar
  insulators},\ }\href@noop {} {\bibfield  {journal} {\bibinfo  {journal}
  {arXiv preprint arXiv:2203.11225}\ } (\bibinfo {year} {2022})}\BibitemShut
  {NoStop}%
\bibitem [{\citenamefont {Holder}\ \emph {et~al.}(2020)\citenamefont {Holder},
  \citenamefont {Kaplan},\ and\ \citenamefont
  {Yan}}]{Yan2020TRBInCurvatureMetricAndMotion}%
  \BibitemOpen
  \bibfield  {author} {\bibinfo {author} {\bibfnamefont {T.}~\bibnamefont
  {Holder}}, \bibinfo {author} {\bibfnamefont {D.}~\bibnamefont {Kaplan}},\
  and\ \bibinfo {author} {\bibfnamefont {B.}~\bibnamefont {Yan}},\ }\bibfield
  {title} {\bibinfo {title} {Consequences of time-reversal-symmetry breaking in
  the light-matter interaction: Berry curvature, quantum metric, and diabatic
  motion},\ }\href {https://doi.org/10.1103/PhysRevResearch.2.033100}
  {\bibfield  {journal} {\bibinfo  {journal} {Phys. Rev. Res.}\ }\textbf
  {\bibinfo {volume} {2}},\ \bibinfo {pages} {033100} (\bibinfo {year}
  {2020})}\BibitemShut {NoStop}%
\bibitem [{\citenamefont {Orenstein}\ \emph {et~al.}(2021)\citenamefont
  {Orenstein}, \citenamefont {Moore}, \citenamefont {Morimoto}, \citenamefont
  {Torchinsky}, \citenamefont {Harter},\ and\ \citenamefont
  {Hsieh}}]{orenstein2021topology}%
  \BibitemOpen
  \bibfield  {author} {\bibinfo {author} {\bibfnamefont {J.}~\bibnamefont
  {Orenstein}}, \bibinfo {author} {\bibfnamefont {J.}~\bibnamefont {Moore}},
  \bibinfo {author} {\bibfnamefont {T.}~\bibnamefont {Morimoto}}, \bibinfo
  {author} {\bibfnamefont {D.}~\bibnamefont {Torchinsky}}, \bibinfo {author}
  {\bibfnamefont {J.}~\bibnamefont {Harter}},\ and\ \bibinfo {author}
  {\bibfnamefont {D.}~\bibnamefont {Hsieh}},\ }\bibfield  {title} {\bibinfo
  {title} {Topology and symmetry of quantum materials via nonlinear optical
  responses},\ }\href@noop {} {\bibfield  {journal} {\bibinfo  {journal}
  {Annual Review of Condensed Matter Physics}\ }\textbf {\bibinfo {volume}
  {12}},\ \bibinfo {pages} {247} (\bibinfo {year} {2021})}\BibitemShut
  {NoStop}%
\bibitem [{\citenamefont {Wu}\ \emph {et~al.}(2017)\citenamefont {Wu},
  \citenamefont {Patankar}, \citenamefont {Morimoto}, \citenamefont {Nair},
  \citenamefont {Thewalt}, \citenamefont {Little}, \citenamefont {Analytis},
  \citenamefont {Moore},\ and\ \citenamefont {Orenstein}}]{wu2017giant}%
  \BibitemOpen
  \bibfield  {author} {\bibinfo {author} {\bibfnamefont {L.}~\bibnamefont
  {Wu}}, \bibinfo {author} {\bibfnamefont {S.}~\bibnamefont {Patankar}},
  \bibinfo {author} {\bibfnamefont {T.}~\bibnamefont {Morimoto}}, \bibinfo
  {author} {\bibfnamefont {N.~L.}\ \bibnamefont {Nair}}, \bibinfo {author}
  {\bibfnamefont {E.}~\bibnamefont {Thewalt}}, \bibinfo {author} {\bibfnamefont
  {A.}~\bibnamefont {Little}}, \bibinfo {author} {\bibfnamefont {J.~G.}\
  \bibnamefont {Analytis}}, \bibinfo {author} {\bibfnamefont {J.~E.}\
  \bibnamefont {Moore}},\ and\ \bibinfo {author} {\bibfnamefont
  {J.}~\bibnamefont {Orenstein}},\ }\bibfield  {title} {\bibinfo {title} {Giant
  anisotropic nonlinear optical response in transition metal monopnictide
  {{Weyl}} semimetals},\ }\href@noop {} {\bibfield  {journal} {\bibinfo
  {journal} {Nature Physics}\ }\textbf {\bibinfo {volume} {13}},\ \bibinfo
  {pages} {350} (\bibinfo {year} {2017})}\BibitemShut {NoStop}%
\bibitem [{\citenamefont {Streda}(1982)}]{streda1982theory}%
  \BibitemOpen
  \bibfield  {author} {\bibinfo {author} {\bibfnamefont {P.}~\bibnamefont
  {Streda}},\ }\bibfield  {title} {\bibinfo {title} {Theory of quantised hall
  conductivity in two dimensions},\ }\href@noop {} {\bibfield  {journal}
  {\bibinfo  {journal} {Journal of Physics C: Solid State Physics}\ }\textbf
  {\bibinfo {volume} {15}},\ \bibinfo {pages} {L717} (\bibinfo {year}
  {1982})}\BibitemShut {NoStop}%
\bibitem [{\citenamefont {Smrcka}\ and\ \citenamefont
  {Streda}(1977)}]{smrcka1977transport}%
  \BibitemOpen
  \bibfield  {author} {\bibinfo {author} {\bibfnamefont {L.}~\bibnamefont
  {Smrcka}}\ and\ \bibinfo {author} {\bibfnamefont {P.}~\bibnamefont
  {Streda}},\ }\bibfield  {title} {\bibinfo {title} {Transport coefficients in
  strong magnetic fields},\ }\href@noop {} {\bibfield  {journal} {\bibinfo
  {journal} {Journal of Physics C: Solid State Physics}\ }\textbf {\bibinfo
  {volume} {10}},\ \bibinfo {pages} {2153} (\bibinfo {year}
  {1977})}\BibitemShut {NoStop}%
\bibitem [{\citenamefont {Malashevich}\ \emph {et~al.}(2010)\citenamefont
  {Malashevich}, \citenamefont {Souza}, \citenamefont {Coh},\ and\
  \citenamefont {Vanderbilt}}]{malashevich2010theory}%
  \BibitemOpen
  \bibfield  {author} {\bibinfo {author} {\bibfnamefont {A.}~\bibnamefont
  {Malashevich}}, \bibinfo {author} {\bibfnamefont {I.}~\bibnamefont {Souza}},
  \bibinfo {author} {\bibfnamefont {S.}~\bibnamefont {Coh}},\ and\ \bibinfo
  {author} {\bibfnamefont {D.}~\bibnamefont {Vanderbilt}},\ }\bibfield  {title}
  {\bibinfo {title} {Theory of orbital magnetoelectric response},\ }\href@noop
  {} {\bibfield  {journal} {\bibinfo  {journal} {New Journal of Physics}\
  }\textbf {\bibinfo {volume} {12}},\ \bibinfo {pages} {053032} (\bibinfo
  {year} {2010})}\BibitemShut {NoStop}%
\bibitem [{\citenamefont {Zhong}\ \emph {et~al.}(2016)\citenamefont {Zhong},
  \citenamefont {Moore},\ and\ \citenamefont {Souza}}]{zhong2016gyrotropic}%
  \BibitemOpen
  \bibfield  {author} {\bibinfo {author} {\bibfnamefont {S.}~\bibnamefont
  {Zhong}}, \bibinfo {author} {\bibfnamefont {J.~E.}\ \bibnamefont {Moore}},\
  and\ \bibinfo {author} {\bibfnamefont {I.}~\bibnamefont {Souza}},\ }\bibfield
   {title} {\bibinfo {title} {Gyrotropic magnetic effect and the magnetic
  moment on the fermi surface},\ }\href@noop {} {\bibfield  {journal} {\bibinfo
   {journal} {Physical review letters}\ }\textbf {\bibinfo {volume} {116}},\
  \bibinfo {pages} {077201} (\bibinfo {year} {2016})}\BibitemShut {NoStop}%
\bibitem [{\citenamefont {Gassner}\ and\ \citenamefont
  {Mele}(2023)}]{Mele2023NonlinearOpticalRegLattice}%
  \BibitemOpen
  \bibfield  {author} {\bibinfo {author} {\bibfnamefont {S.}~\bibnamefont
  {Gassner}}\ and\ \bibinfo {author} {\bibfnamefont {E.~J.}\ \bibnamefont
  {Mele}},\ }\bibfield  {title} {\bibinfo {title} {Regularized lattice theory
  for spatially dispersive nonlinear optical conductivities},\ }\href@noop {}
  {\bibfield  {journal} {\bibinfo  {journal} {arXiv preprint arXiv:2304.13073}\
  } (\bibinfo {year} {2023})},\ \Eprint {https://arxiv.org/abs/2304.13073}
  {arXiv:2304.13073} \BibitemShut {NoStop}%
\bibitem [{\citenamefont {Hoyos}\ and\ \citenamefont
  {Son}(2012)}]{hoyos2012hall}%
  \BibitemOpen
  \bibfield  {author} {\bibinfo {author} {\bibfnamefont {C.}~\bibnamefont
  {Hoyos}}\ and\ \bibinfo {author} {\bibfnamefont {D.~T.}\ \bibnamefont
  {Son}},\ }\bibfield  {title} {\bibinfo {title} {Hall {{Viscosity}} and
  {{Electromagnetic Response}}},\ }\href
  {https://doi.org/10.1103/PhysRevLett.108.066805} {\bibfield  {journal}
  {\bibinfo  {journal} {Phys. Rev. Lett.}\ }\textbf {\bibinfo {volume} {108}},\
  \bibinfo {pages} {066805} (\bibinfo {year} {2012})}\BibitemShut {NoStop}%
\bibitem [{\citenamefont {Bradlyn}\ \emph {et~al.}(2012)\citenamefont
  {Bradlyn}, \citenamefont {Goldstein},\ and\ \citenamefont
  {Read}}]{bradlyn2012kubo}%
  \BibitemOpen
  \bibfield  {author} {\bibinfo {author} {\bibfnamefont {B.}~\bibnamefont
  {Bradlyn}}, \bibinfo {author} {\bibfnamefont {M.}~\bibnamefont {Goldstein}},\
  and\ \bibinfo {author} {\bibfnamefont {N.}~\bibnamefont {Read}},\ }\bibfield
  {title} {\bibinfo {title} {Kubo formulas for viscosity: {{Hall}} viscosity,
  {{Ward}} identities, and the relation with conductivity},\ }\href@noop {}
  {\bibfield  {journal} {\bibinfo  {journal} {Physical Review B}\ }\textbf
  {\bibinfo {volume} {86}},\ \bibinfo {pages} {245309} (\bibinfo {year}
  {2012})}\BibitemShut {NoStop}%
\bibitem [{\citenamefont {Holder}\ \emph {et~al.}(2019)\citenamefont {Holder},
  \citenamefont {Queiroz},\ and\ \citenamefont {Stern}}]{holder2019unified}%
  \BibitemOpen
  \bibfield  {author} {\bibinfo {author} {\bibfnamefont {T.}~\bibnamefont
  {Holder}}, \bibinfo {author} {\bibfnamefont {R.}~\bibnamefont {Queiroz}},\
  and\ \bibinfo {author} {\bibfnamefont {A.}~\bibnamefont {Stern}},\ }\bibfield
   {title} {\bibinfo {title} {Unified description of the classical hall
  viscosity},\ }\href@noop {} {\bibfield  {journal} {\bibinfo  {journal}
  {Physical review letters}\ }\textbf {\bibinfo {volume} {123}},\ \bibinfo
  {pages} {106801} (\bibinfo {year} {2019})}\BibitemShut {NoStop}%
\bibitem [{\citenamefont {Lucas}\ and\ \citenamefont
  {Fong}(2018)}]{lucas2018hydrodynamics}%
  \BibitemOpen
  \bibfield  {author} {\bibinfo {author} {\bibfnamefont {A.}~\bibnamefont
  {Lucas}}\ and\ \bibinfo {author} {\bibfnamefont {K.~C.}\ \bibnamefont
  {Fong}},\ }\bibfield  {title} {\bibinfo {title} {Hydrodynamics of electrons
  in graphene},\ }\href@noop {} {\bibfield  {journal} {\bibinfo  {journal}
  {Journal of Physics: Condensed Matter}\ }\textbf {\bibinfo {volume} {30}},\
  \bibinfo {pages} {053001} (\bibinfo {year} {2018})}\BibitemShut {NoStop}%
\bibitem [{\citenamefont {Sulpizio}\ \emph {et~al.}(2019)\citenamefont
  {Sulpizio}, \citenamefont {Ella}, \citenamefont {Rozen}, \citenamefont
  {Birkbeck}, \citenamefont {Perello}, \citenamefont {Dutta}, \citenamefont
  {{Ben-Shalom}}, \citenamefont {Taniguchi}, \citenamefont {Watanabe},
  \citenamefont {Holder} \emph {et~al.}}]{sulpizio2019visualizing}%
  \BibitemOpen
  \bibfield  {author} {\bibinfo {author} {\bibfnamefont {J.~A.}\ \bibnamefont
  {Sulpizio}}, \bibinfo {author} {\bibfnamefont {L.}~\bibnamefont {Ella}},
  \bibinfo {author} {\bibfnamefont {A.}~\bibnamefont {Rozen}}, \bibinfo
  {author} {\bibfnamefont {J.}~\bibnamefont {Birkbeck}}, \bibinfo {author}
  {\bibfnamefont {D.~J.}\ \bibnamefont {Perello}}, \bibinfo {author}
  {\bibfnamefont {D.}~\bibnamefont {Dutta}}, \bibinfo {author} {\bibfnamefont
  {M.}~\bibnamefont {{Ben-Shalom}}}, \bibinfo {author} {\bibfnamefont
  {T.}~\bibnamefont {Taniguchi}}, \bibinfo {author} {\bibfnamefont
  {K.}~\bibnamefont {Watanabe}}, \bibinfo {author} {\bibfnamefont
  {T.}~\bibnamefont {Holder}}, \emph {et~al.},\ }\bibfield  {title} {\bibinfo
  {title} {Visualizing {{Poiseuille}} flow of hydrodynamic electrons},\
  }\href@noop {} {\bibfield  {journal} {\bibinfo  {journal} {Nature}\ }\textbf
  {\bibinfo {volume} {576}},\ \bibinfo {pages} {75} (\bibinfo {year}
  {2019})}\BibitemShut {NoStop}%
\bibitem [{\citenamefont {Delacr{\'e}taz}\ and\ \citenamefont
  {Gromov}(2017)}]{delacretaz2017transport}%
  \BibitemOpen
  \bibfield  {author} {\bibinfo {author} {\bibfnamefont {L.~V.}\ \bibnamefont
  {Delacr{\'e}taz}}\ and\ \bibinfo {author} {\bibfnamefont {A.}~\bibnamefont
  {Gromov}},\ }\bibfield  {title} {\bibinfo {title} {Transport {{Signatures}}
  of the {{Hall Viscosity}}},\ }\href@noop {} {\bibfield  {journal} {\bibinfo
  {journal} {Physical review letters}\ }\textbf {\bibinfo {volume} {119}},\
  \bibinfo {pages} {226602} (\bibinfo {year} {2017})}\BibitemShut {NoStop}%
\bibitem [{\citenamefont {Zhang}\ and\ \citenamefont
  {Rhim}(2022)}]{zhang2022GeometricSpinHallEffectFromInhomogeneity}%
  \BibitemOpen
  \bibfield  {author} {\bibinfo {author} {\bibfnamefont {A.}~\bibnamefont
  {Zhang}}\ and\ \bibinfo {author} {\bibfnamefont {J.-W.}\ \bibnamefont
  {Rhim}},\ }\bibfield  {title} {\bibinfo {title} {Geometric origin of
  intrinsic spin hall effect in an inhomogeneous electric field},\ }\href
  {https://doi.org/10.1038/s42005-022-00975-3} {\bibfield  {journal} {\bibinfo
  {journal} {Communications Physics}\ }\textbf {\bibinfo {volume} {5}},\
  \bibinfo {pages} {195} (\bibinfo {year} {2022})}\BibitemShut {NoStop}%
\bibitem [{\citenamefont {Kozii}\ \emph {et~al.}(2021)\citenamefont {Kozii},
  \citenamefont {Avdoshkin}, \citenamefont {Zhong},\ and\ \citenamefont
  {Moore}}]{kozii2021intrinsic}%
  \BibitemOpen
  \bibfield  {author} {\bibinfo {author} {\bibfnamefont {V.}~\bibnamefont
  {Kozii}}, \bibinfo {author} {\bibfnamefont {A.}~\bibnamefont {Avdoshkin}},
  \bibinfo {author} {\bibfnamefont {S.}~\bibnamefont {Zhong}},\ and\ \bibinfo
  {author} {\bibfnamefont {J.~E.}\ \bibnamefont {Moore}},\ }\bibfield  {title}
  {\bibinfo {title} {Intrinsic {{Anomalous Hall Conductivity}} in a
  {{Nonuniform Electric Field}}},\ }\href
  {https://doi.org/10.1103/PhysRevLett.126.156602} {\bibfield  {journal}
  {\bibinfo  {journal} {Phys. Rev. Lett.}\ }\textbf {\bibinfo {volume} {126}},\
  \bibinfo {pages} {156602} (\bibinfo {year} {2021})}\BibitemShut {NoStop}%
\bibitem [{\citenamefont {Cao}\ \emph {et~al.}(2018)\citenamefont {Cao},
  \citenamefont {Fatemi}, \citenamefont {Fang}, \citenamefont {Watanabe},
  \citenamefont {Taniguchi}, \citenamefont {Kaxiras},\ and\ \citenamefont
  {{Jarillo-Herrero}}}]{cao2018unconventional}%
  \BibitemOpen
  \bibfield  {author} {\bibinfo {author} {\bibfnamefont {Y.}~\bibnamefont
  {Cao}}, \bibinfo {author} {\bibfnamefont {V.}~\bibnamefont {Fatemi}},
  \bibinfo {author} {\bibfnamefont {S.}~\bibnamefont {Fang}}, \bibinfo {author}
  {\bibfnamefont {K.}~\bibnamefont {Watanabe}}, \bibinfo {author}
  {\bibfnamefont {T.}~\bibnamefont {Taniguchi}}, \bibinfo {author}
  {\bibfnamefont {E.}~\bibnamefont {Kaxiras}},\ and\ \bibinfo {author}
  {\bibfnamefont {P.}~\bibnamefont {{Jarillo-Herrero}}},\ }\bibfield  {title}
  {\bibinfo {title} {Unconventional superconductivity in magic-angle graphene
  superlattices},\ }\href@noop {} {\bibfield  {journal} {\bibinfo  {journal}
  {Nature}\ }\textbf {\bibinfo {volume} {556}},\ \bibinfo {pages} {43}
  (\bibinfo {year} {2018})}\BibitemShut {NoStop}%
\bibitem [{\citenamefont {Bistritzer}\ and\ \citenamefont
  {MacDonald}(2011)}]{bistritzer2011moire}%
  \BibitemOpen
  \bibfield  {author} {\bibinfo {author} {\bibfnamefont {R.}~\bibnamefont
  {Bistritzer}}\ and\ \bibinfo {author} {\bibfnamefont {A.~H.}\ \bibnamefont
  {MacDonald}},\ }\bibfield  {title} {\bibinfo {title} {Moir\'e butterflies in
  twisted bilayer graphene},\ }\href@noop {} {\bibfield  {journal} {\bibinfo
  {journal} {Physical Review B}\ }\textbf {\bibinfo {volume} {84}},\ \bibinfo
  {pages} {035440} (\bibinfo {year} {2011})}\BibitemShut {NoStop}%
\bibitem [{\citenamefont {Ciarrocchi}\ \emph {et~al.}(2022)\citenamefont
  {Ciarrocchi}, \citenamefont {Tagarelli}, \citenamefont {Avsar},\ and\
  \citenamefont {Kis}}]{ciarrocchi2022excitonic}%
  \BibitemOpen
  \bibfield  {author} {\bibinfo {author} {\bibfnamefont {A.}~\bibnamefont
  {Ciarrocchi}}, \bibinfo {author} {\bibfnamefont {F.}~\bibnamefont
  {Tagarelli}}, \bibinfo {author} {\bibfnamefont {A.}~\bibnamefont {Avsar}},\
  and\ \bibinfo {author} {\bibfnamefont {A.}~\bibnamefont {Kis}},\ }\bibfield
  {title} {\bibinfo {title} {Excitonic devices with van der waals
  heterostructures: valleytronics meets twistronics},\ }\href@noop {}
  {\bibfield  {journal} {\bibinfo  {journal} {Nature Reviews Materials}\
  }\textbf {\bibinfo {volume} {7}},\ \bibinfo {pages} {449} (\bibinfo {year}
  {2022})}\BibitemShut {NoStop}%
\bibitem [{\citenamefont {Andrei}\ \emph {et~al.}(2021)\citenamefont {Andrei},
  \citenamefont {Efetov}, \citenamefont {Jarillo-Herrero}, \citenamefont
  {MacDonald}, \citenamefont {Mak}, \citenamefont {Senthil}, \citenamefont
  {Tutuc}, \citenamefont {Yazdani},\ and\ \citenamefont
  {Young}}]{andrei2021marvels}%
  \BibitemOpen
  \bibfield  {author} {\bibinfo {author} {\bibfnamefont {E.~Y.}\ \bibnamefont
  {Andrei}}, \bibinfo {author} {\bibfnamefont {D.~K.}\ \bibnamefont {Efetov}},
  \bibinfo {author} {\bibfnamefont {P.}~\bibnamefont {Jarillo-Herrero}},
  \bibinfo {author} {\bibfnamefont {A.~H.}\ \bibnamefont {MacDonald}}, \bibinfo
  {author} {\bibfnamefont {K.~F.}\ \bibnamefont {Mak}}, \bibinfo {author}
  {\bibfnamefont {T.}~\bibnamefont {Senthil}}, \bibinfo {author} {\bibfnamefont
  {E.}~\bibnamefont {Tutuc}}, \bibinfo {author} {\bibfnamefont
  {A.}~\bibnamefont {Yazdani}},\ and\ \bibinfo {author} {\bibfnamefont {A.~F.}\
  \bibnamefont {Young}},\ }\bibfield  {title} {\bibinfo {title} {The marvels of
  moir{\'e} materials},\ }\href@noop {} {\bibfield  {journal} {\bibinfo
  {journal} {Nature Reviews Materials}\ }\textbf {\bibinfo {volume} {6}},\
  \bibinfo {pages} {201} (\bibinfo {year} {2021})}\BibitemShut {NoStop}%
\bibitem [{\citenamefont {T{\"o}rm{\"a}}\ \emph {et~al.}(2022)\citenamefont
  {T{\"o}rm{\"a}}, \citenamefont {Peotta},\ and\ \citenamefont
  {Bernevig}}]{torma2022superconductivity}%
  \BibitemOpen
  \bibfield  {author} {\bibinfo {author} {\bibfnamefont {P.}~\bibnamefont
  {T{\"o}rm{\"a}}}, \bibinfo {author} {\bibfnamefont {S.}~\bibnamefont
  {Peotta}},\ and\ \bibinfo {author} {\bibfnamefont {B.~A.}\ \bibnamefont
  {Bernevig}},\ }\bibfield  {title} {\bibinfo {title} {Superconductivity,
  superfluidity and quantum geometry in twisted multilayer systems},\
  }\href@noop {} {\bibfield  {journal} {\bibinfo  {journal} {Nature Reviews
  Physics}\ }\textbf {\bibinfo {volume} {4}},\ \bibinfo {pages} {528} (\bibinfo
  {year} {2022})}\BibitemShut {NoStop}%
\bibitem [{\citenamefont {Liu}\ \emph {et~al.}(2020)\citenamefont {Liu},
  \citenamefont {Hao}, \citenamefont {Khalaf}, \citenamefont {Lee},
  \citenamefont {Ronen}, \citenamefont {Yoo}, \citenamefont {Haei~Najafabadi},
  \citenamefont {Watanabe}, \citenamefont {Taniguchi}, \citenamefont
  {Vishwanath} \emph {et~al.}}]{liu2020tunable}%
  \BibitemOpen
  \bibfield  {author} {\bibinfo {author} {\bibfnamefont {X.}~\bibnamefont
  {Liu}}, \bibinfo {author} {\bibfnamefont {Z.}~\bibnamefont {Hao}}, \bibinfo
  {author} {\bibfnamefont {E.}~\bibnamefont {Khalaf}}, \bibinfo {author}
  {\bibfnamefont {J.~Y.}\ \bibnamefont {Lee}}, \bibinfo {author} {\bibfnamefont
  {Y.}~\bibnamefont {Ronen}}, \bibinfo {author} {\bibfnamefont
  {H.}~\bibnamefont {Yoo}}, \bibinfo {author} {\bibfnamefont {D.}~\bibnamefont
  {Haei~Najafabadi}}, \bibinfo {author} {\bibfnamefont {K.}~\bibnamefont
  {Watanabe}}, \bibinfo {author} {\bibfnamefont {T.}~\bibnamefont {Taniguchi}},
  \bibinfo {author} {\bibfnamefont {A.}~\bibnamefont {Vishwanath}}, \emph
  {et~al.},\ }\bibfield  {title} {\bibinfo {title} {Tunable spin-polarized
  correlated states in twisted double bilayer graphene},\ }\href@noop {}
  {\bibfield  {journal} {\bibinfo  {journal} {Nature}\ }\textbf {\bibinfo
  {volume} {583}},\ \bibinfo {pages} {221} (\bibinfo {year}
  {2020})}\BibitemShut {NoStop}%
\bibitem [{\citenamefont {Park}\ \emph {et~al.}(2021)\citenamefont {Park},
  \citenamefont {Cao}, \citenamefont {Watanabe}, \citenamefont {Taniguchi},\
  and\ \citenamefont {Jarillo-Herrero}}]{park2021tunable}%
  \BibitemOpen
  \bibfield  {author} {\bibinfo {author} {\bibfnamefont {J.~M.}\ \bibnamefont
  {Park}}, \bibinfo {author} {\bibfnamefont {Y.}~\bibnamefont {Cao}}, \bibinfo
  {author} {\bibfnamefont {K.}~\bibnamefont {Watanabe}}, \bibinfo {author}
  {\bibfnamefont {T.}~\bibnamefont {Taniguchi}},\ and\ \bibinfo {author}
  {\bibfnamefont {P.}~\bibnamefont {Jarillo-Herrero}},\ }\bibfield  {title}
  {\bibinfo {title} {Tunable strongly coupled superconductivity in magic-angle
  twisted trilayer graphene},\ }\href@noop {} {\bibfield  {journal} {\bibinfo
  {journal} {Nature}\ }\textbf {\bibinfo {volume} {590}},\ \bibinfo {pages}
  {249} (\bibinfo {year} {2021})}\BibitemShut {NoStop}%
\bibitem [{\citenamefont {Zhang}\ \emph {et~al.}(2021)\citenamefont {Zhang},
  \citenamefont {Liu},\ and\ \citenamefont {Fu}}]{zhang2021electronic}%
  \BibitemOpen
  \bibfield  {author} {\bibinfo {author} {\bibfnamefont {Y.}~\bibnamefont
  {Zhang}}, \bibinfo {author} {\bibfnamefont {T.}~\bibnamefont {Liu}},\ and\
  \bibinfo {author} {\bibfnamefont {L.}~\bibnamefont {Fu}},\ }\bibfield
  {title} {\bibinfo {title} {Electronic structures, charge transfer, and charge
  order in twisted transition metal dichalcogenide bilayers},\ }\href@noop {}
  {\bibfield  {journal} {\bibinfo  {journal} {Physical Review B}\ }\textbf
  {\bibinfo {volume} {103}},\ \bibinfo {pages} {155142} (\bibinfo {year}
  {2021})}\BibitemShut {NoStop}%
\bibitem [{\citenamefont {Sung}\ \emph {et~al.}(2020)\citenamefont {Sung},
  \citenamefont {Zhou}, \citenamefont {Scuri}, \citenamefont {Z{\'o}lyomi},
  \citenamefont {Andersen}, \citenamefont {Yoo}, \citenamefont {Wild},
  \citenamefont {Joe}, \citenamefont {Gelly}, \citenamefont {Heo} \emph
  {et~al.}}]{sung2020broken}%
  \BibitemOpen
  \bibfield  {author} {\bibinfo {author} {\bibfnamefont {J.}~\bibnamefont
  {Sung}}, \bibinfo {author} {\bibfnamefont {Y.}~\bibnamefont {Zhou}}, \bibinfo
  {author} {\bibfnamefont {G.}~\bibnamefont {Scuri}}, \bibinfo {author}
  {\bibfnamefont {V.}~\bibnamefont {Z{\'o}lyomi}}, \bibinfo {author}
  {\bibfnamefont {T.~I.}\ \bibnamefont {Andersen}}, \bibinfo {author}
  {\bibfnamefont {H.}~\bibnamefont {Yoo}}, \bibinfo {author} {\bibfnamefont
  {D.~S.}\ \bibnamefont {Wild}}, \bibinfo {author} {\bibfnamefont {A.~Y.}\
  \bibnamefont {Joe}}, \bibinfo {author} {\bibfnamefont {R.~J.}\ \bibnamefont
  {Gelly}}, \bibinfo {author} {\bibfnamefont {H.}~\bibnamefont {Heo}}, \emph
  {et~al.},\ }\bibfield  {title} {\bibinfo {title} {Broken mirror symmetry in
  excitonic response of reconstructed domains in twisted mose2/mose2
  bilayers},\ }\href@noop {} {\bibfield  {journal} {\bibinfo  {journal} {Nature
  Nanotechnology}\ }\textbf {\bibinfo {volume} {15}},\ \bibinfo {pages} {750}
  (\bibinfo {year} {2020})}\BibitemShut {NoStop}%
\bibitem [{\citenamefont {Wang}\ \emph {et~al.}(2022)\citenamefont {Wang},
  \citenamefont {Yu}, \citenamefont {Kwan}, \citenamefont {Jia}, \citenamefont
  {Lei}, \citenamefont {Klemenz}, \citenamefont {Cevallos}, \citenamefont
  {Singha}, \citenamefont {Devakul}, \citenamefont {Watanabe} \emph
  {et~al.}}]{wang2022one}%
  \BibitemOpen
  \bibfield  {author} {\bibinfo {author} {\bibfnamefont {P.}~\bibnamefont
  {Wang}}, \bibinfo {author} {\bibfnamefont {G.}~\bibnamefont {Yu}}, \bibinfo
  {author} {\bibfnamefont {Y.~H.}\ \bibnamefont {Kwan}}, \bibinfo {author}
  {\bibfnamefont {Y.}~\bibnamefont {Jia}}, \bibinfo {author} {\bibfnamefont
  {S.}~\bibnamefont {Lei}}, \bibinfo {author} {\bibfnamefont {S.}~\bibnamefont
  {Klemenz}}, \bibinfo {author} {\bibfnamefont {F.~A.}\ \bibnamefont
  {Cevallos}}, \bibinfo {author} {\bibfnamefont {R.}~\bibnamefont {Singha}},
  \bibinfo {author} {\bibfnamefont {T.}~\bibnamefont {Devakul}}, \bibinfo
  {author} {\bibfnamefont {K.}~\bibnamefont {Watanabe}}, \emph {et~al.},\
  }\bibfield  {title} {\bibinfo {title} {One-dimensional luttinger liquids in a
  two-dimensional moir{\'e} lattice},\ }\href@noop {} {\bibfield  {journal}
  {\bibinfo  {journal} {Nature}\ }\textbf {\bibinfo {volume} {605}},\ \bibinfo
  {pages} {57} (\bibinfo {year} {2022})}\BibitemShut {NoStop}%
\bibitem [{\citenamefont {Liu}\ and\ \citenamefont
  {Dai}(2020)}]{Liu2020AHEInBilayerGraphene}%
  \BibitemOpen
  \bibfield  {author} {\bibinfo {author} {\bibfnamefont {J.}~\bibnamefont
  {Liu}}\ and\ \bibinfo {author} {\bibfnamefont {X.}~\bibnamefont {Dai}},\
  }\bibfield  {title} {\bibinfo {title} {Anomalous {Hall} effect,
  magneto-optical properties, and nonlinear optical properties of twisted
  graphene systems},\ }\href {https://doi.org/10.1038/s41524-020-0299-4}
  {\bibfield  {journal} {\bibinfo  {journal} {npj Computational Materials}\
  }\textbf {\bibinfo {volume} {6}},\ \bibinfo {pages} {57} (\bibinfo {year}
  {2020})}\BibitemShut {NoStop}%
\bibitem [{\citenamefont {Moon}\ and\ \citenamefont
  {Koshino}(2013)}]{moon2013optical}%
  \BibitemOpen
  \bibfield  {author} {\bibinfo {author} {\bibfnamefont {P.}~\bibnamefont
  {Moon}}\ and\ \bibinfo {author} {\bibfnamefont {M.}~\bibnamefont {Koshino}},\
  }\bibfield  {title} {\bibinfo {title} {Optical absorption in twisted bilayer
  graphene},\ }\href@noop {} {\bibfield  {journal} {\bibinfo  {journal}
  {Physical Review B}\ }\textbf {\bibinfo {volume} {87}},\ \bibinfo {pages}
  {205404} (\bibinfo {year} {2013})}\BibitemShut {NoStop}%
\bibitem [{\citenamefont {Ventura}\ \emph {et~al.}(2017)\citenamefont
  {Ventura}, \citenamefont {Passos}, \citenamefont {Lopes~dos Santos},
  \citenamefont {Viana Parente~Lopes},\ and\ \citenamefont
  {Peres}}]{Peres2018GuageCovarianceInNonlinearOpticalResponse}%
  \BibitemOpen
  \bibfield  {author} {\bibinfo {author} {\bibfnamefont {G.~B.}\ \bibnamefont
  {Ventura}}, \bibinfo {author} {\bibfnamefont {D.~J.}\ \bibnamefont {Passos}},
  \bibinfo {author} {\bibfnamefont {J.~M.~B.}\ \bibnamefont {Lopes~dos
  Santos}}, \bibinfo {author} {\bibfnamefont {J.~M.}\ \bibnamefont {Viana
  Parente~Lopes}},\ and\ \bibinfo {author} {\bibfnamefont {N.~M.~R.}\
  \bibnamefont {Peres}},\ }\bibfield  {title} {\bibinfo {title} {Gauge
  covariances and nonlinear optical responses},\ }\href
  {https://doi.org/10.1103/PhysRevB.96.035431} {\bibfield  {journal} {\bibinfo
  {journal} {Phys. Rev. B}\ }\textbf {\bibinfo {volume} {96}},\ \bibinfo
  {pages} {035431} (\bibinfo {year} {2017})}\BibitemShut {NoStop}%
\bibitem [{\citenamefont {Passos}\ \emph {et~al.}(2018)\citenamefont {Passos},
  \citenamefont {Ventura}, \citenamefont {Lopes}, \citenamefont {{dos
  Santos}},\ and\ \citenamefont {Peres}}]{passos2018nonlinear}%
  \BibitemOpen
  \bibfield  {author} {\bibinfo {author} {\bibfnamefont {D.}~\bibnamefont
  {Passos}}, \bibinfo {author} {\bibfnamefont {G.}~\bibnamefont {Ventura}},
  \bibinfo {author} {\bibfnamefont {J.~V.~P.}\ \bibnamefont {Lopes}}, \bibinfo
  {author} {\bibfnamefont {J.~L.}\ \bibnamefont {{dos Santos}}},\ and\ \bibinfo
  {author} {\bibfnamefont {N.}~\bibnamefont {Peres}},\ }\bibfield  {title}
  {\bibinfo {title} {Nonlinear optical responses of crystalline systems:
  {{Results}} from a velocity gauge analysis},\ }\href@noop {} {\bibfield
  {journal} {\bibinfo  {journal} {Physical Review B}\ }\textbf {\bibinfo
  {volume} {97}},\ \bibinfo {pages} {235446} (\bibinfo {year}
  {2018})}\BibitemShut {NoStop}%
\bibitem [{\citenamefont {Zhong}(2018)}]{zhong2018linear}%
  \BibitemOpen
  \bibfield  {author} {\bibinfo {author} {\bibfnamefont {S.}~\bibnamefont
  {Zhong}},\ }\href@noop {} {\emph {\bibinfo {title} {Linear and Nonlinear
  Electromagnetic Responses in Topological Semimetals}}}\ (\bibinfo
  {publisher} {{University of California, Berkeley}},\ \bibinfo {year}
  {2018})\BibitemShut {NoStop}%
\bibitem [{\citenamefont {Vignale}(1991)}]{Vignale1991OrbitalParamagnetism}%
  \BibitemOpen
  \bibfield  {author} {\bibinfo {author} {\bibfnamefont {G.}~\bibnamefont
  {Vignale}},\ }\bibfield  {title} {\bibinfo {title} {Orbital paramagnetism of
  electrons in a two-dimensional lattice},\ }\href
  {https://doi.org/10.1103/PhysRevLett.67.358} {\bibfield  {journal} {\bibinfo
  {journal} {Phys. Rev. Lett.}\ }\textbf {\bibinfo {volume} {67}},\ \bibinfo
  {pages} {358} (\bibinfo {year} {1991})}\BibitemShut {NoStop}%
\bibitem [{\citenamefont {Óscar Pozo~Ocaña}\ and\ \citenamefont
  {Souza}(2023)}]{Souza2023MidpointMethod}%
  \BibitemOpen
  \bibfield  {author} {\bibinfo {author} {\bibnamefont {Óscar Pozo~Ocaña}}\
  and\ \bibinfo {author} {\bibfnamefont {I.}~\bibnamefont {Souza}},\ }\bibfield
   {title} {\bibinfo {title} {{Multipole theory of optical spatial dispersion
  in crystals}},\ }\href {https://doi.org/10.21468/SciPostPhys.14.5.118}
  {\bibfield  {journal} {\bibinfo  {journal} {SciPost Phys.}\ }\textbf
  {\bibinfo {volume} {14}},\ \bibinfo {pages} {118} (\bibinfo {year}
  {2023})}\BibitemShut {NoStop}%
\bibitem [{\citenamefont {Blaha}\ \emph {et~al.}(1990)\citenamefont {Blaha},
  \citenamefont {Schwarz}, \citenamefont {Sorantin},\ and\ \citenamefont
  {Trickey}}]{blaha1990full}%
  \BibitemOpen
  \bibfield  {author} {\bibinfo {author} {\bibfnamefont {P.}~\bibnamefont
  {Blaha}}, \bibinfo {author} {\bibfnamefont {K.}~\bibnamefont {Schwarz}},
  \bibinfo {author} {\bibfnamefont {P.}~\bibnamefont {Sorantin}},\ and\
  \bibinfo {author} {\bibfnamefont {S.}~\bibnamefont {Trickey}},\ }\bibfield
  {title} {\bibinfo {title} {Full-potential, linearized augmented plane wave
  programs for crystalline systems},\ }\href@noop {} {\bibfield  {journal}
  {\bibinfo  {journal} {Computer physics communications}\ }\textbf {\bibinfo
  {volume} {59}},\ \bibinfo {pages} {399} (\bibinfo {year} {1990})}\BibitemShut
  {NoStop}%
\bibitem [{\citenamefont {Zhao}\ \emph {et~al.}(2023)\citenamefont {Zhao},
  \citenamefont {Mai}, \citenamefont {Bradlyn},\ and\ \citenamefont
  {Phillips}}]{zhao2023failure}%
  \BibitemOpen
  \bibfield  {author} {\bibinfo {author} {\bibfnamefont {J.}~\bibnamefont
  {Zhao}}, \bibinfo {author} {\bibfnamefont {P.}~\bibnamefont {Mai}}, \bibinfo
  {author} {\bibfnamefont {B.}~\bibnamefont {Bradlyn}},\ and\ \bibinfo {author}
  {\bibfnamefont {P.}~\bibnamefont {Phillips}},\ }\bibfield  {title} {\bibinfo
  {title} {Failure of topological invariants in strongly correlated matter},\
  }\href@noop {} {\bibfield  {journal} {\bibinfo  {journal} {Phys. Rev. Lett.}\
  }\textbf {\bibinfo {volume} {131}},\ \bibinfo {pages} {106601} (\bibinfo
  {year} {2023})}\BibitemShut {NoStop}%
\bibitem [{\citenamefont {Karplus}\ and\ \citenamefont
  {Schwinger}(1948)}]{karplus1948note}%
  \BibitemOpen
  \bibfield  {author} {\bibinfo {author} {\bibfnamefont {R.}~\bibnamefont
  {Karplus}}\ and\ \bibinfo {author} {\bibfnamefont {J.}~\bibnamefont
  {Schwinger}},\ }\bibfield  {title} {\bibinfo {title} {A note on saturation in
  microwave spectroscopy},\ }\href@noop {} {\bibfield  {journal} {\bibinfo
  {journal} {Physical Review}\ }\textbf {\bibinfo {volume} {73}},\ \bibinfo
  {pages} {1020} (\bibinfo {year} {1948})}\BibitemShut {NoStop}%
\bibitem [{\citenamefont {Vanderbilt}(2018)}]{vanderbilt2018berry}%
  \BibitemOpen
  \bibfield  {author} {\bibinfo {author} {\bibfnamefont {D.}~\bibnamefont
  {Vanderbilt}},\ }\href@noop {} {\emph {\bibinfo {title} {Berry {{Phases}} in
  {{Electronic Structure Theory}}: {{Electric Polarization}}, {{Orbital
  Magnetization}} and {{Topological Insulators}}}}}\ (\bibinfo  {publisher}
  {{Cambridge University Press}},\ \bibinfo {year} {2018})\BibitemShut
  {NoStop}%
\bibitem [{\citenamefont {Thonhauser}\ and\ \citenamefont
  {Vanderbilt}(2006)}]{thonhauser2006insulator}%
  \BibitemOpen
  \bibfield  {author} {\bibinfo {author} {\bibfnamefont {T.}~\bibnamefont
  {Thonhauser}}\ and\ \bibinfo {author} {\bibfnamefont {D.}~\bibnamefont
  {Vanderbilt}},\ }\bibfield  {title} {\bibinfo {title}
  {Insulator/{{Chern-Insulator}} transition in the {{Haldane}} model},\
  }\href@noop {} {\bibfield  {journal} {\bibinfo  {journal} {Physical Review
  B}\ }\textbf {\bibinfo {volume} {74}},\ \bibinfo {pages} {235111} (\bibinfo
  {year} {2006})}\BibitemShut {NoStop}%
\bibitem [{\citenamefont {Thouless}(1984)}]{thouless1984wannier}%
  \BibitemOpen
  \bibfield  {author} {\bibinfo {author} {\bibfnamefont {{\relax
  DJ}.}~\bibnamefont {Thouless}},\ }\bibfield  {title} {\bibinfo {title}
  {Wannier functions for magnetic sub-bands},\ }\href@noop {} {\bibfield
  {journal} {\bibinfo  {journal} {Journal of Physics C: Solid State Physics}\
  }\textbf {\bibinfo {volume} {17}},\ \bibinfo {pages} {L325} (\bibinfo {year}
  {1984})}\BibitemShut {NoStop}%
\bibitem [{\citenamefont {Peskin}\ and\ \citenamefont
  {Schroeder}(1995)}]{peskin1995introduction}%
  \BibitemOpen
  \bibfield  {author} {\bibinfo {author} {\bibfnamefont {M.}~\bibnamefont
  {Peskin}}\ and\ \bibinfo {author} {\bibfnamefont {D.}~\bibnamefont
  {Schroeder}},\ }\href {https://books.google.com/books?id=i35LALN0GosC} {\emph
  {\bibinfo {title} {An {{Introduction}} to {{Quantum Field Theory}}}}},\
  Advanced Book Classics\ (\bibinfo  {publisher} {{Avalon Publishing}},\
  \bibinfo {year} {1995})\BibitemShut {NoStop}%
\bibitem [{\citenamefont {Kaplan}\ \emph {et~al.}(2023)\citenamefont {Kaplan},
  \citenamefont {Holder},\ and\ \citenamefont
  {Yan}}]{Yan2023UnifyingSemiclassicsNonlinear}%
  \BibitemOpen
  \bibfield  {author} {\bibinfo {author} {\bibfnamefont {D.}~\bibnamefont
  {Kaplan}}, \bibinfo {author} {\bibfnamefont {T.}~\bibnamefont {Holder}},\
  and\ \bibinfo {author} {\bibfnamefont {B.}~\bibnamefont {Yan}},\ }\bibfield
  {title} {\bibinfo {title} {{Unifying semiclassics and quantum perturbation
  theory at nonlinear order}},\ }\href
  {https://doi.org/10.21468/SciPostPhys.14.4.082} {\bibfield  {journal}
  {\bibinfo  {journal} {SciPost Phys.}\ }\textbf {\bibinfo {volume} {14}},\
  \bibinfo {pages} {082} (\bibinfo {year} {2023})}\BibitemShut {NoStop}%
\bibitem [{\citenamefont {Coleman}(2015)}]{coleman2015introduction}%
  \BibitemOpen
  \bibfield  {author} {\bibinfo {author} {\bibfnamefont {P.}~\bibnamefont
  {Coleman}},\ }\href@noop {} {\emph {\bibinfo {title} {Introduction to
  Many-Body Physics}}}\ (\bibinfo  {publisher} {{Cambridge University Press}},\
  \bibinfo {year} {2015})\BibitemShut {NoStop}%
\bibitem [{\citenamefont {Bruus}\ and\ \citenamefont
  {Flensberg}(2004)}]{bruus2004manybody}%
  \BibitemOpen
  \bibfield  {author} {\bibinfo {author} {\bibfnamefont {H.}~\bibnamefont
  {Bruus}}\ and\ \bibinfo {author} {\bibfnamefont {K.}~\bibnamefont
  {Flensberg}},\ }\href@noop {} {\emph {\bibinfo {title} {Many-{{Body Quantum
  Theory}} in {{Condensed Matter Physics}}: {{An Introduction}}}}},\ \bibinfo
  {edition} {1st}\ ed.\ (\bibinfo  {publisher} {{Oxford University Press}},\
  \bibinfo {address} {{Oxford, England}},\ \bibinfo {year} {2004})\BibitemShut
  {NoStop}%
\bibitem [{\citenamefont {Ting}\ \emph {et~al.}(2018)\citenamefont {Ting},
  \citenamefont {Chen}, \citenamefont {Sohn},\ and\ \citenamefont
  {Derbes}}]{Coleman2018QFTBook}%
  \BibitemOpen
  \bibfield  {author} {\bibinfo {author} {\bibfnamefont {Y.-S.}\ \bibnamefont
  {Ting}}, \bibinfo {author} {\bibfnamefont {B.~G.-G.}\ \bibnamefont {Chen}},
  \bibinfo {author} {\bibfnamefont {R.}~\bibnamefont {Sohn}},\ and\ \bibinfo
  {author} {\bibfnamefont {D.}~\bibnamefont {Derbes}},\ }\href@noop {} {\emph
  {\bibinfo {title} {Quantum Field Theory: Lectures by Sydney Coleman}}}\
  (\bibinfo  {publisher} {World Scientific Publishing Co},\ \bibinfo {year}
  {2018})\BibitemShut {NoStop}%
\bibitem [{\citenamefont {McKay}\ and\ \citenamefont
  {Bradlyn}(2021)}]{McKay2021CDWWeyl}%
  \BibitemOpen
  \bibfield  {author} {\bibinfo {author} {\bibfnamefont {R.~C.}\ \bibnamefont
  {McKay}}\ and\ \bibinfo {author} {\bibfnamefont {B.}~\bibnamefont
  {Bradlyn}},\ }\bibfield  {title} {\bibinfo {title} {Optical response from
  charge-density waves in weyl semimetals},\ }\bibfield  {journal} {\bibinfo
  {journal} {Physical Review B}\ }\textbf {\bibinfo {volume} {104}},\ \href
  {https://doi.org/10.1103/physrevb.104.155120} {10.1103/physrevb.104.155120}
  (\bibinfo {year} {2021})\BibitemShut {NoStop}%
\bibitem [{\citenamefont {Altland}\ and\ \citenamefont
  {Simons}(2010)}]{altland2010condensed}%
  \BibitemOpen
  \bibfield  {author} {\bibinfo {author} {\bibfnamefont {A.}~\bibnamefont
  {Altland}}\ and\ \bibinfo {author} {\bibfnamefont {B.~D.}\ \bibnamefont
  {Simons}},\ }\href@noop {} {\emph {\bibinfo {title} {Condensed Matter Field
  Theory}}}\ (\bibinfo  {publisher} {{Cambridge university press}},\ \bibinfo
  {year} {2010})\BibitemShut {NoStop}%
\bibitem [{\citenamefont {Fenton}\ and\ \citenamefont
  {Psaltakis}(1983)}]{fenton1983infrared}%
  \BibitemOpen
  \bibfield  {author} {\bibinfo {author} {\bibfnamefont {E.~W.}\ \bibnamefont
  {Fenton}}\ and\ \bibinfo {author} {\bibfnamefont {G.~C.}\ \bibnamefont
  {Psaltakis}},\ }\bibfield  {title} {\bibinfo {title} {Infrared conductivity
  from spin and charge density waves: ({{TMTSF}}) {{2X}}},\ }\href
  {https://doi.org/10.1016/0038-1098(83)90064-9} {\bibfield  {journal}
  {\bibinfo  {journal} {Solid State Communications}\ }\textbf {\bibinfo
  {volume} {47}},\ \bibinfo {pages} {767} (\bibinfo {year} {1983})}\BibitemShut
  {NoStop}%
\bibitem [{\citenamefont {Takano}(1982)}]{takano1982diagrammatical}%
  \BibitemOpen
  \bibfield  {author} {\bibinfo {author} {\bibfnamefont {K.}~\bibnamefont
  {Takano}},\ }\bibfield  {title} {\bibinfo {title} {Diagrammatical
  {{Approach}} to {{Functional Integral Method}} in a {{One-Dimensional Peierls
  System}}},\ }\href {https://doi.org/10.1143/PTP.68.1} {\bibfield  {journal}
  {\bibinfo  {journal} {Progress of Theoretical Physics}\ }\textbf {\bibinfo
  {volume} {68}},\ \bibinfo {pages} {1} (\bibinfo {year} {1982})},\ \bibinfo
  {note} {\_eprint:
  https://academic.oup.com/ptp/article-pdf/68/1/1/5317559/68-1-1.pdf}\BibitemShut
  {NoStop}%
\bibitem [{Note1()}]{Note1}%
  \BibitemOpen
  \bibinfo {note} {Note that we can generalize this rule to allow for
  calculation of nonzero-$\protect \mathbf {G}$ Umklapp response in
  Eq.~\protect \textup {\hbox {\mathsurround \z@ \protect \normalfont
  (\ignorespaces \ref {eq:responsewithnonzeroG}\unskip \@@italiccorr )}} by
  imposing momentum conservation only modulo a reciprocal lattice
  vector}\BibitemShut {NoStop}%
\bibitem [{\citenamefont {Husain}\ \emph {et~al.}(2023)\citenamefont {Husain},
  \citenamefont {Huang}, \citenamefont {Mitrano}, \citenamefont {Rak},
  \citenamefont {Rubeck}, \citenamefont {Guo}, \citenamefont {Yang},
  \citenamefont {Sow}, \citenamefont {Maeno}, \citenamefont {Uchoa},
  \citenamefont {Chiang}, \citenamefont {Batson}, \citenamefont {Phillips},\
  and\ \citenamefont {Abbamonte}}]{husain2023pines}%
  \BibitemOpen
  \bibfield  {author} {\bibinfo {author} {\bibfnamefont {A.~A.}\ \bibnamefont
  {Husain}}, \bibinfo {author} {\bibfnamefont {E.~W.}\ \bibnamefont {Huang}},
  \bibinfo {author} {\bibfnamefont {M.}~\bibnamefont {Mitrano}}, \bibinfo
  {author} {\bibfnamefont {M.~S.}\ \bibnamefont {Rak}}, \bibinfo {author}
  {\bibfnamefont {S.~I.}\ \bibnamefont {Rubeck}}, \bibinfo {author}
  {\bibfnamefont {X.}~\bibnamefont {Guo}}, \bibinfo {author} {\bibfnamefont
  {H.}~\bibnamefont {Yang}}, \bibinfo {author} {\bibfnamefont {C.}~\bibnamefont
  {Sow}}, \bibinfo {author} {\bibfnamefont {Y.}~\bibnamefont {Maeno}}, \bibinfo
  {author} {\bibfnamefont {B.}~\bibnamefont {Uchoa}}, \bibinfo {author}
  {\bibfnamefont {T.~C.}\ \bibnamefont {Chiang}}, \bibinfo {author}
  {\bibfnamefont {P.~E.}\ \bibnamefont {Batson}}, \bibinfo {author}
  {\bibfnamefont {P.~W.}\ \bibnamefont {Phillips}},\ and\ \bibinfo {author}
  {\bibfnamefont {P.}~\bibnamefont {Abbamonte}},\ }\bibfield  {title} {\bibinfo
  {title} {Pines’ demon observed as a 3d acoustic plasmon in sr$_2$ruo$_4$},\
  }\bibfield  {journal} {\bibinfo  {journal} {Nature}\ }\href
  {https://doi.org/10.1038/s41586-023-06318-8} {10.1038/s41586-023-06318-8}
  (\bibinfo {year} {2023})\BibitemShut {NoStop}%
\bibitem [{\citenamefont {Kadanoff}\ and\ \citenamefont
  {Martin}(1963)}]{kadanoff1963hydrodynamic}%
  \BibitemOpen
  \bibfield  {author} {\bibinfo {author} {\bibfnamefont {L.~P.}\ \bibnamefont
  {Kadanoff}}\ and\ \bibinfo {author} {\bibfnamefont {P.~C.}\ \bibnamefont
  {Martin}},\ }\bibfield  {title} {\bibinfo {title} {Hydrodynamic equations and
  correlation functions},\ }\href@noop {} {\bibfield  {journal} {\bibinfo
  {journal} {Annals of Physics}\ }\textbf {\bibinfo {volume} {24}},\ \bibinfo
  {pages} {419} (\bibinfo {year} {1963})}\BibitemShut {NoStop}%
\bibitem [{\citenamefont {{Mendez-Valderrama}}\ \emph
  {et~al.}(2023)\citenamefont {{Mendez-Valderrama}}, \citenamefont {Mao},\ and\
  \citenamefont {Chowdhury}}]{mendez2023theory}%
  \BibitemOpen
  \bibfield  {author} {\bibinfo {author} {\bibfnamefont {{\relax
  JF}.}~\bibnamefont {{Mendez-Valderrama}}}, \bibinfo {author} {\bibfnamefont
  {D.}~\bibnamefont {Mao}},\ and\ \bibinfo {author} {\bibfnamefont
  {D.}~\bibnamefont {Chowdhury}},\ }\bibfield  {title} {\bibinfo {title} {A
  theory for the low-energy optical sum-rule in moir{\textbackslash}'e
  graphene},\ }\href@noop {} {\bibfield  {journal} {\bibinfo  {journal} {arXiv
  preprint arXiv:2312.03819}\ } (\bibinfo {year} {2023})},\ \Eprint
  {https://arxiv.org/abs/2312.03819} {arxiv:2312.03819} \BibitemShut {NoStop}%
\bibitem [{\citenamefont {Weyl}(1929)}]{Weyl1929OriginalWeylFermionPaper}%
  \BibitemOpen
  \bibfield  {author} {\bibinfo {author} {\bibfnamefont {H.}~\bibnamefont
  {Weyl}},\ }\bibfield  {title} {\bibinfo {title} {Gravitation and the
  electron},\ }\href {https://doi.org/10.1073/pnas.15.4.323} {\bibfield
  {journal} {\bibinfo  {journal} {Proceedings of the National Academy of
  Sciences}\ }\textbf {\bibinfo {volume} {15}},\ \bibinfo {pages} {323}
  (\bibinfo {year} {1929})},\ \Eprint
  {https://arxiv.org/abs/https://www.pnas.org/doi/pdf/10.1073/pnas.15.4.323}
  {https://www.pnas.org/doi/pdf/10.1073/pnas.15.4.323} \BibitemShut {NoStop}%
\bibitem [{\citenamefont {Steiner}\ \emph {et~al.}(2017)\citenamefont
  {Steiner}, \citenamefont {Andreev},\ and\ \citenamefont
  {Pesin}}]{Steiner2017AnomalousHallWSM}%
  \BibitemOpen
  \bibfield  {author} {\bibinfo {author} {\bibfnamefont {J.}~\bibnamefont
  {Steiner}}, \bibinfo {author} {\bibfnamefont {A.}~\bibnamefont {Andreev}},\
  and\ \bibinfo {author} {\bibfnamefont {D.}~\bibnamefont {Pesin}},\ }\bibfield
   {title} {\bibinfo {title} {Anomalous hall effect in type-i weyl metals},\
  }\bibfield  {journal} {\bibinfo  {journal} {Physical Review Letters}\
  }\textbf {\bibinfo {volume} {119}},\ \href
  {https://doi.org/10.1103/physrevlett.119.036601}
  {10.1103/physrevlett.119.036601} (\bibinfo {year} {2017})\BibitemShut
  {NoStop}%
\bibitem [{\citenamefont {Burkov}(2014)}]{burkov2014anomalous}%
  \BibitemOpen
  \bibfield  {author} {\bibinfo {author} {\bibfnamefont {A.~A.}\ \bibnamefont
  {Burkov}},\ }\bibfield  {title} {\bibinfo {title} {Anomalous {{Hall Effect}}
  in {{Weyl Metals}}},\ }\href {https://doi.org/10.1103/PhysRevLett.113.187202}
  {\bibfield  {journal} {\bibinfo  {journal} {Physical Review Letters}\
  }\textbf {\bibinfo {volume} {113}},\ \bibinfo {pages} {187202} (\bibinfo
  {year} {2014})},\ \bibinfo {note} {arXiv: 1406.3033}\BibitemShut {NoStop}%
\bibitem [{\citenamefont {Huang}\ \emph {et~al.}(2017)\citenamefont {Huang},
  \citenamefont {Zhou},\ and\ \citenamefont
  {Shen}}]{Huang2017MultifoldWSMAnomHall}%
  \BibitemOpen
  \bibfield  {author} {\bibinfo {author} {\bibfnamefont {Z.-M.}\ \bibnamefont
  {Huang}}, \bibinfo {author} {\bibfnamefont {J.}~\bibnamefont {Zhou}},\ and\
  \bibinfo {author} {\bibfnamefont {S.-Q.}\ \bibnamefont {Shen}},\ }\bibfield
  {title} {\bibinfo {title} {Topological responses from chiral anomaly in
  multi-weyl semimetals},\ }\bibfield  {journal} {\bibinfo  {journal} {Physical
  Review B}\ }\textbf {\bibinfo {volume} {96}},\ \href
  {https://doi.org/10.1103/physrevb.96.085201} {10.1103/physrevb.96.085201}
  (\bibinfo {year} {2017})\BibitemShut {NoStop}%
\bibitem [{\citenamefont {McCormick}\ \emph
  {et~al.}(2017{\natexlab{a}})\citenamefont {McCormick}, \citenamefont
  {Kimchi},\ and\ \citenamefont {Trivedi}}]{mccormick2017minimal}%
  \BibitemOpen
  \bibfield  {author} {\bibinfo {author} {\bibfnamefont {T.~M.}\ \bibnamefont
  {McCormick}}, \bibinfo {author} {\bibfnamefont {I.}~\bibnamefont {Kimchi}},\
  and\ \bibinfo {author} {\bibfnamefont {N.}~\bibnamefont {Trivedi}},\
  }\bibfield  {title} {\bibinfo {title} {Minimal models for topological
  {{Weyl}} semimetals},\ }\href@noop {} {\bibfield  {journal} {\bibinfo
  {journal} {Physical Review B}\ }\textbf {\bibinfo {volume} {95}},\ \bibinfo
  {pages} {075133} (\bibinfo {year} {2017}{\natexlab{a}})}\BibitemShut
  {NoStop}%
\bibitem [{\citenamefont {McCormick}\ \emph
  {et~al.}(2017{\natexlab{b}})\citenamefont {McCormick}, \citenamefont
  {McKay},\ and\ \citenamefont {Trivedi}}]{mccormick2017tiltedweyl}%
  \BibitemOpen
  \bibfield  {author} {\bibinfo {author} {\bibfnamefont {T.~M.}\ \bibnamefont
  {McCormick}}, \bibinfo {author} {\bibfnamefont {R.~C.}\ \bibnamefont
  {McKay}},\ and\ \bibinfo {author} {\bibfnamefont {N.}~\bibnamefont
  {Trivedi}},\ }\bibfield  {title} {\bibinfo {title} {Semiclassical theory of
  anomalous transport in type-{II} topological weyl semimetals},\ }\bibfield
  {journal} {\bibinfo  {journal} {Physical Review B}\ }\textbf {\bibinfo
  {volume} {96}},\ \href {https://doi.org/10.1103/physrevb.96.235116}
  {10.1103/physrevb.96.235116} (\bibinfo {year}
  {2017}{\natexlab{b}})\BibitemShut {NoStop}%
\bibitem [{\citenamefont {Yang}\ \emph {et~al.}(2011)\citenamefont {Yang},
  \citenamefont {Lu},\ and\ \citenamefont {Ran}}]{Ran2011WeylModel}%
  \BibitemOpen
  \bibfield  {author} {\bibinfo {author} {\bibfnamefont {K.-Y.}\ \bibnamefont
  {Yang}}, \bibinfo {author} {\bibfnamefont {Y.-M.}\ \bibnamefont {Lu}},\ and\
  \bibinfo {author} {\bibfnamefont {Y.}~\bibnamefont {Ran}},\ }\bibfield
  {title} {\bibinfo {title} {Quantum hall effects in a weyl semimetal: Possible
  application in pyrochlore iridates},\ }\href
  {https://doi.org/10.1103/PhysRevB.84.075129} {\bibfield  {journal} {\bibinfo
  {journal} {Phys. Rev. B}\ }\textbf {\bibinfo {volume} {84}},\ \bibinfo
  {pages} {075129} (\bibinfo {year} {2011})}\BibitemShut {NoStop}%
\bibitem [{\citenamefont {Sharma}\ \emph {et~al.}(2016)\citenamefont {Sharma},
  \citenamefont {Goswami},\ and\ \citenamefont
  {Tewari}}]{Tewari2016WeylMagnetothermalConductivity}%
  \BibitemOpen
  \bibfield  {author} {\bibinfo {author} {\bibfnamefont {G.}~\bibnamefont
  {Sharma}}, \bibinfo {author} {\bibfnamefont {P.}~\bibnamefont {Goswami}},\
  and\ \bibinfo {author} {\bibfnamefont {S.}~\bibnamefont {Tewari}},\
  }\bibfield  {title} {\bibinfo {title} {Nernst and magnetothermal conductivity
  in a lattice model of weyl fermions},\ }\href
  {https://doi.org/10.1103/PhysRevB.93.035116} {\bibfield  {journal} {\bibinfo
  {journal} {Phys. Rev. B}\ }\textbf {\bibinfo {volume} {93}},\ \bibinfo
  {pages} {035116} (\bibinfo {year} {2016})}\BibitemShut {NoStop}%
\bibitem [{\citenamefont {Wieder}\ \emph {et~al.}(2020)\citenamefont {Wieder},
  \citenamefont {Lin},\ and\ \citenamefont
  {Bradlyn}}]{Bradlyn2020AxionicWSMCDW}%
  \BibitemOpen
  \bibfield  {author} {\bibinfo {author} {\bibfnamefont {B.~J.}\ \bibnamefont
  {Wieder}}, \bibinfo {author} {\bibfnamefont {K.-S.}\ \bibnamefont {Lin}},\
  and\ \bibinfo {author} {\bibfnamefont {B.}~\bibnamefont {Bradlyn}},\
  }\bibfield  {title} {\bibinfo {title} {Axionic band topology in
  inversion-symmetric weyl-charge-density waves},\ }\bibfield  {journal}
  {\bibinfo  {journal} {Physical Review Research}\ }\textbf {\bibinfo {volume}
  {2}},\ \href {https://doi.org/10.1103/physrevresearch.2.042010}
  {10.1103/physrevresearch.2.042010} (\bibinfo {year} {2020})\BibitemShut
  {NoStop}%
\bibitem [{\citenamefont {Cano}\ \emph {et~al.}(2019)\citenamefont {Cano},
  \citenamefont {Bradlyn},\ and\ \citenamefont
  {Vergniory}}]{Cano2019MultifoldWeyl}%
  \BibitemOpen
  \bibfield  {author} {\bibinfo {author} {\bibfnamefont {J.}~\bibnamefont
  {Cano}}, \bibinfo {author} {\bibfnamefont {B.}~\bibnamefont {Bradlyn}},\ and\
  \bibinfo {author} {\bibfnamefont {M.~G.}\ \bibnamefont {Vergniory}},\
  }\bibfield  {title} {\bibinfo {title} {Multifold nodal points in magnetic
  materials},\ }\href {https://doi.org/10.1063/1.5124314} {\bibfield  {journal}
  {\bibinfo  {journal} {{APL} Materials}\ }\textbf {\bibinfo {volume} {7}},\
  \bibinfo {pages} {101125} (\bibinfo {year} {2019})}\BibitemShut {NoStop}%
\bibitem [{\citenamefont {Hoyos}(2014)}]{hoyos2014hall}%
  \BibitemOpen
  \bibfield  {author} {\bibinfo {author} {\bibfnamefont {C.}~\bibnamefont
  {Hoyos}},\ }\bibfield  {title} {\bibinfo {title} {Hall viscosity, topological
  states and effective theories},\ }\href@noop {} {\bibfield  {journal}
  {\bibinfo  {journal} {International Journal of Modern Physics B}\ }\textbf
  {\bibinfo {volume} {28}},\ \bibinfo {pages} {1430007} (\bibinfo {year}
  {2014})}\BibitemShut {NoStop}%
\bibitem [{\citenamefont {Sharma}(2016)}]{sharma2016topological}%
  \BibitemOpen
  \bibfield  {author} {\bibinfo {author} {\bibfnamefont {G.}~\bibnamefont
  {Sharma}},\ }\href@noop {} {\emph {\bibinfo {title} {Topological Phenomena in
  Metals and Superconductors}}}\ (\bibinfo  {publisher} {Clemson University},\
  \bibinfo {year} {2016})\BibitemShut {NoStop}%
\bibitem [{\citenamefont {Messica}\ \emph {et~al.}(2022)\citenamefont
  {Messica}, \citenamefont {Gutman},\ and\ \citenamefont
  {Ostrovsky}}]{messica2022anomalousWSM}%
  \BibitemOpen
  \bibfield  {author} {\bibinfo {author} {\bibfnamefont {Y.}~\bibnamefont
  {Messica}}, \bibinfo {author} {\bibfnamefont {D.~B.}\ \bibnamefont
  {Gutman}},\ and\ \bibinfo {author} {\bibfnamefont {P.~M.}\ \bibnamefont
  {Ostrovsky}},\ }\bibfield  {title} {\bibinfo {title} {Anomalous hall effect
  in disordered weyl semimetals},\ }\href@noop {} {\bibfield  {journal}
  {\bibinfo  {journal} {arXiv preprint arXiv:2211.03185}\ } (\bibinfo {year}
  {2022})},\ \Eprint {https://arxiv.org/abs/2211.03185} {arXiv:2211.03185}
  \BibitemShut {NoStop}%
\bibitem [{\citenamefont {Kanagaraj}\ \emph {et~al.}(2022)\citenamefont
  {Kanagaraj}, \citenamefont {Ning},\ and\ \citenamefont
  {He}}]{kanagaraj2022topologicalWSM}%
  \BibitemOpen
  \bibfield  {author} {\bibinfo {author} {\bibfnamefont {M.}~\bibnamefont
  {Kanagaraj}}, \bibinfo {author} {\bibfnamefont {J.}~\bibnamefont {Ning}},\
  and\ \bibinfo {author} {\bibfnamefont {L.}~\bibnamefont {He}},\ }\bibfield
  {title} {\bibinfo {title} {Topological {Co3Sn2S2} magnetic weyl semimetal:
  {From} fundamental understanding to diverse fields of study},\ }\href
  {https://doi.org/10.1016/j.revip.2022.100072} {\bibfield  {journal} {\bibinfo
   {journal} {Reviews in Physics}\ }\textbf {\bibinfo {volume} {8}},\ \bibinfo
  {pages} {100072} (\bibinfo {year} {2022})}\BibitemShut {NoStop}%
\bibitem [{\citenamefont {Buccheri}\ \emph {et~al.}(2022)\citenamefont
  {Buccheri}, \citenamefont {Egger},\ and\ \citenamefont
  {De~Martino}}]{Francesco2022QuadradicConductanceVsChemPot}%
  \BibitemOpen
  \bibfield  {author} {\bibinfo {author} {\bibfnamefont {F.}~\bibnamefont
  {Buccheri}}, \bibinfo {author} {\bibfnamefont {R.}~\bibnamefont {Egger}},\
  and\ \bibinfo {author} {\bibfnamefont {A.}~\bibnamefont {De~Martino}},\
  }\bibfield  {title} {\bibinfo {title} {Transport, refraction, and interface
  arcs in junctions of weyl semimetals},\ }\href
  {https://doi.org/10.1103/PhysRevB.106.045413} {\bibfield  {journal} {\bibinfo
   {journal} {Phys. Rev. B}\ }\textbf {\bibinfo {volume} {106}},\ \bibinfo
  {pages} {045413} (\bibinfo {year} {2022})}\BibitemShut {NoStop}%
\bibitem [{\citenamefont {Osterloh}\ \emph {et~al.}(1994)\citenamefont
  {Osterloh}, \citenamefont {Oppeneer}, \citenamefont {Sticht},\ and\
  \citenamefont {Kubler}}]{Osterloh1994MOKE}%
  \BibitemOpen
  \bibfield  {author} {\bibinfo {author} {\bibfnamefont {I.}~\bibnamefont
  {Osterloh}}, \bibinfo {author} {\bibfnamefont {P.~M.}\ \bibnamefont
  {Oppeneer}}, \bibinfo {author} {\bibfnamefont {J.}~\bibnamefont {Sticht}},\
  and\ \bibinfo {author} {\bibfnamefont {J.}~\bibnamefont {Kubler}},\
  }\bibfield  {title} {\bibinfo {title} {A theoretical study of the
  magneto-optical {Kerr} effect in {FeX} ({X}={Co},{Ni},{Pd},{Pt})},\ }\href
  {https://doi.org/10.1088/0953-8984/6/1/028} {\bibfield  {journal} {\bibinfo
  {journal} {Journal of Physics: Condensed Matter}\ }\textbf {\bibinfo {volume}
  {6}},\ \bibinfo {pages} {285} (\bibinfo {year} {1994})}\BibitemShut {NoStop}%
\bibitem [{\citenamefont {Armitage}(2014)}]{armitage2014constraints}%
  \BibitemOpen
  \bibfield  {author} {\bibinfo {author} {\bibfnamefont {N.}~\bibnamefont
  {Armitage}},\ }\bibfield  {title} {\bibinfo {title} {Constraints on jones
  transmission matrices from time-reversal invariance and discrete spatial
  symmetries},\ }\href@noop {} {\bibfield  {journal} {\bibinfo  {journal}
  {Physical Review B}\ }\textbf {\bibinfo {volume} {90}},\ \bibinfo {pages}
  {035135} (\bibinfo {year} {2014})}\BibitemShut {NoStop}%
\bibitem [{\citenamefont {Xia}\ \emph {et~al.}(2006)\citenamefont {Xia},
  \citenamefont {Maeno}, \citenamefont {Beyersdorf}, \citenamefont {Fejer},\
  and\ \citenamefont {Kapitulnik}}]{xia2006high}%
  \BibitemOpen
  \bibfield  {author} {\bibinfo {author} {\bibfnamefont {J.}~\bibnamefont
  {Xia}}, \bibinfo {author} {\bibfnamefont {Y.}~\bibnamefont {Maeno}}, \bibinfo
  {author} {\bibfnamefont {P.~T.}\ \bibnamefont {Beyersdorf}}, \bibinfo
  {author} {\bibfnamefont {M.}~\bibnamefont {Fejer}},\ and\ \bibinfo {author}
  {\bibfnamefont {A.}~\bibnamefont {Kapitulnik}},\ }\bibfield  {title}
  {\bibinfo {title} {High resolution polar kerr effect measurements of sr 2 ruo
  4: Evidence for broken time-reversal symmetry in the superconducting state},\
  }\href@noop {} {\bibfield  {journal} {\bibinfo  {journal} {Physical review
  letters}\ }\textbf {\bibinfo {volume} {97}},\ \bibinfo {pages} {167002}
  (\bibinfo {year} {2006})}\BibitemShut {NoStop}%
\bibitem [{\citenamefont {Xia}\ \emph {et~al.}(2007)\citenamefont {Xia},
  \citenamefont {Schemm}, \citenamefont {Deutscher}, \citenamefont {Kivelson},
  \citenamefont {Bonn}, \citenamefont {Hardy}, \citenamefont {Liang},
  \citenamefont {Siemons}, \citenamefont {Koster}, \citenamefont {Fejer} \emph
  {et~al.}}]{xia2007polar}%
  \BibitemOpen
  \bibfield  {author} {\bibinfo {author} {\bibfnamefont {J.}~\bibnamefont
  {Xia}}, \bibinfo {author} {\bibfnamefont {E.}~\bibnamefont {Schemm}},
  \bibinfo {author} {\bibfnamefont {G.}~\bibnamefont {Deutscher}}, \bibinfo
  {author} {\bibfnamefont {S.}~\bibnamefont {Kivelson}}, \bibinfo {author}
  {\bibfnamefont {D.}~\bibnamefont {Bonn}}, \bibinfo {author} {\bibfnamefont
  {W.}~\bibnamefont {Hardy}}, \bibinfo {author} {\bibfnamefont
  {R.}~\bibnamefont {Liang}}, \bibinfo {author} {\bibfnamefont
  {W.}~\bibnamefont {Siemons}}, \bibinfo {author} {\bibfnamefont
  {G.}~\bibnamefont {Koster}}, \bibinfo {author} {\bibfnamefont
  {M.}~\bibnamefont {Fejer}}, \emph {et~al.},\ }\bibfield  {title} {\bibinfo
  {title} {Polar kerr effect measurements of yba2cu3o6+ x: Evidence for broken
  symmetry near the pseudogap temperature},\ }\href@noop {} {\bibfield
  {journal} {\bibinfo  {journal} {arXiv preprint arXiv:0711.2494}\ } (\bibinfo
  {year} {2007})}\BibitemShut {NoStop}%
\bibitem [{\citenamefont {Schemm}\ \emph {et~al.}(2014)\citenamefont {Schemm},
  \citenamefont {Gannon}, \citenamefont {Wishne}, \citenamefont {Halperin},\
  and\ \citenamefont {Kapitulnik}}]{schemm2014observation}%
  \BibitemOpen
  \bibfield  {author} {\bibinfo {author} {\bibfnamefont {E.}~\bibnamefont
  {Schemm}}, \bibinfo {author} {\bibfnamefont {W.}~\bibnamefont {Gannon}},
  \bibinfo {author} {\bibfnamefont {C.}~\bibnamefont {Wishne}}, \bibinfo
  {author} {\bibfnamefont {W.}~\bibnamefont {Halperin}},\ and\ \bibinfo
  {author} {\bibfnamefont {A.}~\bibnamefont {Kapitulnik}},\ }\bibfield  {title}
  {\bibinfo {title} {Observation of broken time-reversal symmetry in the
  heavy-fermion superconductor upt3},\ }\href@noop {} {\bibfield  {journal}
  {\bibinfo  {journal} {Science}\ }\textbf {\bibinfo {volume} {345}},\ \bibinfo
  {pages} {190} (\bibinfo {year} {2014})}\BibitemShut {NoStop}%
\bibitem [{\citenamefont {Kapitulnik}\ \emph {et~al.}(2009)\citenamefont
  {Kapitulnik}, \citenamefont {Xia}, \citenamefont {Schemm},\ and\
  \citenamefont {Palevski}}]{kapitulnik2009polar}%
  \BibitemOpen
  \bibfield  {author} {\bibinfo {author} {\bibfnamefont {A.}~\bibnamefont
  {Kapitulnik}}, \bibinfo {author} {\bibfnamefont {J.}~\bibnamefont {Xia}},
  \bibinfo {author} {\bibfnamefont {E.}~\bibnamefont {Schemm}},\ and\ \bibinfo
  {author} {\bibfnamefont {A.}~\bibnamefont {Palevski}},\ }\bibfield  {title}
  {\bibinfo {title} {Polar kerr effect as probe for time-reversal symmetry
  breaking in unconventional superconductors},\ }\href@noop {} {\bibfield
  {journal} {\bibinfo  {journal} {New Journal of Physics}\ }\textbf {\bibinfo
  {volume} {11}},\ \bibinfo {pages} {055060} (\bibinfo {year}
  {2009})}\BibitemShut {NoStop}%
\bibitem [{\citenamefont {Saykin}\ \emph {et~al.}(2023)\citenamefont {Saykin},
  \citenamefont {Farhang}, \citenamefont {Kountz}, \citenamefont {Chen},
  \citenamefont {Ortiz}, \citenamefont {Shekhar}, \citenamefont {Felser},
  \citenamefont {Wilson}, \citenamefont {Thomale}, \citenamefont {Xia} \emph
  {et~al.}}]{saykin2023high}%
  \BibitemOpen
  \bibfield  {author} {\bibinfo {author} {\bibfnamefont {D.~R.}\ \bibnamefont
  {Saykin}}, \bibinfo {author} {\bibfnamefont {C.}~\bibnamefont {Farhang}},
  \bibinfo {author} {\bibfnamefont {E.~D.}\ \bibnamefont {Kountz}}, \bibinfo
  {author} {\bibfnamefont {D.}~\bibnamefont {Chen}}, \bibinfo {author}
  {\bibfnamefont {B.~R.}\ \bibnamefont {Ortiz}}, \bibinfo {author}
  {\bibfnamefont {C.}~\bibnamefont {Shekhar}}, \bibinfo {author} {\bibfnamefont
  {C.}~\bibnamefont {Felser}}, \bibinfo {author} {\bibfnamefont {S.~D.}\
  \bibnamefont {Wilson}}, \bibinfo {author} {\bibfnamefont {R.}~\bibnamefont
  {Thomale}}, \bibinfo {author} {\bibfnamefont {J.}~\bibnamefont {Xia}}, \emph
  {et~al.},\ }\bibfield  {title} {\bibinfo {title} {High resolution polar kerr
  effect studies of csv 3 sb 5: Tests for time-reversal symmetry breaking below
  the charge-order transition},\ }\href@noop {} {\bibfield  {journal} {\bibinfo
   {journal} {Physical Review Letters}\ }\textbf {\bibinfo {volume} {131}},\
  \bibinfo {pages} {016901} (\bibinfo {year} {2023})}\BibitemShut {NoStop}%
\bibitem [{\citenamefont {Gong}\ \emph {et~al.}(2017)\citenamefont {Gong},
  \citenamefont {Kargarian}, \citenamefont {Stern}, \citenamefont {Yue},
  \citenamefont {Zhou}, \citenamefont {Jin}, \citenamefont {Galitski},
  \citenamefont {Yakovenko},\ and\ \citenamefont {Xia}}]{gong2017time}%
  \BibitemOpen
  \bibfield  {author} {\bibinfo {author} {\bibfnamefont {X.}~\bibnamefont
  {Gong}}, \bibinfo {author} {\bibfnamefont {M.}~\bibnamefont {Kargarian}},
  \bibinfo {author} {\bibfnamefont {A.}~\bibnamefont {Stern}}, \bibinfo
  {author} {\bibfnamefont {D.}~\bibnamefont {Yue}}, \bibinfo {author}
  {\bibfnamefont {H.}~\bibnamefont {Zhou}}, \bibinfo {author} {\bibfnamefont
  {X.}~\bibnamefont {Jin}}, \bibinfo {author} {\bibfnamefont {V.~M.}\
  \bibnamefont {Galitski}}, \bibinfo {author} {\bibfnamefont {V.~M.}\
  \bibnamefont {Yakovenko}},\ and\ \bibinfo {author} {\bibfnamefont
  {J.}~\bibnamefont {Xia}},\ }\bibfield  {title} {\bibinfo {title}
  {Time-reversal symmetry-breaking superconductivity in epitaxial
  bismuth/nickel bilayers},\ }\href@noop {} {\bibfield  {journal} {\bibinfo
  {journal} {Science advances}\ }\textbf {\bibinfo {volume} {3}},\ \bibinfo
  {pages} {e1602579} (\bibinfo {year} {2017})}\BibitemShut {NoStop}%
\bibitem [{\citenamefont {Haldane}(1988)}]{haldane1988HaldaneModel}%
  \BibitemOpen
  \bibfield  {author} {\bibinfo {author} {\bibfnamefont {F.~D.~M.}\
  \bibnamefont {Haldane}},\ }\bibfield  {title} {\bibinfo {title} {Model for a
  {Quantum} {Hall} {Effect} without {Landau} {Levels}: {Condensed}-{Matter}
  {Realization} of the "{Parity} {Anomaly}"},\ }\href
  {https://doi.org/10.1103/PhysRevLett.61.2015} {\bibfield  {journal} {\bibinfo
   {journal} {Physical Review Letters}\ }\textbf {\bibinfo {volume} {61}},\
  \bibinfo {pages} {2015} (\bibinfo {year} {1988})}\BibitemShut {NoStop}%
\bibitem [{\citenamefont {Mannaï}\ and\ \citenamefont
  {Haddad}(2020)}]{Manna2020StrainHaldaneModel}%
  \BibitemOpen
  \bibfield  {author} {\bibinfo {author} {\bibfnamefont {M.}~\bibnamefont
  {Mannaï}}\ and\ \bibinfo {author} {\bibfnamefont {S.}~\bibnamefont
  {Haddad}},\ }\bibfield  {title} {\bibinfo {title} {Strain tuned topology in
  the haldane and the modified haldane models},\ }\href
  {https://doi.org/10.1088/1361-648x/ab73a1} {\bibfield  {journal} {\bibinfo
  {journal} {Journal of Physics: Condensed Matter}\ }\textbf {\bibinfo {volume}
  {32}},\ \bibinfo {pages} {225501} (\bibinfo {year} {2020})}\BibitemShut
  {NoStop}%
\bibitem [{\citenamefont {Filusch}\ and\ \citenamefont
  {Fehske}(2023)}]{Filusch2023FlatHaldaneModel}%
  \BibitemOpen
  \bibfield  {author} {\bibinfo {author} {\bibfnamefont {A.}~\bibnamefont
  {Filusch}}\ and\ \bibinfo {author} {\bibfnamefont {H.}~\bibnamefont
  {Fehske}},\ }\bibfield  {title} {\bibinfo {title} {Singular flat bands in the
  modified {Haldane}-{Dice} model},\ }\href
  {https://doi.org/10.1016/j.physb.2023.414848} {\bibfield  {journal} {\bibinfo
   {journal} {Physica B: Condensed Matter}\ }\textbf {\bibinfo {volume}
  {659}},\ \bibinfo {pages} {414848} (\bibinfo {year} {2023})}\BibitemShut
  {NoStop}%
\bibitem [{\citenamefont {Wright}(2013)}]{wright2013BuckledHaldaneModel}%
  \BibitemOpen
  \bibfield  {author} {\bibinfo {author} {\bibfnamefont {A.~R.}\ \bibnamefont
  {Wright}},\ }\bibfield  {title} {\bibinfo {title} {Realising {Haldane}'s
  vision for a {Chern} insulator in buckled lattices},\ }\href
  {https://doi.org/10.1038/srep02736} {\bibfield  {journal} {\bibinfo
  {journal} {Scientific Reports}\ }\textbf {\bibinfo {volume} {3}},\ \bibinfo
  {pages} {2736} (\bibinfo {year} {2013})}\BibitemShut {NoStop}%
\bibitem [{\citenamefont {Po}\ \emph {et~al.}(2019)\citenamefont {Po},
  \citenamefont {Zou}, \citenamefont {Senthil},\ and\ \citenamefont
  {Vishwanath}}]{Po2019tightbindingModelsForBilayerGraphene}%
  \BibitemOpen
  \bibfield  {author} {\bibinfo {author} {\bibfnamefont {H.~C.}\ \bibnamefont
  {Po}}, \bibinfo {author} {\bibfnamefont {L.}~\bibnamefont {Zou}}, \bibinfo
  {author} {\bibfnamefont {T.}~\bibnamefont {Senthil}},\ and\ \bibinfo {author}
  {\bibfnamefont {A.}~\bibnamefont {Vishwanath}},\ }\bibfield  {title}
  {\bibinfo {title} {Faithful {Tight}-binding {Models} and {Fragile} {Topology}
  of {Magic}-angle {Bilayer} {Graphene}},\ }\href
  {https://doi.org/10.1103/PhysRevB.99.195455} {\bibfield  {journal} {\bibinfo
  {journal} {Physical Review B}\ }\textbf {\bibinfo {volume} {99}},\ \bibinfo
  {pages} {195455} (\bibinfo {year} {2019})},\ \bibinfo {note}
  {arXiv:1808.02482 [cond-mat]}\BibitemShut {NoStop}%
\bibitem [{\citenamefont {Brey}\ \emph {et~al.}(2020)\citenamefont {Brey},
  \citenamefont {Stauber}, \citenamefont {Slipchenko},\ and\ \citenamefont
  {Martín-Moreno}}]{brey2020PlasmonicsInBilayerGraphene}%
  \BibitemOpen
  \bibfield  {author} {\bibinfo {author} {\bibfnamefont {L.}~\bibnamefont
  {Brey}}, \bibinfo {author} {\bibfnamefont {T.}~\bibnamefont {Stauber}},
  \bibinfo {author} {\bibfnamefont {T.}~\bibnamefont {Slipchenko}},\ and\
  \bibinfo {author} {\bibfnamefont {L.}~\bibnamefont {Martín-Moreno}},\
  }\bibfield  {title} {\bibinfo {title} {Plasmonic {Dirac} {Cone} in {Twisted}
  {Bilayer} {Graphene}},\ }\href
  {https://doi.org/10.1103/PhysRevLett.125.256804} {\bibfield  {journal}
  {\bibinfo  {journal} {Physical Review Letters}\ }\textbf {\bibinfo {volume}
  {125}},\ \bibinfo {pages} {256804} (\bibinfo {year} {2020})},\ \bibinfo
  {note} {arXiv:2006.11763 [cond-mat]}\BibitemShut {NoStop}%
\bibitem [{\citenamefont {Kang}\ and\ \citenamefont
  {Vafek}(2018)}]{Kang2018WannierStatesInBilayerGraphene}%
  \BibitemOpen
  \bibfield  {author} {\bibinfo {author} {\bibfnamefont {J.}~\bibnamefont
  {Kang}}\ and\ \bibinfo {author} {\bibfnamefont {O.}~\bibnamefont {Vafek}},\
  }\bibfield  {title} {\bibinfo {title} {Symmetry, {Maximally} {Localized}
  {Wannier} {States}, and a {Low}-{Energy} {Model} for {Twisted} {Bilayer}
  {Graphene} {Narrow} {Bands}},\ }\href
  {https://doi.org/10.1103/PhysRevX.8.031088} {\bibfield  {journal} {\bibinfo
  {journal} {Physical Review X}\ }\textbf {\bibinfo {volume} {8}},\ \bibinfo
  {pages} {031088} (\bibinfo {year} {2018})}\BibitemShut {NoStop}%
\bibitem [{\citenamefont {Faulstich}\ \emph {et~al.}(2022)\citenamefont
  {Faulstich}, \citenamefont {Stubbs}, \citenamefont {Zhu}, \citenamefont
  {Soejima}, \citenamefont {Dilip}, \citenamefont {Zhai}, \citenamefont {Kim},
  \citenamefont {Zaletel}, \citenamefont {Chan},\ and\ \citenamefont
  {Lin}}]{Faulstich2022QuantumChemistryBilayerGraphene}%
  \BibitemOpen
  \bibfield  {author} {\bibinfo {author} {\bibfnamefont {F.~M.}\ \bibnamefont
  {Faulstich}}, \bibinfo {author} {\bibfnamefont {K.~D.}\ \bibnamefont
  {Stubbs}}, \bibinfo {author} {\bibfnamefont {Q.}~\bibnamefont {Zhu}},
  \bibinfo {author} {\bibfnamefont {T.}~\bibnamefont {Soejima}}, \bibinfo
  {author} {\bibfnamefont {R.}~\bibnamefont {Dilip}}, \bibinfo {author}
  {\bibfnamefont {H.}~\bibnamefont {Zhai}}, \bibinfo {author} {\bibfnamefont
  {R.}~\bibnamefont {Kim}}, \bibinfo {author} {\bibfnamefont {M.~P.}\
  \bibnamefont {Zaletel}}, \bibinfo {author} {\bibfnamefont {G.~K.-L.}\
  \bibnamefont {Chan}},\ and\ \bibinfo {author} {\bibfnamefont
  {L.}~\bibnamefont {Lin}},\ }\href {http://arxiv.org/abs/2211.09243} {\bibinfo
  {title} {Interacting models for twisted bilayer graphene: a quantum chemistry
  approach}} (\bibinfo {year} {2022})\BibitemShut {NoStop}%
\bibitem [{\citenamefont {Choi}\ \emph {et~al.}(2011)\citenamefont {Choi},
  \citenamefont {Lee}, \citenamefont {Kim}, \citenamefont {Kang}, \citenamefont
  {Shin}, \citenamefont {Kwak}, \citenamefont {Kang}, \citenamefont {Lee},
  \citenamefont {Park},\ and\ \citenamefont
  {Min}}]{Choi2011HighRefractiveIndex}%
  \BibitemOpen
  \bibfield  {author} {\bibinfo {author} {\bibfnamefont {M.}~\bibnamefont
  {Choi}}, \bibinfo {author} {\bibfnamefont {S.~H.}\ \bibnamefont {Lee}},
  \bibinfo {author} {\bibfnamefont {Y.}~\bibnamefont {Kim}}, \bibinfo {author}
  {\bibfnamefont {S.}~\bibnamefont {Kang}}, \bibinfo {author} {\bibfnamefont
  {J.}~\bibnamefont {Shin}}, \bibinfo {author} {\bibfnamefont {M.}~\bibnamefont
  {Kwak}}, \bibinfo {author} {\bibfnamefont {K.-Y.}\ \bibnamefont {Kang}},
  \bibinfo {author} {\bibfnamefont {Y.-H.}\ \bibnamefont {Lee}}, \bibinfo
  {author} {\bibfnamefont {N.}~\bibnamefont {Park}},\ and\ \bibinfo {author}
  {\bibfnamefont {B.}~\bibnamefont {Min}},\ }\bibfield  {title} {\bibinfo
  {title} {A terahertz metamaterial with unnaturally high refractive index},\
  }\href {https://doi.org/10.1038/nature09776} {\bibfield  {journal} {\bibinfo
  {journal} {Nature}\ }\textbf {\bibinfo {volume} {470}},\ \bibinfo {pages}
  {369} (\bibinfo {year} {2011})}\BibitemShut {NoStop}%
\bibitem [{\citenamefont {Luttinger}(1964)}]{luttinger1964theory}%
  \BibitemOpen
  \bibfield  {author} {\bibinfo {author} {\bibfnamefont {J.~M.}\ \bibnamefont
  {Luttinger}},\ }\bibfield  {title} {\bibinfo {title} {Theory of {{Thermal
  Transport Coefficients}}},\ }\href
  {https://doi.org/10.1103/PhysRev.135.A1505} {\bibfield  {journal} {\bibinfo
  {journal} {Phys. Rev.}\ }\textbf {\bibinfo {volume} {135}},\ \bibinfo {pages}
  {A1505} (\bibinfo {year} {1964})}\BibitemShut {NoStop}%
\bibitem [{\citenamefont {Cooper}\ \emph {et~al.}(1997)\citenamefont {Cooper},
  \citenamefont {Halperin},\ and\ \citenamefont
  {Ruzin}}]{cooper1997thermoelectric}%
  \BibitemOpen
  \bibfield  {author} {\bibinfo {author} {\bibfnamefont {N.}~\bibnamefont
  {Cooper}}, \bibinfo {author} {\bibfnamefont {B.}~\bibnamefont {Halperin}},\
  and\ \bibinfo {author} {\bibfnamefont {I.}~\bibnamefont {Ruzin}},\ }\bibfield
   {title} {\bibinfo {title} {Thermoelectric response of an interacting
  two-dimensional electron gas in a quantizing magnetic field},\ }\href@noop {}
  {\bibfield  {journal} {\bibinfo  {journal} {Physical Review B}\ }\textbf
  {\bibinfo {volume} {55}},\ \bibinfo {pages} {2344} (\bibinfo {year}
  {1997})}\BibitemShut {NoStop}%
\bibitem [{\citenamefont {Scheibner}\ \emph {et~al.}(2019)\citenamefont
  {Scheibner}, \citenamefont {Souslov}, \citenamefont {Banerjee}, \citenamefont
  {Surowka}, \citenamefont {Irvine},\ and\ \citenamefont
  {Vitelli}}]{scheibner2019odd}%
  \BibitemOpen
  \bibfield  {author} {\bibinfo {author} {\bibfnamefont {C.}~\bibnamefont
  {Scheibner}}, \bibinfo {author} {\bibfnamefont {A.}~\bibnamefont {Souslov}},
  \bibinfo {author} {\bibfnamefont {D.}~\bibnamefont {Banerjee}}, \bibinfo
  {author} {\bibfnamefont {P.}~\bibnamefont {Surowka}}, \bibinfo {author}
  {\bibfnamefont {W.~T.~M.}\ \bibnamefont {Irvine}},\ and\ \bibinfo {author}
  {\bibfnamefont {V.}~\bibnamefont {Vitelli}},\ }\href@noop {} {\emph {\bibinfo
  {title} {Odd Elasticity}}}\ (\bibinfo {year} {2019})\ \bibinfo {note}
  {\_eprint: 1902.07760}\BibitemShut {NoStop}%
\bibitem [{\citenamefont {Mauri}\ and\ \citenamefont
  {Louie}(1996)}]{Louie1996MagneticSusceptibilityOfInsulators}%
  \BibitemOpen
  \bibfield  {author} {\bibinfo {author} {\bibfnamefont {F.}~\bibnamefont
  {Mauri}}\ and\ \bibinfo {author} {\bibfnamefont {S.~G.}\ \bibnamefont
  {Louie}},\ }\bibfield  {title} {\bibinfo {title} {Magnetic susceptibility of
  insulators from first principles},\ }\href
  {https://doi.org/10.1103/PhysRevLett.76.4246} {\bibfield  {journal} {\bibinfo
   {journal} {Phys. Rev. Lett.}\ }\textbf {\bibinfo {volume} {76}},\ \bibinfo
  {pages} {4246} (\bibinfo {year} {1996})}\BibitemShut {NoStop}%
\bibitem [{\citenamefont {Bustamante}\ \emph {et~al.}(2021)\citenamefont
  {Bustamante}, \citenamefont {Wu}, \citenamefont {Fermon}, \citenamefont
  {Pannetier-Lecoeur}, \citenamefont {Wakamura}, \citenamefont {Watanabe},
  \citenamefont {Taniguchi}, \citenamefont {Pellegrin}, \citenamefont
  {Bernard}, \citenamefont {Daddinounou}, \citenamefont {Bouchiat},
  \citenamefont {Guéron}, \citenamefont {Ferrier}, \citenamefont
  {Montambaux},\ and\ \citenamefont
  {Bouchiat}}]{Bouchiat2021DiamagnetismInDirac}%
  \BibitemOpen
  \bibfield  {author} {\bibinfo {author} {\bibfnamefont {J.~V.}\ \bibnamefont
  {Bustamante}}, \bibinfo {author} {\bibfnamefont {N.~J.}\ \bibnamefont {Wu}},
  \bibinfo {author} {\bibfnamefont {C.}~\bibnamefont {Fermon}}, \bibinfo
  {author} {\bibfnamefont {M.}~\bibnamefont {Pannetier-Lecoeur}}, \bibinfo
  {author} {\bibfnamefont {T.}~\bibnamefont {Wakamura}}, \bibinfo {author}
  {\bibfnamefont {K.}~\bibnamefont {Watanabe}}, \bibinfo {author}
  {\bibfnamefont {T.}~\bibnamefont {Taniguchi}}, \bibinfo {author}
  {\bibfnamefont {T.}~\bibnamefont {Pellegrin}}, \bibinfo {author}
  {\bibfnamefont {A.}~\bibnamefont {Bernard}}, \bibinfo {author} {\bibfnamefont
  {S.}~\bibnamefont {Daddinounou}}, \bibinfo {author} {\bibfnamefont
  {V.}~\bibnamefont {Bouchiat}}, \bibinfo {author} {\bibfnamefont
  {S.}~\bibnamefont {Guéron}}, \bibinfo {author} {\bibfnamefont
  {M.}~\bibnamefont {Ferrier}}, \bibinfo {author} {\bibfnamefont
  {G.}~\bibnamefont {Montambaux}},\ and\ \bibinfo {author} {\bibfnamefont
  {H.}~\bibnamefont {Bouchiat}},\ }\bibfield  {title} {\bibinfo {title}
  {Detection of graphene's divergent orbital diamagnetism at the dirac point},\
  }\href {https://doi.org/10.1126/science.abf9396} {\bibfield  {journal}
  {\bibinfo  {journal} {Science}\ }\textbf {\bibinfo {volume} {374}},\ \bibinfo
  {pages} {1399} (\bibinfo {year} {2021})}\BibitemShut {NoStop}%
\bibitem [{\citenamefont {Oriekhov}\ \emph {et~al.}(2021)\citenamefont
  {Oriekhov}, \citenamefont {Gusynin},\ and\ \citenamefont
  {Loktev}}]{Loktev2021OrbitalSuscpInGraphene}%
  \BibitemOpen
  \bibfield  {author} {\bibinfo {author} {\bibfnamefont {D.~O.}\ \bibnamefont
  {Oriekhov}}, \bibinfo {author} {\bibfnamefont {V.~P.}\ \bibnamefont
  {Gusynin}},\ and\ \bibinfo {author} {\bibfnamefont {V.~M.}\ \bibnamefont
  {Loktev}},\ }\bibfield  {title} {\bibinfo {title} {Orbital susceptibility of
  t-graphene: Interplay of high-order van hove singularities and dirac cones},\
  }\href {https://doi.org/10.1103/PhysRevB.103.195104} {\bibfield  {journal}
  {\bibinfo  {journal} {Phys. Rev. B}\ }\textbf {\bibinfo {volume} {103}},\
  \bibinfo {pages} {195104} (\bibinfo {year} {2021})}\BibitemShut {NoStop}%
\bibitem [{\citenamefont {Gliozzi}\ \emph {et~al.}(2022)\citenamefont
  {Gliozzi}, \citenamefont {Lin},\ and\ \citenamefont
  {Hughes}}]{hughes2022quadrupolemagneticmoment}%
  \BibitemOpen
  \bibfield  {author} {\bibinfo {author} {\bibfnamefont {J.}~\bibnamefont
  {Gliozzi}}, \bibinfo {author} {\bibfnamefont {M.}~\bibnamefont {Lin}},\ and\
  \bibinfo {author} {\bibfnamefont {T.~L.}\ \bibnamefont {Hughes}},\ }\bibfield
   {title} {\bibinfo {title} {Orbital magnetic quadrupole moment in higher
  order topological phases},\ }\href@noop {} {\bibfield  {journal} {\bibinfo
  {journal} {arXiv preprint arXiv:2211.08438}\ } (\bibinfo {year} {2022})},\
  \Eprint {https://arxiv.org/abs/2211.08438} {arXiv:2211.08438} \BibitemShut
  {NoStop}%
\bibitem [{\citenamefont {Ceresoli}\ \emph {et~al.}(2006)\citenamefont
  {Ceresoli}, \citenamefont {Thonhauser}, \citenamefont {Vanderbilt},\ and\
  \citenamefont {Resta}}]{ceresoli2006orbital}%
  \BibitemOpen
  \bibfield  {author} {\bibinfo {author} {\bibfnamefont {D.}~\bibnamefont
  {Ceresoli}}, \bibinfo {author} {\bibfnamefont {T.}~\bibnamefont
  {Thonhauser}}, \bibinfo {author} {\bibfnamefont {D.}~\bibnamefont
  {Vanderbilt}},\ and\ \bibinfo {author} {\bibfnamefont {R.}~\bibnamefont
  {Resta}},\ }\bibfield  {title} {\bibinfo {title} {Orbital magnetization in
  crystalline solids: Multi-band insulators, chern insulators, and metals},\
  }\href@noop {} {\bibfield  {journal} {\bibinfo  {journal} {Physical Review
  B}\ }\textbf {\bibinfo {volume} {74}},\ \bibinfo {pages} {024408} (\bibinfo
  {year} {2006})}\BibitemShut {NoStop}%
\bibitem [{\citenamefont {Seleznev}\ and\ \citenamefont
  {Vanderbilt}(2022)}]{vanderbilt2022surfaceorbitalmagnetization}%
  \BibitemOpen
  \bibfield  {author} {\bibinfo {author} {\bibfnamefont {D.}~\bibnamefont
  {Seleznev}}\ and\ \bibinfo {author} {\bibfnamefont {D.}~\bibnamefont
  {Vanderbilt}},\ }\bibfield  {title} {\bibinfo {title} {Towards a theory of
  surface orbital magnetization},\ }\href@noop {} {\bibfield  {journal}
  {\bibinfo  {journal} {arXiv preprint arXiv:2210.08736}\ } (\bibinfo {year}
  {2022})},\ \Eprint {https://arxiv.org/abs/2210.08736} {arXiv:2210.08736}
  \BibitemShut {NoStop}%
\bibitem [{\citenamefont {Bohm}(1949)}]{bohm1949BlochsTheorem}%
  \BibitemOpen
  \bibfield  {author} {\bibinfo {author} {\bibfnamefont {D.}~\bibnamefont
  {Bohm}},\ }\bibfield  {title} {\bibinfo {title} {Note on a {Theorem} of
  {Bloch} {Concerning} {Possible} {Causes} of superconductivity},\ }\href
  {https://doi.org/10.1103/PhysRev.75.502} {\bibfield  {journal} {\bibinfo
  {journal} {Physical Review}\ }\textbf {\bibinfo {volume} {75}},\ \bibinfo
  {pages} {502} (\bibinfo {year} {1949})}\BibitemShut {NoStop}%
\bibitem [{\citenamefont {Watanabe}(2022)}]{watanabe2022bloch}%
  \BibitemOpen
  \bibfield  {author} {\bibinfo {author} {\bibfnamefont {H.}~\bibnamefont
  {Watanabe}},\ }\bibfield  {title} {\bibinfo {title} {Bloch theorem in the
  presence of an additional conserved charge},\ }\href@noop {} {\bibfield
  {journal} {\bibinfo  {journal} {Physical Review Research}\ }\textbf {\bibinfo
  {volume} {4}},\ \bibinfo {pages} {013043} (\bibinfo {year}
  {2022})}\BibitemShut {NoStop}%
\bibitem [{\citenamefont {Zeng}\ \emph {et~al.}(2021)\citenamefont {Zeng},
  \citenamefont {Nandy},\ and\ \citenamefont
  {Tewari}}]{Tewari2021NonlinearBerryCurvature}%
  \BibitemOpen
  \bibfield  {author} {\bibinfo {author} {\bibfnamefont {C.}~\bibnamefont
  {Zeng}}, \bibinfo {author} {\bibfnamefont {S.}~\bibnamefont {Nandy}},\ and\
  \bibinfo {author} {\bibfnamefont {S.}~\bibnamefont {Tewari}},\ }\bibfield
  {title} {\bibinfo {title} {Nonlinear transport in weyl semimetals induced by
  berry curvature dipole},\ }\href
  {https://doi.org/10.1103/PhysRevB.103.245119} {\bibfield  {journal} {\bibinfo
   {journal} {Phys. Rev. B}\ }\textbf {\bibinfo {volume} {103}},\ \bibinfo
  {pages} {245119} (\bibinfo {year} {2021})}\BibitemShut {NoStop}%
\bibitem [{\citenamefont {Rees}\ \emph {et~al.}(2020)\citenamefont {Rees},
  \citenamefont {Manna}, \citenamefont {Lu}, \citenamefont {Morimoto},
  \citenamefont {Borrmann}, \citenamefont {Felser}, \citenamefont {Moore},
  \citenamefont {Torchinsky},\ and\ \citenamefont
  {Orenstein}}]{rees2020helicitydependent}%
  \BibitemOpen
  \bibfield  {author} {\bibinfo {author} {\bibfnamefont {D.}~\bibnamefont
  {Rees}}, \bibinfo {author} {\bibfnamefont {K.}~\bibnamefont {Manna}},
  \bibinfo {author} {\bibfnamefont {B.}~\bibnamefont {Lu}}, \bibinfo {author}
  {\bibfnamefont {T.}~\bibnamefont {Morimoto}}, \bibinfo {author}
  {\bibfnamefont {H.}~\bibnamefont {Borrmann}}, \bibinfo {author}
  {\bibfnamefont {C.}~\bibnamefont {Felser}}, \bibinfo {author} {\bibfnamefont
  {J.}~\bibnamefont {Moore}}, \bibinfo {author} {\bibfnamefont {D.~H.}\
  \bibnamefont {Torchinsky}},\ and\ \bibinfo {author} {\bibfnamefont
  {J.}~\bibnamefont {Orenstein}},\ }\bibfield  {title} {\bibinfo {title}
  {Helicity-dependent photocurrents in the chiral {{Weyl}} semimetal
  {{RhSi}}},\ }\href@noop {} {\bibfield  {journal} {\bibinfo  {journal}
  {Science advances}\ }\textbf {\bibinfo {volume} {6}},\ \bibinfo {pages}
  {eaba0509} (\bibinfo {year} {2020})}\BibitemShut {NoStop}%
\bibitem [{\citenamefont {Kim}\ \emph {et~al.}(2023)\citenamefont {Kim},
  \citenamefont {Lv}, \citenamefont {Sun}, \citenamefont {Zhao}, \citenamefont
  {Bielinski}, \citenamefont {Murzabekova}, \citenamefont {Qu}, \citenamefont
  {Duncan}, \citenamefont {Nguyen}, \citenamefont {Trigo} \emph
  {et~al.}}]{kim2023observation}%
  \BibitemOpen
  \bibfield  {author} {\bibinfo {author} {\bibfnamefont {S.}~\bibnamefont
  {Kim}}, \bibinfo {author} {\bibfnamefont {Y.}~\bibnamefont {Lv}}, \bibinfo
  {author} {\bibfnamefont {X.-Q.}\ \bibnamefont {Sun}}, \bibinfo {author}
  {\bibfnamefont {C.}~\bibnamefont {Zhao}}, \bibinfo {author} {\bibfnamefont
  {N.}~\bibnamefont {Bielinski}}, \bibinfo {author} {\bibfnamefont
  {A.}~\bibnamefont {Murzabekova}}, \bibinfo {author} {\bibfnamefont
  {K.}~\bibnamefont {Qu}}, \bibinfo {author} {\bibfnamefont {R.~A.}\
  \bibnamefont {Duncan}}, \bibinfo {author} {\bibfnamefont {Q.~L.}\
  \bibnamefont {Nguyen}}, \bibinfo {author} {\bibfnamefont {M.}~\bibnamefont
  {Trigo}}, \emph {et~al.},\ }\bibfield  {title} {\bibinfo {title} {Observation
  of a massive phason in a charge-density-wave insulator},\ }\href@noop {}
  {\bibfield  {journal} {\bibinfo  {journal} {Nature Materials}\ }\textbf
  {\bibinfo {volume} {22}},\ \bibinfo {pages} {429} (\bibinfo {year}
  {2023})}\BibitemShut {NoStop}%
\bibitem [{\citenamefont {Weber}\ \emph {et~al.}(2007)\citenamefont {Weber},
  \citenamefont {Orenstein}, \citenamefont {Bernevig}, \citenamefont {Zhang},
  \citenamefont {Stephens},\ and\ \citenamefont
  {Awschalom}}]{weber2007nondiffusive}%
  \BibitemOpen
  \bibfield  {author} {\bibinfo {author} {\bibfnamefont {C.~P.}\ \bibnamefont
  {Weber}}, \bibinfo {author} {\bibfnamefont {J.}~\bibnamefont {Orenstein}},
  \bibinfo {author} {\bibfnamefont {B.~A.}\ \bibnamefont {Bernevig}}, \bibinfo
  {author} {\bibfnamefont {S.-C.}\ \bibnamefont {Zhang}}, \bibinfo {author}
  {\bibfnamefont {J.}~\bibnamefont {Stephens}},\ and\ \bibinfo {author}
  {\bibfnamefont {D.~D.}\ \bibnamefont {Awschalom}},\ }\bibfield  {title}
  {\bibinfo {title} {Nondiffusive spin dynamics in a two-dimensional electron
  gas},\ }\href@noop {} {\bibfield  {journal} {\bibinfo  {journal} {Physical
  review letters}\ }\textbf {\bibinfo {volume} {98}},\ \bibinfo {pages}
  {076604} (\bibinfo {year} {2007})}\BibitemShut {NoStop}%
\bibitem [{\citenamefont {Koralek}\ \emph {et~al.}(2009)\citenamefont
  {Koralek}, \citenamefont {Weber}, \citenamefont {Orenstein}, \citenamefont
  {Bernevig}, \citenamefont {Zhang}, \citenamefont {Mack},\ and\ \citenamefont
  {Awschalom}}]{koralek2009emergence}%
  \BibitemOpen
  \bibfield  {author} {\bibinfo {author} {\bibfnamefont {J.~D.}\ \bibnamefont
  {Koralek}}, \bibinfo {author} {\bibfnamefont {C.~P.}\ \bibnamefont {Weber}},
  \bibinfo {author} {\bibfnamefont {J.}~\bibnamefont {Orenstein}}, \bibinfo
  {author} {\bibfnamefont {B.~A.}\ \bibnamefont {Bernevig}}, \bibinfo {author}
  {\bibfnamefont {S.-C.}\ \bibnamefont {Zhang}}, \bibinfo {author}
  {\bibfnamefont {S.}~\bibnamefont {Mack}},\ and\ \bibinfo {author}
  {\bibfnamefont {D.}~\bibnamefont {Awschalom}},\ }\bibfield  {title} {\bibinfo
  {title} {Emergence of the persistent spin helix in semiconductor quantum
  wells},\ }\href@noop {} {\bibfield  {journal} {\bibinfo  {journal} {Nature}\
  }\textbf {\bibinfo {volume} {458}},\ \bibinfo {pages} {610} (\bibinfo {year}
  {2009})}\BibitemShut {NoStop}%
\bibitem [{\citenamefont {Torchinsky}\ \emph {et~al.}(2013)\citenamefont
  {Torchinsky}, \citenamefont {Mahmood}, \citenamefont {Bollinger},
  \citenamefont {Bo{\v{z}}ovi{\'c}},\ and\ \citenamefont
  {Gedik}}]{torchinsky2013fluctuating}%
  \BibitemOpen
  \bibfield  {author} {\bibinfo {author} {\bibfnamefont {D.~H.}\ \bibnamefont
  {Torchinsky}}, \bibinfo {author} {\bibfnamefont {F.}~\bibnamefont {Mahmood}},
  \bibinfo {author} {\bibfnamefont {A.~T.}\ \bibnamefont {Bollinger}}, \bibinfo
  {author} {\bibfnamefont {I.}~\bibnamefont {Bo{\v{z}}ovi{\'c}}},\ and\
  \bibinfo {author} {\bibfnamefont {N.}~\bibnamefont {Gedik}},\ }\bibfield
  {title} {\bibinfo {title} {Fluctuating charge-density waves in a cuprate
  superconductor},\ }\href@noop {} {\bibfield  {journal} {\bibinfo  {journal}
  {Nature materials}\ }\textbf {\bibinfo {volume} {12}},\ \bibinfo {pages}
  {387} (\bibinfo {year} {2013})}\BibitemShut {NoStop}%
\bibitem [{\citenamefont {Rouxel}\ \emph {et~al.}(2021)\citenamefont {Rouxel},
  \citenamefont {Fainozzi}, \citenamefont {Mankowsky}, \citenamefont
  {R{\"o}sner}, \citenamefont {Seniutinas}, \citenamefont {Mincigrucci},
  \citenamefont {Catalini}, \citenamefont {Foglia}, \citenamefont {Cucini},
  \citenamefont {D{\"o}ring} \emph {et~al.}}]{rouxel2021hard}%
  \BibitemOpen
  \bibfield  {author} {\bibinfo {author} {\bibfnamefont {J.~R.}\ \bibnamefont
  {Rouxel}}, \bibinfo {author} {\bibfnamefont {D.}~\bibnamefont {Fainozzi}},
  \bibinfo {author} {\bibfnamefont {R.}~\bibnamefont {Mankowsky}}, \bibinfo
  {author} {\bibfnamefont {B.}~\bibnamefont {R{\"o}sner}}, \bibinfo {author}
  {\bibfnamefont {G.}~\bibnamefont {Seniutinas}}, \bibinfo {author}
  {\bibfnamefont {R.}~\bibnamefont {Mincigrucci}}, \bibinfo {author}
  {\bibfnamefont {S.}~\bibnamefont {Catalini}}, \bibinfo {author}
  {\bibfnamefont {L.}~\bibnamefont {Foglia}}, \bibinfo {author} {\bibfnamefont
  {R.}~\bibnamefont {Cucini}}, \bibinfo {author} {\bibfnamefont
  {F.}~\bibnamefont {D{\"o}ring}}, \emph {et~al.},\ }\bibfield  {title}
  {\bibinfo {title} {Hard x-ray transient grating spectroscopy on bismuth
  germanate},\ }\href@noop {} {\bibfield  {journal} {\bibinfo  {journal}
  {Nature Photonics}\ }\textbf {\bibinfo {volume} {15}},\ \bibinfo {pages}
  {499} (\bibinfo {year} {2021})}\BibitemShut {NoStop}%
\bibitem [{\citenamefont {Bencivenga}\ \emph {et~al.}(2019)\citenamefont
  {Bencivenga}, \citenamefont {Mincigrucci}, \citenamefont {Capotondi},
  \citenamefont {Foglia}, \citenamefont {Naumenko}, \citenamefont {Maznev},
  \citenamefont {Pedersoli}, \citenamefont {Simoncig}, \citenamefont
  {Caporaletti}, \citenamefont {Chiloyan} \emph
  {et~al.}}]{bencivenga2019nanoscale}%
  \BibitemOpen
  \bibfield  {author} {\bibinfo {author} {\bibfnamefont {F.}~\bibnamefont
  {Bencivenga}}, \bibinfo {author} {\bibfnamefont {R.}~\bibnamefont
  {Mincigrucci}}, \bibinfo {author} {\bibfnamefont {F.}~\bibnamefont
  {Capotondi}}, \bibinfo {author} {\bibfnamefont {L.}~\bibnamefont {Foglia}},
  \bibinfo {author} {\bibfnamefont {D.}~\bibnamefont {Naumenko}}, \bibinfo
  {author} {\bibfnamefont {A.}~\bibnamefont {Maznev}}, \bibinfo {author}
  {\bibfnamefont {E.}~\bibnamefont {Pedersoli}}, \bibinfo {author}
  {\bibfnamefont {A.}~\bibnamefont {Simoncig}}, \bibinfo {author}
  {\bibfnamefont {F.}~\bibnamefont {Caporaletti}}, \bibinfo {author}
  {\bibfnamefont {V.}~\bibnamefont {Chiloyan}}, \emph {et~al.},\ }\bibfield
  {title} {\bibinfo {title} {Nanoscale transient gratings excited and probed by
  extreme ultraviolet femtosecond pulses},\ }\href@noop {} {\bibfield
  {journal} {\bibinfo  {journal} {Science advances}\ }\textbf {\bibinfo
  {volume} {5}},\ \bibinfo {pages} {eaaw5805} (\bibinfo {year}
  {2019})}\BibitemShut {NoStop}%
\bibitem [{\citenamefont {Takasan}\ \emph {et~al.}(2021)\citenamefont
  {Takasan}, \citenamefont {Morimoto}, \citenamefont {Orenstein},\ and\
  \citenamefont {Moore}}]{Moore2021SecondHarmonicGenerationinWeyl}%
  \BibitemOpen
  \bibfield  {author} {\bibinfo {author} {\bibfnamefont {K.}~\bibnamefont
  {Takasan}}, \bibinfo {author} {\bibfnamefont {T.}~\bibnamefont {Morimoto}},
  \bibinfo {author} {\bibfnamefont {J.}~\bibnamefont {Orenstein}},\ and\
  \bibinfo {author} {\bibfnamefont {J.~E.}\ \bibnamefont {Moore}},\ }\bibfield
  {title} {\bibinfo {title} {Current-induced second harmonic generation in
  inversion-symmetric dirac and weyl semimetals},\ }\href
  {https://doi.org/10.1103/PhysRevB.104.L161202} {\bibfield  {journal}
  {\bibinfo  {journal} {Phys. Rev. B}\ }\textbf {\bibinfo {volume} {104}},\
  \bibinfo {pages} {L161202} (\bibinfo {year} {2021})}\BibitemShut {NoStop}%
\bibitem [{\citenamefont {Lee}\ \emph {et~al.}(2020)\citenamefont {Lee},
  \citenamefont {Yap}, \citenamefont {Tai}, \citenamefont {Xu}, \citenamefont
  {Zhang},\ and\ \citenamefont {Gong}}]{Gong2020HigherHarmonicGeneration}%
  \BibitemOpen
  \bibfield  {author} {\bibinfo {author} {\bibfnamefont {C.~H.}\ \bibnamefont
  {Lee}}, \bibinfo {author} {\bibfnamefont {H.~H.}\ \bibnamefont {Yap}},
  \bibinfo {author} {\bibfnamefont {T.}~\bibnamefont {Tai}}, \bibinfo {author}
  {\bibfnamefont {G.}~\bibnamefont {Xu}}, \bibinfo {author} {\bibfnamefont
  {X.}~\bibnamefont {Zhang}},\ and\ \bibinfo {author} {\bibfnamefont
  {J.}~\bibnamefont {Gong}},\ }\bibfield  {title} {\bibinfo {title} {Enhanced
  higher harmonic generation from nodal topology},\ }\href
  {https://doi.org/10.1103/PhysRevB.102.035138} {\bibfield  {journal} {\bibinfo
   {journal} {Phys. Rev. B}\ }\textbf {\bibinfo {volume} {102}},\ \bibinfo
  {pages} {035138} (\bibinfo {year} {2020})}\BibitemShut {NoStop}%
\bibitem [{\citenamefont {Cea}\ \emph {et~al.}(2016)\citenamefont {Cea},
  \citenamefont {Castellani},\ and\ \citenamefont
  {Benfatto}}]{Benfatto2016ThirdHarmonicsInSuperconductors}%
  \BibitemOpen
  \bibfield  {author} {\bibinfo {author} {\bibfnamefont {T.}~\bibnamefont
  {Cea}}, \bibinfo {author} {\bibfnamefont {C.}~\bibnamefont {Castellani}},\
  and\ \bibinfo {author} {\bibfnamefont {L.}~\bibnamefont {Benfatto}},\
  }\bibfield  {title} {\bibinfo {title} {Nonlinear optical effects and
  third-harmonic generation in superconductors: Cooper pairs versus higgs mode
  contribution},\ }\href {https://doi.org/10.1103/PhysRevB.93.180507}
  {\bibfield  {journal} {\bibinfo  {journal} {Phys. Rev. B}\ }\textbf {\bibinfo
  {volume} {93}},\ \bibinfo {pages} {180507} (\bibinfo {year}
  {2016})}\BibitemShut {NoStop}%
\bibitem [{\citenamefont {Gooth}\ \emph {et~al.}(2019)\citenamefont {Gooth},
  \citenamefont {Bradlyn}, \citenamefont {Honnali}, \citenamefont {Schindler},
  \citenamefont {Kumar}, \citenamefont {Noky}, \citenamefont {Qi},
  \citenamefont {Shekhar}, \citenamefont {Sun}, \citenamefont {Wang},
  \citenamefont {Bernevig},\ and\ \citenamefont {Felser}}]{gooth2019axionic}%
  \BibitemOpen
  \bibfield  {author} {\bibinfo {author} {\bibfnamefont {J.}~\bibnamefont
  {Gooth}}, \bibinfo {author} {\bibfnamefont {B.}~\bibnamefont {Bradlyn}},
  \bibinfo {author} {\bibfnamefont {S.}~\bibnamefont {Honnali}}, \bibinfo
  {author} {\bibfnamefont {C.}~\bibnamefont {Schindler}}, \bibinfo {author}
  {\bibfnamefont {N.}~\bibnamefont {Kumar}}, \bibinfo {author} {\bibfnamefont
  {J.}~\bibnamefont {Noky}}, \bibinfo {author} {\bibfnamefont {Y.}~\bibnamefont
  {Qi}}, \bibinfo {author} {\bibfnamefont {C.}~\bibnamefont {Shekhar}},
  \bibinfo {author} {\bibfnamefont {Y.}~\bibnamefont {Sun}}, \bibinfo {author}
  {\bibfnamefont {Z.}~\bibnamefont {Wang}}, \bibinfo {author} {\bibfnamefont
  {B.~A.}\ \bibnamefont {Bernevig}},\ and\ \bibinfo {author} {\bibfnamefont
  {C.}~\bibnamefont {Felser}},\ }\bibfield  {title} {\bibinfo {title} {Axionic
  charge-density wave in the {{Weyl}} semimetal ({{TaSe4}}){{2I}}},\ }\href
  {https://doi.org/10.1038/s41586-019-1630-4} {\bibfield  {journal} {\bibinfo
  {journal} {Nature}\ }\textbf {\bibinfo {volume} {575}},\ \bibinfo {pages}
  {315} (\bibinfo {year} {2019})}\BibitemShut {NoStop}%
\bibitem [{\citenamefont {Shi}\ \emph {et~al.}(2021)\citenamefont {Shi},
  \citenamefont {Wieder}, \citenamefont {Meyerheim}, \citenamefont {Sun},
  \citenamefont {Zhang}, \citenamefont {Li}, \citenamefont {Shen},
  \citenamefont {Qi}, \citenamefont {Yang}, \citenamefont {Jena} \emph
  {et~al.}}]{shi2021chargedensitywave}%
  \BibitemOpen
  \bibfield  {author} {\bibinfo {author} {\bibfnamefont {W.}~\bibnamefont
  {Shi}}, \bibinfo {author} {\bibfnamefont {B.~J.}\ \bibnamefont {Wieder}},
  \bibinfo {author} {\bibfnamefont {H.}~\bibnamefont {Meyerheim}}, \bibinfo
  {author} {\bibfnamefont {Y.}~\bibnamefont {Sun}}, \bibinfo {author}
  {\bibfnamefont {Y.}~\bibnamefont {Zhang}}, \bibinfo {author} {\bibfnamefont
  {Y.}~\bibnamefont {Li}}, \bibinfo {author} {\bibfnamefont {L.}~\bibnamefont
  {Shen}}, \bibinfo {author} {\bibfnamefont {Y.}~\bibnamefont {Qi}}, \bibinfo
  {author} {\bibfnamefont {L.}~\bibnamefont {Yang}}, \bibinfo {author}
  {\bibfnamefont {J.}~\bibnamefont {Jena}}, \emph {et~al.},\ }\bibfield
  {title} {\bibinfo {title} {A charge-density-wave topological semimetal},\
  }\href@noop {} {\bibfield  {journal} {\bibinfo  {journal} {Nature Physics}\
  }\textbf {\bibinfo {volume} {17}},\ \bibinfo {pages} {381} (\bibinfo {year}
  {2021})}\BibitemShut {NoStop}%
\bibitem [{\citenamefont {Ament}\ \emph {et~al.}(2011)\citenamefont {Ament},
  \citenamefont {Van~Veenendaal}, \citenamefont {Devereaux}, \citenamefont
  {Hill},\ and\ \citenamefont {Van Den~Brink}}]{ament2011resonant}%
  \BibitemOpen
  \bibfield  {author} {\bibinfo {author} {\bibfnamefont {L.~J.}\ \bibnamefont
  {Ament}}, \bibinfo {author} {\bibfnamefont {M.}~\bibnamefont
  {Van~Veenendaal}}, \bibinfo {author} {\bibfnamefont {T.~P.}\ \bibnamefont
  {Devereaux}}, \bibinfo {author} {\bibfnamefont {J.~P.}\ \bibnamefont
  {Hill}},\ and\ \bibinfo {author} {\bibfnamefont {J.}~\bibnamefont {Van
  Den~Brink}},\ }\bibfield  {title} {\bibinfo {title} {Resonant inelastic
  {{X-ray}} scattering studies of elementary excitations},\ }\href@noop {}
  {\bibfield  {journal} {\bibinfo  {journal} {Reviews of Modern Physics}\
  }\textbf {\bibinfo {volume} {83}},\ \bibinfo {pages} {705} (\bibinfo {year}
  {2011})}\BibitemShut {NoStop}%
\bibitem [{\citenamefont {Forsythe}\ \emph {et~al.}(2018)\citenamefont
  {Forsythe}, \citenamefont {Zhou}, \citenamefont {Watanabe}, \citenamefont
  {Taniguchi}, \citenamefont {Pasupathy}, \citenamefont {Moon}, \citenamefont
  {Koshino}, \citenamefont {Kim},\ and\ \citenamefont
  {Dean}}]{forsythe2018band}%
  \BibitemOpen
  \bibfield  {author} {\bibinfo {author} {\bibfnamefont {C.}~\bibnamefont
  {Forsythe}}, \bibinfo {author} {\bibfnamefont {X.}~\bibnamefont {Zhou}},
  \bibinfo {author} {\bibfnamefont {K.}~\bibnamefont {Watanabe}}, \bibinfo
  {author} {\bibfnamefont {T.}~\bibnamefont {Taniguchi}}, \bibinfo {author}
  {\bibfnamefont {A.}~\bibnamefont {Pasupathy}}, \bibinfo {author}
  {\bibfnamefont {P.}~\bibnamefont {Moon}}, \bibinfo {author} {\bibfnamefont
  {M.}~\bibnamefont {Koshino}}, \bibinfo {author} {\bibfnamefont
  {P.}~\bibnamefont {Kim}},\ and\ \bibinfo {author} {\bibfnamefont {C.~R.}\
  \bibnamefont {Dean}},\ }\bibfield  {title} {\bibinfo {title} {Band structure
  engineering of 2d materials using patterned dielectric superlattices},\
  }\href@noop {} {\bibfield  {journal} {\bibinfo  {journal} {Nature
  nanotechnology}\ }\textbf {\bibinfo {volume} {13}},\ \bibinfo {pages} {566}
  (\bibinfo {year} {2018})}\BibitemShut {NoStop}%
\bibitem [{\citenamefont {Li}\ \emph {et~al.}(2021)\citenamefont {Li},
  \citenamefont {Dietrich}, \citenamefont {Forsythe}, \citenamefont
  {Taniguchi}, \citenamefont {Watanabe}, \citenamefont {Moon},\ and\
  \citenamefont {Dean}}]{li2021anisotropic}%
  \BibitemOpen
  \bibfield  {author} {\bibinfo {author} {\bibfnamefont {Y.}~\bibnamefont
  {Li}}, \bibinfo {author} {\bibfnamefont {S.}~\bibnamefont {Dietrich}},
  \bibinfo {author} {\bibfnamefont {C.}~\bibnamefont {Forsythe}}, \bibinfo
  {author} {\bibfnamefont {T.}~\bibnamefont {Taniguchi}}, \bibinfo {author}
  {\bibfnamefont {K.}~\bibnamefont {Watanabe}}, \bibinfo {author}
  {\bibfnamefont {P.}~\bibnamefont {Moon}},\ and\ \bibinfo {author}
  {\bibfnamefont {C.~R.}\ \bibnamefont {Dean}},\ }\bibfield  {title} {\bibinfo
  {title} {Anisotropic band flattening in graphene with one-dimensional
  superlattices},\ }\href@noop {} {\bibfield  {journal} {\bibinfo  {journal}
  {Nature Nanotechnology}\ }\textbf {\bibinfo {volume} {16}},\ \bibinfo {pages}
  {525} (\bibinfo {year} {2021})}\BibitemShut {NoStop}%
\bibitem [{\citenamefont {Barcons~Ruiz}\ \emph {et~al.}(2022)\citenamefont
  {Barcons~Ruiz}, \citenamefont {Herzig~Sheinfux}, \citenamefont {Hoffmann},
  \citenamefont {Torre}, \citenamefont {Agarwal}, \citenamefont {Kumar},
  \citenamefont {Vistoli}, \citenamefont {Taniguchi}, \citenamefont {Watanabe},
  \citenamefont {Bachtold} \emph {et~al.}}]{barcons2022engineering}%
  \BibitemOpen
  \bibfield  {author} {\bibinfo {author} {\bibfnamefont {D.}~\bibnamefont
  {Barcons~Ruiz}}, \bibinfo {author} {\bibfnamefont {H.}~\bibnamefont
  {Herzig~Sheinfux}}, \bibinfo {author} {\bibfnamefont {R.}~\bibnamefont
  {Hoffmann}}, \bibinfo {author} {\bibfnamefont {I.}~\bibnamefont {Torre}},
  \bibinfo {author} {\bibfnamefont {H.}~\bibnamefont {Agarwal}}, \bibinfo
  {author} {\bibfnamefont {R.~K.}\ \bibnamefont {Kumar}}, \bibinfo {author}
  {\bibfnamefont {L.}~\bibnamefont {Vistoli}}, \bibinfo {author} {\bibfnamefont
  {T.}~\bibnamefont {Taniguchi}}, \bibinfo {author} {\bibfnamefont
  {K.}~\bibnamefont {Watanabe}}, \bibinfo {author} {\bibfnamefont
  {A.}~\bibnamefont {Bachtold}}, \emph {et~al.},\ }\bibfield  {title} {\bibinfo
  {title} {Engineering high quality graphene superlattices via ion milled
  ultra-thin etching masks},\ }\href@noop {} {\bibfield  {journal} {\bibinfo
  {journal} {Nature Communications}\ }\textbf {\bibinfo {volume} {13}},\
  \bibinfo {pages} {6926} (\bibinfo {year} {2022})}\BibitemShut {NoStop}%
\bibitem [{\citenamefont {Zeng}\ \emph {et~al.}(2024)\citenamefont {Zeng},
  \citenamefont {Wolf}, \citenamefont {Huang}, \citenamefont {Wei},
  \citenamefont {Ghorashi}, \citenamefont {MacDonald},\ and\ \citenamefont
  {Cano}}]{zeng2024gate}%
  \BibitemOpen
  \bibfield  {author} {\bibinfo {author} {\bibfnamefont {Y.}~\bibnamefont
  {Zeng}}, \bibinfo {author} {\bibfnamefont {T.~M.}\ \bibnamefont {Wolf}},
  \bibinfo {author} {\bibfnamefont {C.}~\bibnamefont {Huang}}, \bibinfo
  {author} {\bibfnamefont {N.}~\bibnamefont {Wei}}, \bibinfo {author}
  {\bibfnamefont {S.~A.~A.}\ \bibnamefont {Ghorashi}}, \bibinfo {author}
  {\bibfnamefont {A.~H.}\ \bibnamefont {MacDonald}},\ and\ \bibinfo {author}
  {\bibfnamefont {J.}~\bibnamefont {Cano}},\ }\bibfield  {title} {\bibinfo
  {title} {Gate-tunable topological phases in superlattice modulated bilayer
  graphene},\ }\href@noop {} {\bibfield  {journal} {\bibinfo  {journal} {arXiv
  preprint arXiv:2401.04321}\ } (\bibinfo {year} {2024})}\BibitemShut {NoStop}%
\bibitem [{\citenamefont {Sun}\ \emph {et~al.}(2023)\citenamefont {Sun},
  \citenamefont {Ghorashi}, \citenamefont {Watanabe}, \citenamefont
  {Taniguchi}, \citenamefont {Camino}, \citenamefont {Cano},\ and\
  \citenamefont {Du}}]{sun2023signature}%
  \BibitemOpen
  \bibfield  {author} {\bibinfo {author} {\bibfnamefont {J.}~\bibnamefont
  {Sun}}, \bibinfo {author} {\bibfnamefont {S.~A.~A.}\ \bibnamefont
  {Ghorashi}}, \bibinfo {author} {\bibfnamefont {K.}~\bibnamefont {Watanabe}},
  \bibinfo {author} {\bibfnamefont {T.}~\bibnamefont {Taniguchi}}, \bibinfo
  {author} {\bibfnamefont {F.}~\bibnamefont {Camino}}, \bibinfo {author}
  {\bibfnamefont {J.}~\bibnamefont {Cano}},\ and\ \bibinfo {author}
  {\bibfnamefont {X.}~\bibnamefont {Du}},\ }\bibfield  {title} {\bibinfo
  {title} {Signature of correlated insulator in electric field controlled
  superlattice},\ }\href@noop {} {\bibfield  {journal} {\bibinfo  {journal}
  {arXiv preprint arXiv:2306.06848}\ } (\bibinfo {year} {2023})},\ \Eprint
  {https://arxiv.org/abs/2306.06848} {arxiv:2306.06848} \BibitemShut {NoStop}%
\bibitem [{\citenamefont {Ghorashi}\ \emph {et~al.}(2023)\citenamefont
  {Ghorashi}, \citenamefont {Dunbrack}, \citenamefont {Abouelkomsan},
  \citenamefont {Sun}, \citenamefont {Du},\ and\ \citenamefont
  {Cano}}]{ghorashi2023topological}%
  \BibitemOpen
  \bibfield  {author} {\bibinfo {author} {\bibfnamefont {S.~A.~A.}\
  \bibnamefont {Ghorashi}}, \bibinfo {author} {\bibfnamefont {A.}~\bibnamefont
  {Dunbrack}}, \bibinfo {author} {\bibfnamefont {A.}~\bibnamefont
  {Abouelkomsan}}, \bibinfo {author} {\bibfnamefont {J.}~\bibnamefont {Sun}},
  \bibinfo {author} {\bibfnamefont {X.}~\bibnamefont {Du}},\ and\ \bibinfo
  {author} {\bibfnamefont {J.}~\bibnamefont {Cano}},\ }\bibfield  {title}
  {\bibinfo {title} {Topological and stacked flat bands in bilayer graphene
  with a superlattice potential},\ }\href@noop {} {\bibfield  {journal}
  {\bibinfo  {journal} {Physical Review Letters}\ }\textbf {\bibinfo {volume}
  {130}},\ \bibinfo {pages} {196201} (\bibinfo {year} {2023})}\BibitemShut
  {NoStop}%
\bibitem [{\citenamefont {Messiah}(2014)}]{messiah2014quantum}%
  \BibitemOpen
  \bibfield  {author} {\bibinfo {author} {\bibfnamefont {A.}~\bibnamefont
  {Messiah}},\ }\href@noop {} {\emph {\bibinfo {title} {Quantum mechanics}}}\
  (\bibinfo  {publisher} {Courier Corporation},\ \bibinfo {year}
  {2014})\BibitemShut {NoStop}%
\bibitem [{\citenamefont {Catarina}\ \emph {et~al.}(2020)\citenamefont
  {Catarina}, \citenamefont {Peres},\ and\ \citenamefont
  {Fernández-Rossier}}]{Catarina2020KerrIn2DDirac}%
  \BibitemOpen
  \bibfield  {author} {\bibinfo {author} {\bibfnamefont {G.}~\bibnamefont
  {Catarina}}, \bibinfo {author} {\bibfnamefont {N.~M.~R.}\ \bibnamefont
  {Peres}},\ and\ \bibinfo {author} {\bibfnamefont {J.}~\bibnamefont
  {Fernández-Rossier}},\ }\bibfield  {title} {\bibinfo {title}
  {Magneto-optical {Kerr} effect in spin split two-dimensional massive {Dirac}
  materials},\ }\href {https://doi.org/10.1088/2053-1583/ab6781} {\bibfield
  {journal} {\bibinfo  {journal} {2D Materials}\ }\textbf {\bibinfo {volume}
  {7}},\ \bibinfo {pages} {025011} (\bibinfo {year} {2020})}\BibitemShut
  {NoStop}%
\bibitem [{\citenamefont {Kargarian}\ \emph {et~al.}(2015)\citenamefont
  {Kargarian}, \citenamefont {Randeria},\ and\ \citenamefont
  {Trivedi}}]{Kargarian2015KerrAndFaradayInWeyl}%
  \BibitemOpen
  \bibfield  {author} {\bibinfo {author} {\bibfnamefont {M.}~\bibnamefont
  {Kargarian}}, \bibinfo {author} {\bibfnamefont {M.}~\bibnamefont
  {Randeria}},\ and\ \bibinfo {author} {\bibfnamefont {N.}~\bibnamefont
  {Trivedi}},\ }\bibfield  {title} {\bibinfo {title} {Theory of {Kerr} and
  {Faraday} rotations and linear dichroism in {Topological} {Weyl}
  {Semimetals}},\ }\href {https://doi.org/10.1038/srep12683} {\bibfield
  {journal} {\bibinfo  {journal} {Scientific Reports}\ }\textbf {\bibinfo
  {volume} {5}},\ \bibinfo {pages} {12683} (\bibinfo {year}
  {2015})}\BibitemShut {NoStop}%
\bibitem [{\citenamefont {Oppeneer}(1999)}]{Oppeneer1999MOKEinFerromagnets}%
  \BibitemOpen
  \bibfield  {author} {\bibinfo {author} {\bibfnamefont {P.~M.}\ \bibnamefont
  {Oppeneer}},\ }\href {https://doi.org/10.13140/2.1.3171.4083} {\emph
  {\bibinfo {title} {Theory of the {Magneto}-{Optical} {Kerr} {Effect} in
  {Ferromagnetic} {Compounds}}}}\ (\bibinfo  {publisher} {Technische
  Universität Dresden},\ \bibinfo {year} {1999})\BibitemShut {NoStop}%
\bibitem [{\citenamefont {Ghosh}\ \emph {et~al.}(2023)\citenamefont {Ghosh},
  \citenamefont {Sahoo},\ and\ \citenamefont {Nandy}}]{Ghosh2023KerrInWeyl}%
  \BibitemOpen
  \bibfield  {author} {\bibinfo {author} {\bibfnamefont {S.}~\bibnamefont
  {Ghosh}}, \bibinfo {author} {\bibfnamefont {A.}~\bibnamefont {Sahoo}},\ and\
  \bibinfo {author} {\bibfnamefont {S.}~\bibnamefont {Nandy}},\ }\bibfield
  {title} {\bibinfo {title} {Theoretical investigations on {Kerr} and {Faraday}
  rotations in topological multi-{Weyl} {Semimetals}},\ }\href@noop {}
  {\bibfield  {journal} {\bibinfo  {journal} {arXiv preprint arXiv:2209.11217}\
  } (\bibinfo {year} {2023})},\ \Eprint {https://arxiv.org/abs/2209.11217}
  {arXiv:2209.11217} \BibitemShut {NoStop}%
\bibitem [{\citenamefont {Creed}(2020)}]{Creed2020MethodsforKerrDetection}%
  \BibitemOpen
  \bibfield  {author} {\bibinfo {author} {\bibfnamefont {J.}~\bibnamefont
  {Creed}},\ }\emph {\bibinfo {title} {Methods for {Sensitive} {Detection} of
  {Magneto} {Optic} {Kerr} {Effect}}},\ \href
  {https://doi.org/10.31979/etd.8brz-y9ww} {\bibinfo {type} {Master of
  {Science}}},\ \bibinfo  {school} {San Jose State University}, \bibinfo
  {address} {San Jose, CA, USA} (\bibinfo {year} {2020})\BibitemShut {NoStop}%
\bibitem [{\citenamefont {Jackson}(1999)}]{Jackson1999ClassicalEMBook}%
  \BibitemOpen
  \bibfield  {author} {\bibinfo {author} {\bibfnamefont {J.~D.}\ \bibnamefont
  {Jackson}},\ }\href {http://cdswebcern.ch/record/490457} {\emph {\bibinfo
  {title} {Classical electrodynamics}}},\ \bibinfo {edition} {3rd}\ ed.\
  (\bibinfo  {publisher} {Wiley},\ \bibinfo {address} {New York, {NY}},\
  \bibinfo {year} {1999})\BibitemShut {NoStop}%
\bibitem [{\citenamefont {Wolski}(2014)}]{Wolski2014theoryOfEMFields}%
  \BibitemOpen
  \bibfield  {author} {\bibinfo {author} {\bibfnamefont {A.}~\bibnamefont
  {Wolski}},\ }\bibfield  {title} {\bibinfo {title} {Theory of electromagnetic
  fields},\ }\href@noop {} {\bibfield  {journal} {\bibinfo  {journal} {arXiv
  preprint arXiv:1111.4354}\ } (\bibinfo {year} {2014})},\ \Eprint
  {https://arxiv.org/abs/1111.4354} {arXiv:1111.4354} \BibitemShut {NoStop}%
\bibitem [{\citenamefont {Ebert}(1996)}]{Ebert1996MOKEInTransitionMetals}%
  \BibitemOpen
  \bibfield  {author} {\bibinfo {author} {\bibfnamefont {H.}~\bibnamefont
  {Ebert}},\ }\bibfield  {title} {\bibinfo {title} {Magneto-optical effects in
  transition metal systems},\ }\href
  {https://doi.org/10.1088/0034-4885/59/12/003} {\bibfield  {journal} {\bibinfo
   {journal} {Reports on Progress in Physics}\ }\textbf {\bibinfo {volume}
  {59}},\ \bibinfo {pages} {1665} (\bibinfo {year} {1996})}\BibitemShut
  {NoStop}%
\bibitem [{\citenamefont {Higo}\ \emph {et~al.}(2018)\citenamefont {Higo},
  \citenamefont {Man}, \citenamefont {Gopman}, \citenamefont {Wu},
  \citenamefont {Koretsune}, \citenamefont {Erve}, \citenamefont {Kabanov},
  \citenamefont {Rees}, \citenamefont {Li}, \citenamefont {Suzuki},
  \citenamefont {Patankar}, \citenamefont {Ikhlas}, \citenamefont {Chien},
  \citenamefont {Arita}, \citenamefont {Shull}, \citenamefont {Orenstein},\
  and\ \citenamefont {Nakatsuji}}]{Higo2018LargeMOKE}%
  \BibitemOpen
  \bibfield  {author} {\bibinfo {author} {\bibfnamefont {T.}~\bibnamefont
  {Higo}}, \bibinfo {author} {\bibfnamefont {H.}~\bibnamefont {Man}}, \bibinfo
  {author} {\bibfnamefont {D.~B.}\ \bibnamefont {Gopman}}, \bibinfo {author}
  {\bibfnamefont {L.}~\bibnamefont {Wu}}, \bibinfo {author} {\bibfnamefont
  {T.}~\bibnamefont {Koretsune}}, \bibinfo {author} {\bibfnamefont {O.~M. J.
  v.~t.}\ \bibnamefont {Erve}}, \bibinfo {author} {\bibfnamefont {Y.~P.}\
  \bibnamefont {Kabanov}}, \bibinfo {author} {\bibfnamefont {D.}~\bibnamefont
  {Rees}}, \bibinfo {author} {\bibfnamefont {Y.}~\bibnamefont {Li}}, \bibinfo
  {author} {\bibfnamefont {M.-T.}\ \bibnamefont {Suzuki}}, \bibinfo {author}
  {\bibfnamefont {S.}~\bibnamefont {Patankar}}, \bibinfo {author}
  {\bibfnamefont {M.}~\bibnamefont {Ikhlas}}, \bibinfo {author} {\bibfnamefont
  {C.~L.}\ \bibnamefont {Chien}}, \bibinfo {author} {\bibfnamefont
  {R.}~\bibnamefont {Arita}}, \bibinfo {author} {\bibfnamefont {R.~D.}\
  \bibnamefont {Shull}}, \bibinfo {author} {\bibfnamefont {J.}~\bibnamefont
  {Orenstein}},\ and\ \bibinfo {author} {\bibfnamefont {S.}~\bibnamefont
  {Nakatsuji}},\ }\bibfield  {title} {\bibinfo {title} {Large magneto-optical
  {Kerr} effect and imaging of magnetic octupole domains in an
  antiferromagnetic metal},\ }\href {https://doi.org/10.1038/s41566-017-0086-z}
  {\bibfield  {journal} {\bibinfo  {journal} {Nature Photonics}\ }\textbf
  {\bibinfo {volume} {12}},\ \bibinfo {pages} {73} (\bibinfo {year}
  {2018})}\BibitemShut {NoStop}%
\bibitem [{\citenamefont {McGee}(1991)}]{McGee1991MOKETheory}%
  \BibitemOpen
  \bibfield  {author} {\bibinfo {author} {\bibfnamefont {N.~W.~E.}\
  \bibnamefont {McGee}},\ }\emph {\bibinfo {title} {The {Magneto}-{Optical}
  {Kerr} {Effect}: {Theory}, {Measurement} and {Application}}},\ \href
  {https://research.tue.nl/en/studentTheses/the-magneto-optical-kerr-effect-theory-measurement-and-application}
  {\bibinfo {type} {Master}},\ \bibinfo  {school} {Eindhoven University of
  Technology}, \bibinfo {address} {Eindhoven, NB, Netherlands} (\bibinfo {year}
  {1991})\BibitemShut {NoStop}%
\bibitem [{\citenamefont {Shah}\ \emph {et~al.}(2021)\citenamefont {Shah},
  \citenamefont {Akbar}, \citenamefont {Sajid},\ and\ \citenamefont
  {Sabieh~Anwar}}]{Shah2021FaradayAndKerrInThinFilms}%
  \BibitemOpen
  \bibfield  {author} {\bibinfo {author} {\bibfnamefont {M.}~\bibnamefont
  {Shah}}, \bibinfo {author} {\bibfnamefont {A.}~\bibnamefont {Akbar}},
  \bibinfo {author} {\bibfnamefont {M.}~\bibnamefont {Sajid}},\ and\ \bibinfo
  {author} {\bibfnamefont {M.}~\bibnamefont {Sabieh~Anwar}},\ }\bibfield
  {title} {\bibinfo {title} {Transitional {Faraday} and {Kerr} effect in
  hybridized topological insulator thin films},\ }\href
  {https://doi.org/10.1364/OME.413973} {\bibfield  {journal} {\bibinfo
  {journal} {Optical Materials Express}\ }\textbf {\bibinfo {volume} {11}},\
  \bibinfo {pages} {525} (\bibinfo {year} {2021})}\BibitemShut {NoStop}%
\bibitem [{\citenamefont {Hwang}\ and\ \citenamefont
  {Sarma}(2007)}]{Hwang2007PlasmonDispersionDirac}%
  \BibitemOpen
  \bibfield  {author} {\bibinfo {author} {\bibfnamefont {E.~H.}\ \bibnamefont
  {Hwang}}\ and\ \bibinfo {author} {\bibfnamefont {S.~D.}\ \bibnamefont
  {Sarma}},\ }\bibfield  {title} {\bibinfo {title} {Dielectric function,
  screening, and plasmons in two-dimensional graphene},\ }\bibfield  {journal}
  {\bibinfo  {journal} {Physical Review B}\ }\textbf {\bibinfo {volume} {75}},\
  \href {https://doi.org/10.1103/physrevb.75.205418}
  {10.1103/physrevb.75.205418} (\bibinfo {year} {2007})\BibitemShut {NoStop}%
\bibitem [{\citenamefont {Mishra}\ \emph {et~al.}(2012)\citenamefont {Mishra},
  \citenamefont {Ashraf},\ and\ \citenamefont
  {Sharma}}]{Mishra2012PlasmonDispersionDirac}%
  \BibitemOpen
  \bibfield  {author} {\bibinfo {author} {\bibfnamefont {K.~N.}\ \bibnamefont
  {Mishra}}, \bibinfo {author} {\bibfnamefont {S.}~\bibnamefont {Ashraf}},\
  and\ \bibinfo {author} {\bibfnamefont {A.~C.}\ \bibnamefont {Sharma}},\
  }\bibfield  {title} {\bibinfo {title} {Ground state properties of monolayer
  doped graphene},\ }\href {https://doi.org/10.4236/wjcmp.2012.21006}
  {\bibfield  {journal} {\bibinfo  {journal} {World Journal of Condensed Matter
  Physics}\ }\textbf {\bibinfo {volume} {02}},\ \bibinfo {pages} {36} (\bibinfo
  {year} {2012})}\BibitemShut {NoStop}%
\bibitem [{\citenamefont {Jani}\ \emph {et~al.}(2003)\citenamefont {Jani},
  \citenamefont {Kolorenc},\ and\ \citenamefont
  {Picka}}]{Jani2003DensityDensityPlasmon}%
  \BibitemOpen
  \bibfield  {author} {\bibinfo {author} {\bibfnamefont {V.}~\bibnamefont
  {Jani}}, \bibinfo {author} {\bibfnamefont {J.}~\bibnamefont {Kolorenc}},\
  and\ \bibinfo {author} {\bibfnamefont {V.}~\bibnamefont {Picka}},\ }\bibfield
   {title} {\bibinfo {title} {Density and current response functions in
  strongly disordered electron systems: diffusion, electrical conductivity and
  {Einstein} relation},\ }\href {https://doi.org/10.1140/epjb/e2003-00258-4}
  {\bibfield  {journal} {\bibinfo  {journal} {The European Physical Journal B -
  Condensed Matter}\ }\textbf {\bibinfo {volume} {35}},\ \bibinfo {pages} {77}
  (\bibinfo {year} {2003})}\BibitemShut {NoStop}%
\end{thebibliography}%
\end{document}